\documentclass[%
 reprint,  
superscriptaddress,
floatfix,
 amsmath,amssymb,
 aps,
doi=true,
rmp,
floatfix,
raggedbottom,
]{revtex4-1}
\usepackage[colorlinks=true,linkcolor=blue,urlcolor=blue,citecolor=blue]{hyperref}
\usepackage{amsthm,amssymb,amsmath,epsfig,graphics,color,verbatim,booktabs,multirow,rotating}
\AtBeginDocument{
\heavyrulewidth=.08em
\lightrulewidth=.05em
\cmidrulewidth=.03em
\belowrulesep=.65ex
\belowbottomsep=0pt
\aboverulesep=.4ex
\abovetopsep=0pt
\cmidrulesep=\doublerulesep
\cmidrulekern=.5em
\defaultaddspace=.5em
}

\usepackage{sidecap}
\usepackage[normalem]{ulem} 
\usepackage{url} 
\usepackage{amsmath}
\usepackage{amsfonts}
\usepackage{dsfont}
\usepackage{relsize}
\usepackage{wrapfig}
\usepackage{subfigure}
\usepackage[scr]{rsfso}
\usepackage{upgreek}
\usepackage{xfrac}
\usepackage[usenames,dvipsnames,table,xcdraw]{xcolor}
\usepackage[margin=.58in]{geometry}
\usepackage{multirow}
\usepackage{colortbl}

\usepackage[percent]{overpic}
\usepackage{chngcntr}
\usepackage{bm}
\counterwithin{figure}{section}
\counterwithin{table}{section}

\usepackage{lipsum} 
\usepackage{enumitem} 
\usepackage{bbm}
\usepackage{longtable}

\usepackage[]{nicefrac}

\usepackage[export]{adjustbox}

\usepackage{array}
\newcolumntype{L}[1]{>{\raggedright\let\newline\\\arraybackslash\hspace{0pt}}m{#1}}
\newcolumntype{C}[1]{>{\centering\let\newline\\\arraybackslash\hspace{0pt}}m{#1}}
\newcolumntype{R}[1]{>{\raggedleft\let\newline\\\arraybackslash\hspace{0pt}}m{#1}}

\usepackage{upgreek}

\newcommand{\appropto}{\mathrel{\vcenter{
  \offinterlineskip\halign{\hfil$##$\cr
    \propto\cr\noalign{\kern2pt}\sim\cr\noalign{\kern-2pt}}}}}
    
\usepackage[T1]{fontenc}

\usepackage{chngcntr}
\counterwithout{figure}{section}
\counterwithout{table}{section}

\usepackage[outline]{contour}

\definecolor{dandelion}{rgb}{0.94, 0.88, 0.19}
\definecolor{thepurple}{rgb}{0.65, 0.24, 0.59}
\definecolor{theorange}{rgb}{0.95, 0.40, 0.13}
\definecolor{thegreen}{rgb}{0.05, 0.50, 0.25}
\definecolor{mathematicagreen}{rgb}{0.0, 0.80, 0.0}
\definecolor{igoraquamarine}{rgb}{0.0, 0.6, 0.6}
\definecolor{igorgreen}{rgb}{0.0, 0.6, 0.0}
\definecolor{springgreen}{rgb}{0.0, 0.8, 0.0}
\definecolor{skyblue}{rgb}{0.0, 0.6, 1.0}
\definecolor{black}{rgb}{0.0 0.0 0.0}

\definecolor{inkscapeblue}{rgb}{0.0 0.0 1.0}
\definecolor{inkscapepurple}{rgb}{0.5 0.0 0.5}
\definecolor{inkscapegreen}{rgb}{0.0 0.5 0.0}
\definecolor{inkscapered}{rgb}{1 0.0 0.0}
\definecolor{black}{rgb}{0.0 0.0 0.0}

\usepackage{physics}

\usepackage{xspace}

\usepackage[acronym]{glossaries}
\makeglossaries

\newcommand{\ignore}[1]{ }
\newcommand{\why}[1]{\textcolor{RubineRed}{?}}

\newcommand{\NVm}{NV$^-$\xspace}
\newcommand{\NVo}{NV$^0$\xspace}
\newcommand{\NVp}{NV$^+$\xspace}
\newcommand{\Nfif}{$^{15}$N\xspace}
\newcommand{\Nfor}{$^{14}$N\xspace}

\newcommand{\Pone}{$^{31}$P\xspace}
\newcommand{\Ttwo}{$T_2$\xspace}
\newcommand{\Tone}{$T_1$\xspace}
\newcommand{\Ttwostar}{$T^*_2$\xspace}

\newcommand{\Cthir}{$^{13}\text{C}$\xspace}
\newcommand{\Hone}{$^{1}\text{H}$\xspace}
\newcommand{\Htwo}{$^{2}\text{H}$\xspace}
\newcommand{\Fnine}{$^{19}\text{F}$\xspace}
\newcommand{\Aiso}{$A_{\text{iso}}$\xspace}
\newcommand{\Tonerou}{$T_{1\rho}$\xspace}
\newcommand{\nTHz}{nT$\cdot$Hz$^{-1/2}$\xspace}
\newcommand{\sqrtHz}{$\cdot$Hz$^{-1/2}$\xspace}
\newcommand{\uB}{$\mu_{\text{B}}$\xspace}
\newcommand{\mTnm}{mT$\cdot$nm$^{-1}$\xspace}

\newcommand{\celsius}{$^{\circ}$C\xspace}
\newcommand{\sm}{single-molecule\xspace}
\newcommand{\SM}{Single-molecule\xspace}
\newcommand{\degree}{$^{\circ}$\xspace}
\newcommand{\td}{two-dimensional\xspace}
\newcommand{\Td}{Two-dimensional\xspace}
\newcommand{\Texp}{T_{\text{exp}}}

\newcommand{\gNV}{\gamma_{\text{NV}} }

\newcommand{\se}{Sec.\xspace}
\newcommand{\app}{Appendix\xspace}
\newcommand{\eq}{Eq.\xspace}

\newacronym{AFM}{AFM}{atomic force microscopy} 
\newcommand{\afm}{\gls*{AFM}\xspace}
\newcommand{\AFM}{\Gls*{AFM}\xspace}

\newacronym{cw}{cw}{continuous-wave}

\newacronym{cwODMR}{cw-ODMR}{continuous-wave optically detected magnetic resonance} 
\newcommand{\cwODMR}{\gls*{cwODMR}\xspace}
\newcommand{\CwODMR}{\Gls*{cwODMR}\xspace}

\newacronym{COSY}{COSY}{correlated spectroscopy} 
\newcommand{\cosy}{\gls*{COSY}\xspace}

\newacronym{CPMG}{CPMG}{Carr-Purcell-Meiboom-Gill} 
\newcommand{\cpmg}{\gls*{CPMG}\xspace}

\newacronym{CVD}{CVD}{chemical vapor deposition} 
\newcommand{\cvd}{\gls*{CVD}\xspace}

\newacronym{DD}{DD}{dynamical decoupling} 
\newcommand{\dyde}{\gls*{DD}\xspace}
\newcommand{\DyDe}{\Gls*{DD}\xspace}

\newacronym{DEER}{DEER}{double electron-electron resonance} 
\newcommand{\deer}{\gls*{DEER}\xspace}
\newcommand{\DEER}{\Gls*{DEER}\xspace}

\newacronym{DNP}{DNP}{dynamical nuclear polarization} 
\newcommand{\dnp}{\gls*{DNP}\xspace}
\newcommand{\DNP}{\Gls*{DNP}\xspace}

\newacronym{DQ}{DQ}{double-quantum} 
\newcommand{\dq}{\gls*{DQ}\xspace}

\newacronym{EBL}{EBL}{electron beam lithography} 
\newcommand{\ebl}{\gls*{EBL}\xspace}

\newacronym{EDMR}{EDMR}{electrically detected magnetic resonance} 

\newcommand{\EDMR}{\Gls*{EDMR}\xspace}

\newacronym{ENDOR}{ENDOR}{electron nuclear double resonance} 
\newcommand{\enor}{\gls*{ENDOR}\xspace}

\newacronym{EPR}{EPR}{electron paramagnetic resonance} 
\newcommand{\epr}{\gls*{EPR}\xspace}
\newcommand{\EPR}{\Gls*{EPR}\xspace}

\newacronym{FIB}{FIB}{focused ion beam} 

\newcommand{\FIB}{\Gls*{FIB}\xspace}

\newacronym{FRET}{FRET}{fluorescence resonance energy transfer} 
\newcommand{\fret}{\gls*{FRET}\xspace}

\newacronym{HH}{HH}{Hartmann-Hahn} 
\newcommand{\hh}{\gls*{HH}\xspace}
\newcommand{\HH}{\Gls*{HH}\xspace}

\newacronym{HHDR}{HHDR}{Hartmann-Hahn double resonance} 
\newcommand{\hhdr}{\gls*{HHDR}\xspace}

\newacronym{ICP}{ICP}{inductively coupled plasma} 

\newcommand{\ICP}{\Gls*{ICP}\xspace}

\newacronym{ISC}{ISC}{inter-system crossing} 
\newcommand{\isc}{\gls*{ISC}\xspace}

\newacronym{MFM}{MFM}{magnetic force microscopy} 
\newcommand{\mfm}{\gls*{MFM}\xspace}
\newcommand{\MFM}{\Gls*{MFM}\xspace}

\newacronym{MRFM}{MRFM}{magnetic resonance force microscopy} 
\newcommand{\mrfm}{\gls*{MRFM}\xspace}
\newcommand{\MRFM}{\Gls*{MRFM}\xspace}

\newacronym{MRI}{MRI}{magnetic resonance imaging} 
\newcommand{\mri}{\gls*{MRI}\xspace}

\newacronym{MW}{MW}{microwave}

\newacronym{NMR}{NMR}{nuclear magnetic resonance} 
\newcommand{\nmr}{\gls*{NMR}\xspace}
\newcommand{\NMR}{\Gls*{NMR}\xspace}

\newacronym{NQR}{NQR}{nuclear quadrupole resonance} 
\newcommand{\nqr}{\gls*{NQR}\xspace}
\newcommand{\NQR}{\Gls*{NQR}\xspace}

\newacronym{NV}{NV}{nitrogen-vacancy} 
\newcommand{\NV}{\Gls*{NV}\xspace}
\newcommand{\nv}{\gls*{NV}\xspace}

\newacronym{ODMR}{ODMR}{optically detected magnetic resonance}

\newacronym{ONP}{ONP}{optically nuclear polarization} 
\newcommand{\onp}{\gls*{ONP}\xspace}

\newacronym{PL}{PL}{photoluminescence} 
\newcommand{\phl}{\gls*{PL}\xspace}
\newcommand{\PhL}{\Gls*{PL}\xspace}

\newacronym{PSF}{PSF}{point spread function} 
\newcommand{\psf}{\gls*{PSF}\xspace}

\newacronym{QND}{QND}{quantum non-demolition}

\newacronym{Qdyne}{Qdyne}{quantum heterodyne} 
\newcommand{\qdyne}{\gls*{Qdyne}\xspace}
\newcommand{\Qdyne}{\Gls*{Qdyne}\xspace}

\newacronym{RIE}{RIE}{eactive ion etching} 
\newcommand{\rie}{\gls*{RIE}\xspace}
\newcommand{\RIE}{\Gls*{RIE}\xspace}

\newacronym{RF}{RF}{radio frequency}

\newacronym{SCC}{SCC}{spin-to-charge} 
\newcommand{\scc}{\gls*{SCC}\xspace}
\newcommand{\SCC}{\Gls*{SCC}\xspace}

\newacronym{SEM}{SEM}{scanning electron microscopy} 
\newcommand{\sem}{\gls*{SEM}\xspace}

\newacronym{SNR}{SNR}{signal-to-noise ratio}

\newacronym{SIL}{SIL}{solid immersion lens} 
\newcommand{\sil}{\gls*{SIL}\xspace}
\newcommand{\SIL}{\Gls*{SIL}\xspace}

\newacronym{SRIM}{SRIM}{stopping and range of ions in matter}
\newcommand{\srim}{\gls*{SRIM}\xspace}

\newacronym{SQS}{SQS}{solid-state quantum sensor}
\newcommand{\sqs}{\gls*{SQS}\xspace}
\newcommand{\SQS}{\Gls*{SQS}\xspace}

\newacronym{SSMR}{SSMR}{spin-based single-molecule magnetic resonance technologies}
\newcommand{\ssmr}{\gls*{SSMR}\xspace}
\newcommand{\SSMR}{\Gls*{SSMR}\xspace}

\newacronym{STM}{STM}{scanning tunneling microscopy} 
\newcommand{\stm}{\gls*{STM}\xspace}

\newacronym{SQ}{SQ}{single-quantum} 
\newcommand{\sq}{\gls*{SQ}\xspace}

\newacronym{TEM}{TEM}{transmission electron microscopy} 
\newcommand{\tem}{\gls*{TEM}\xspace}
\newcommand{\TEM}{\Gls*{TEM}\xspace}

\newacronym{UDD}{UDD}{Uhrig dynamical decoupling} 
\newcommand{\udd}{\gls*{UDD}\xspace}

\newacronym{UHV}{UHV}{ultra-high vacuum} 
\newcommand{\uhv}{\gls*{UHV}\xspace}

\newacronym{ZFEPR}{ZF-EPR}{zero-field EPR} 
\newcommand{\zfepr}{\gls*{ZFEPR}\xspace}

\newacronym{ZPL}{ZPL}{zero phonon line} 
\newcommand{\zpl}{\gls*{ZPL}\xspace}

\footnotetext{$^{\ast}$ djf@ustc.edu.cn}
\footnotetext{$^{\dag}$ fzshi@ustc.edu.cn}
\footnotetext{$^{\ddag}$ kongxi@nju.edu.cn}
\footnotetext{$^{\S}$ fedor.jelezko@uni-ulm.de}
\footnotetext{$^{\P}$ j.wrachtrup@pi3.uni-stuttgart.de}

\begin{document}

\title{Single-molecule Scale  Magnetic Resonance Spectroscopy using Nitrogen-Vacancy Centers in Diamond}

\author
  {\large{Jiangfeng Du,$^{1,2,8, \ast}$
  Fazhan Shi,$^{1,2,7, \dag}$
  Xi Kong,$^{3, \ddag}$
  Fedor Jekezko,$^{4, \S}$
  J\"{o}rg Wrachtrup,$^{5,6, \P}$}
  \\
  \normalsize{$^{1}$\emph{CAS Key Laboratory of Microscale Magnetic Resonance and School of Physical Sciences,\\ University of Science and Technology of China, Hefei 230026, China}}\\
  {$^{2}$\emph{Hefei National Laboratory, University of Science and Technology of China, Hefei 230088, China}}\\
  {$^{3}$\emph{The State Key Laboratory of Solid State Microstructures and Department of Physics,\\ Nanjing University, 210093 Nanjing, China}}\\
  {$^{4}$\emph{Institute for Quantum Optics and Center for Integrated Quantum Science and Technology (IQST),\\ Ulm University, Ulm, Germany}}\\
  {$^{5}$\emph{3rd Institute of Physics and Institute for Integrated Quantum Science and Technology  (IQST),\\ University Stuttgart, Stuttgart, Germany}}\\
  {$^{6}$\emph{Max Planck Institute for Solid State Research, Stuttgart, Germany}}\\
  {$^{7}$\emph{School of Biomedical Engineering and Suzhou Institute for Advanced Research,\\ University of Science and Technology of China, Suzhou 215123, China}}\\
  {$^{8}$\emph{School of Physics, Zhejiang University, Hangzhou 310027, China}}
}

\date{\today}

\begin{abstract}
    \SM  technology stands as a powerful tool, enabling the characterization of intricate structural and dynamic information that would otherwise remain concealed within the averaged behaviors of numerous molecules. This technology finds extensive application across diverse fields including physics, chemistry, biology, and medicine. Quantum sensing, particularly leveraging nitrogen-vacancy (NV) centers within diamond structures, presents a promising avenue for \sm magnetic resonance, offering prospects for sensing and imaging technology at the \sm level. Notably, while significant strides have been made in \sm scale magnetic resonance using NV centers over the past two decades, current approaches still exhibit limitations in magnetic sensitivity, spectral resolution, and spatial resolution. Particularly, the full reconstruction of three-dimensional positions of nuclear spins within single molecules remains an unattained goal.
    This review provides a comprehensive overview of the current state-of-the-art in \sm scale magnetic resonance, encompassing an analysis of various relevant techniques involving NV centers. Additionally, it explores the optimization of technical parameters associated with these methods. This detailed analysis serves as a foundation for the development of new technologies and the exploration of potential applications.  
\end{abstract}

\maketitle

\tableofcontents

\renewcommand{\thetable}{\arabic{table}}   
\renewcommand{\thefigure}{\arabic{figure}}

\section{Introduction and background}\label{introduction}

The advent of single-molecule technology has ushered in a new era of understanding fundamental biological, chemical, and physical phenomena. Conventional ensemble methods interrogate the behavior from a large number of molecules collectively and average out the individual properties and rare states.  Unlike conventional ensemble methods, single-molecule technology unveils insights into individual properties and rare states. Presently, this technology encompasses a wide array of techniques, spanning optical, electrical, mechanical, and magnetic methods like super-resolution microscopy, \fret, \stm, \tem, nanopores, atomic force microscopy, magnetic resonance force, among others. These techniques facilitate the detection of various facets of single molecules, including imaging, motion, structure, and chemical environment. Nonetheless, the simultaneous multimodal detection of diverse information using these methods, especially within living conditions, remains a significant challenge.

\begin{figure}[hbtp]
\begin{overpic}[width=1\columnwidth]{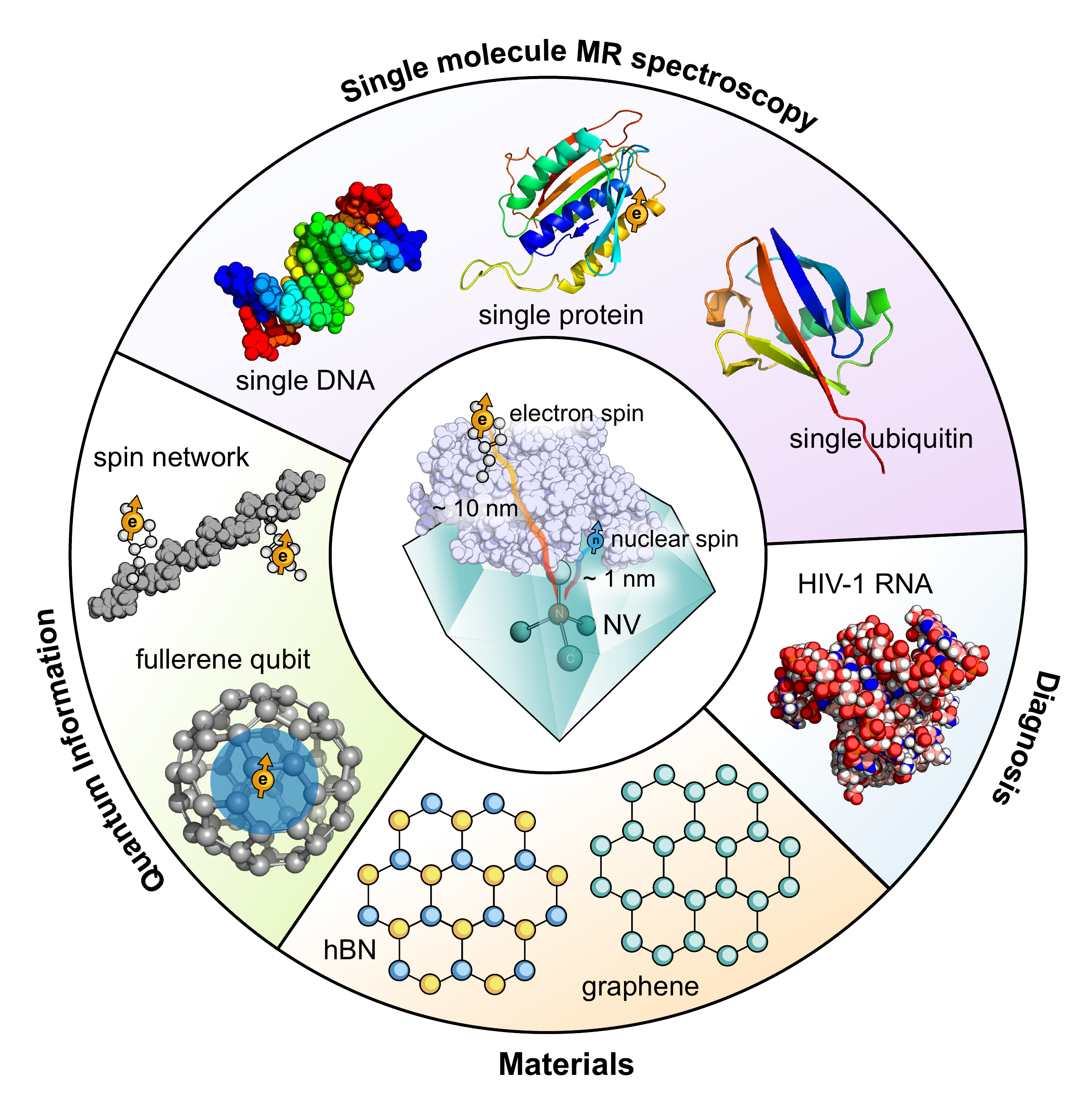} 
\end{overpic}
\caption[NV intro]{
The NV center is a point defect in diamond, comprising a substitutional nitrogen atom adjacent to a vacancy. It exhibits sensitivity to various physical quantities, especially to magnetic fields. As a solid-state quantum sensor operating under ambient temperature, the NV center can leverage quantum sensing protocols to enhance sensitivity, spectral resolution, and spatial resolution. This versatile quantum sensor finds applications in single-molecule studies, including magnetic spectroscopy of single proteins \cite{shi2015Singleprotein,Lovchinsky2016} and single DNA molecules \cite{shi2018SingleDNA}, detection of single qubits in spin networks \cite{schlipf2017molecular} and doped fullerenes \cite{pinto2020Readout}, characterization of atomic-thin materials like monolayer hexagonal boron nitride (hBN) \cite{Lovchinsky2017magnetic} and graphene \cite{hao2023Sensing}, and diagnosis of HIV-1 RNA \cite{millerSpinenhancedNanodiamondBiosensing2020}.
  } \label{fig:NV_intro} 
\end{figure}

In recent years, solid-state quantum sensing has emerged as a progressive technology that not only presents new possibilities but has also garnered considerable attention.
\SQS, exemplified by the NV center, demonstrate quantum properties akin to those of traditional atomic systems. Nevertheless, their exceptional characteristics empower high-sensitivity quantum sensing across diverse environments. The NV center materializes as a point defect created by a nitrogen atom positioned adjacent to a vacancy within the diamond crystal (Fig. \ref{fig:NV_intro}). It exists in three distinct charge states: \NVm, \NVo, and \NVp. Among these states, the negatively charged \NVm center, featuring a spin-1 ground state, is predominantly harnessed for applications in quantum sensing and quantum information. In this context, ``NV'' denotes NV$^-$ unless explicitly specified otherwise in subsequent text. Functioning as a sensor, the NV center offers straightforward initialization and readout procedures, necessitating only an approximate 10-$\mu$W laser at a frequency of 532 nm, obviating the requirement for specialized narrowband lasers. Under optical excitation, the spin-induced spontaneous emission facilitates the attainment of both spin-related fluorescence contrast and optical spin initialization of the \nv center's $m_s$=0 ground state. The state of the NV sensor can be readily identified, initialized, and extracted using confocal microscopy.

The NV sensor operates under ambient temperature, atmospheric pressure, and ambient magnetic fields, eliminating the requirement for cryogenic or vacuum systems, as well as high-level applied bias magnetic fields. Despite being at ambient temperature, the \NVm defect still sustains a long-lived spin ground state, with a longitudinal relaxation time extending up to 10 ms and coherent time reaching several milliseconds. Moreover, the spin freedom of this defect exhibits sensitivity to variations in magnetic fields, electric fields, strain, and temperature \cite{Doherty2013the}, enabling the NV center to function as a multimodal sensor. Moreover, diamond possesses chemical inertness, offering exceptional biocompatibility for NV sensors. These characteristics enable the sensor to be located approximately 1 nm from the field source \cite{Lovchinsky2016,muller2014Nuclear}, facilitating magnetic field imaging with sub-nanoscale spatial resolution and achieving high sensitivity in single-molecule measurements. NV sensors have been successfully employed for various applications, including single-molecule ESR, single-molecule NMR, single-DNA structure measurement, spin network detection, two-dimensional material detection, spin-qubit detection, and single-virus diagnosis. NV sensors can simultaneously detect magnetic fields, electric fields, strain, pH value, and temperature changes in the surrounding environment.

This review aims to present the current progress, limitations, and outlooks of single-molecule techniques based on NV. The review is intended to serve as an introductory resource for students and researchers new to this field and as a fundamental reference for researchers already involved in related studies.

\subsection{Single-molecule technology introduction}\label{single_molecule}

\begin{figure*}[hbtp]
\begin{overpic}[width=1\textwidth]{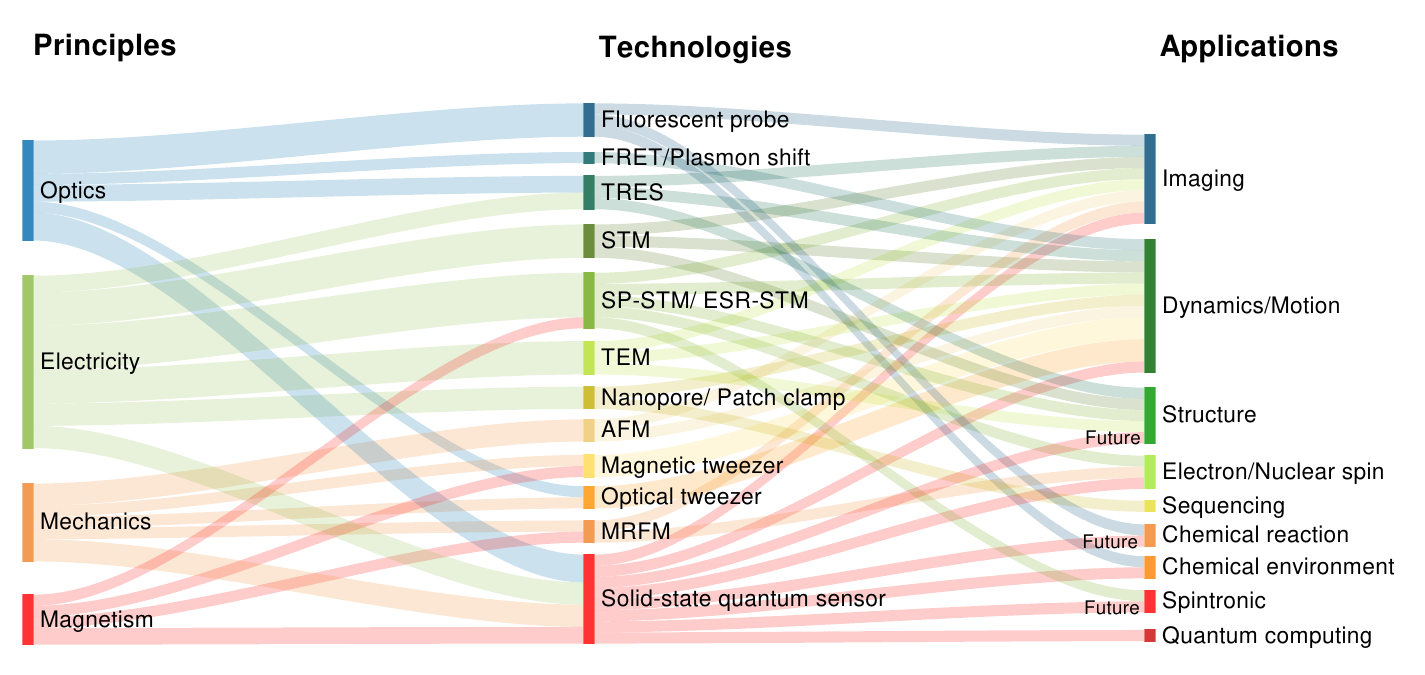} 
\end{overpic}
\caption[NV Center]{
The breakthrough of single-molecule techniques has allowed access to unprecedented physical, chemical, and biological details of molecules, ranging from tracking the movement of individual molecules to observing the vibration and conformational changes of single bonds within them. These measurements yield a single-molecule scale view of chemical reactions, physical processes, and intermolecular interactions, offering insight into the behavior and properties of individual molecules and offering novel understanding for interdisciplinary fields. Herein, we present a comprehensive overview of various types of single-molecule measurements, categorized based on the principles employed in the techniques; moreover, several of these techniques combine multiple principles, which are also addressed herein. The first column of color blocks corresponds to the four major physical principles: optics, electricity, mechanics, and magnetism. The second column corresponds to single-molecule techniques, connected by lines to their associated underlying principles. The last column highlights the applications of single-molecule techniques. The solid-state quantum sensor, which is linked to all four fundamental physical principles and boasts high sensitivity, spectral resolution, and spatial resolution, finds widespread applications (``future'' marks potential applications). 
  } \label{fig:smtech} 
\end{figure*}

Single-molecule technology serves as a conduit for delving into the behavior of individual molecules, unlocking realms that were previously unattainable. Across diverse disciplines, numerous potent technologies (Fig. \ref{fig:smtech}) have emerged, enabling the exploration and resolution of questions deemed ``unanswerable'' before. These methodologies predominantly fall into four main categories, each rooted in distinct measurement principles: optical, electric, mechanical, and magnetic measurements (Fig. \ref{fig:smtech}).

Optical microscopy is the oldest mode of microscopy, and optical methods are used to probe physical location or chemical environment of molecules. Despite the diffraction limit restricting spatial resolution, super-resolution microscopy techniques, such as stimulated emission depletion (STED) \cite{hell1994Breaking} and stochastic optical reconstruction microscopy (STORM) \cite{moerner2003Methods,betzig2006Imaging,rust2006Subdiffractionlimit}, have enabled determination of labeled fluorophore positions beyond the diffraction limit, down to approximately 10 nm. In particular, smaller distance scales can be measured via the fluorescence resonance energy transfer (FRET) \cite{roy2008practical,lerner2018dynamic} between fluorophores. In particular, FRET and plasmon shift enable the measurement of conformal changes of single molecules and local chemical environment \cite{alivisatos2004Use,giepmans2006Fluorescent,zhang2002Creating}.

Electrons with short wavelengths and small mass are suitable for single-molecule measurements with extremely short distance scales. 
The resolution of \tem \cite{claridge2005Directed,feldkamp2006Rational,fu2004Discrete,giljohann2010Gold} depends on the wavelength of the particle, allowing for the derivation of atomic-scale structures.
Due to their small mass, electrons exhibit quantum tunneling behavior, enabling \stm \cite{durkan2002Electronic,messina2007Spin,moore2008Functional,wiesendanger2009Spin,choi2017Atomicscale} us to detect surfaces and molecule structures with a spatial resolution of less than 0.1 nm. Apart from static measurements of molecules, the electric method has also been applied to observe the motions \cite{kwon2005Unidirectional,shirai2006SurfaceRolling,taranovskyy2010Quantitative,wong2007Molecule} and conformational changes \cite{choi2006Conformational,feringa2000Chiroptical,liljeroth2007CurrentInduced,loppacher2003Direct,moresco2001Conformational} of molecules. Nanopore, which allows only one molecule to move through at a time \cite{dekker2007Solidstate,branton2008potential}, is small enough to enable direct, real-time analysis of long DNA or ribonucleic acid (RNA) fragments. The single molecule driven by electric field, for example DNA molecules with typical conditions, moves through the nanopore with a residence time of approximately 1 $\mu$s per base \cite{meller2001VoltageDriven}. The characteristic ionic current is recorded, and the corresponding nucleotide sequence is determined.

 Single molecules can be measured via the forces between a sharp tip and the molecules.
Magnetic \cite{neuman2008Singlemolecule} and optical tweezers \cite{ashkin1986Observation,grier2003revolution,crampton2010Unravelling} are utilized for observing force-induced motion and dynamic and conformational changes in single molecules within a force range of 0.1--1000 pN and $\sim$ nm scale. \AFM \cite{giessibl2003Advances,neuman2008Singlemolecule,crampton2010Unravelling} enables the imaging of single molecules by facilitating measurements at larger force scales and distance ranges. In this regard, the \afm method is commonly used for imaging the topography of the molecule rather than identifying the chemical species. 
 
To analyze various aspects of single molecules, new technologies are developed by combining different physical principles. The combination of electric and optical methods, tip-enhanced Raman scattering (TERS) \cite{shi2017Advances,sonntag2014Recent}, yields a valuable ``fingerprint'' for chemical recognition in vibrational spectroscopy. \SM Raman spectroscopy has been demonstrated \cite{ichimura2009Subnanometric,steidtner2008TipEnhanced} and resolves the inner structure of a single molecule with chemical recognition \cite{zhang2013Chemical,shi2017Advances}. Combining electric and magnetic methods, electron spin resonance STM (ESR-STM) \cite{durkan2002Electronic,manassen1989Direct} and spin-polarized STM (SP-STM) \cite{wiesendanger2009Spin,atodiresei2010Design,brede2010Spin,loth2010Measurementa} allow chemical recognition through \sm ESR spectroscopy. Moreover, combining magnetic and mechanical methods, \mrfm measures the force experienced by polarized spins \cite{rugar2004Single,mamin2007Nuclear} in a gradient magnetic field. Furthermore, \MRFM allows for both mechanical detection \cite{rugar2004Single} and imaging \cite{mamin2007Nuclear} of the electron and nuclear spins of different chemical species. These techniques require careful instrumental design and operation at ultrahigh vacuum conditions and cryogenic temperatures.

Single molecule techniques also enable chemical recognition under ambient condition. The atomic-scale \sqs \cite{Degen2017},  the \nv center in diamond, combines optical and magnetic methods \cite{gruber1997Scanning} to enable quantum sensing for a variety of physical quantities, including force \cite{ovartchaiyapong2014Dynamica,teissier2014Straina}, magnetic field \cite{Balasubramanian2008,maze2008Nanoscale,Taylor2008}, electric field \cite{dolde2011Electricfield}, and temperature \cite{Kucsko2013,neumann2013HighPrecisiona}. The \ssmr utilizes \epr and \nmr spectroscopy for chemical recognition of a single molecule \cite{shi2015Singleprotein,Lovchinsky2016,sushkov2014Magnetic} while evaluating various other aspects, including imaging with sub-nanometer resolution \cite{grinolds2014Subnanometre,Arai2015}, sensing biological dynamics \cite{feng2021Association,igarashi2020Trackinga,mcguinness2011Quantuma,xia2019Nanometerprecision}, chemical environment \cite{fujisaku2019PH,hall2010Monitoringa,rendler2017Opticala,steinert2013Magneticb,ziem2013Highlya,Karaveli2016}, and chemical reactions \cite{peronamartinez2020Nanodiamond} under ambient conditions. Moreover, this characterization technique has potential applications in diagnostics \cite{millerSpinenhancedNanodiamondBiosensing2020,li2022SARSCoV2,chen2022Immunomagnetic}, quantum information \cite{Neumann2010b,schlipf2017molecular} and spintronics \cite{du2017Control,thiel2019Probing,wang2022Noninvasive,yan2022Quantum,zhou2021Magnon}.
The great potential lies the combination of multidisciplinary applications, non-labeling quantum sensing with high resolution contextual information, allowing for the correlation of various types of information regarding a single molecule under ambient conditions.

The single molecule techniques enable imaging of molecular structure, translational and rotational motion, conformational changes, molecular vibration, chemical environment, and chemical functionality. In turn, the ability of biology and chemistry to design and synthesize new molecules, from small molecules to supra-molecular assemblies, offers a wide range of possibilities for realizing practical quantum information applications using hybrid systems. One of the central tasks for scalable quantum technologies require an unprecedented combination of precision and complexity for designing stable structures of well-controllable quantum systems on the nanoscale.

A well-controlled spin-labeled biological network \cite{schlipf2017molecular} offers an exceptional opportunity to combine quantum spin systems with well-developed bio-programmable assembly techniques \cite{dai2016Optical,dey2021DNA}. \SSMR has the potential to read and potentially control the spin network, yielding a solid foundation for achieving the hybrid scalable quantum spin systems. Another important goal in physics are single-molecule electronics, which use individual molecules as electronic component for microelectronic industry \cite{aradhya2013Singlemolecule,chen2007Measurement,heath2009Molecular,rascon-ramos2015Binding,tao2006Electron,xu2003Measurement}. Single-molecule magnets \cite{leuenberger2001Quantum,wernsdorfer1999Quantum} also have potential applications in high density storage and quantum computation.
Additionally, \ssmr can provide more convenient and loose conditions for the readout of single-molecule electronics and single-molecule magnets.

\begin{figure*}[hbtp] 
\begin{overpic}[width=0.87\textwidth]{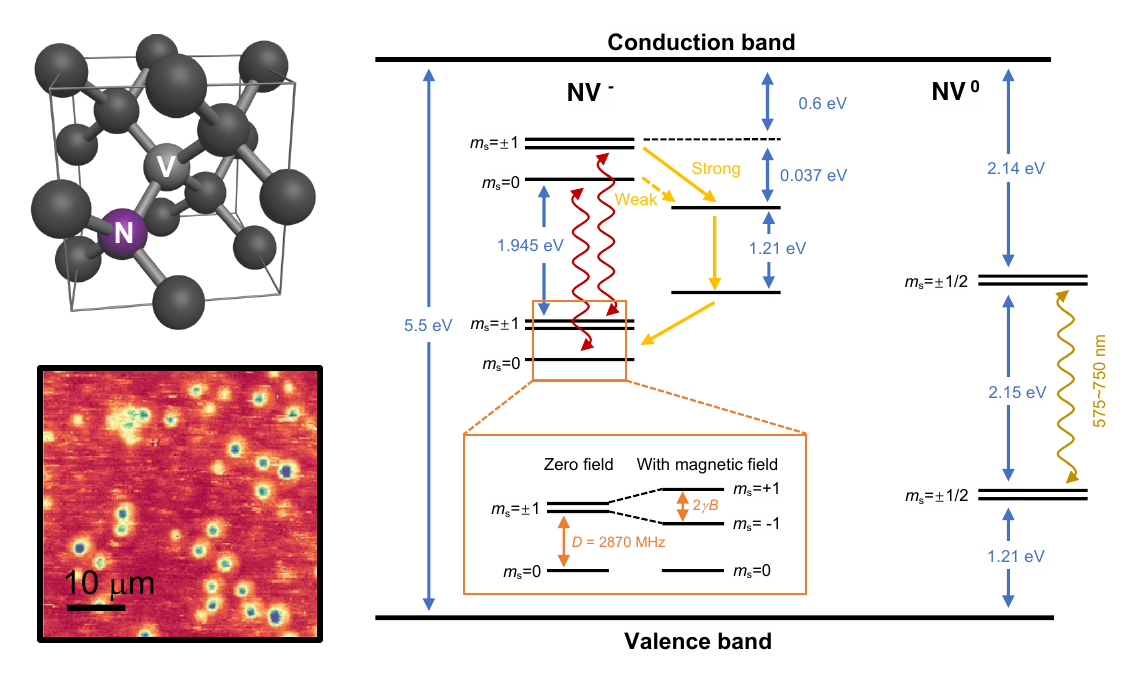}
	\put (-1, 54) {(a)}
	\put (-1, 27) {(b)}
	\put (31, 51) {(c)}
	\put (76, 51) {(d)}
\end{overpic}
\caption[NV Center]{ The \NV center. (a) An NV center in hexoctahedral diamond lattice. The NV center comprises a nitrogen atom (purple ball labeled as N) and an adjacent lattice vacancy (grey ball labeled as V). The principal axis lies along the connection line between nitrogen and the vacancy. (b) The scanning optical confocal imaging of the NV centers.
(c-d) The energy level configurations of the \NVm and \NVo centers in diamond. The NV center is situated within the conduction and valence band of the diamond lattice. (c) The negatively charged \NVm center energy level. The energy difference between the excited and the ground state is 1.945 eV which is corresponding to 637 nm zero phonon line. Refer to the orange box below for the detailed energy level of the ground state.
The ground state of \NVm is a triplet spin-1 system, with a zero-field splitting $D=2.87$ GHz between the electronic spin levels $m_s=0$ and $m_s=\pm 1$. Furthermore, applying a magnetic field $B$ causes an energy splitting between the $\ket{m_s=\pm 1}$ levels. The spin state can be detected through fluorescence intensity due to a spin-dependent transition, which results from an intersystem crossing between the excited state and meta-stable level. (d) The neutrally charged \NVo center energy level. The energy difference between the excited and the ground state is 2.15 eV, which corresponds to the 575-nm zero phonon line and 575--750 nm sideband.
}   \label{fig:NV}
\end{figure*}
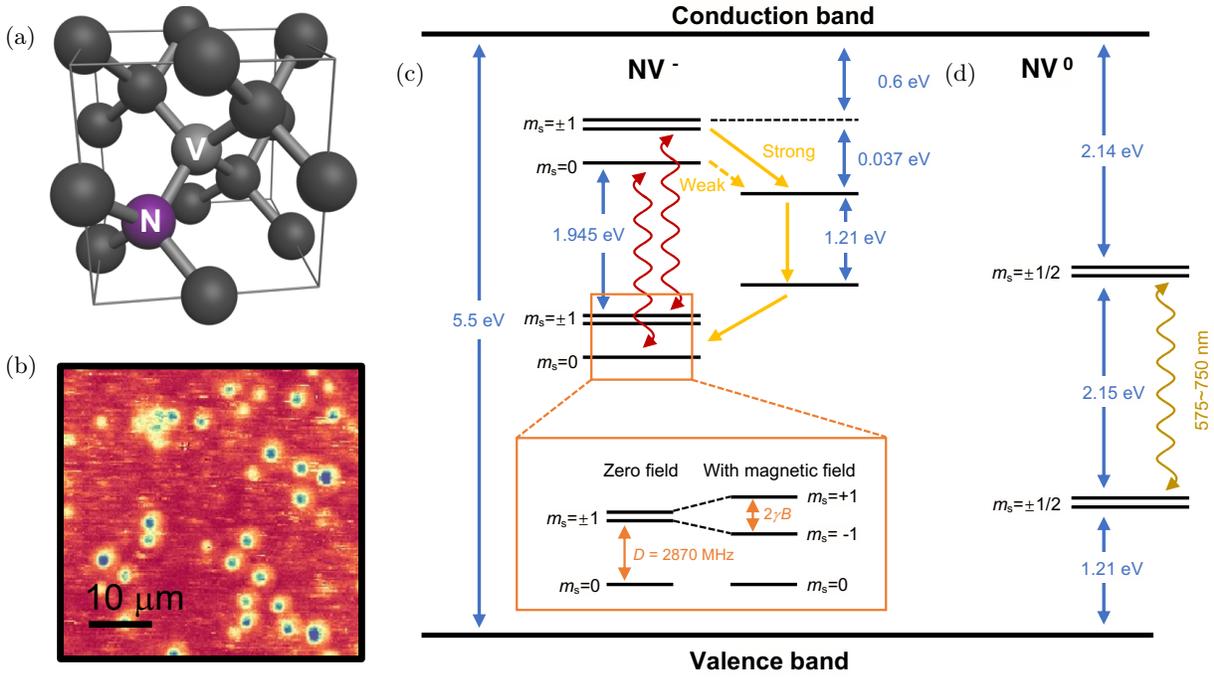

\subsection{NV-diamond sensing overview }\label{NV_intro} 

Quantum sensing \cite{Degen2017} technology utilizes quantum resources, such as quantum coherence, quantum entanglement, and coherent control of quantum systems to obtain precision measurements. The use of quantum metrology techniques and quantum resources allows the surpassing of the standard quantum limit. This technology is based on techniques developed in the fields of atomic physics and magnetic resonance techniques, including atomic clocks \cite{ludlow2015Optical}, atomic vapor magnetometers \cite{mitchell2020Colloquium}, superconducting quantum interference devices \cite{greenberg1998Application}, and solid-state NV centers \cite{Barry2020Sensitivity}. Among these modalities, the \nv center serves as a unique platform for quantum sensing, due to its substantial coherence time under ambient conditions.

The single NV center in diamond is first observed by confocal scanning microscopy \cite{gruber1997Scanning}. It is an optically active point defect in diamond, and its spin state can be initialized and measured using optical excitation \cite{jelezko2004Observation,jelezko2004Observationa}. In particular, the ultra-long spin relaxation times \cite{Balasubramanian2009,BarGill2013,herbschleb2019Ultralong} yield a unique opportunity to realize high sensitivity at the nano-scale or even at the molecular scale. The use of NV centers as nanoscale magnetic sensors was first proposed and experimentally demonstrated  \cite{Degen2008,Taylor2008,Balasubramanian2008,maze2008Nanoscale} in 2008. The NV center has emerged as a promising quantum sensing platform in the following decades, as reported in \nmr \cite{Staudacher2013nuclear,mamin2013nanoscale}, \epr \cite{shi2015Singleprotein}, as well as in the fields of condensed matter physics \cite{casola2018Probing} and biology \cite{millerSpinenhancedNanodiamondBiosensing2020}.

The NV center is a point defect comprising a substitutional nitrogen adjacent to a vacancy in diamond (Fig.~\ref{fig:NV}(a,b)). There are different charge states of NV center, NV$^-$, NV$^0$, and NV$^+$. The negatively charged state \NV exhibits a triplet ground state and a spin-dependent \isc, which allows for measurement of the spin state based on the spin-dependent fluorescence (Fig.~\ref{fig:NV}(c)). The NV center possesses long spin relaxation time at ambient temperature, with the best longitudinal relaxation times $T_1\sim 7$ ms, typically 0.3 -- 5 ms \cite{rosskopf2014investigation,Myers2017} with NV center depth under 10 nm, the best coherence time $T_2$ approaching 2.4 ms, typically 5 -- 200 $\mu$s under 10 nm \cite{Lovchinsky2016,myers2014probing,rosskopf2014investigation,Romach2015}, and the best dephasing time $T_2^*$ approaching 1.5 ms \cite{herbschleb2019Ultralong}, typically 3 -- 50 $\mu$s with NV center depth under 10 nm limited by $T_2$. 
The NV center's long coherence time at ambient temperature makes it advantageous to utilize its quantum coherence to measure physical quantities, including the magnetic fields \cite{Balasubramanian2008,maze2008Nanoscale,Barry2020Sensitivity}, electric fields \cite{dolde2011Electricfield}, strain fields or pressure fields \cite{Hsieh2018imaging,lesik2019Magnetic,yip2019Measuring}, and temperature \cite{Doherty2013the,Kucsko2013}. The quantum feature of \NV center in diamond can be utilized to increase sensitivity, which can be enhanced by designing tailored protocols suited for use with the NV center. Additionally, the NV center can be engineered specifically for use as a quantum sensor.

The \NV center exhibits quantum sensing across a vast range of environments, with experiments performed spanning from ultra-low temperatures ($\sim$ 10 mK) \cite{zhu2011Coherent} to extreme heat ($\sim$ 1000 K) \cite{liu2019Coherenta}, from \uhv \cite{rosskopf2014investigation} to high pressure conditions ($\sim$ 30 GPa) \cite{Hsieh2018imaging,lesik2019Magnetic,yip2019Measuring}, and magnetic fields from zero up to $\sim$ 3 T \cite{aslam2017Nanoscale}. Additionally, the diamond material that houses the NV center provides exceptional biocompatibility and low toxicity, which permits its successful integration into living biological cells \cite{mcguinness2011Quantuma,millerSpinenhancedNanodiamondBiosensing2020,Kucsko2013}. 
Through the ability to precisely manufacture NV sensors within 2 nm of the diamond surface \cite{muller2014Nuclear,Lovchinsky2016}, their superior sensitivity and spatial resolution allows for detection of entities at the \sm level; however, reconstruction of the three-dimensional position of nuclear spins on single molecule has not been achieved.

\subsection{The NV spin control technique}\label{NV}

The \NV center exhibits a spin-1 triplet electronic spin ground state. The main axis of the NV center is along the vector joining the nitrogen atom and the vacancy. In particular, the spin-1 ground-state Hamiltonian of the \NV center is comprised of three essential components, the spin Hamiltonian with an external magnetic field $H_0$, the interaction between the \NV spin and adjacent nitrogen nuclear spin $H_{\text{nuclear}}$, the electron spin interaction with the electric fields and strain $H_{\text{electrical|strain}}$, 
\begin{align}\label{eqn:NV_ham}
   H  = & \underbrace{\mathcal{D} S_{z}^2+\gamma_{\text{NV}}\vb{B}_0\cdot\vb{S}}_{H_{\text{0}}}\\\nonumber
   &  +A_{\parallel}S_{z}I_{z,\text{N}}\\\nonumber
   &   \underbrace{+A_{\perp}(S_{x}I_{x,\text{N}}+S_{y}I_{y,\text{N}})+\mathcal{P}I_z^2-\gamma_{\text{N}}(\vb{B}_0\cdot\vb{I}_{\text{N}})}_{H_{\text{nuclear}}}\\\nonumber
   &  +d_{\parallel}(E_z+\delta_z)\qty(S_{z}^2-2/3)\\\nonumber
   &  \underbrace{+d_{\perp}(E_x+\delta_x)\qty(S_{y}^2-S_{x}^2)+d_{\perp}(E_y+\delta_y)\qty(S_{x}S_{y}+S_{y}S_{x})}_{H_{\text{electrical|strain}}}.
\end{align}
$H_0$ consists of the zero-field-splitting term with $\mathcal{D}$, resulting from the spin--spin interaction, which breaks the symmetry, and $\gamma_{\mathrm{NV}} \mathbf{B}_{0} \cdot \mathbf{S}$, which describes the interaction between the \NV electron spin and the external magnetic field $\vb{B}_0$, where 
$\gamma_{\text{NV}}$ represents the gyromagnetic ratio. $H_{\text{nuclear}}$ accounts for terms related to the \NV center's nitrogen nuclear spin ($I_{N}=1$ for $^{14}$N and $I_{N}=1/2$ for $^{15}$N), where $A_{\parallel}$, $A_{\perp}$ are the parallel and perpendicular hyperfine coupling strengths respectively, $\mathcal{P}$ denotes the nuclear electric quadrupole coupling, and $\gamma_{\text{N}}$ indicates the gyromagnetic ratio of the respective nitrogen nuclear isotope. $H_{\text{electrical|strain}}$ represents the response to the electric field and the strain field, where $d_{\parallel}$ and $d_{\perp}$ are the ground state components of the electric dipole moment. Moreover, $\vb{S}=(S_x,S_y,S_z)$ is the \NV electronic spin operator, $\vb{I}_{N}=(I_{x,N},I_{y,N},I_{z,N})$ is the nitrogen nuclear spin operator, $\vb{E}=(E_x,E_y,E_z)$ is the electrical field vector, and $\bm{\delta}=(\delta_x,\delta_y,\delta_z)$ is the strain field vector. Values of the parameters $\mathcal{D}$, $\gamma_{\text{NV}}$, $\gamma_{\text{N}}$, $A_{\parallel}$, $A_{\perp}$, $\mathcal{P}$, $d_{\parallel}$, and $d_{\perp}$ are listed in \app \ref{sec:symbols}.

\subsubsection{Coherent control and readout of NV center}

\NV  center exhibits spin state-dependent fluorescence emission \cite{Doherty2013the} in the 600--800 nm range. Under the 532-nm laser excitation, the \NV system cycles between the ground and excited states and shelves to the metastable singlet manifold preferentially for $m_s=\pm 1$ spin states ($\ket{\pm 1})$ via the triplet-to-singlet ISC. The singlet-to-triplet ISC is less spin selective compared to the triplet-to-singlet ISC, leading to a ground-state spin polarization to $m_s=0$ spin state ($\ket{0}$) after the laser illumination \cite{Doherty2013the,waldherr2011Dark,robledo2011Spin}. 

\begin{figure}[htp]
\begin{overpic}[width=0.8\columnwidth]{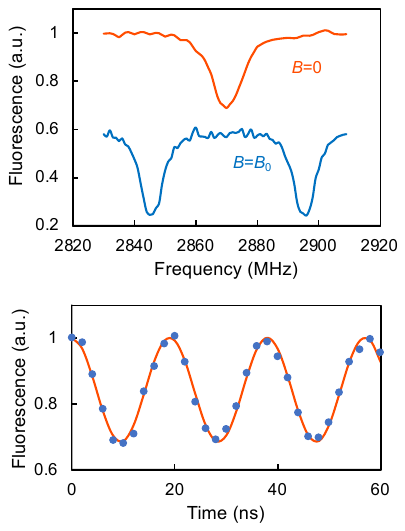}
\put (-6, 94) { (a) }
\put (-6, 42) { (b) }
\end{overpic}
\caption[cw]{  ODMR methods. (a) The \cwODMR spectra of a single NV center. The spectrum is observed by sweeping the applied cw microwave frequency while laser pumping is applied. The fluorescence decreases at the resonant frequency. The upper curve is obtained under zero external field where the $m_s=\pm 1$ states are degenerate. The lower curve is obtained under $B_0$ field with $2\gamma_eB_0/2\pi$ splitting. (b) The Rabi oscillation of single NV center. By resonant microwave pulse, the \NV center spin oscillates between the $m_s=0$ and $m_s=1$ spin states. }\label{fig:cw}
\end{figure}

\CwODMR is a common tool to characterize spin-related energy level transitions \cite{wrachtrupOpticalDetectionMagnetic1993,gruber1997Scanning} and also a powerful tool for magnetometry \cite{Balasubramanian2008,lesage2013Optical,Kucsko2013,Barry2016}. The \NV spin is driven with continuous laser and microwave irradiation simultaneously. The \NV spin is driven to $\ket{\pm 1}$ state when the microwave is tuned to be resonant with one of the $\ket{0}\leftrightarrow \ket{\pm 1}$ transitions, shown as the fluorescence reduction dip in Fig. \ref{fig:cw}(a). The \NV exhibits an intrinsic zero-field-splitting transition frequency, $D$ = 2.87 GHz, and its zero-field spectrum appears as a single dip. Moreover, the Zeeman effect causes the spectrum to split into double dips under magnetic field $B_0$.

Pulsed techniques are more sensitive sensing approaches wherein the laser is turned off during the operational period. The long coherence time $\sim$ ms \cite{Balasubramanian2009,BarGill2013} at ambient temperature allows complicated magnetic resonance sequences, developed over decades in the field of magnetic resonance \cite{Slichter1990}. Coherent control is realized by a resonant microwave pulse applied during the no optical period known as Rabi oscillation \cite{Slichter1990}. A typical example of the spin oscillation between $\ket{0}$ and $\ket{1}$ for a single \NV at ambient temperature is shown in Fig. \ref{fig:cw}(b).

\begin{figure}[tbhp]
\begin{overpic}[width=1\columnwidth]{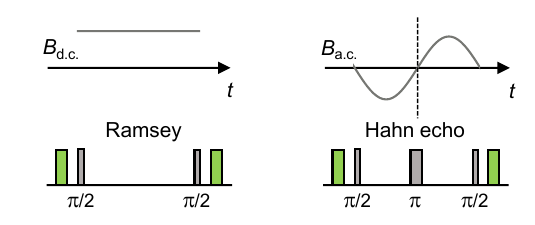}
\put (0, 38) {(a)}
\put (48, 38) {(b)}
\end{overpic}
\caption[NV Center]{  The control protocols for quantum sensing for d.c. and a.c. target magnetic field. (a) Pulse sequence for the d.c. magnetometry using Ramsey fringe with a sequence of $\pi/2-\tau-\pi/2$. (b) Pulse sequence $\pi/2-\tau-\pi-\tau-\pi/2$ for the a.c. magnetometry. A $\pi$ is inserted in the middle of the sequence to eliminate unwanted environmental noise and accumulate desired magnetic field with the right frequency $\omega=\pi/\tau$. }  \label{fig:sensing_seq}
\end{figure}

Ramsey interferometry measurement \cite{lee2002quantum,Taylor2008} serves as a powerful tool for measuring static magnetic fields $B_{\text{d.c.}}$ (Fig. \ref{fig:sensing_seq}(a)). A $\pi/2$ microwave pulse is applied to either convert the initialized state to a superposition state or transform the interacted state to a final state. The static magnetic field, or any static energy shift, introduces a relative phase $\phi$ between $\ket{0}$ and $\ket{1}$ states. This relative phase is given by $\phi=\gamma_eB_{\text{d.c.}}t_s$, where $t_s$ denotes the coherent evolution time during the sensing protocol between the sensor and the signal $B_{\text{d.c.}}$. 
Furthermore, the Hahn echo control sequence (Fig. \ref{fig:sensing_seq}(b)) is utilized to measure the alternating current (a.c.) field while eliminating static environmental fluctuations. An additional microwave $\pi$ pulse is introduced in the middle of the Ramsey sequence. Under the Hahn echo control, the time-dependent oscillating magnetic field $B(t)=B_{\text{a.c.}}\cos (\omega t+\varphi_0)$ during sensing interrogation time $t_s=2\tau$ contributes to an overall additive phase shift. The accumulated phase is given by $\phi=\gamma_e[\int_0^{\tau}B(t)\dd t-\int_{\tau}^{2\tau}B(t)\dd t]$ and reaches its maximum when the sequence is resonant and in phase, i.e., when $\tau=\pi/\omega$ and $\varphi_0=0$. The interrogation time $t_s$ can be extended from \Ttwostar to \Ttwo, thereby improving the sensitivity and spectral resolution. Moreover, the sensitivity and spectral resolution can be further improved by utilizing more complex sequences, such as the \dyde (\se \ref{sec:deer} and \ref{sec:NMR:DD}), correlation spectroscopy, and the \qdyne method (\se \ref{sec:correlation}).

\subsubsection{Quantum sensing } \label{sec:quantum_sensing}

\begin{figure}[htb]
\begin{overpic}[width=0.89\columnwidth]{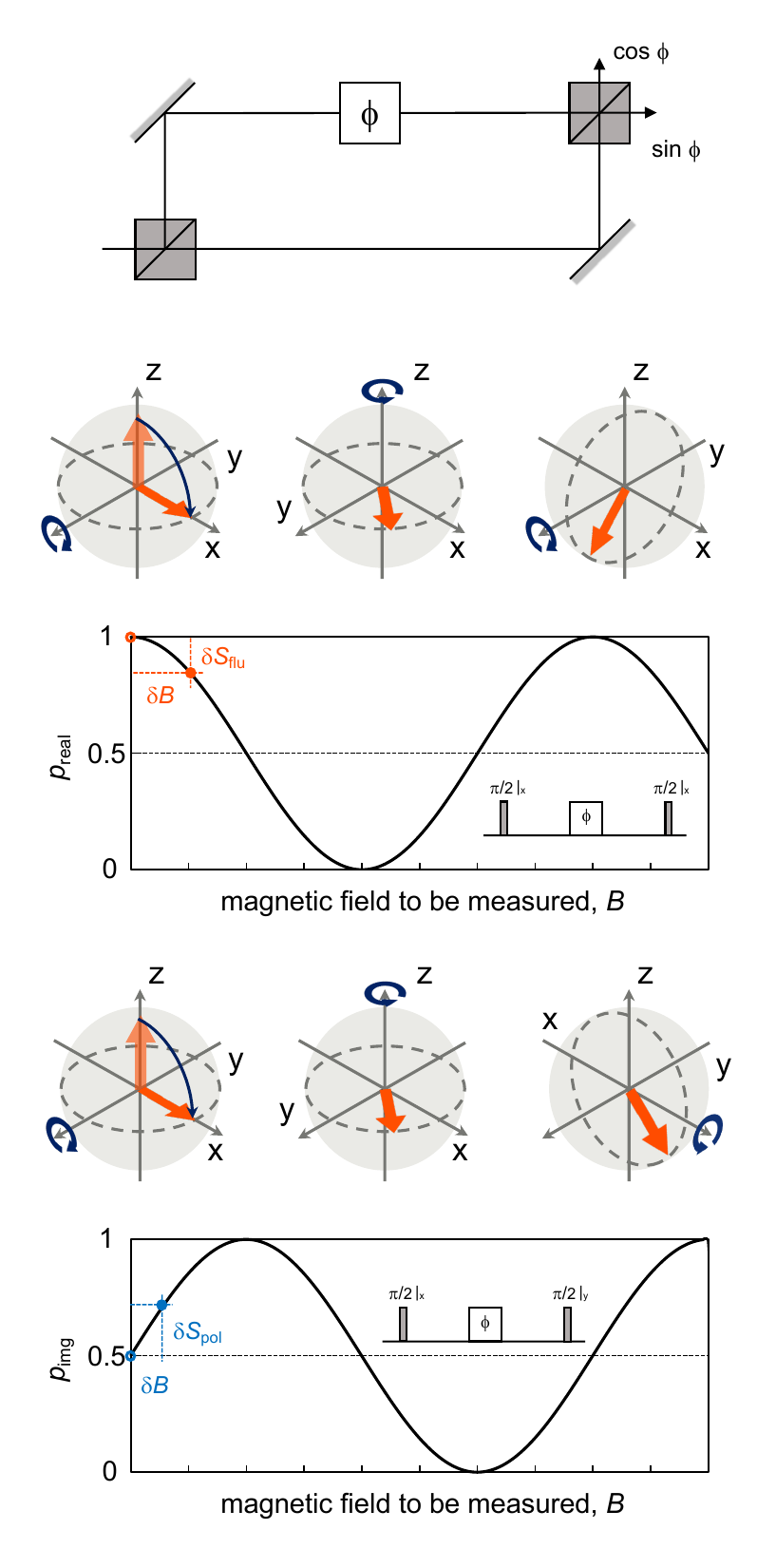}
\put (0, 95) { (a) }
\put (0, 76) { (b) }
\put (0, 59) { (c) }
\put (0, 37) { (d) }
\put (0, 21) { (e) }
\end{overpic}
\caption[quantum sensing]{  Quantum sensing. (a) The concept of quantum sensing is illustrated through an interferometer with two beam-splitters (corresponding to $\pi/2$ pulse) and a phase accumulation operation ($\phi$). (b) The spin evolution in the Bloch sphere. The spin is initially rotated by a $\pi/2$ pulse along the $x$-axis to establish coherence on the $x-y$ plane. Subsequently, a second $\pi/2$ pulse along the $y$-axis is used to project the real component of the spin coherence onto the $z$-axis. (c) Measurement of the signal for the real component of spin coherence versus the magnetic field $B$. (d) Similar to (b) but with the second $\pi/2$ pulse along the $x$-axis. Signal measurement for the imaginary component of the coherence. (e) Signal measurement for the imaginary component of the spin coherence versus the magnetic field $B$.}  \label{fig:sensing}
\end{figure}
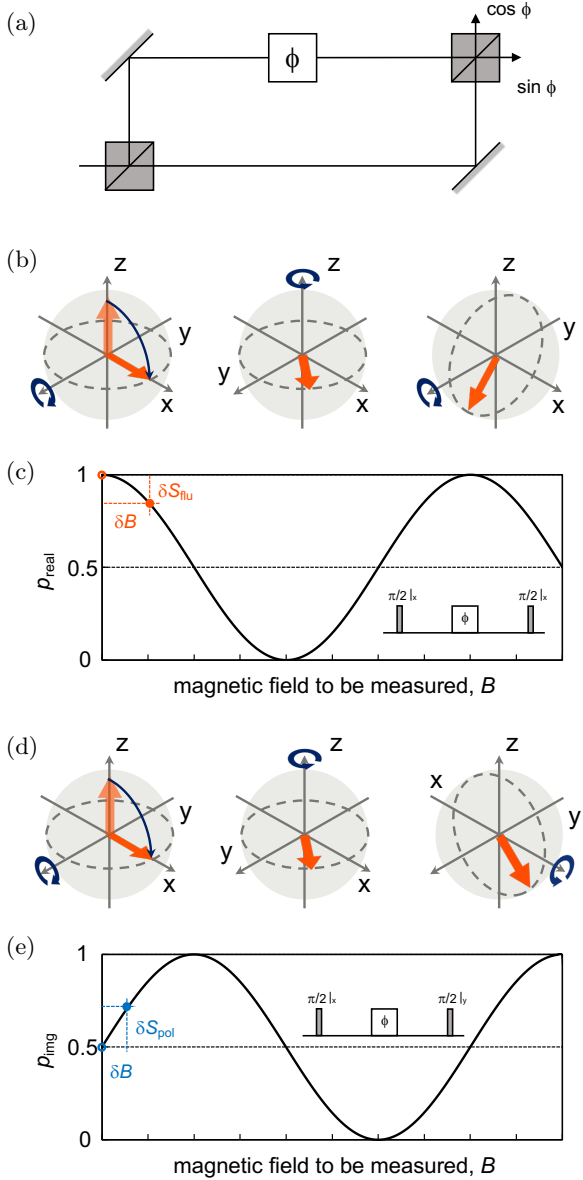

Quantum sensing involves using a quantum object to measure a physical quantities or utilizing quantum coherence and quantum entanglement to perform measurements \cite{Degen2017}. It always follows a generic protocol, sensor initialization, signal and sensor coupling, sensor readout, and signal estimation \cite{Degen2017}. The canonical approach is illustrated as a quantum interferometer \cite{lee2002quantum} in Fig. \ref{fig:sensing}(a). The Hamiltonian for the quantum sensor can be described as \cite{Degen2017},
\begin{equation}
H(t)=H_{0}+H_{V}(t)+H_{\text {control }}(t), \label{eqn:sensing_hamiltonian}
\end{equation}
where $H_0$ is the NV sensor Hamiltonian, $H_{V}(t)$ is the response Hamiltonian of the quantum sensor coupled with physical quantity $V(t)$, and $H_{\text {control }}(t)$ is the control Hamiltonian. The control Hamiltonian $H_{\text {control }}(t)$ can be implemented through electromagnetic field control. By optimizing $H_{\text {control }}(t)$ protocols, the target physical quantity $V(t)$ can be readout precisely. As a quantum interferometer, the \NV quantum sensor is firstly initialized into $\ket{0}$ by a 532-nm laser. A $\pi/2$ pulse along $x$-axis converts the quantum sensor to a superposition state (Fig. \ref{fig:sensing}(b,d)),
\begin{equation}\label{eq:superposition}
\ket{+}=\frac{1}{\sqrt{2}}\qty(\ket{0}+\ket{1}).
\end{equation}
Subsequently, the sensor evolves for time duration $t$ under the interaction with the target sample. As a result of the interaction, the superposition state picks up a phase $\phi$ dependent on the physical quantity $V(t)$. Without loss of generality, the accumulated phase is $\phi=\int_0^{\tau}g_sV(t)\dd t$ if the response Hamiltonian $H_{V}(t)=g_sV(t)S_z$, where $g_s$ is the response coefficient of the physical quantity $V(t)$. The state of the quantum sensor evolves to 
\begin{equation}
\ket{\psi(t)}=\frac{1}{\sqrt{2}}\qty(\ket{0}+e^{-i\phi}\ket{1}).
\end{equation}
The spin coherence is then defined as $\xi=e^{-i\phi}$. The final state is either transformed to 
\begin{equation}
\ket{\psi_{\text{real}}}=e^{-i\phi/2}\qty(\cos\frac{\phi}{2}\ket{0}+i\sin\frac{\phi}{2}\ket{1})
\end{equation}
through the real component sensing protocol ($U_{\text{real}}$) with an ending $\pi/2$ pulse in $x$-axis or 
\begin{equation}
\ket{\psi_{\text{img}}}=e^{i\pi/2}\qty(\cos\qty(\frac{\phi}{2}-\frac{\pi}{4})\ket{0}+\cos\qty(\frac{\phi}{2}+\frac{\pi}{4})\ket{1})
\end{equation}
through the imaginary component sensing protocol ($U_{\text{img}}$) with an ending $\pi/2$ pulse in $y$-axis. The final signal is  determined by the projection probability on $\ket{0}$. The results of the real component and imaginary component readout  are (Fig. \ref{fig:sensing}(c,e))
\begin{align}
p_{\text{real}}=\braket{0}{\psi_{\text{real}}}^2=\frac{1}{2}(1+\Re \xi) , \label{eqn:real_read}  \\
p_{\text{img}}=\braket{0}{\psi_{\text{img}}}^2=\frac{1}{2}(1+\Im \xi) \label{eqn:img_read}
\end{align}
respectively. 

\begin{figure}[hbtp]
\begin{overpic}[width=0.9\columnwidth]{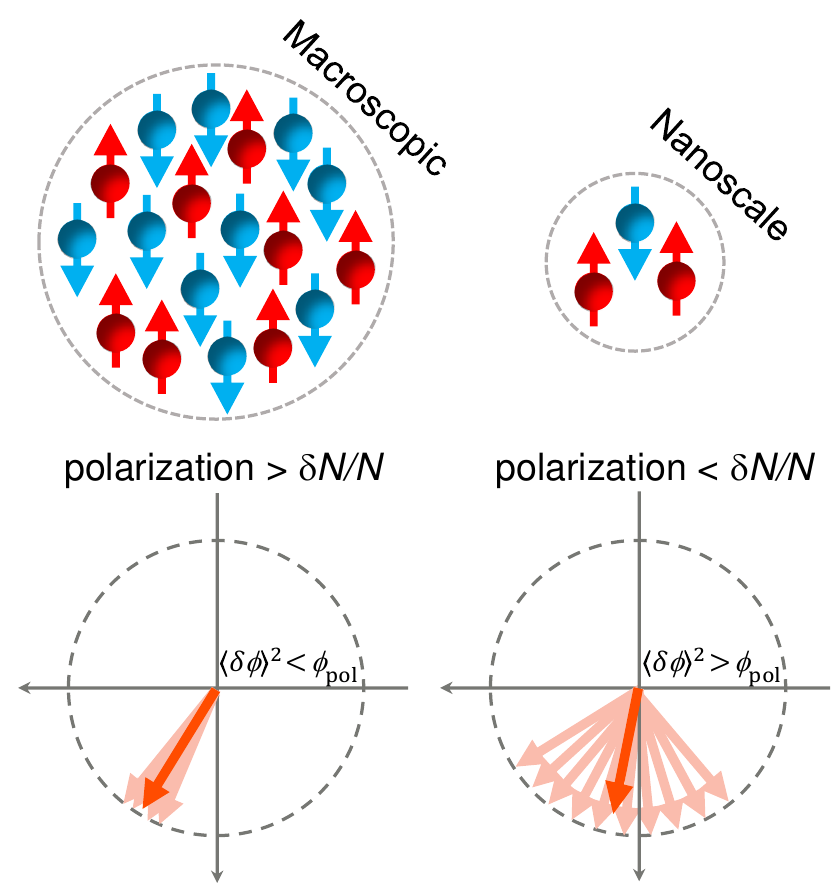}
\put (0, 94) { (a) }
\put (49, 94) { (b) }
\end{overpic}
\caption[quantum sensing]{ Measurement of macroscopic spin and nanoscale spin system. (a) The fluctuation of the macroscopic ensemble spin system, characterized by $\delta N/N\sim 1/\sqrt{N}$, is always smaller than the polarization. The real component readout signal, which detects the fluctuation $\sim \langle \delta\phi\rangle^2$, is smaller than the imaginary component readout signal, $\sim \phi_{\text{pol}}=\langle \phi\rangle$, which is proportional to the polarization. (b) Conversely, the fluctuation in a few-spin system, characterized by $\delta N/N\sim 1/\sqrt{N}$, can be larger than the polarization. In such cases, the fluctuation signal, $\langle \delta \phi\rangle^2$, can surpass the polarization signal. Further details are reported in Fig. \ref{fig:nmr_pol_flu}. } \label{fig:sensing_real_img}
\end{figure}

The fluctuation of field $V(t)$ is the main source of the decoherence in the NV center spin. The decoherence could reduce both the sensitivity and spectral resolution of the quantum sensing (see details in \se \ref{sec:decoherence}). Nevertheless, it is possible to utilize the decoherence for the noise field measurements. The correlation function $S(t)$ characterize the stochastic processes of the fluctuations $\Delta V(t)$ and is defined as,
\begin{equation}
S(t)=g_s^2\langle \Delta V(0)\Delta V(t)\rangle. \label{eqn:PSD_define}
\end{equation}
The spectral density of noise $S(\omega)=\mathscr{F}S(t)$ is obtained by Fourier transform of $S(t)$.
The relaxation of the NV center spin depends on the control pulse sequence, which provides a filter function $F_t(\omega)$ in the frequency domain (see details in \se \ref{sec:decoherence:sensor_control}), and the noise spectral density $S(\omega)$. The coherence is given by \cite{klauder1962Spectral},
\begin{equation}
\xi(t)=e^{-\chi(t)/2}, \label{eqn:decoherence_define}
\end{equation}  where the decoherence function $\chi(t)$ is
\begin{equation}
\chi(t)=\frac{1}{\pi} \int_{-\infty}^{\infty} \dd \omega S(\omega) \frac{F_{t}(\omega)}{\omega^{2}}. \label{eqn:deocherence}
\end{equation}

Different measurement methods are required for different detection tasks depending on the target sample. Take the spin detection as example, the corresponding physical quantity are $V(t)=\langle I_x\rangle$ and $V(t)=\langle I_z\rangle$. The strengths of statistical fluctuation $\sim 1/\sqrt{N}$ are different for macroscopic and microscopic spin ensembles, as shown in Fig. \ref{fig:sensing_real_img}. Notably, macroscopic spin ensembles generally exhibit smaller fluctuation than the thermal polarization or hyperpolarization. Conversely, microscopic spin ensembles can have fluctuation with magnitude greater than the polarization. The readout results (Eq. \ref{eqn:real_read} and \ref{eqn:img_read}) dependent on the accumulated phase $\phi\ll 1$ are 
\begin{align}
\langle p_{\text{real}}\rangle & \approx 1-\frac{{\langle\phi\rangle}^2}{4}-\frac{\delta\phi^2}{4},  \label{eqn:real_result} \\
\langle p_{\text{img}}\rangle & \approx \frac{1}{2}+\frac{{\langle\phi\rangle}}{2}, \label{eqn:img_result}
\end{align}
where ${\langle\phi\rangle}=\/g_st_s P$, $\delta\phi=g_st_s /\sqrt{N}$ and $P$ is the polarization of the target sample spin ensemble. The real component readout is an efficient method to use when quantum fluctuation predominate, while the imaginary component readout is efficient for polarization-dominant scenarios (as illustrated in Fig. \ref{fig:sensing_real_img}).  

Detection of a single spin presents a different scenario. With a sufficient level of sensitivity, the target spin state can be readout with single shot \cite{Degen2017,Neumann2010}. This possibility depends upon the spin--sensor coupling strength $g_s$, the target spin relaxation rate $\gamma$, and the sensor readout noise $\sigma_R$. When the accumulated measurement signal suffices to surpass the readout noise, $g_s/2\gamma\gg\sigma_R$, before the target spin relaxes, the single shot readout protocol \cite{Neumann2010,Maurer2012,dreau2013SingleShot,liu2017SingleShot} accurately determines the target spin state with high fidelity. The NV sensor accumulates a phase that is determined by $\phi= \langle I_z \rangle{g_s}{t_s} $, and the imaginary component readout result is $S_{\text{img}}=1/2\,\sin {g_s}{t_s}$ (Eq. \ref{eqn:real_read}, excluding the constant term). Similar conclusions apply to tracking measurements of $I_x(t)$ \cite{cujia2019Tracking}. The signal strength is insufficient to exceed the readout noise before the target spin relaxes if ${g_s}/{2}\gamma \ll \sigma_R$. Only the fluctuation signal ($S_{\text{real}}$) can be observed, which is given by $(1-\cos {g_s}{t_s})/2 \approx {g_s}^2{t_s}^2/4 \ll S_{\text{img}}$.

\section{Quantum sensing approaches and research progresses}\label{sec:progress}

\begin{figure*}[hbtp]
\begin{overpic}[width=0.8\textwidth]{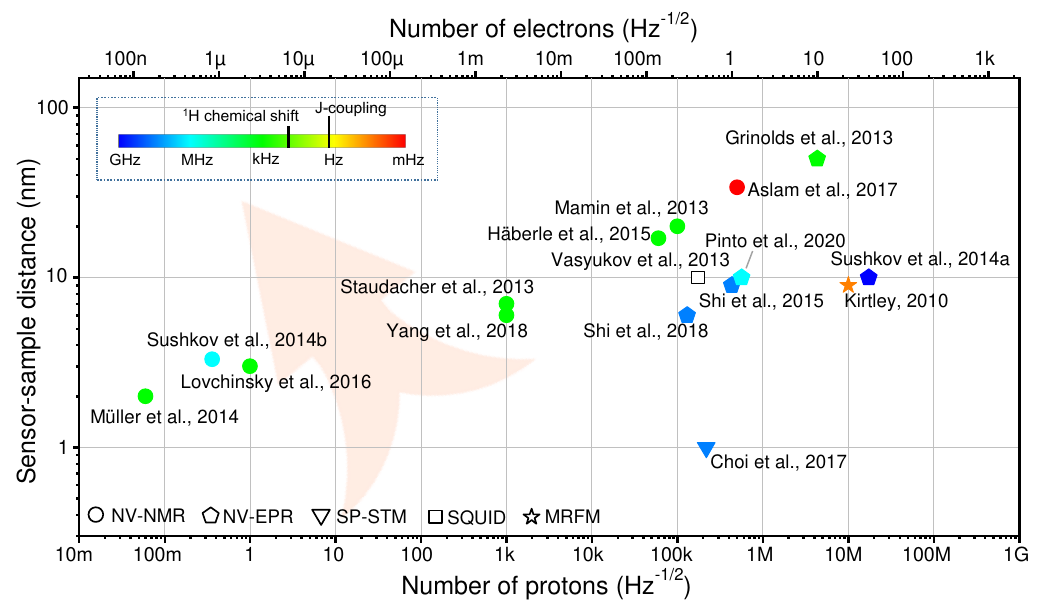} 
\end{overpic}
\caption[NV Center]{ The reported spin number detection sensitivity $\eta_{\text{spin}}$ by different magnetic resonance technologies versus the sensor--sample distance. The spin detection sensitivity $\eta_{\text{spin}}$ represents the minimum detectable spin number (standard deviation) per Hz$^{1/2}$. For the works without specific reported values, we estimated the spin sensitivity of electrons/protons per Hz$^{1/2}$ for different detection methods by estimating the magnetic field generated by spins or by comparing the signal standard deviation with the experimental time.
The applications of \ssmr are restricted to two regions, one which has high spin number sensitivity at short sensor--sample distances, and the other which has low spin number sensitivity at far sensor--sample distances.
Our ultimate research goal is to achieve single spin sensitivity for far sensor--sample distance conditions. The details are presented in Table \ref{tab:sensitivity_distance}.
 }  \label{sensitivity_distance}
\end{figure*}

This chapter introduces the measurement protocols used in \ssmr in two sections, focusing on \epr and \nmr. The \epr section summarizes the methods for detecting electron spin, including \deer \cite{shi2015Singleprotein,shi2018SingleDNA,schlipf2017molecular,pinto2020Readout}, quantum relaxometry \cite{kong2018Nanoscale,lillie2017Environmentally,sushkov2014alloptical,steinert2013Magnetic,ziem2013Highly,ermakova2013Detection} and other methods. The NMR section includes \dyde \cite{zhaoSensingSingleRemote2012,Staudacher2013nuclear,shi2014Sensing,muller2014Nuclear}, \enor \cite{mamin2013nanoscale,aslam2017Nanoscale}, \hhdr \cite{londonDetectingPolarizingNuclear2013,shagievaMicrowaveAssistedCrossPolarizationNuclear2018}, two-dimensional \nmr \cite{yang2018Detection,kong2020Artificial,abobeih2019Atomicscale,smits2019Twodimensional} and other methods. Works with the capability to detect single-molecule are highlighted, including single nuclear spin sensitivity NMR \cite{muller2014Nuclear}, single proton NMR \cite{sushkov2014Magnetic}, all-optical single-molecule electron sensing \cite{sushkov2014alloptical}, single protein \EPR \cite{shi2015Singleprotein}, single-molecule NMR \cite{Lovchinsky2016}, single-molecule liquid \EPR \cite{shi2018SingleDNA}, and single endofullerene \EPR \cite{pinto2020Readout}. Although \ssmr have advanced significantly, research has primarily focused on two regions, i.e., short sensor--sample distances with high spin number sensitivity or long sensor--sample distances with low spin number sensitivity (Fig. \ref{sensitivity_distance}), especially for the single-molecule NMR at the micron scale. 
On the one hand, \ssmr aims to achieve high sensitivity at short distance. On the other hand, it is also critical to aim for high spin sensitivity at far sensor--sample distances (the arrow in Fig. \ref{sensitivity_distance}).

\begin{table*}[]
\caption[NV Center]{Spin number sensitivities for different sensing methods in Fig. \ref{sensitivity_distance}. Values are assumed directly from the cited references. Other values are estimated from the experimental curves or experimental descriptions. In cases where values are nonexistent or not applicable, we denoted these values using the symbol ``---''.} \label{tab:sensitivity_distance}
\begin{tabular}{clcccc}
\toprule
 \toprule
 \parbox{6em}{Method}         &\parbox{6em}{References}    & \parbox{8em}{Sensitivity\\$\mu_B/\sqrt{\text{Hz}}$ } & \parbox{8em}{Sensitivity\\proton/$\sqrt{\text{Hz}}$  } & \parbox{10em}{Sensor--sample distance   (nm)}   & \parbox{8em}{Spectral resolution  } \\ \midrule
 MRFM      &  \cite{kirtley2010Fundamental}      &   23        & ---   & 9                           & 0.21 Hz                 \\[0.1in]
 SQUID      & \cite{vasyukov2013scanning}     & 0.4        & ---    & 10                         &        ---          \\[0.1in]
SP-STM      &  \cite{choi2017Atomicscale}         &   0.5        & ---   & $\sim$ 1                           & 10 MHz                 \\[0.1in]
NV-EPR     &  \cite{grinolds2013Nanoscale}     &    10          & ---    & 50                          & $\sim$ 10 kHz                 \\
  &     \cite{sushkov2014alloptical}      & 40          & ---   & 10                          & ---                 \\
   &   \cite{shi2015Singleprotein}            & 1          & ---     & 9                           & 10 MHz                \\ 
   &  \cite{shi2018SingleDNA}         &     0.3           & ---       & 6                          & 20 MHz                 \\
   &  \cite{pinto2020Readout}         & 1.3         & ---    & 10                          & 1 MHz               \\ [0.1in]
  NV-NMR      &  \cite{Staudacher2013nuclear}        & ---     & $\sim 10^{3}$     & 7                          & 20 kHz                 \\ 
       &  \cite{mamin2013nanoscale}        & ---     & $\sim 10^{5}$     & 20                          & 20 kHz                 \\ 
     &   \cite{sushkov2014Magnetic}     &     ---   & 0.36        & 3.3                         & $\sim$ MHz                \\
   &   \cite{muller2014Nuclear}       & ---   & 0.06         & 2 to 3                           & 10 kHz                 \\
   &     \cite{haberle2015Nanoscale}     & ---     & 6$\times 10^{4}$        & 17                          & 20 kHz                  \\
 &     \cite{Lovchinsky2016}  & ---   & 1           & 3                           & 5 kHz                  \\
   &   \cite{aslam2017Nanoscale}         & ---     & 5$\times 10^{5}$       & 34                          & 1 mHz                  \\
     &   \cite{yang2018Detection}         & ---     & $\sim 10^{3}$       & 6                          & 5 kHz                  \\
 \bottomrule
\end{tabular}
\end{table*}

\subsection{Electron paramagnetic  resonance (EPR)}

\EPR is crucial for determining the structure, dynamics, and spatial distribution of paramagnetic species \cite{borbat2001Electron}. Although most biological molecules are non-paramagnetic and thus cannot be measured by this technique, labeling biological molecules with a small spin-bearing moiety, such as nitroxide spin labels, enables EPR to acquire a broad range of structural and dynamical information. Compared to non-labeling NMR techniques, EPR technique is much more sensitive because the electron has a much greater magnetic moment compared to a nucleus. Moreover, the \NV electron spin and the target labeled electron spin exhibit magnetic dipolar interaction. Denoting the spins by $\mathbf{S}_{\text NV}$ (\NV sensor spin), $\mathbf{S}_e$ (target electron spin), and $\mathbf{I}$ (nuclear spin of the target electron spin), the combined system Hamiltonian is expressed as follows:

\begin{align}\label{eqn:esr_ham}
    H  = & \underbrace{\mathcal{D} S_{z,\text{NV}}^2+\gamma_{\text{NV}}\mathbf{B}_0 \cdot \mathbf{S}_{\text{NV}}}_{H_0}\\\nonumber
   &  +\underbrace{\frac{\mu_0}{4\pi}\gamma_{\text NV}\gamma_e\hbar\frac{\mathbf{S}_{\text{NV}}\cdot\mathbf{S}_{e}-3(\mathbf{S}_{\text{NV}}\cdot\hat{\mathbf{r}})(\mathbf{S}_{e}\cdot\hat{\mathbf{r}})}{r^3}}_{H_{\text{NV},\text{ele}}}\\\nonumber
   &  +\underbrace{\gamma_e\mathbf{B}_0\cdot\mathbf{S}_{e}+\mathbf{S}_e\cdot \mathbb{A}\cdot\mathbf{I}-\gamma_n\mathbf{B}_0\cdot\mathbf{I}+\mathbf{I}\cdot \mathbb{Q}\cdot\mathbf{I}}_{H_{\text{ele}}},
\end{align}
where $\mathbf{B}_0$ is the external static field, $\mathbb{A}$ is the hyperfine tensor, $\mathbb{Q}$ is the nuclear quadrupole coupling tensor, $\mathbf{r}$ and $\hat{\mathbf{r}}$ is the distance and the direction between \NV and target electron spin. Also  $\gamma_{\text{NV}}$, $\gamma_e$ and $\gamma_n$ are the gyromagnetic ratios of the NV, target electron and nuclear spin, respectively. 

\subsubsection{Double electron-electron resonance (DEER)} \label{sec:deer}
\DEER is one of the most  widely used techniques to characterize the coupling or the distance between two electron spins \cite{schweiger2001Principles}. Under the secular approximation, the effective coupling is
\begin{equation}
H_{\text{dip}}=\frac{\mu_0}{4\pi}\frac{\gamma_{\text NV}\gamma_{\text{tar}}\hbar}{r^3}(1-3\cos^2\theta),\label{Hdip}
\end{equation}
where $r$ denotes the distance between the NV center and the target electron spin, $\theta$ is the angle between NV-target-spin connection direction and the external magnetic field. By resolving the coupling strength, we detect the label spin using \ssmr.

\begin{figure}[htp]
\begin{overpic}[width=1\columnwidth]{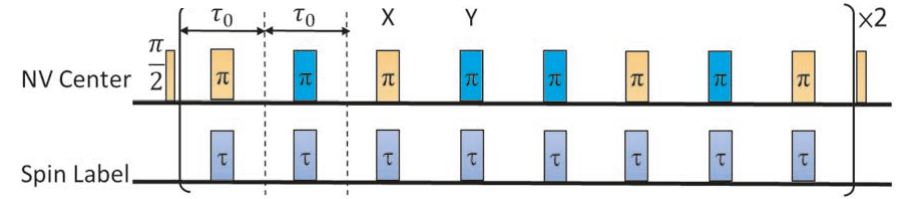}
\end{overpic}
\begin{overpic}[width=0.15\columnwidth]{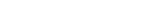}
\end{overpic}
\begin{overpic}[width=1\columnwidth]{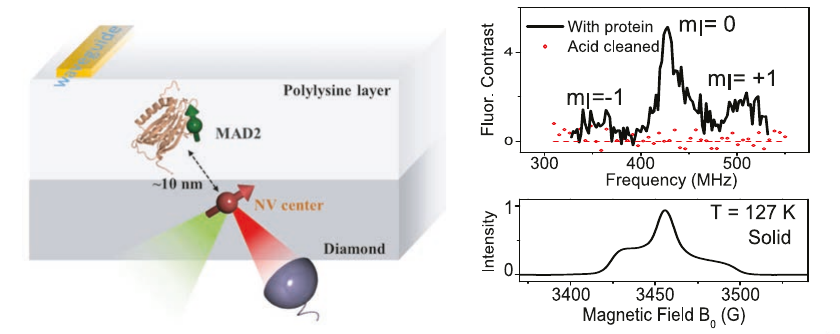}
  \put (0, 36) {(b)}
  \put (53, 36) {(c)}
  \put (53, 16) {(d)}
  \put (0, 63) {(a)}
\end{overpic}
\caption[DEER_sequence]{ Single-molecule \epr realized by \deer method. (a) The \deer pulse control sequence to probe the coupling between the \NV spin and the labeled electron spin on single protein. The pulse sequence indicates the timing and order of pulses applied to the spins. The microwave frequency is resonant with the \NV spin, and the radio frequency is resonant with the labelled electron spins. A \dyde sequence is applied on the \NV spin while synchronous series $\pi$ pulses are applied simultaneously with the microwave $\pi$ pulses. Synchronously driving the \NV electron spin and the target spin allows for preservation of the coupling between the \NV spin and the target spin, while eliminating most non-synchronous magnetic noise. (b) The protein is labeled using nitroxide spin labels. Moreover, the protein is reliably immobilized on the diamond surface close to the \NV center by embedding in a polylysine layer. Microwave radiation is delivered through a coplanar waveguide, and the fluorescence is collected using a confocal system. (c) The single spin \epr spectra under ambient conditions. The spectrum disappears after removing the protein via acid cleaning. (d) The ensemble ESR spectrum of protein molecules in a frozen buffer solution with glycerine at 127 K. The figures are adapted from \cite{shi2015Singleprotein}. }  \label{fig:deer_seq}
\end{figure}

The quantum sensing control protocols are shown in Fig. \ref{fig:deer_seq}(b). The \NV sensor is firstly initialized to {$\ket{0}$} state. Then an initial $\pi/2$ pulse prepares the sensor in a coherent superposition state. Afterwards a series of \DyDe $\pi$ pulses $(\tau/2-\pi-\tau/2)^N$ with resonant frequency $\omega_{\mathrm{NV}}=\mathcal{D}\pm \gamma_eB_0$, are applied to NV electron spin, while the target labeled electron spin is flipped synchronously with $\pi$ pulses. Detecting the single-electron spin label is challenging because the coherence time of shallow NV centers is usually limited by surface noise (see \se \ref{sec:decoherence}), which obscures the coupling between the spin label and the NV center. The \DyDe control pulses elongate the coherence time of NV electron spin, thereby facilitating single electron spin sensitivity. The signal preserves during the evolution, and other noise is eliminated when the driving frequency matches the resonant frequency of the target electron spin. The final signal is observed by the real component readout as \cite{sushkov2014Magnetic,shi2015Singleprotein,schlipf2017molecular},
\begin{align}\label{eqn:deer_signal}
S_{\text{DEER}}=\frac{1}{2}\left[1+\prod_n \cos \left(H_{n,\text{dip}} N\tau\right)\right],
\end{align}
where $H_{n,\text{dip}}$ is the effective coupling of the $n$th labeled electron spin. 

Although single electron spins inside diamond or on diamond surfaces have been sensed \cite{grinolds2011Quantum,grinolds2013Nanoscale,grinolds2014Subnanometre}, the first electron spin resonance on a single protein is realized in  \cite{shi2015Singleprotein}. Single nitroxide spin labeled MAD2 (mitotic arrest deficient-2) molecules are chosen as an essential spindle checkpoint protein \cite{martin-lluesma2002Rolea,rothman1992Molecular} and are reliably immobilized on the diamond surface by embedding them in a polylysine layer, as shown in Fig. \ref{fig:deer_seq}(a). The positions to the single NV centers rely on statistical proximity. For this purpose, the protein surface concentration is optimized  and various NV centers are tested for dipolar interaction with single spin labels.

The single-molecule \epr spectrum is obtained by sweeping the driving microwave field frequency (Fig. \ref{fig:deer_seq}(c)). The spin coherence transfer will occur when the driving field frequency is resonant with the target electron spin. Three clearly resolved spectral peaks indicate that the detected spin is the nitroxide-labeled electron spin on the protein.
In the spectrum of a solid-state ensemble EPR, the random orientation of the molecular spin principal axis leads to the broadening of spectral peaks, as depicted in Fig. 9(c). However, in the case of the single molecule EPR, the fast tumbling of molecules averages out the anisotropic terms of the hyperfine interaction. Single spin \epr allows reliable determination of anisotropic hyperfine coupling, which is crucial for structural and dynamical information.

\begin{figure*}[htp]
\begin{minipage}[h]{0.27\textwidth}
\centering
\begin{overpic}[width=\textwidth]{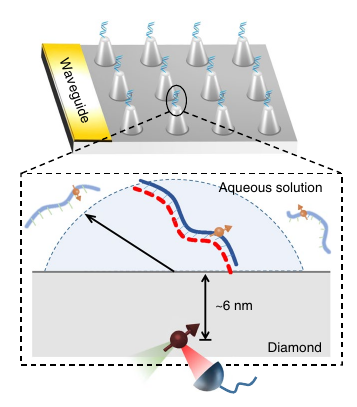}
\put (-5, 95) { (a) }
\put (-5, -3) { (b) }
\put (89, 95) { (c) }
\put (89, 8) { (d) }
\put (190, 95) { (e) }
\put (190, 4) { (f) }
\end{overpic}
\begin{overpic}[width=\textwidth]{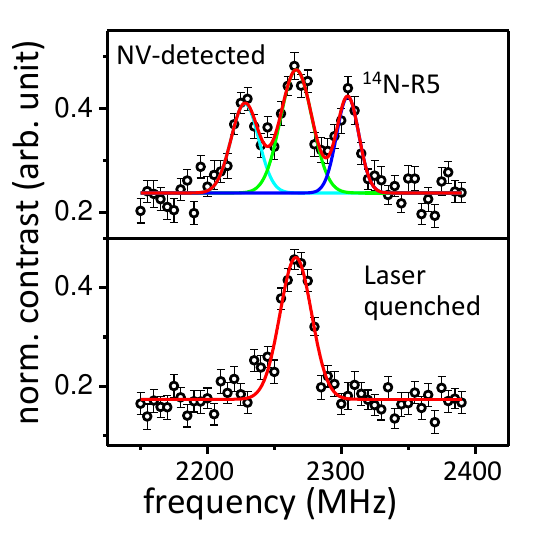}
\end{overpic}
\end{minipage}
\begin{minipage}[h]{0.3\textwidth}
\begin{overpic}[width=\textwidth]{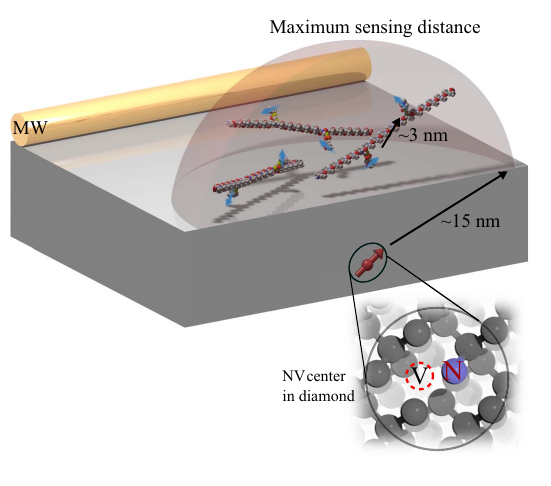}
\end{overpic}
\begin{overpic}[width=\textwidth]{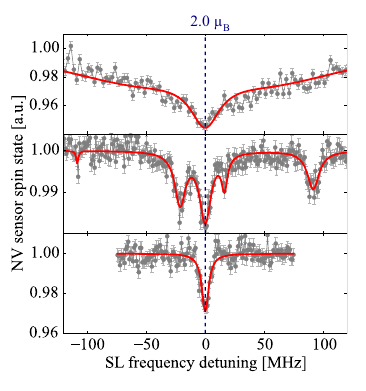}
\end{overpic}
\end{minipage}
\begin{minipage}[h]{0.35\textwidth}
\begin{overpic}[width=0.85\textwidth]{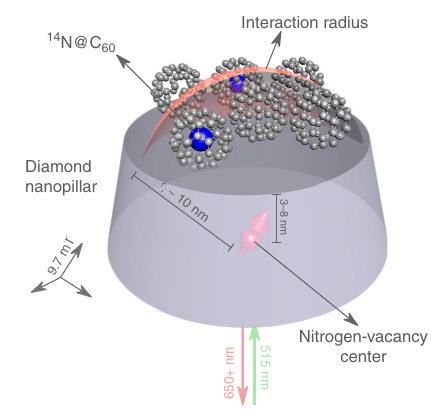}
\end{overpic}
\begin{overpic}[width=\textwidth]{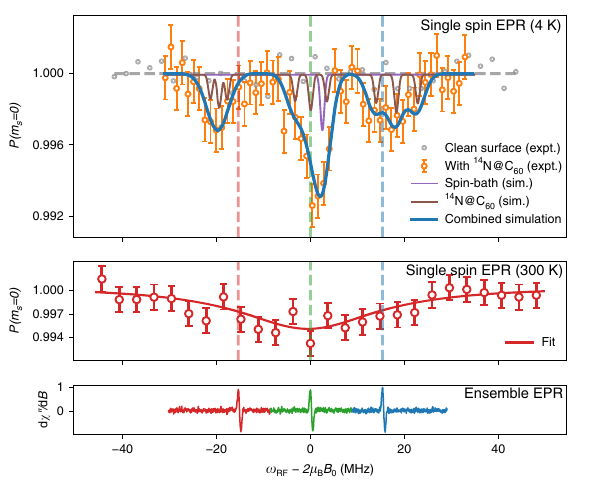}
\end{overpic}
\end{minipage}
\caption[NV Center]{ Single-molecule EPR. (a) Single-molecule EPR in liquid. The diamond is attached with DNA on the diamond pillar surface. The surface-tethered DNA strand (indicated by a red dashed line) hybridizes with a strand (indicated by a blue solid line) that contains a spin label (indicated by an orange arrow), leading to the localization of a spin-labeled duplex within the detection volume of an NV center (indicated by a dark red arrow).
(b) Top: the single electron spin EPR. The splitting features the \Nfor-R5 spin label. Bottom: the \Nfor-R5 spin label electron spin is quenched in the detected spectrum after the laser irradiation. (a-b) are adapted from \cite{shi2018SingleDNA}. (c) The EPR spectra of multiple electron spin have been measured at $T$ = 4.2 K in ultrahigh vacuum. The target sample, doubly spin-labeled polyprolines, is coupled with shallow implanted NV center. The hemisphere shown in the figure represents the sensing range of the NV sensor and is approximately the same size as the depth of the NV sensor. (d) Measured spectra on the sensor. From top to bottom: blank spectrum without spin-labeled peptides on surface, spectrum of diluted polyproline labels on diamond surface (labelled and unlabelled ratio 1:10) and spectrum of the cleaned diamond surface after above measurement. (c,d) are adapted from \cite{schlipf2017molecular}. (e) The N@C$_{60}$ diluted with empty C$_{60}$ cages are detected by the NV sensor. (f) From top to bottom. Top, the EPR spectrum of 
N@C$_{60}$ on diamond surface at low temperature. The solid blue line indicates the simulation of single N@C$_{60}$ electron spin EPR spectrum. The dashed vertical lines are the positions of ensemble EPR hyperfine components. Middle, spectrum under  ambient temperature, the linewidth is largely broadening. Down, ensemble EPR. (e,f) are adapted from \cite{pinto2020Readout}. }  \label{fig:single_molecule_epr_three}
\end{figure*}

The single-molecule \epr \cite{shi2015Singleprotein} is demonstrated with a single spin-labeled protein immobilized in a solid layer. However, the majority of biomolecules function in aqueous solution under ambient temperature, where they undergo a range of motions. NV detection of single-molecules under near-physiological conditions presents considerable additional challenges compared with studies performed with a stationary solid phase. Chemical tethering scheme \cite{Lovchinsky2016,shi2018SingleDNA} is executed to confine the spin-labeled molecule (Fig. \ref{fig:single_molecule_epr_three}(a)), a DNA duplex, within $\sim$ 10 nm from \ a shallow NV center \cite{shi2018SingleDNA}. With the implementation of a diamond pillar array design (see Sec \ref{sec:diam_engineer}), the detection time for an \epr spectrum reduces by approximately one order compared with \cite{shi2015Singleprotein}. This strategy enables multiple \epr spectra before the labels to be quenched via laser irradiation. The resolved \Aiso values from the spectra vary by $\sim$ 12\%, reflecting heterogeneity among the individual molecules (Fig. \ref{fig:single_molecule_epr_three}(b)), which may reflect heterogeneity of the polarity profile at the individual DNA duplexes; nevertheless, several other factors (e.g., spin label dynamics and local electrostatics) cannot be completely ruled out. The liquid-state single molecule EPR spectra indicate that analysis of an NV center-based single-molecule spectra can render information on the local environment and motion dynamics of a bio-molecule, such as variations in DNA conformation and interactions between DNA and proteins or ligands.

One of the crucial obstacles for ambient temperature single molecule EPR is the laser induced spin label quenching \cite{shi2018SingleDNA,schlipf2017molecular}. By conducting the experiment at liquid helium temperature (4.2 K) and in a UHV environment (<$10^{-8}$ Pa), the photo-bleaching of the electron spin label is eliminated. Common methanethiosulfonate spin label (MTSSL) bearing peptides as a network spin is covered on the diamond containing the NV sensor (Fig. \ref{fig:single_molecule_epr_three}(c)). Subsequently, the collective readout and coherent manipulation of very few ($\leqslant$ 6) of these $S$ = 1/2 electron spin systems and access their direct dipolar coupling tensor is demonstrated as shown in Fig. \ref{fig:single_molecule_epr_three}(d) \cite{schlipf2017molecular}. Another spin system, ${ }^{14}\mathrm{N} @ \mathrm{C}_{60}$ fullerene cages \cite{olyanich2013manipulation,ozmaian2016Diffusion,smith1998Encapsulated}, with the capability to self-assemble on monolayer graphyne sheets, is also measured and controlled using similar technology as shown in Fig. \ref{fig:single_molecule_epr_three}(e,f) \cite{pinto2020Readout}.

\subsubsection{Quantum relaxometry} \label{sec:relaxometry}

Quantum technology often relies on the preservation of quantum states; however, relaxation can lead to dissipation of quantum states, thereby hindering quantum technology. 
Magnetic field fluctuations generated by spins are common in nanoscale biology. Quantum relaxation provides an unprecedented opportunity for nanoscale measurements, as it provides valuable information regarding processes that generate these fluctuations, rather than presenting an obstacle to quantum technology. 
Conventional techniques include magnetic resonance force microscopy \cite{rugar2004Single}, atomic vapor magnetometers \cite{kominis2003subfemtotesla} or SQUIDs \cite{greenberg1998Application}, lack the ability to combine high sensitivity, nanoscale spatial resolution, and biologically living conditions, which limits their scope of application. In contrast, \sqs yields an opportunity in this field. By measuring relaxation, \ssmr allows sensing of samples labeled with Mn$^{2+}$ \cite{ziem2013Highly}, Fe$^{3+}$ \cite{ziem2013Highly,ermakova2013Detection,schafer-nolte2014Tracking,wang2019Nanoscale,radu2020Dynamic}, Cu$^{2+}$ \cite{simpson2017Electron}, La$^{3+}$ \cite{radu2020Dynamic}, Gd$^{3+}$ \cite{sushkov2014alloptical,rendler2017Opticala,gorrini2019Fast}, pH value \cite{fujisaku2019PH}, and free radicals \cite{barton2020Nanoscale,peronamartinez2020Nanodiamond,vanderlaan2018Using} involved in chemical reactions.

\begin{figure}[htp]
\begin{overpic}[width=0.65\columnwidth]{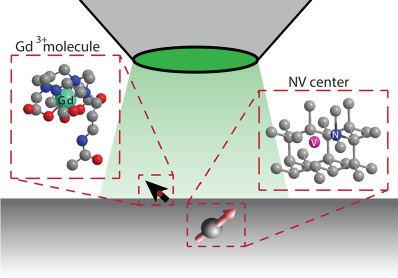}
\end{overpic}
\begin{overpic}[width=0.69\columnwidth]{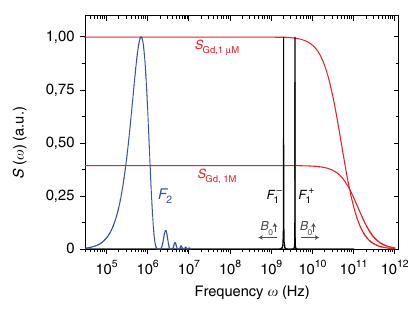}
	\put (-10, 136) { (a) }
	\put (-10, 68) { (b) }
	\put (-10, 0) { (c) }
\end{overpic}
\begin{overpic}[width=0.65\columnwidth]{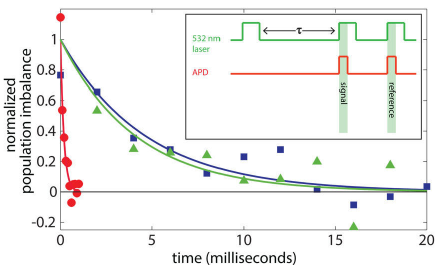}
\end{overpic}
\caption[NV Center]{  Schematic of single-molecule quantum relaxometry. (a) A single shallow NV center is used as an \sqs. A single-molecule labelled as ion (Gd$^{3+}$) is attached on the surface of the diamond with a shallow NV center. The NV center is shed with laser, and fluorescence is observed using a confocal microscope. The figure is adapted from \cite{sushkov2014alloptical}. (b) 
The noise spectral density for ions (Gd$^{3+}$) is shown with different concentrations (red curves), demonstrating the broadening effect of coupling ($f_{\text{dipole}}$) at higher concentrations. Although the filter function of \dyde (blue, $F_2$) is limited to $\sim$ MHz fluctuations, $T_1$-relaxometry can probe a wide frequency range up to GHz with two sensitivity windows ($F_1^+$ and $F_1^-$ for $m_s=\pm 1$ transition), shown in black. $F_1^{\pm}$ can be Zeeman-shifted via $B_0$, enabling the experimental detection of $S_{\text{ion}}(\omega)$. (c) Quantum relaxometry measurement protocols. The fluorescence is measured after initialization and a waiting time of $\tau$. The presence of a nearby target molecule can increase the relaxation of the NV sensor, represented by the red line. Figures (b-c) are adapted from \cite{steinert2013Magnetic}. } \label{fig:ion}
\end{figure}

The target single non-fluorescent molecule can be attached covalently on the diamond surface (Fig. \ref{fig:ion}(a)). The sensing protocol is performed under ambient condition with only optical method involved in the experiment. The paramagnetic component is introduced with metallic ions such as Mn$^{2+}$, Fe$^{3+}$, Cu$^{2+}$, La$^{3+}$, Gd$^{3+}$. The paramagnetic ion spins generate a fast fluctuating magnetic field that is described by the magnetic fluctuation spectral noise density $S(\omega)$ (Fig. \ref{fig:ion}(b)),
\begin{equation}
S(\omega)=\frac{\gNV^2\langle B^2_{\perp,\text{ion}}\rangle}{\pi} \frac{f_{\mathrm{ion}}}{f_{\mathrm{ion}}^{2}+\left(\omega-\omega_{0}\right)^{2}},\label{eq:psd_ion}
\end{equation}
where $f_{\text{ion}}$ is the relaxation rate of the ion and $\omega_0$ represents the Larmor precession of the ion. The fluctuation strength $\langle B^2_{\perp,\text{ion}}\rangle$ depends on the coupling strength between the NV center spin and the ion spin \cite{steinert2013Magnetic},
\begin{equation}
\langle B^2_{\perp,\text{ion}}\rangle=\frac{21 \cdot 10^{3} \pi N_{\mathrm{A}} c_{\mathrm{ion}}}{16 d^{3}}\left(\frac{\mu_{0} \hbar}{4 \pi}  \gamma_{\mathrm{ion}}\right)^{2}, \label{eqn:ion_field}  
\end{equation}
where $c_{\text{ion}}$ represents the ion concentration, $N_A$ denotes the Avogadro constant, $\gamma_{\mathrm{ion}}$ is the gyromagnetic ratio of ion, and $d$ is the sensor depth.

The rate of \sqs relaxation \cite{steinert2013Magnetic,ziem2013Highly} for the ion can be calculated using the following equation:
\begin{equation}
f_{\text{ion}}=f_{\text{dip}}+f_{\text{vib}}+f_{\text{trans}}+f_{\text{rot}}. \label{eqn:ion_linewidth}
\end{equation} 
 Herein, $f_{\text{dipole}}= c_{\text{ion}}\cdot 77 $ GHz $\text{M}^{-1}$ is a concentration-dependent fluctuation rate that is induced by the dipolar coupling between ion spins.
The magnetic fluctuations undergo a significant broadening, $f_{\text{vib}}$, due to intrinsic relaxation from surrounding chemical environment. The broadening $f_{\text{vib}}$  is determined by the ensemble EPR spectroscopy. The motion of spins in aqueous solution also induces the spin relaxation. The fluctuation rate of the effective fluctuation field due to translational spatial diffusion is thus \cite{steinert2013Magnetic}
\begin{equation}
f_{\text {trans }}= D_{\text {diff }}\left(\frac{3}{4 d}\right)^{2},
\end{equation}
where $D_{\text{diff}}$ is the diffusion coefficient. Furthermore, molecular rotation yields a fluctuation rate governed by Stokes's law with 
\begin{equation}
f_{\text {rot }}= \frac{k_{B} T}{4 d^{3} \varrho},
\end{equation}
where $\varrho$ is the fluid viscosity. The motion induced spin relaxation is on the order of $\sim$ 0.1 GHz in water at ambient temperature \cite{steinert2013Magnetic}.

Thus, the metallic ion labeled molecules on a diamond surface is sensed by a single shallow NV center acting as a magnetometer. In the presence of nearby paramagnetic molecules, a relaxation process occurs naturally if the noise spectrum suffices to cover the filter function of the NV center. This type of measurement does not require any microwave controls, and thus, is purely optical in nature. In the absence of any quantum controls, the evolution of NV center follows a normal relaxation process, where the population of $\ket{0}$ state is described by 
\begin{equation}
  P(\ket{0}) = \frac{1}{6} \left(2+ e^{-\Gamma_1^- t}+ e^{-\Gamma_1^+ t}+ 2e^{-(\Gamma_1^-+\Gamma_1^+) t}\right),
\end{equation}
The longitudinal relaxation rate $\Gamma_1^{\pm}$ is given by \cite{simpson2017Electron}
\begin{equation}
  \Gamma_1(B_0) = \Gamma_{1,\text{int}} +\int_{-\infty}^{\infty}\frac{\Gamma_2}{2(\Gamma_2^2+(\omega_{\text{NV}}(B_0)-\omega)^2)}S(\omega)
   \text{d} \omega,
\end{equation}where $\Gamma_{1,\text{int}} = 1/T_1$ is the intrinsic relaxation rate, 
$\Gamma_2 = 1/T_2^{*}$ is the dephasing rate of the NV center, $\omega_{\text{NV}}^{\pm}(B_0)=\mathcal{D} {\pm} \gamma_{\text{NV}}B_0$ is the transition frequency,
and $S(\omega)$ is the magentic fluctuation spectral noise density of the ion (Eq. \ref{eq:psd_ion}). The maximum relaxation rate occurs at the crossover point, and $\Gamma_1(B_0)$ represents the \epr spectrum based on quantum relaxometry.

All-optical quantum relaxometry has achieved single molecule sensitivity. A single paramagnetic molecule is labeled with a gadolinium ion (Gd$^{3+}$) chelated by an amine-terminated organic ligand \cite{sushkov2014alloptical}. 
The surface density of molecules could be controlled by varying the concentration of the Gd$^{3+}$ molecules during the reaction. The procedure can be used to covalently attach any water-soluble amine-terminated molecule to the diamond surface with controlled surface coverage. Fig. \ref{fig:ion}(c) depicts the change in \NV spin relaxation rate ($T_1$) caused by a single molecule \cite{sushkov2014alloptical}. The advantage of all-optical techniques is their convenience, which may have implications for studying a broad range of biochemical molecules and processes without requiring microwave or radio-frequency control \cite{sushkov2014alloptical,ermakova2013Detection,kaufmann2013Detection,simpson2017Electron,rendler2017Opticala,steinert2013Magnetic,ziem2013Highly,barton2020Nanoscale}. 

\begin{figure}[htp]
	\begin{overpic}[width=1\columnwidth]{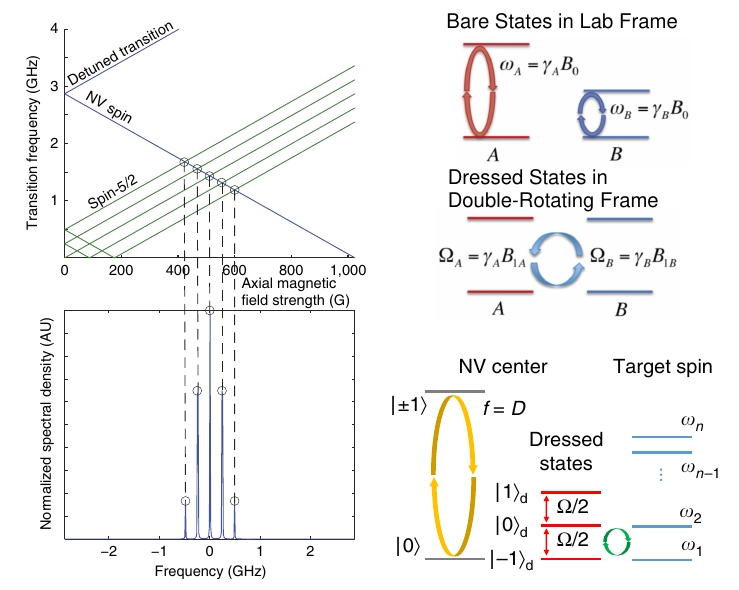}
		\put (0, 76) {(a)}
		\put (53, 76) {(b)}
		\put (53, 31) {(c)}
	\end{overpic}
	\caption[NV Center]{ Three  types of EPR spectroscopic methods based on electron relaxometry. (a) Sweeping magnetic field \cite{hall2016detection}. Resonance  occurs at the energy level crossover. The figure is adapted from \cite{hall2016detection}. (b) Tuning energy levels of dressed states by continuously driving \cite{Belthangady2013Dressed}. Resonance  occurs when the two driving powers match. The figure is adapted \cite{Belthangady2013Dressed}. (c)   \sqs is driven to the dressed states \cite{kong2018Nanoscale}. Resonance occurs when the Rabi frequency matches the energy splitting of the bare states. The figure is adapted from \cite{kong2018Nanoscale}.}  \label{fig:electron_relaxametry}
\end{figure}

The relaxation signal originates from the spectral overlap between the NV sensor filter function and the noise frequency of the target spin. Adjusting the magnetic field is one of the most convenient ways to study EPR spectroscopy. Adjusting the magnetic field can accordingly change the energy levels of both the NV sensor and the target, allowing for the detection of the characteristic spectra of several paramagnetic species. For instance, in detecting the electron spin label of a nitroxide free radical, the characteristic frequency of the noise signal increases almost linearly with the external magnetic field, at approximately $\sim \gamma_{\text{e}} B_0$, while for the NV center, the transition frequency changes with the magnetic field, as $\mathcal{D} - \gamma_{\text{NV}}B_0$. A crossover occurs at $ \mathcal{D}/(\gamma_{\text{NV}}+ \gamma_{\text{tar}})$, based upon which the $g$-factor of the target spin is obtained.

The spins with different resonant frequencies  cannot effectively couple in the laboratory frame. In addition to adjusting the magnetic field, another approach involves utilizing the dressed state method for \epr spectroscopy, with the Hamiltonian given by
\begin{equation}
  H_{\text{NV,tar}} = \Omega_{\text{NV}} S_{x,\text{NV}} + \Omega_{\text{tar}} S_{x,\text{tar}} + H_{\text{dip}}S_{z,\text{NV}}S_{z,\text{tar}}.
\end{equation}
Energy transfer between two resonantly-driven spins can occur with  $\Omega_{\text{NV}}=\Omega_{\text{tar}}$, similar to the Hartmann--Hahn matching condition. 
In contrast to the aforementioned magnetic-field-adjusting method, this method can be applied to all magnetic fields if the corresponding driving field strength is available. The driving on the target is not necessary when the driving field on the NV sensor is sufficiently strong \cite{kong2018Nanoscale} (Fig. \ref{fig:electron_relaxametry}(c)). Subsequently, as the resonance condition, the Rabi frequency of NV sensor matches the bare transition frequency of the target spin. 
This method is suitable for cases wherein the target spin is not easy to control or is unstable, akin to a molecule exhibiting rotational dynamics.

\subsection{Nuclear magnetic resonance (NMR)} \label{sec:NMR}

\begin{figure*}[htp]
\begin{overpic}[width=1\textwidth]{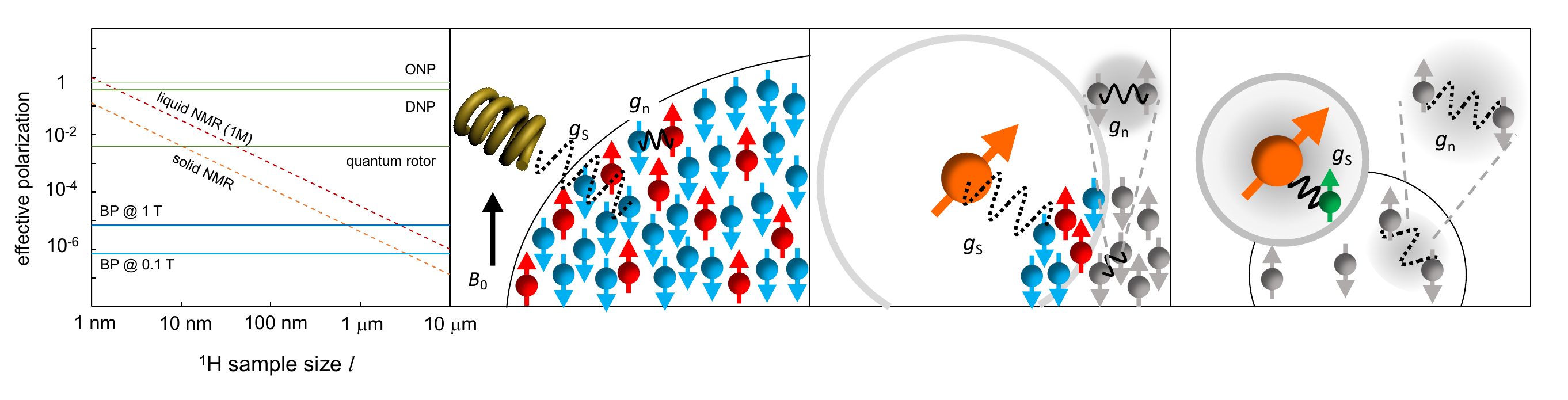} 
	\put (7, 21) {(a)}
	\put (30, 21) {(b)}
	\put (53, 21) {(c)}
	\put (76, 21) {(d)}
  \put (36, 1.2) {$S \propto \frac{N \gamma_n B_{0}}{k_{B} T} {g_s} {t_s}$}
  \put (60, 1.2) {$S\propto N{g_s^2}{t_s^2}$}
  \put (76, 1.2) {$S\propto\left\{\begin{array}{ll} {g_s}{t_s}, & \text{if }{t_s} <T_{\text{1,tar}} \\ {g_s^2}{t_s^2}, & \text{if }{t_s} >T_{\text{1,tar}} \end{array}\right.$}
\end{overpic}
\caption[NV Center]{ Polarization-dependent and fluctuation-dependent NMR. (a) Statistical fluctuations (indicated by effective polarization in figure) may exceed the thermal or enhanced nuclear spin polarization depending on the sample size. The typical polarization levels reached by ONP \cite{tateishi2014Room}, DNP \cite{ardenkjaer-larsen2003Increase} and quantum rotor polarization \cite{roy2015Enhancement} are shown as light green, green, and dark green lines, respectively. The Boltzmann polarization (BP) of proton under 0.1 T and 1 T are displayed as dark blue and blue lines, respectively. The estimated effective polarization of liquid (1 M protons concentration) and solid samples (proton density 50 nm$^{-3}$) are shown as the dark red and red dashed curves depending on the sample size $l$, respectively. (b) Polarization-dependent NMR. For ensembles with large spins ($\gtrsim 10^{12}$ nuclear spins), the signal $S$, which corresponds to the time-averaged nuclear induction magnetic field, is proportional to the sample polarization. Under high-temperature conditions, the sample magnetization can be calculated as $N\gamma_nB_0/k_{\text B}T$, where $N$ is the number of spins, $B_0$ is the external magnetic field, $T$ is the temperature and $k_{\text B}$ is the Boltzmann constant. (c) For detection of small ensembles, the signal $S$ is proportional to the statistical fluctuation. The effective sample magnetization is ${\propto}\sqrt{N} $ if the statistical fluctuations exceed polarization. The signal is observed by the real-component readout; thus, $S\propto N{g_s^2}t^2$. (d) Detection of individual nuclear. The optimal strength signal $S$ depends on the nuclear spin relaxation time $T_{1,\text{tar}}$. }  \label{fig:nmr_pol_flu}
\end{figure*}

NMR spectroscopy is a powerful tool to determine the chemical makeup of macromolecules and for resolving the structures of organic compounds and biological macromolecules \cite{wuthrich2001way,yves2010Physics}. However, the weak interaction strengths between sample nuclear spins and the conventional NMR probe and the low polarization of the sample nuclear spins require macroscopic sample quantities to investigate and are difficult to apply directly in the study of nanoscale molecules. 
However, conventional NMR offers limited sensitivity due to the weak interaction strength between sample spins and the inductive detectors, and the low-thermal polarization of the sample nuclear spins Fig. \ref{fig:nmr_pol_flu}(b). In addition, the maximum magnetic field gradient restricts the spatial resolution of NMR imaging methods \cite{glover2002Limits}. Thus, the conventional NMR requires macroscopic sample quantities to investigate and are difficult to apply directly to the study of nanoscale molecules.
By miniaturizing the detector to approach the sample closer, we can enhance sensor--sample coupling. In particular, the signal of conventional \NMR is proportional to the magnetic polarization $P^{\text{nuclear}}$ of the nuclear spins \cite{Levitt2008,Abragam1983},
\begin{align}
P^{\text{nuclear}}  \approx \frac{\gamma_n B_0 \hbar}{2 k_{\text B} T},
\end{align}
where $k_{\text B}$ is the Boltzmann constant and $T$ is the temperature. The nuclear spin polarization can be further increased by hyperpolarization approaches such as optically-induced polarization \cite{tateishi2014Room}, dynamic nuclear polarization \cite{ardenkjaer-larsen2003Increase,gajan2014Hybrid,griesinger2012Dynamic}, and quantum-rotor-induced polarization \cite{meier2013LongLived,roy2015Enhancement} (Fig. \ref{fig:nmr_pol_flu}(a)). The statistical fluctuation $\sqrt{N}$ (Fig. \ref{fig:nmr_pol_flu}(c)) is always significantly weaker than the thermal polarization or the hyperpolarization for conventional NMR techniques \cite{mccoy1989Nuclear,sleator1985Nuclearspin,muller2006Nuclear}. However, recently developed sensors, including \NV sensors, the superconducting quantum interference device (SQUID) sensor \cite{vasyukov2013scanning}, and the \mfm sensor have the capability to decrease the sensor--sample distance to within 100 nm \cite{mitchell2020Colloquium}. This proximity allows the statistical fluctuations to invariably exceed the thermal nuclear spin polarization (Fig. \ref{fig:nmr_pol_flu}(a)). Thus, for the NV sensor-based nanoscale or even single-molecule NMR (Fig. \ref{fig:nmr_pol_flu}(c)), the detection of statistical fluctuations signal can potentially enhance the sensitivity. The sensitivity of the protocol depends critically on the stand-off distance between the NV sensor and the target nuclear spins.  

Let's consider a target sample containing proton nuclei—perhaps a liquid sample with a concentration of 1 M (1 mol/L) protons or a solid sample featuring a proton density of 50 nm$^{-3}$. At spatial scales around 10--100 nm, the statistical fluctuation exceeds both the Boltzmann polarization (equivalent to 1 T at ambient temperature) and the nuclear polarization calculated through the quantum rotor method (Fig. \ref{fig:nmr_pol_flu}(a)). These fluctuations approach the polarization achieved by \onp or \dnp methods, especially on the spatial scale of approximately nanometers, even reaching the scale of a single molecule or single spin. The NV center presents a unique opportunity to position a single quantum sensor in close proximity to the target sample, comprising a few nuclear spins. This spatial proximity enables robust coupling between the quantum sensor and the nuclear spins of the sample. Additionally, the NV center generates a dipolar field from its electron spin, serving as a field gradient for \mri. With the ability to observe nuclear spin polarization or fluctuations, the \sqs using NV centers proves ideal for conducting nanoscale and \sm \nmr spectroscopy.

In the context of NV-based NMR detection, a hyperfine interaction exists between the \NV sensor and the nuclei. Notably, the Fermi contact interaction between the \NV and nuclear spins becomes negligible for distances exceeding 2 nm \cite{gali2008initio,nizovtsev2018Nonflipping13C}. Consequently, the dynamics of the NV-nuclear spin system are governed by the following Hamiltonian:
\begin{widetext}
\begin{align}\label{eqn:nmr_ham}
    H  = & \underbrace{\mathcal{D} S_{z,\text{NV}}^2+\gamma_{\text NV} \mathbf{B}_{0} \cdot \mathbf{S}_{\text{NV}}}_{H_0} +\underbrace{\sum_j\frac{\mu_0\gamma_{\text NV}\gamma_n\hbar}{4\pi}\left[\left(-\frac{8\pi}{3}\rho_s(\mathbf{r}_j)+\frac{1}{r_j^3}\right)\mathbf{S}_{\text{NV}}\cdot\mathbf{I}_j-\frac{3(\mathbf{S}_{\text{NV}}\cdot\mathbf{r}_j)(\mathbf{I}_n\cdot\mathbf{r}_j)}{r^5_j}\right]}_{H_{\text{NV},\text{nuclear}}}\\\nonumber
   &  \underbrace{-\gamma_{\mathrm{n}} \mathbf{B}_{0} \cdot \sum_{j} \mathbf{I}_{j} +\sum_{i<j}  \frac{\mu_{0} \gamma_{\mathrm{n}}^{2}\hbar }{4 \pi r_{i j}^{3}}  \left[\mathbf{I}_{i} \cdot \mathbf{I}_{j}-\frac{3\left(\mathbf{I}_{i} \cdot \mathbf{r}_{i j}\right)\left(\mathbf{r}_{i j} \cdot \mathbf{I}_{j}\right)}{r_{i j}^{2}}\right]  }_{H_{\text{nuclear}}},
\end{align}
\end{widetext}
where $\mathbf{I}_j=(I_{x,j},I_{y,j},I_{z,j})$ is the spin operator for $j^{\text{th}}$ nuclear, $\gamma_{n}$ is the gyro-magnetic ratio of the $j^{\text{th}}$ nuclear spin, $\mathbf{r}_j$ is the distance between the \NV sensor and the $j^{\text{th}}$ nuclear spin, $\rho_s(\mathbf{r}_j)$ is the electron wavefunction density of NV on the $j^{\text{th}}$ nuclear spin, and $\mathbf{r}_{ij}$ represents the distance between the $i^{\text{th}}$ and $j^{\text{th}}$ nuclear spins. In the interaction picture of the Hamiltonian $H_0$ and rotating wave approximation, where the coupling between the nuclear spin and the NV center is far slower than the rotating frame frequency, the full Hamiltonian Eq. \ref{eqn:nmr_ham} can be simplified as follows,
\begin{equation}\label{eqn:nmr_ham_equ}
H=\omega_{L} \sum_{j}^{N} {I}_{j}^{z}+{S}_{z} \sum_{j}^{N}\left(a_{\|,j} {I}_{z,j}+a_{\perp ,j}  {I}_{\perp,j}\right),
\end{equation}
where $\omega_{L}=\gamma_nB_0$ is the Larmor frequency of the nuclear spin, and
\begin{gather}
a_{\|}=\frac{\mu_{0} \gamma_{\mathrm{e}} \gamma_{\mathrm{n}} \hbar}{4 \pi r^{3}}\left(3 \cos ^{2} \theta-1\right)+a_{\text {iso }} \\
a_{\perp}=\frac{\mu_{0} \gamma_{\mathrm{e}} \gamma_{\mathrm{n}} \hbar}{4 \pi r^{3}} 3 \sin \theta \cos \theta ,
\end{gather}
are the parallel and perpendicular components of the hyperfine coupling respectively, where $a_{\text {iso }} $ is the isotopic component of hyperfine interaction.

The direct measurement of nuclear spin by an NV center poses challenges due to the energy mismatch between the resonant frequencies of the nuclear spin ($\omega_L=\gamma_nB$) and the \NV spin ($\omega_e=\mathcal{D}\pm \gamma_eB$). To address this, various techniques are employed to adjust these resonant frequencies. One strategy involves driving the NV center's electron spin using either chopped microwave pulses (\dyde, see \se \ref{sec:NMR:DD}) or continuous microwaves (\HH, see \se \ref{sec:NMR:HH}). Coherence transfer between the electron spin and the nuclear spin occurs when the driven frequency of the electron spin aligns with the nuclear Larmor frequency. An alternative technique, ENDOR, synchronously flips the nuclear spin with the electric spin of the NV center (see \se \ref{sec:NMR:ENDOR}). Beyond one-dimensional NMR, two-dimensional NMR is also employed to identify correlations between nuclei of different species exhibiting different frequencies in conjunction with the NV center (see \se \ref{sec:NMR:twod}).

\subsubsection{Dynamical decoupling (DD) } \label{sec:NMR:DD}
DD sequences offer a combination of quantum sensing with the safeguarding of NV quantum states against environmental decoherence. By aligning the control frequency of flipping the \NV electron spin with that of the nuclear spin, coherence transfer becomes achievable.

\begin{figure}[htp]
\begin{overpic}[width=0.98\columnwidth]{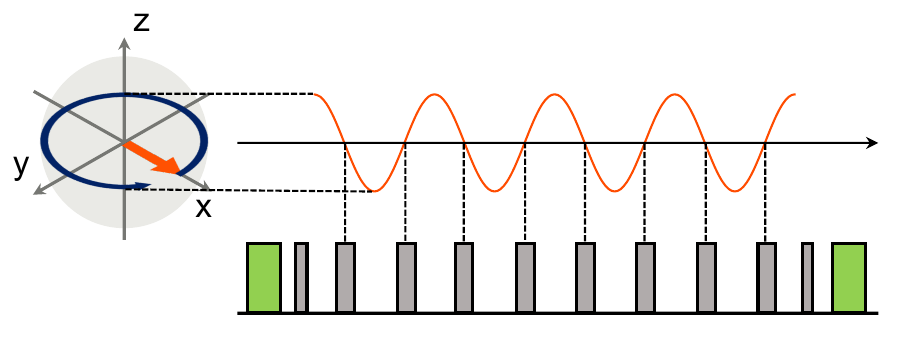}
\end{overpic}
\begin{overpic}[width=0.42\columnwidth]{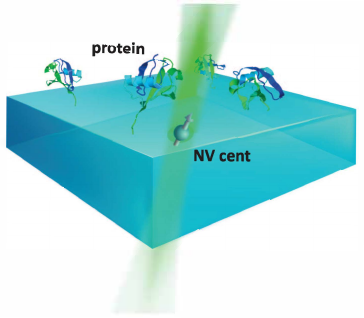}
\end{overpic}
\begin{overpic}[width=0.48\columnwidth]{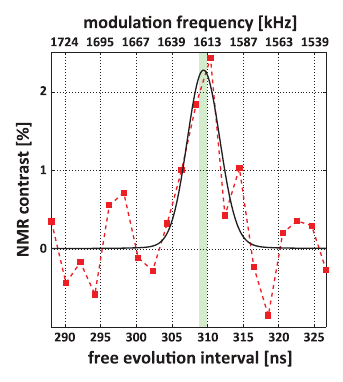}
  \put (-90, 165) { (a) }
  \put (-90, 83) { (b) }
  \put (-4, 83) { (c) }
\end{overpic}
\caption[NV Center]{ The NV based DD sensing protocol. (a) The \dyde sequence. The pulse sequence involves two green laser pulses that perform the initialization and readout of the \NV spin state (green blocks) and resonant microwave pulses (grey blocks) that start and end with $\pi/2$ pulses. In between, periodic $\pi$ pulses (grey blocks) are introduced with elements that repeat as $\tau/2-\pi-\tau-\pi-\tau/2$. (b) The schematic for the single-molecule NMR with NV sensor. The single proteins are covalently bonding on the diamond surface. (c) \Htwo NMR spectrum of an isotopically enriched ubiquitin protein (\Htwo at > 98\% abundance) at external magnetic field  247.3 mT, using the XY8-507 sequence and Gaussian fit (black solid line).
 Figures (b,c) are adapted from \cite{Lovchinsky2016}. }  \label{fig:nmr_seq}
\end{figure}

Initially, the \NV sensor is initialized into the $\ket{0}$ state, followed by a $\pi/2$ pulse along the $y$-axis, transforming the sensor into the superposition state $\ket{+}$, in line with the quantum sensing protocol described in \se \ref{sec:quantum_sensing}. Subsequently, a sequence of resonant $\pi$ microwave pulses is applied to the NV sensor, utilizing a fundamental unit such as $\tau/2-\pi-\tau-\pi-\tau/2$ (refer to Table \ref{tab:dd_sequence} for different \dyde sequences), as depicted in Fig. \ref{fig:nmr_seq}(a). Here, $\tau$ represents the free evolution time.

The collective evolution of the NV-nuclear spin system results in a rotation of the nuclear spin by an angle $\phi$ around an axis ${\mathbf{h}}_{m_s}$ (detailed in \app \ref{sec:sta_fluc_signal}), contingent upon the initial \NV spin state $m_s$ \cite{taminiauDetectionControlIndividual2012,zhaoSensingSingleRemote2012,kolkowitz2012Sensing,zhao2012Decoherence}. This rotation occurs when the free evolution time satisfies the resonant condition:
\begin{equation}
\tau=\frac{\pi}{\omega_L+a_{\parallel}/2},
\end{equation}
the two axes, ${\mathbf{h}}_{0}$ and ${\mathbf{h}}_{1}$, align in antiparallel directions (see \app \ref{sec:sta_fluc_signal} for detail description of ${\mathbf{h}}_{m_s}$) and the rotation angle is
\begin{equation}
\phi\approx\frac{a_{\perp}}{\pi}N\tau,
\end{equation}
with high field condition $\omega_L\gg a_{\|},a_{\perp}$.

Both statistical fluctuations and spin polarizations can be measured using the \dyde method. In particular, the statistical fluctuations can be measured by the real component readout protocol (Eq. \ref{eqn:real_read}), and the spin polarization can be measured by imaginary component readout protocol (Eq. \ref{eqn:img_read}). The relative microwave phase $\vartheta$ between the first and second $\pi/2$ pulses determines the specific protocols used (Fig. \ref{fig:sensing}). When the relative phase $\vartheta=0$, the final population, dependent on the statistical fluctuation of target spins, is observed by the real component readout (Eq. \ref{eqn:real_read}). Under the small-signal approximation, the result is calculated as (\app \ref{sec:sta_fluc_signal}) 
\begin{equation}
p_{\text{real}} \approx 1 - \frac{1}{4}\sum_{i=1} \left(\frac{a_{\perp, i} N \tau}{\pi}\right)^{2},
\end{equation}
which is independent of the polarization of nuclear spins. When the relative phase $\vartheta=\pi/2$, the final population, dependent on the polarization of target spins, observed by the imaginary component readout. The result is calculated as
\begin{equation}
p_{\text{img}} \approx \frac{1}{2} + \frac{1}{2}\sum_{i} P_{\mathrm{n},i} \cdot\frac{a_{\perp, i} N\tau}{\pi},
\end{equation}
where $P_{\mathrm{n},i}$ is the polarization of the $i$th nuclear spin.

Herein, we consider the practical applications of using a near-surface NV sensor to probe nuclear spins in a  sample. Suppose that nuclear spins are evenly distributed in the sample with the density $\rho_{\text{n}}$ and placed upon the diamond surface,  while the NV sensor resides at a distance $d$ from the surface which is perpendicular to the [100] crystal axis. Under this circumstance, the fluctuation signal $S_{\text{fluc}}$ for sensing statistical fluctuated nuclear spins can be calculated as 
\begin{equation}
\label{eqn:S_fluc}
S_{\text{fluc}}  \approx \frac{5\pi\rho_{\text{n}}}{96d^3} \left(\frac{\mu_0\hbar\gamma_n\gamma_e}{4\pi^2}\cdot N\tau\right)^2.
\end{equation}
If nuclear spins is uniformly polarized with the polarization $P$, the polarization signal $S_{\text{pol}}$ can be acquired as
\begin{equation}
\label{eqn:S_pola}
S_{\text{pol}}  \approx P\cdot\frac{\sqrt{2}\pi\rho_{\text{n}}}{3}\cdot\frac{\mu_0\hbar\gamma_n\gamma_e}{4\pi^2}\cdot N\tau.
\end{equation}

The relatively weaker coupling between the nuclear spin and the NV sensor, owing to the former's lower gyromagnetic ratio, demands heightened sensitivity. In addition to minimizing the sample-sensor distance by covalently bonding them (Fig. \ref{fig:nmr_seq}(b)), further enhancements in the readout fidelity and the decoherence time of NV centers near the surface are imperative. To achieve this, an auxiliary nuclear spin is employed to store sensing results in the \NV spin state, enabling repetitive readouts without state resetting through optical pumping \cite{Jiang2009,Neumann2010}. Notably, this approach enhances readout fidelity by nearly tenfold (Fig. \ref{fig:ancilla-assisted_readout}(b)). However, repetitive readout reaches saturation when the readout time aligns with the sensing time, emphasizing the criticality of enhancing the decoherence time to augment sensitivity. Strategies involving wet oxidative chemistry combined with annealing at 465$^{\circ}$C (Tab. \ref{tab:surface}) in a dry oxygen environment have been utilized to boost the coherence times of NV centers near the surface by over an order of magnitude \cite{Lovchinsky2016}. This comprehensive enhancement in spin number sensitivity exceeds 500 times, rendering the sensitivity adequate to potentially detect a single proton within a second. Nonetheless, the direct detection of a single proton has not yet been experimentally realized. An illustrative spectrum, as shown in Fig. \ref{fig:nmr_seq}(c), serves as an example of the achieved results. For detailed discussions, refer to \se \ref{sec:anc_ass_read} regarding the readout techniques and \se \ref{sec:decoherence} along with Tab. \ref{tab:surface} concerning decoherence time reduction.

\subsubsection{Electron nuclear double resonance (ENDOR)} \label{sec:NMR:ENDOR}

\begin{figure}[htp]
\begin{overpic}[width=0.95\columnwidth]{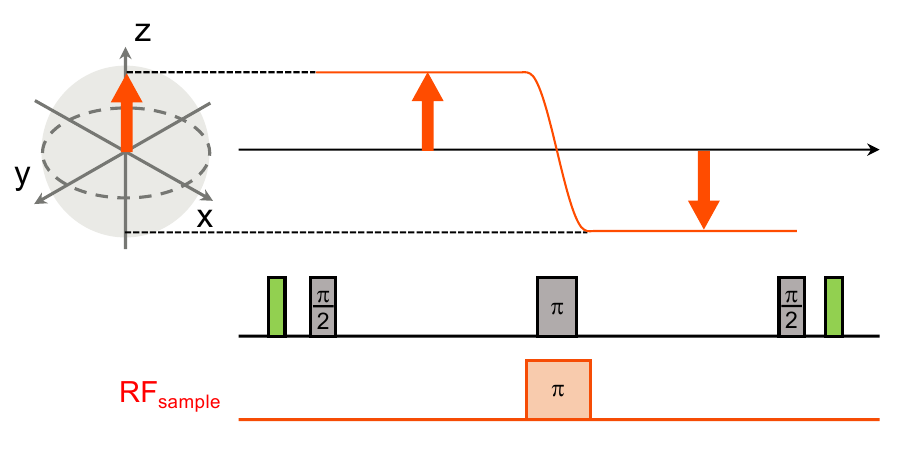}
\end{overpic}
\caption[NV Center]{ ENODR sequence. The sequence is based on the Hahn-echo sequence of the NV center spin. At the midpoint of the sequence, a resonant $\pi$ pulse is applied to both the nuclear and electron spins. }  \label{fig:endor_seq}
\end{figure}

ENDOR, a widely adopted technique, plays a pivotal role in dissecting interactions within \epr spectra. Its traditional application lies in unraveling the molecular and electron structures of paramagnetic species \cite{schweiger2001Principles}. The detection of nuclear spin by the electron spin involves a specific sequence, as depicted in Fig. \ref{fig:endor_seq}. Initially, the electron spin is initialized to the $\ket{0}$ state and then flipped to a $\ket{+}$ superposition state via a $\pi/2$ microwave pulse. Following this, microwave and radio-frequency $\pi$ pulses simultaneously flip both the electron and nuclear spin states. This meticulously designed process nullifies environmental noise on the \NV spin while conserving the coupling between the \NV spin and nuclear spin. Subsequently, the accumulated signal is detected via the second $\pi/2$ pulse (refer to Eq. \ref{eqn:real_read},\ref{eqn:img_read}). 
With the relative phase $\vartheta=0$, the fluctuation signal by real component readout is accordingly expressed as (\app \ref{sec:sta_fluc_signal}),
\begin{equation}
p_{\text{real}} \approx 1 - \frac{1}{4}\sum_{i=1} \left(\frac{a_{\|, i} N \tau}{2}\right)^{2},
\end{equation}
which is independent of the polarization of the nuclear spins. With the relative phase $\vartheta=\pi/2$, the polarization signal by imaginary readout is expressed as
\begin{equation}
p_{\text{img}} \approx \frac{1}{2} + \frac{1}{2}\sum_{i} P_{\mathrm{n},i} \left(\frac{a_{\|, i} N\tau}{2}\right),
\end{equation}
where $P_{\mathrm{n},i}$ is the polarization of $i$th nuclear spin. The ENDOR method is limited by the pulse length and the operation fidelity of the RF pulse. Thus, applying ENDOR sequences with multiple $\pi$ pulses is a challenging task. Typically, in practice, the NV center used in the ENDOR method requires a coherence time greater than approximately 100 $\mu$s \cite{aslam2017Nanoscale,mamin2013nanoscale}.

We consider the same sensing scenario as \dyde, but the diamond surface is perpendicular to the [111] crystal axis. Accordingly, the fluctuation signal $S_{\text{pol}}$ and the polarization signal $S_{\text{pol}}$ are calculated as
\begin{gather}
S_{\text{fluc}}  \approx \frac{\pi\rho_{\text{n}}}{16d^3} \left(\frac{\mu_0\hbar\gamma_n\gamma_e}{8\pi}\cdot N\tau\right)^2 \label{eqn:S_fluc}
\end{gather}
and
\begin{gather}
S_{\text{pol}}  \approx P\cdot\frac{2\pi\rho_{\text{n}}}{3}\left(\frac{\mu_0\hbar\gamma_n\gamma_e}{8\pi}\cdot N\tau\right) . \label{eqn:S_pola}
\end{gather}

\subsubsection{Hartmann-Hahn double resonance (HHDR) } \label{sec:NMR:HH}

\begin{figure}[bhtp]
\begin{overpic}[width=0.92\columnwidth]{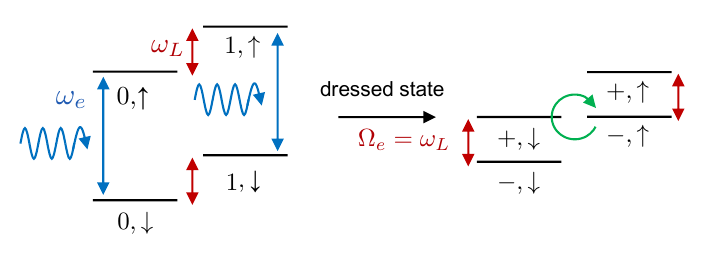}
\end{overpic}
\begin{overpic}[width=0.95\columnwidth]{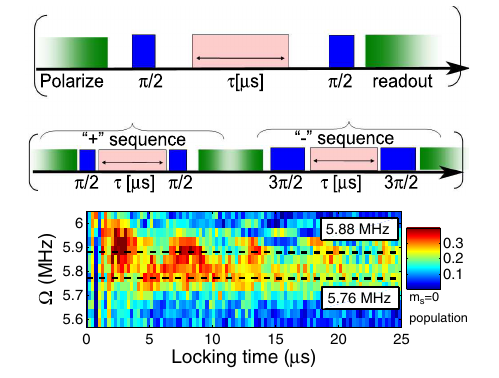}
\put (-2, 105) { (a) }
\put (-2, 72) { (b) }
\put (-2, 49) { (c) }
\put (-2, 33) { (d) }
\end{overpic}
\caption[NV Center]{ Detection and polarization of nuclear spins using \hh. (a) Depicts the energy-level configuration involving the \NV center electron spin and a single spin-1/2 nuclear spin. The hyperfine coupling between the electron spin and the nuclear spin is suppressed due to the energy mismatch between the electron spin ($\omega_e$, labeled blue) and the nuclear spin ($\omega_L$, labeled red). When the electron spin driving frequency $\Omega_e$ aligns with the nuclear spin frequency $\omega_L$, the energy level diagram is described using dressed states $\ket{\pm}$ (shown on the right side of the figure). This removal of energy mismatch allows the flip-flop transition between $\ket{+,\downarrow}$ and $\ket{-,\uparrow}$. (b) Demonstrates the \hh sequence: a 532 nm laser polarizes the \NV electron spin (highlighted in green), with microwave application in between where the $x$ pulse is denoted in blue and the $y$ pulse in pink. (c) Illustrates the alternating \hh sequence designed to avoid polarization of nuclear spins. (d) Displays experimentally observed population of the $\ket{0}$ state concerning the electron spin driving frequency $\Omega$ and sensing time $\tau$. Both isolating nuclear spin and the nuclear spin bath are observed and labeled with dashed lines. Adapted from \cite{londonDetectingPolarizingNuclear2013}. }  \label{fig:HHDR}
\end{figure}

The \hh method is also applicable to nuclear spin sensing \cite{facchi2004Unification,fanchini2007Continuously,cai2013Diamondbased}. In \hh, a continuous, resonant microwave field drives the \NV sensor. Continuous driving isolates the \NV sensor from its environment, making it insensitive to surrounding spins. However, specific frequency components can be selected in \nmr via the \hh resonant condition \cite{cai2013Diamondbased,hartmann1962Nuclear,londonDetectingPolarizingNuclear2013}. Moreover, this technique enables the use of alternating spin-lock sequences to directly polarize the target nuclei. For NV-based nuclear magnetic sensing, \hhdr occurs when the \NV electron spin is driven at a Rabi frequency $\Omega_e$, aligning with the Larmor frequency $\omega_L$ of the nuclear spin (Fig. \ref{fig:HHDR}(a)). Under resonant microwave driving, the entire system transitions from the bare basis to the dressed state basis, with the energy level splitting $\Omega_e$ equating the nuclear spin Larmor frequency $\omega_L$ (Fig. \ref{fig:HHDR}(a)). The energy of the $\ket{+, \downarrow}$ state and the $\ket{-, \uparrow}$ state becomes equal and coupled, evolving coherently together. The remaining states $\ket{+, \uparrow}$ and $\ket{-, \downarrow}$ are decoupled from the joint dynamics (Fig. \ref{fig:HHDR}(b)). The probability of finding the dressed NV spin state, initially set to the state $\ket{+}$, in the opposite state $\ket{-}$ after time ${t_s}$ is
\begin{equation}
p(\tau)=\sin^2\left(   {\frac{\left|\mathbf{a}_{\text {hyp }}\right|t_s}{8}}\sin \theta\right),
\end{equation}
which is dependent on the coupling strength $\mathbf{a}_{\text{hyp}}=(a_{xz},a_{yz},a_{zz})$ and $\theta$, the angle between external magnetic field $B_0$ and $\mathbf{a}_{\text {hyp }}$. The \NV spin is reset to $\ket{+}$ at the beginning of every experiment following the initialization laser and the first $\pi/2$ pulse (Fig. \ref{fig:HHDR}(b)), resulting in $\ket{+,\uparrow}$ as the trapped state. After the several iterations of the experiment, the nuclear spin polarizes to the $\ket{\uparrow}$ state. To prevent polarization, the sensing sequence alternates between resetting the \NV spin state to $\ket{+}$ and $\ket{-}$ (Fig. \ref{fig:HHDR}(c)). An example spectrum is shown in Fig. \ref{fig:HHDR}(d) by sweeping the \NV spin driven frequency $\Omega_e$. Both isolating nuclear spin and the nuclear spin bath are observed. There is a high demand for stable microwave power due to the dependence of the spectrum on $\Omega_e$ \cite{londonDetectingPolarizingNuclear2013}. 

Apart from its application in nuclear spin sensing, \hhdr has found use in polarizing external nuclear spins \cite{fernandez-acebal2018Hyperpolarizationa,shagievaMicrowaveAssistedCrossPolarizationNuclear2018}. The NV center can achieve up to 95\% polarization \cite{vandersar2012Decoherenceprotected} via a square laser pulse and up to 97.7\% using a chopped laser pulse sequence at ambient temperatures \cite{xie2021Beating}. This polarization process occurs within approximately $\sim$ $\mu$s using a $\sim$ 0.1 mW focused 532 nm laser beam ($\sim$ 10 kW/cm$^2$), while the NV center's repolarization time can extend to the $\sim$ ms time scale. Leveraging the \hh resonance method, the NV center becomes a promising candidate for \DNP. While demonstrations have primarily focused on individual NV centers, achieving hyperpolarization of micron-scale samples based on NV ensembles has also been successful \cite{healey2021Polarization}, achieving a best polarization transfer rate of $\approx$ 7500 spins per second per NV center. Anisotropic interactions between the NV electron spin and target nuclear spins are typically averaged out due to molecular motion in fluids and short correlation times. Consequently, the cross-relaxation method is commonly employed in DNP for liquid samples. However, methods for the solid-state counterpart prove to be less efficient \cite{vanbentum2016Solid,wisniewski2016Solid}. Polarizing highly diffusive liquids (e.g., water) poses challenges due to the limitations in effective dipolar coupling strength, with most current experiments relying on viscous liquids (e.g., oil).

\subsubsection{Two-dimensional NMR } \label{sec:NMR:twod} 

\begin{figure}[htp]
	\begin{minipage}[h]{1\columnwidth}
	\centering
	\begin{overpic}[width=\textwidth]{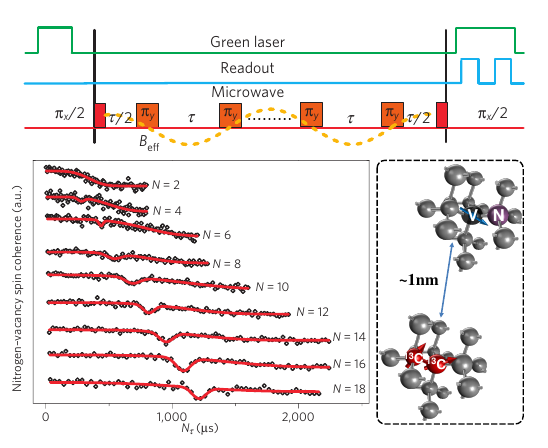}
		\put (0, 75) { (a) }
		\put (0, 49) { (b) }
		\put (70, 49) { (c) }
	\end{overpic}
\end{minipage}
\caption[NV Center]{  Resolving the interaction between nuclear spin pairs. (a) Depicts the \dyde pulse sequence designed for detecting \Cthir-\Cthir nuclear spin pairs. (b) Illustrates dip features in the coherence of the NV sensor under DD control with varying numbers of $\pi$ pulses. Notably, as the number of $\pi$ pulses ($N$) increased from 2 (top curve) to 18 (bottom curve), the dips induced by a pair of nuclear spins emerged and became increasingly prominent. (c) Shows the configurations employed for detecting \Cthir-\Cthir pairs using the NV center. Figures (a-c) have been adapted from \cite{shi2014Sensing}.  }  \label{fig:13Cpair}
\end{figure}

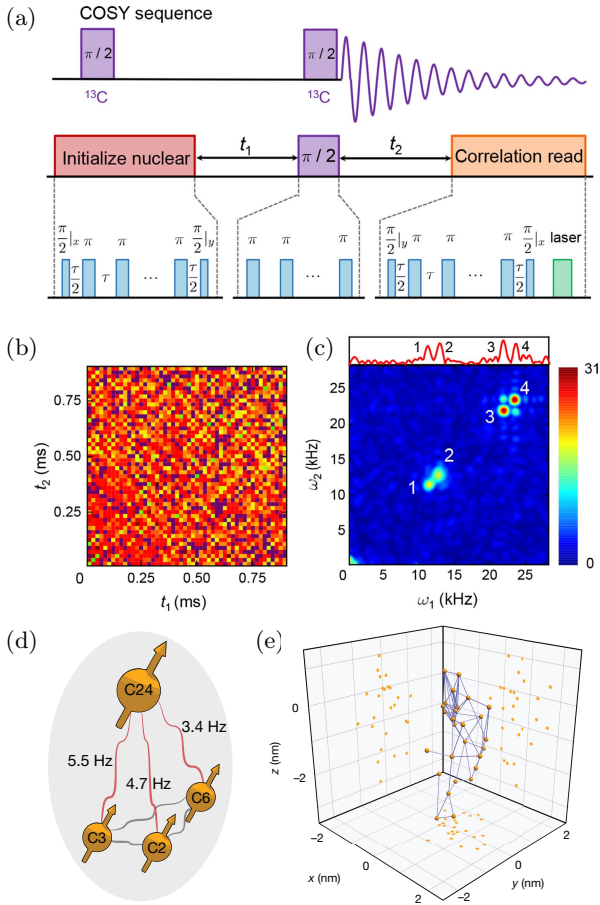
\begin{figure}[htp]
\begin{minipage}[h]{0.85\columnwidth}
	\begin{overpic}[width=\textwidth]{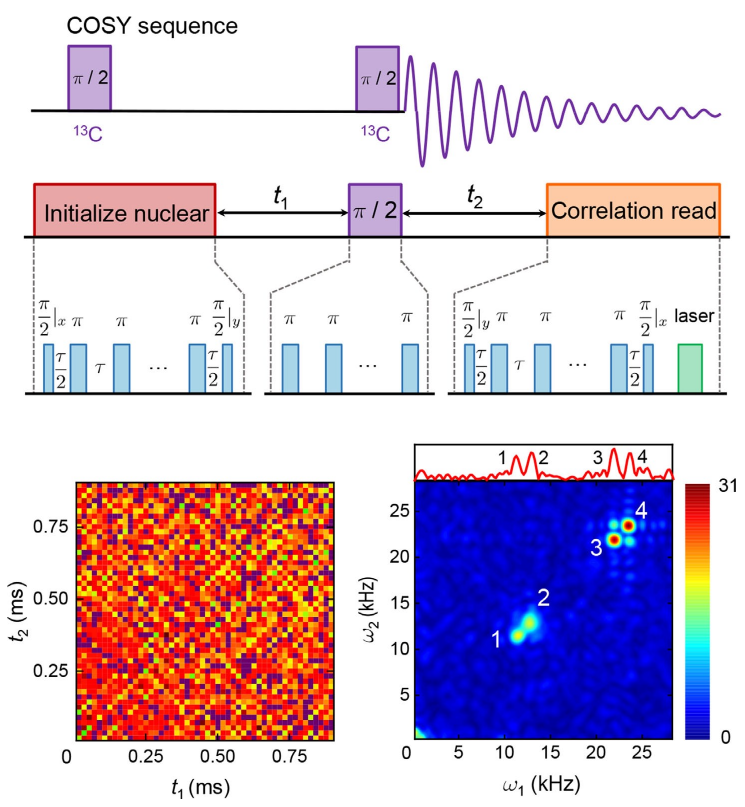}
		\put (-5, 95) { (a) }
		\put (-5, 42) { (b) }
		\put (43, 42) { (c) }
		\put (-5, -5) { (d) }
		\put (35, -5) { (e) }
	\end{overpic}
	\begin{overpic}[width=\textwidth]{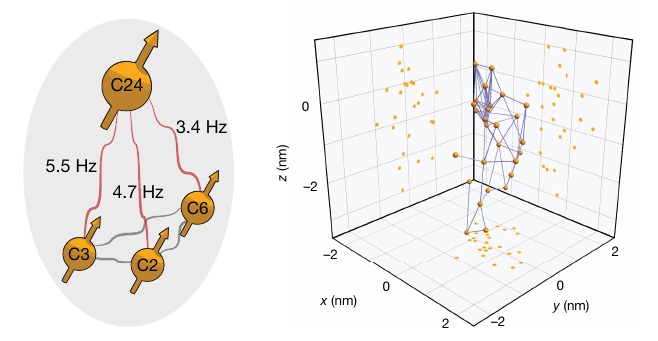}
\end{overpic}
\end{minipage}
\caption[NV Center]{  \Td \nmr. (a) The experimental pulse sequence of \td \nmr along with the structural analysis of the coupled nuclear spin system. This sequence represents an experimental implementation of nanoscale homonuclear correlation spectroscopy \td NMR on the NV center. (b) Sweeps the interrogation times, $t_1$ and $t_2$, from 4 $\mu$s to 0.9 ms with 50 samplings. The time-domain spectroscopy is transformed into frequency-domain spectroscopy in (c) through \td FFT transformation. (c) The \td NMR spectrum revealing cross peaks between the third and fourth peaks in the one-dimensional spectrum (upper inset), indicating that they belong to a coupled spin system. Figures (a--c) have been adapted from \cite{yang2019Structural}.
(d) shows MRI of nuclear spins, resolving the three-dimensional structure by detecting the distance between each nuclear spin dependent on the dipolar coupling strength.
(e) depicts the three-dimensional structure of the nuclear spins within the diamond using the diamond-lattice method. Blue lines indicate couplings greater than 3 Hz, illustrating the connectivity of the cluster. Figures (d,e) are adapted from \cite{abobeih2019Atomicscale}.
}  \label{fig:2dnmr}
\end{figure}

Molecular structure analysis is a cornerstone of biology, chemistry and medicine. \Td \NMR is essential in molecular determination. The one-dimensional nanoscale and \sm \NMR have been previously implemented as mentioned above. 
Although nanoscale one-dimensional NMR and observation of the interaction within a pair of nuclear spins in diamond have been performed using \dyde methods \cite{shi2014Sensing,yang2018Detection}, two-dimensional NMR techniques are still necessary to resolve complex molecular structure information since it can provide a clear and comprehensive picture of intra-molecular interactions (Fig. \ref{fig:13Cpair}(a,b)). Multi-dimensional NMR techniques reveal more spectral information, while measuring times of two-dimensional NMR increase quadratically with experimental sampling numbers.

Two-dimensional nanoscale \nmr spectroscopy based on the NV center has been developed and experimentally demonstrated using coupled $^{13}$C nuclei in diamond \cite{yang2019Structural, kong2020Artificial}. The sequence closely resembles the \cosy sequence in conventional \nmr methods, as depicted in Fig. \ref{fig:2dnmr}(a). The initial \dyde sequence initializes the nuclear spins, followed by a free evolution period $t_1$. Subsequently, another \dyde sequence, equivalent to a $\pi / 2$ pulse on the nuclear spins, is executed. After another free evolution period $t_2$, the final dynamic decoupling sequence transfers the nuclear correlation information onto the NV center. By sweeping the durations $t_1$ and $t_2$, a correlation map (Fig. \ref{fig:2dnmr}(b)) and a \td FFT spectrum (Fig. \ref{fig:2dnmr}(c)) are generated. The presence of cross peaks signifies a coupled nuclear system, allowing for further analysis to determine the precise positions of the nuclear spins and the distances between them. In a different study \cite{abobeih2019Atomicscale}, a much larger cluster comprising 27 coupled $^{13}$C nuclear spins in diamond was detected. This work involved a multidimensional spectroscopy method that isolated individual nuclear-nuclear spin interactions with high spectral resolution (< 80 mHz) and high accuracy (2 mHz). These interactions encode the composition and inter-connectivity of the cluster, enabling the extraction of the three-dimensional structure of the nuclear-spin cluster with sub-angstrom resolution. Moreover, a micron-scale two-dimensional NMR experiment has also demonstrated its capability for structural analysis with chemical resolution \cite{smits2019Twodimensional}.

\section{Detection sensitivity}\label{sensitivity}
\subsection{Introduction}

The pursuit of \sm sciences demands a delicate balance between high sensitivity and spatial resolution, a challenge underscored by the energy resolution limit (ERL) ($E_R = \hbar$) \cite{mitchell2020Colloquium}. Achieving single-molecule spatial resolution with NV centers necessitates a closer sample-probe distance, approximately within several nanometers. However, this proximity amplifies noise from the diamond surface, substantially reducing the coherence times of NV centers. Addressing this predicament remains pivotal in advancing NV-based single-molecule technologies, an ongoing investigation requiring further dedicated efforts.

Within this chapter, we characterize the sensitivities of NV sensors—magnetic field sensitivity and spin number sensitivity, while dissecting the pivotal factors involved, such as readout noise and decoherence. These elements are subjected to meticulous analysis, and the subsequent sections synthesize approaches to enhance them. \se \ref{sec:readout} scrutinizes NV center readout noise and its mitigation strategies. This includes leveraging photonic structures to augment collection efficiency, employing resonant readout at cryogenic temperatures, employing ancilla-assisted repetitive readout, and utilizing spin-to-charge conversion and photoelectric readout techniques.
\se \ref{sec:decoherence} explores the mechanisms behind the decoherence of near-surface NV centers and proposes corresponding methods to extend coherence times. Finally, \se \ref{sec:NMR:amp} introduces sensing protocols designed to amplify nuclear spin signals, an essential strategy to surmount the challenge posed by the low gyromagnetic ratio of nuclear spins.

\subsection{Magnetic field sensitivity and spin number sensitivity}
As discussed in \se \ref{sec:quantum_sensing}, a small magnetic field $B$ can be detected by causing the spin coherence of the NV center to rotate or shorten during a sensing sequence. For sensing a fluctuated field with the variance $(\delta B)^2$, the coherence is shortened, and the sensitivity $\eta_\text{fluc}$ is determined by reading its real component
\begin{equation}
\eta_{\text{fluc}} \approx  \frac{\sigma_S}{|\pdv*{S}{(\delta B)^2}|} \sqrt{T_{\text{exp}}}.
\end{equation}
For sensing a polarized field with the mean $\bar B$, the coherence is rotated and the sensitivity $\eta_\text{pol}$ is given by  imaginary component readout
\begin{equation}
\eta_{\text{pol}} \approx \frac{\sigma_S}{|\pdv*{S}{\bar B}|} \sqrt{T_{\text{exp}}},
\end{equation}
where $\sigma_S$ is the readout noise (see \se \ref{sec:readout}), and the time for a single experimental cycle $T_{\text{exp}}$ is the sum of the time for phase accumulation $T_{\text{accu}}$, state initialization $T_{\text{ini}}$, and readout $T_{\text{read}}$. $\pdv*{S}{\bar B}$ and $\pdv*{S}{(\delta B)^2}$ are the slopes of the signal versus the mean $\bar B$ and the variance $(\delta B)^2$ respectively, resulting in the dimensions of T$\cdot$Hz$^{-1/2}$ and T$^2\cdot$Hz$^{-1/2}$ (Hz$^{-1/2}$ denotes one-second integration) for these two kinds of magnetic field sensing.

By comprehensively considering multiple experimental limitations, including spin decoherence, initialization and readout errors, and duty cycle, the sensitivities $\eta_{\text{fluc}}$ and $\eta_{\text{pol}}$ for an individual NV center can be accurately formulated as
\begin{equation}
\eta_{\text{fluc}} \approx \frac{2}{\gamma_{\text{NV}}^2\sqrt{T_{\text{accu}}^{3}}} \cdot \frac{1}{\xi_{T_{\text{accu}}} F_{\text{read}} F_{\text{ini}}} \cdot \sqrt{1+\frac{T_{\rm ir}}{T_{\text{accu}}}}\label{equ:mag_sensitivity_fluc}
\end{equation}
and
\begin{equation}
\eta_{\text{pol}} \approx \frac{1}{\gamma_{\text{NV}}\sqrt{T_{\text{accu}}}} \cdot \frac{1}{\xi_{T_{\text{accu}}} F_{\text{read}} F_{\text{ini}}} \cdot \sqrt{1+\frac{T_{\rm ir}}{T_{\text{accu}}}},\label{equ:mag_sensitivity_pola}
\end{equation}
where  $\xi_{T_{\text{accu}}}$ is the remained spin coherence at $T_{\text{accu}}$.
The initialization and the readout are both imperfect, with $F_{\text{ini}}$ and $F_{\text{read}}$ denoting their respective fidelities, and occupy considerable time $T_{\text{ir}}=T_{\text{ini}}+T_{\text{read}}$, leading to a reduced duty cycle. 

Herein, we consider the practical parameters utilized in the experiment to estimate the approximate sensitivity. The magnetic sensitivity hinges significantly on two critical factors: readout fidelity and coherence time, to be thoroughly addressed in Section \ref{sec:readout} and Section \ref{sec:decoherence}, respectively.

Regarding the optical readout of the NV spin state, the readout fidelity $F_r$ remains relatively low, approximately $\sim 4\%$, attributable to typical fluorescence collection efficiencies (detailed in Section \ref{sec:collection_efficiency}). During the initialization of the NV center through 532-nm laser illumination, it exists in two charge states. Around 70\% constitutes the useful NV$^-$ state, while the rest remains as the less useful NV$^0$ state \cite{xie2021Beating}. In instances of shallower NV centers, this percentage of the NV$^-$ state further decreases \cite{bluvstein2019identifying}.

For near-surface NV centers exposed to the noise emanating from the diamond surface, the coherence time undergoes a dramatic reduction, plummeting to tens of microseconds \cite{ofori2012spin}, a stark contrast to several milliseconds for NV centers nestled deep within bulk diamond material \cite{herbschleb2019Ultralong}. Substituting these aforementioned values into Eq. \ref{equ:mag_sensitivity_fluc} and Eq. \ref{equ:mag_sensitivity_pola} gives the sensitivities $\eta_{\text{pol}} \approx$ 0.1 $\mu$T$\cdot$Hz$^{-1/2}$ and $\eta_{\text{fluc}} \approx$ (0.2 $\mu$T)$^2\cdot$Hz$^{-1/2}$.

When it comes to magnetic sensing at the single-molecule scale, the magnetic field under detection emanates from the target spins. As such, the primary emphasis lies in detecting the minimum number of spins, integrated over one second, known as spin number sensitivity. Let us consider the scenario where the target spins are positioned at a distance $d$ from a [100]-oriented NV center, leading to the sensitivities $\eta_{\text{fluc}}^{\text{spin}}$ and $\eta_{\text{pola}}^{\text{spin}}$, given by 

\begin{equation}
\label{equ:spin_sensitivity_fluc}
\eta_{\text{fluc}}^{\text{spin}} \approx \left(\frac{4\pi^2}{\mu_0\hbar\gamma_{\text{NV}}\gamma_{\text{tar}}}\right)^2 \cdot \frac{d^6}{\sqrt{T_{\text{accu}}^{3}}}\cdot \frac{1}{\xi_{T_{\text{accu}}} F_{\text{read}} F_{\text{ini}}} \cdot \sqrt{1+\frac{T_{\rm ir}}{T_{\text{accu}}}}
\end{equation}
for sensing  statistical fluctuated spins and
\begin{equation}
\label{equ:spin_sensitivity_pola}
\eta_{\text{pol}}^{\text{spin}} \approx \frac{2\sqrt{2}\pi^2}{\mu_0\hbar\gamma_{\text{NV}}\gamma_{\text{tar}}} \cdot \frac{d^3}{\sqrt{T_{\text{accu}}}} \cdot \frac{1}{P \xi_{T_{\text{accu}}} F_{\text{read}} F_{\text{ini}}} \cdot \sqrt{1+\frac{T_{\rm ir}}{T_{\text{accu}}}}
\end{equation}
for sensing polarized spins with the spin polarization $P$, where $\hbar$ is the reduced Plank constant, $\mu_0$ is the vacuum permeability, and $\gamma_\text{tar}$ is the gyromagnetic ratio of the target spins. 
Eq. \ref{equ:spin_sensitivity_fluc} and Eq. \ref{equ:spin_sensitivity_pola} suggest that the target--sensor distance $d$ and the gyromagnetic ratio of the target spins $\gamma_\text{tar}$ are crucial determinants of the spin number sensitivity, along with the factors denoted in Eq. \ref{equ:mag_sensitivity_fluc} and Eq. \ref{equ:mag_sensitivity_pola}.
Given that the gyromagnetic ratio of the electron spin is approximately three orders of magnitude larger than that of nuclear spins, detecting an individual electron spin outside the diamond is relatively straightforward. Utilizing extremely shallow NV centers (< 5 nm) boasting sub-millisecond spin coherence already provides sufficient sensitivity to detect a single nuclear spin \cite{Lovchinsky2016,muller2014Nuclear}, although it still falls short of resolving a single nuclear spin. Furthermore, single nuclear spins can also be detected by leveraging an electron spin as a reporter (see \se \ref{sec:NMR:amp}).

\subsection{Readout noise}\label{sec:readout}

\begin{figure}[htp]
\begin{overpic}[width=0.9\columnwidth]{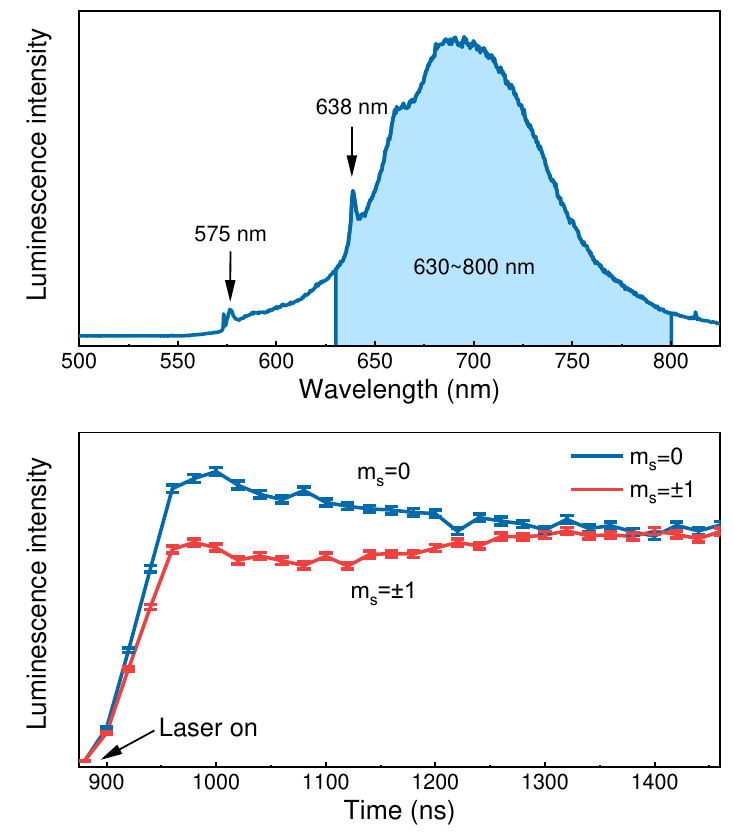}
	\put (-2, 99) {(a)}
	\put (-2, 48) {(b)}
\end{overpic}
\caption[NV Center]{Level diagrams and PL of the NV center. 
(a) The PL spectrum of the NV center, where the laser wavelength for spin initialization is typically 532 nm. Peaks at 575 nm and 638 nm correspond to the ZPL of the \NVo (\NVm) charge state. The range 630--800 nm corresponds to the phonon sideband of the \NVm charge state. (b) The \NVm spin states exhibit spin-dependent fluorescence wherein \NVm spins prepared in the $m_s=0$ state emit a higher rate of photons compared to those prepared in the $m_s=\pm 1$ state. 
} \label{fig:read_process}
\end{figure}

The green laser (typically 532 nm) excites the NV electronic state incoherently via a strong phonon sideband and emits \phl in its 600--800 nm phonon sideband (Fig. \ref{fig:read_process}(a)) at a rate of $\gamma\approx 70$ $\mu$s$^{-1}$ \cite{gupta2016Efficient,robledo2011Spin,fuchs2010Excitedstate}. Unlike the resonant excitation for narrow-linewidth atoms \cite{Itano1993} or color centers under cryogenic temperatures \cite{togan2010Quantum,toganLaserCoolingRealtime2011}, the spin-dependent fluorescence detection for the NV center under ambient temperature, arises through the non-radiative \isc transitions \cite{goldman2015Stateselective,goldman2015PhononInduced,thiering2018Theory} from the spin-triplet excited states to the spin-singlet states \cite{rogers2008Infrared}, which occurs preferentially for the $m_s=\pm 1$ spin states (Fig. \ref{fig:NV}(c)). The spin-singlet ground state is long-lived $\sim$ 200 ns \cite{acosta2010Optical,robledo2011Spin,gupta2016Efficient} and causes $m_s=\pm 1$ the spin state to exhibit a reduced fluorescence intensity, as visualized in Fig. \ref{fig:cw}(a). The imperfect spin-state selectivity of the ISC process limits the spin polarization of the NV center with $\sim$ 90 \% in the $m_s=0$ of the ground state \cite{rong2015Experimental,robledo2011Spin,Doherty2013the} after optical illumination via a laser pulse. However, chopped laser sequences have shown experimental improvement in polarization to 97.7\% \cite{xie2021Beating}.

The spin-dependent PL contrast holds critical significance for non-resonant readout procedures. However, the contrast between $m_s=0$ and $m_s=\pm 1$ dissipates within approximately 300 nanoseconds (referred to as the readout window), constrained by the lifetime of the spin-singlet ground state (Fig. \ref{fig:read_process}(b)). Consequently, the available fluorescence photons, denoted as $n_1$ for the initial state of $m_s=\pm 1$, is relatively scarce compared to the initial state of $m_s=0$ represented as $n_0$ (\se \ref{sec:collection_efficiency}).
In situations where the overall fluorescence collection efficiency is low, 
this scarcity only permits a probabilistic determination of the \NVm spin states. Evaluating the overall readout efficiency involves a comparison between the phase uncertainties obtained from imperfect readouts and those from ideal projective measurements for the imaginary-component readout \cite{Degen2017}, resulting in the ratio of these two uncertainties as
\begin{equation}
\label{equ:readout_noise}
F_{\text{read}} = \left(1+\frac{2(n_0+n_1)}{(n_0-n_1)^2}\right)^{-1/2} = \left(1+\frac{1}{n_{\text{avg}} C^2}\right)^{-1/2}.
\end{equation}
The parameter $n_{\text{avg}}$ represents the average value of $(n_0+n_1)/2$, while $C=(n_0-n_1)/(n_0+n_1)$ stands for the PL contrast. This PL contrast precisely mirrors the readout fidelity utilized in Eq. \ref{equ:mag_sensitivity_fluc} and Eq. \ref{equ:mag_sensitivity_pola} (noting that it differs from the readout fidelity used in projective measurements). The value of $F_{\text{read}}$ must exist within the range of 0 to 1, where $F_{\text{read}}=1$ signifies ideal projective measurements. Expressing the real-component readout involves complexity, particularly when considering decoherence. However, fortunately, it closely aligns with the value derived from Eq. \ref{equ:readout_noise} (usually within a 10\% deviation), thereby allowing the utilization of Eq. \ref{equ:readout_noise} in Eq. \ref{equ:mag_sensitivity_fluc} and Eq. \ref{equ:spin_sensitivity_fluc} \cite{Lovchinsky2016}. Further details regarding the derivation of Eq. \ref{equ:readout_noise} are presented in \app \ref{sec:noise_cal}.

\subsubsection{Photoluminescence enhancement}\label{sec:collection_efficiency}

Enhancing PL becomes evidently crucial, as highlighted in Eq. \ref{equ:readout_noise}, in order to augment the readout fidelity and sensitivity. However, the high refractive index of the diamond host ($n \approx 2.4$) results in significant limitations due to total internal reflection at the diamond surface, severely constraining the collection efficiency of emitted fluorescence photons from the NV center. For a single NV center in a diamond slab observed through a confocal microscope using laser power below the PL-saturation level for magnetometry, the typical count rate averages around 200 kcounts/s (kilo-counts per second) \cite{Lovchinsky2016,Neumann2010}. This translates to an average photon number of $n_{\text{avg}} \approx 0.05$ per single experiment. With an NV center exhibiting a PL contrast of $C \approx 0.18$ (implying $n_1 \approx 0.7 n_0$), the resulting readout fidelity stands at approximately $\sim$ 4\%.

\begin{figure}[htp]
\begin{overpic}[width=0.95\columnwidth]{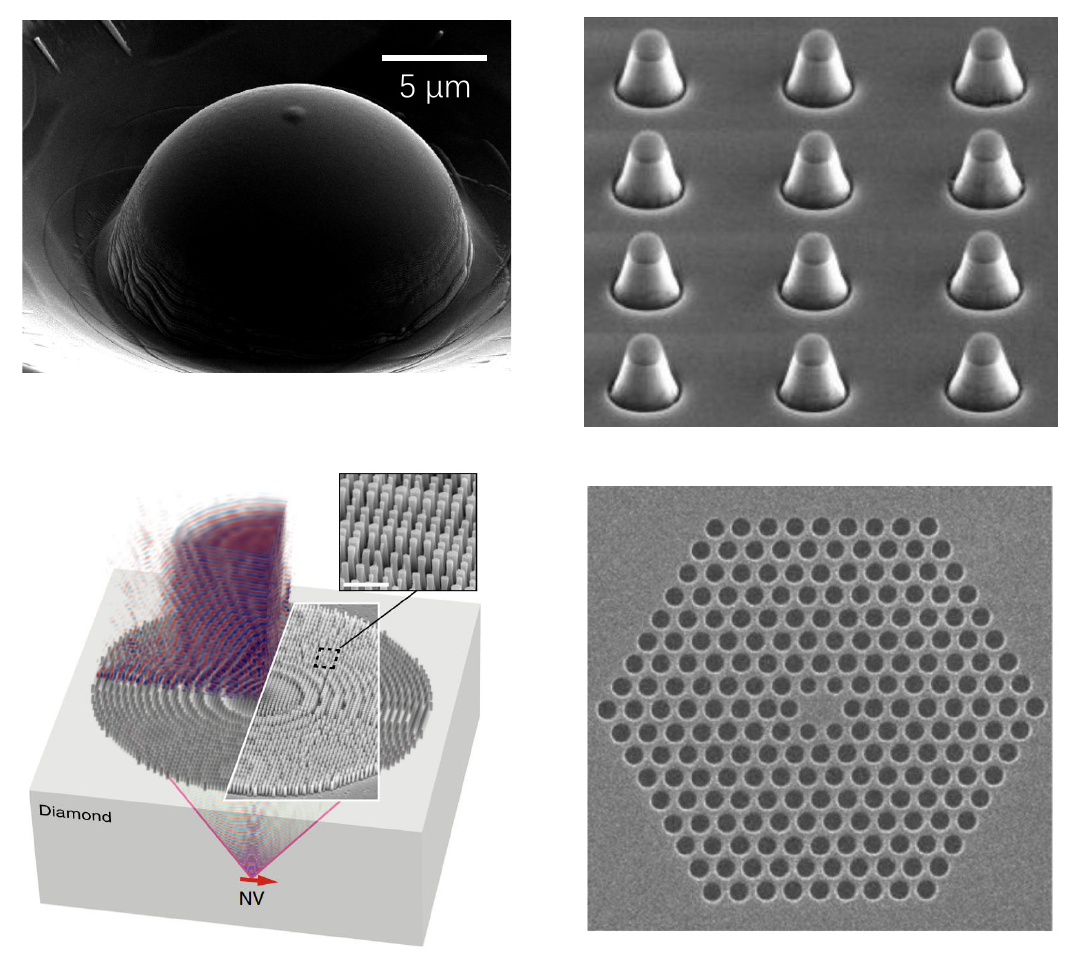}
	\put (-5, 84) {(a)}
	\put (49, 84) {(b)}
	\put (-5, 40) {(c)}
	\put (49, 40) {(d)}
\end{overpic}
\caption[NV Center]{ Photonic structures designed to enhance fluorescence collection efficiency. (a) SIL, a half-ball structure created on the diamond with the NV center situated at its center. (b) An array of diamond nanopillars utilized as photonic waveguides, guiding fluorescence photons from top to bottom. (c) Immersion metalens structures fabricated on the diamond surface to collimate fluorescence emitted by the NV center. Figure adapted from \cite{huang2019monolithic}. (d) A photonic crystal cavity produced on a thin diamond membrane. Figure adapted from \cite{riedrich-moller2015Nanoimplantation}.} \label{fig:nano_photonic}
\end{figure}
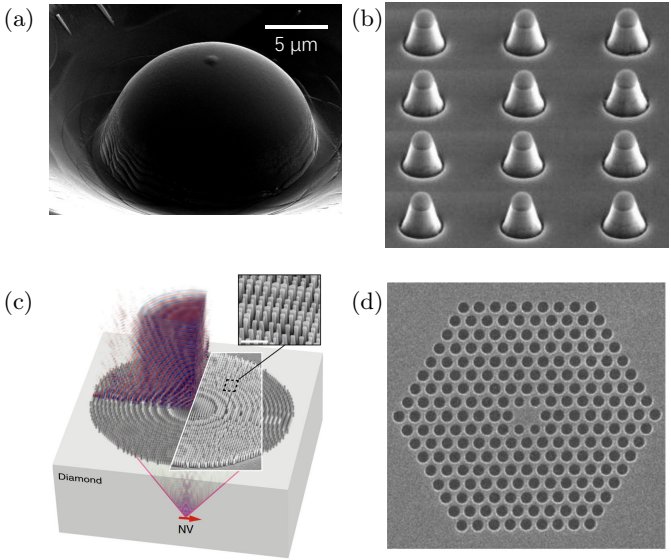

The process of fabricating photonic structures on diamonds represents a significant advancement in enhancing collection efficiency. Diamond, being harder compared to other commonly used optical materials like Si, Si$_3$N$_4$, SiC, and $\alpha$-Al$_2$O$_3$ \cite{rath2015Diamond}, poses challenges in fabrication. However, recent progress in nanolithography has introduced several top-down fabrication techniques, offering viable solutions. One such structure, the \sil, created via focused ion beams, forms a roughly 10 $\mu$m hemisphere housing an NV center at its center (Fig. \ref{fig:nano_photonic}(a)). Fluorescence photons emitted perpendicular to the \sil surface mitigate losses from total internal reflection \cite{hadden2010Strongly,marseglia2011Nanofabricated,siyushev2010Monolithic}. The utilization of \sil structures has shown saturation count rates reaching about 1.2 Mcounts/s (mega-counts per second) \cite{robledo2011Highfidelity}. Furthermore, additional enhancements in fluorescence can be achieved through antireflective coatings \cite{yeung2012anti}. 

The diamond nanopillar, possessing a diameter of approximately 100 nm and a length of several micrometers, is crafted through nanolithography techniques like electron beam lithography (\ebl) and reactive ion etching (\rie). Functioning as an optical waveguide, it houses an embedded NV center (Fig. \ref{fig:nano_photonic}(b)). These nanopillars offer improved convenience for utilization as scanning probes \cite{maletinsky2012robust}. Leveraging self-aligned patterning techniques, the NV center can be precisely positioned at the nanopillar's top center, achieving a count rate of approximately 4 Mcounts/s \cite{wang2022Selfaligned}. Furthermore, variations in nanopillar design have been explored, including inverted nanocones reaching about 3 Mcounts/s \cite{jeon2020Bright}, parabolic profiles attaining approximately 4 Mcounts/s \cite{wan2018Efficient}, and truncated parabolic profiles yielding around 2 Mcounts/s \cite{hedrich2020Parabolica}.

Additionally, metalenses crafted with a Fresnel lens phase profile, formed by etched nanopillars on the diamond surface, have been instrumental in imaging NV centers \cite{huang2019monolithic}. Unlike the \sil structures, the metalens design focuses the emitted light (Fig. \ref{fig:nano_photonic}(c)), negating the requirement for a free-space objective. This stands as a promising approach to directly couple NV centers with optical fibers. Furthermore, coupling to an optical cavity shortens the lifetime of NV excited states, markedly enhancing the count rate. In Fig. \ref{fig:nano_photonic}(d), the photonic crystal cavity, developed from an ultrapure and single crystal diamond membrane, links the implanted NV centers' broadband fluorescence to a cavity mode, showcasing Purcell enhancement of spontaneous emission \cite{riedrich-moller2015Nanoimplantation}.

\subsubsection{Resonant readout}

\begin{figure}[htp]
\begin{overpic}[width=0.95\columnwidth]{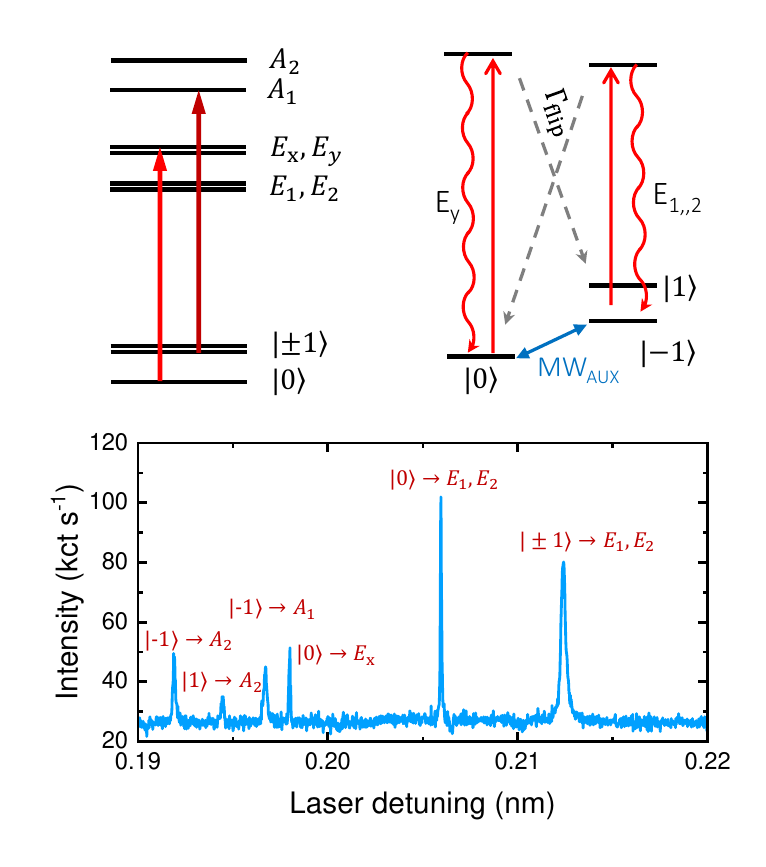}
    \put (2, 95) {(a)}
    \put (43, 95) {(b)}
    \put (2, 49) {(c)}
\end{overpic}
\caption[NV Center]{Resonant detection protocol. (a) The level structure of the NV center under cryogenic temperatures.  The red arrow represents the resonant readout, and the dark red arrow indicates the population pumping. (b) Energy levels  used to prepare and measure the NV electron spin state. The transitions are labeled according to the symmetry of the excited states. Dashed lines indicate spin-non-conserving decay paths.
(c) PL excitation spectrum of the NV center (the wavelength is given relative to 637.2 nm). }  \label{fig:resonant_read}
\end{figure}

The excited state of the \NV triplet consists of six sublevels $\{A_1,A_2,E_x,E_y,E_1,E_2\}$ \cite{maze2011Properties,togan2010Quantum}, as displayed in Fig. \ref{fig:resonant_read}(a). The spin-orbit interaction splits the states into three branches $\{A_1, A_2\}$, $\{E_x, E_y\}$ and $\{E_1, E_2\}$. The spin--spin interaction shifts states with different spin projections and splits $A_1$ and $A_2$ states, and the strain and electric fields additionally split the remaining two branches, namely $\{E_x, E_y\}$ and $\{E_1, E_2\}$ \cite{maze2011Properties,tamarat2008Spinflip,batalov2009Low}. However, at ambient temperatures, the stochastic phonon-mediated transitions average the orbital states, resulting in a simplified excited-state level structure \cite{fu2009Observation}. At low temperatures, the narrow linewidths of the excited states (shown in Fig. \ref{fig:resonant_read}(b)) allow for resonant excitation of the spin-selective optical transitions, enabling coherent coupling between the NV spin and the photon \cite{togan2010Quantum,buckley2010SpinLight} and all-optical coherent control \cite{bassett2014Ultrafasta,yale2013Allopticala}. Under the $A_1$ transition (Fig. \ref{fig:resonant_read}(a)), the spin state of the \NV center can be initialized to $\ket{0}$ with a high fidelity of $99.7\pm 0.1\%$ \cite{robledo2011Highfidelity}. Under the $E_x$ transition, the dark states $\ket{\pm 1}$ emit minimal fluorescence photons, leading to a PL contrast $C\approx 1$ for the readout fidelity in Eq. \ref{equ:readout_noise}. Additionally, the $E_x$ transition, in low-strain NV centers, preserves spin and constitutes a cycle transition for continuous fluorescence photon emission \cite{robledo2011Highfidelity}, as pioneered in the field of cold atoms \cite{happer1972Optical,olmschenk2007Manipulationa,zoller1987Quantum}. However, the cycle transition in NV centers is not as ideal as in cold atoms due to spin mixing and phonon-induced transitions within the excited states. These limitations constrain the spin relaxation time to approximately $10$ microseconds and restrict the number of collected photons to $\sim$ 10 \cite{robledo2011Highfidelity,bernien2013Heralded,hensen2015Loopholefree,humphreys2018Deterministic}. The highest readout fidelity achieved is approximately 97.7 \% \cite{humphreys2018Deterministic}.

\subsubsection{Ancilla-assisted readout} \label{sec:anc_ass_read}

\begin{figure*}[htp]
\begin{overpic}[width=0.8\textwidth]{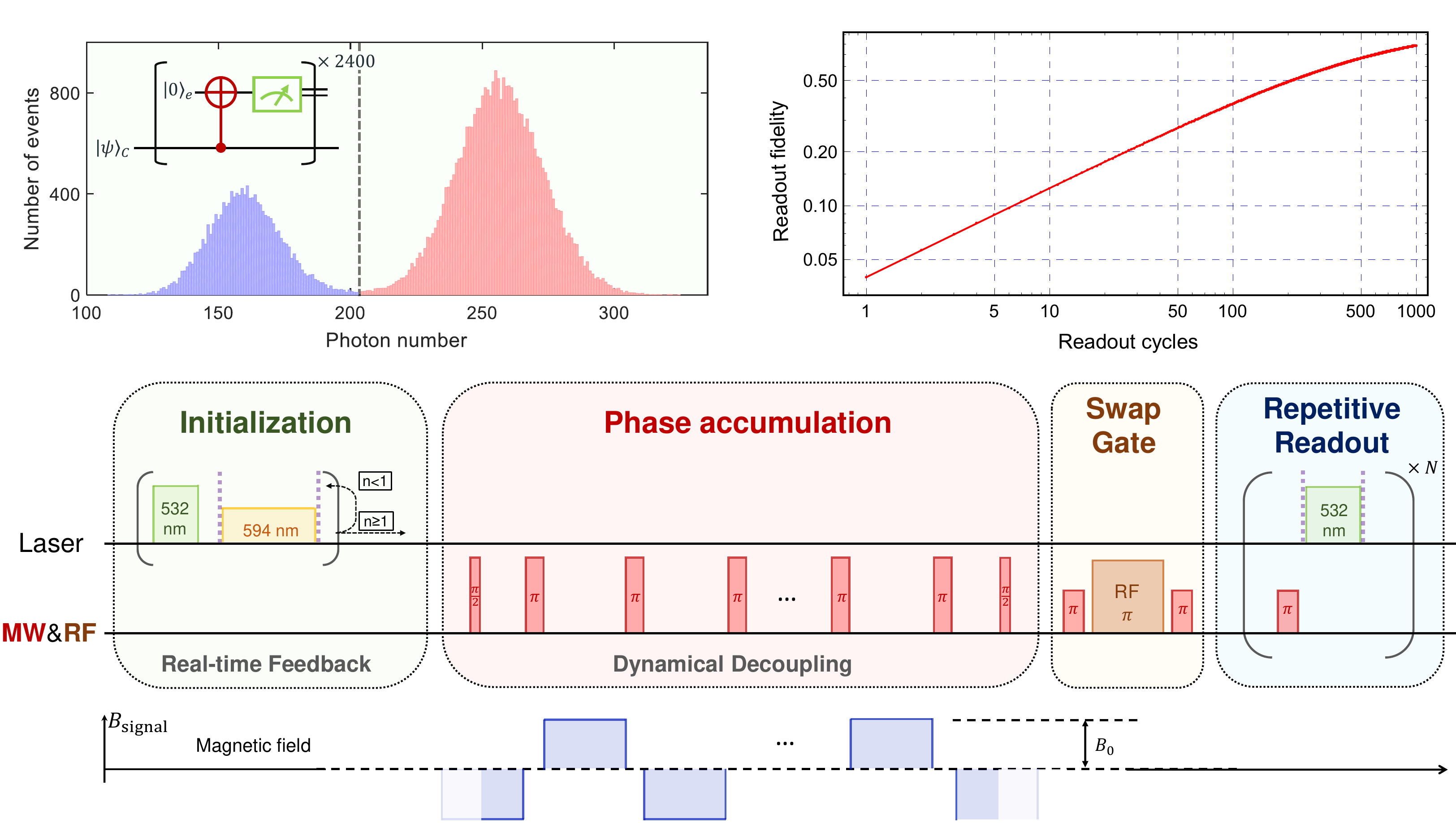}
	\put (0, 54) {(a)}
	\put (50, 54) {(b)}
	\put (0, 30) {(c)}
\end{overpic}
\caption[NV Center]{(a) An example of a photon number histogram of projective measurement for an adjacent nuclear spin with the fidelity $\sim$ 0.99. The figure is adapted from \cite{xie2021Beating}. (b) The fidelity improves with the number of repetitive readout (the last element in (c)).
(c) A complete ancilla-assisted sensing protocol for detecting a small magnetic field with sub-nanotesla sensitivity. It consists of four parts: NV$^-$ charge and spin state initialization by real-time feedback, phase accumulation by dynamical decoupling sequences, SWAP gates, and repetitive readout. The figure is adapted from \cite{zhao2022Subnanotesla}.}  \label{fig:ancilla-assisted_readout}
\end{figure*}

Under ambient temperature, the average number $n_{\text{avg}}$ of collected photons of a single \NV per single measurement is 0.01--0.1 depending on the collection efficiency. As expressed in Eq. \ref{equ:readout_noise}, the photon shot noise will dominate in the overall readout noise when $n_{\text{avg}}\ll 1$. This noise emerges as a significant obstacle for achieving high-fidelity readout at ambient temperature. An alternative method to increase the readout fidelity involves utilizing an adjacent nuclear spin whose spin state is long-lived during laser illumination. The readout process is implemented by transferring the NV spin state to the nuclear spin through a SWAP gate, followed by thousands of readouts of the nuclear spin (the last two parts in Fig. \ref{fig:ancilla-assisted_readout}(c)). The readout fidelity is enhanced by the repetitive readout procedure, as shown in Fig. \ref{fig:ancilla-assisted_readout}(b). The method was first demonstrated with a neighboring \Cthir nuclear spin as the ancilla \cite{Jiang2009}. The projective measurement (fidelity $>0.707$) of the nuclear spin was later achieved with \Nfor nuclear spin \cite{Neumann2010,xie2021Beating} and \Cthir nuclear spin \cite{Maurer2012,dreau2013SingleShot,liu2017SingleShot,unden2016quantum,xie2021Beating} at ambient temperature.

Typical two-peak statistics of photon number after repetitive readout are displayed in Fig. \ref{fig:ancilla-assisted_readout}(a), as the evidence of realizing the projective measurement for a nuclear spin. Nevertheless, we can infer from Eq. \ref{equ:mag_sensitivity_fluc} and Eq. \ref{equ:mag_sensitivity_pola} that several cycles in repetitive readout will require considerable time and in turn degrade the overall sensitivity $\eta$. Therefore, the number of readout cycles will be optimized based on the time for phase accumulation $T_{\text{accu}}$ in a real sensing circumstance (Fig. \ref{fig:ancilla-assisted_readout}(c)). The complete sensing protocol with sub-nanotesla sensitivity (Fig. \ref{fig:ancilla-assisted_readout}(c)) is comprised of four procedures:  
\begin{enumerate}[label=(\roman*)]
\item The NV sensor is firstly initiated to the NV$^-$  state by a real-time feedback and  the $m_s=0$ spin state; 
\item \DyDe sequences are applied to encoding the detected magnetic field into the phase of the \NV spin; 
\item The \NV spin state is mapped onto the nuclear spin by a SWAP gate; 
\item The nuclear spin is readout repetitively for $N$ times to improve the readout fidelity.
\end{enumerate}
The SWAP gate consists of CNOT$e|n$-CNOT$n|e$-CNOT$e|n$ gates, where CNOT$e|n$ (CNOT$n|e$) denotes a nuclear(electronic)-controlled electronic (nuclear) spin NOT gate. The CNOT$e|n$ gate with 99.92\% fidelity achieved in \cite{xie202399} is realized using an MW selective $\pi$ pulse, while the CNOT$n|e$ gate utilizes an RF selective $\pi$ pulse. All unitary operations, including single-qubit gates in DD sequences, CNOT$e|n$ gate, and CNOT$n|e$ gate, can be achieved with near-unit fidelity, and thus do not impair the magnetic sensitivity of the NV sensor (not included in Eq. \ref{equ:mag_sensitivity_fluc} and Eq. \ref{equ:mag_sensitivity_pola}).

By employing real-time feedback and chopped laser illumination, the \NVm charge state and the $m_s=0$ spin state can be prepared with fidelities of 98.94(3)\% and 97.74(18)\%, respectively \cite{xie2021Beating}. Thus, $F_{\text{ini}}$ approximates 1 and slightly lowers the sensitivity. By using $^{12}$C-purified diamond and DD sequences, the interrogation time $T_{\text{accu}}$ can be extended beyond $\sim$ 2 ms with the remained coherence $\xi_{T_{\text{accu}}}$$\approx$0.5. In particular, the readout fidelity can be improved to approximately 80\% with a reasonable duty cycle. The time for initialization and readout may be nearly equal to the interrogation time, which reduces the sensitivity by a factor of $\sim$1.4 (the term $\sqrt{1+T_{\rm ir}/T_{\text{accu}}}$). In accordance with Eq. \ref{equ:mag_sensitivity_fluc} and Eq. \ref{equ:mag_sensitivity_pola}, magnetic field sensitivities achieve $\eta_{\text{pol}} \approx$ 0.5 nT$\cdot$Hz$^{-1/2}$ \cite{zhao2022Subnanotesla} and $\eta_{\text{fluc}} \approx$ (1.6 nT)$^2\cdot$Hz$^{-1/2}$.

\subsubsection{Spin-to-charge conversion (SCC)  and photoelectric readout} \label{sec:SCC}

The charge state of the NV center possesses a lengthy relaxation time and can be optically measured with high fidelity even under ambient temperatures \cite{waldherr2011Dark}. Consequently, the \scc readout presents an alternative to ancilla-assisted readout (\se \ref{sec:anc_ass_read}). In this \scc readout method, the \NV spin state is transferred to its charge state (either the neutral state \NVo or the negative state \NVm), allowing for high-fidelity readout through the long-lived charge state \cite{shields2015Efficient,Hopper2016,irber2021Robust,zhang2021Highfidelity}.
Achieving high-fidelity readout of the charge state requires a considerably longer readout duration compared to resonant readout or nuclear-spin-assisted repetitive readout, resulting in more collected photons and reduced readout noise.

The excitation of \NVo (\zpl at 575 nm) demands a higher photon energy than that of \NVm (\zpl at 637 nm). A laser operating within the wavelength range between these two ZPLs efficiently excites the lower-energy \NVm optical transition over \NVo. For instance, under low-power illumination at 594 nm and at ambient temperature, the \NVm photon count rate could be 28 times higher than that of \NVo \cite{shields2015Efficient}. In NV centers located deeply within bulk diamond, the dominant charge state relaxation occurs via a two-photon process and scales quadratically with the laser power, while fluorescence scales linearly \cite{waldherr2011Dark}. Consequently, employing lower laser power allows for more collected photons and higher charge readout fidelity, albeit at the cost of extended readout time, also constrained by the detector's dark counts \cite{aslam2013Photoinduced}. The scenario becomes more intricate for shallow NV centers. The presence of local electron traps near the diamond surface influences the NV center's charge state, introducing a linearly proportional dependence on the laser power for the ionization rate \cite{dhomkar2018Charge,bluvstein2019identifying}.

\begin{figure*}[htp]
\begin{minipage}[h]{0.45\textwidth}
\begin{overpic}[width=0.48\textwidth]{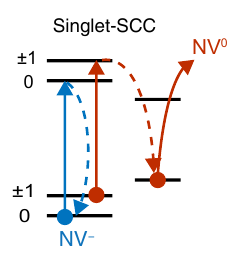}
\end{overpic}
\begin{overpic}[width=0.48\textwidth]{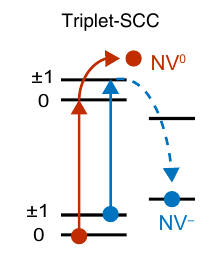}
\end{overpic}
\begin{overpic}[width=1\textwidth]{Figure/blank}
\end{overpic}
\begin{overpic}[width=1\textwidth]{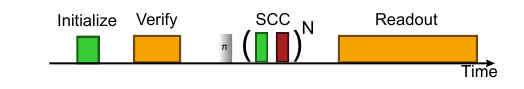}
\end{overpic}
\begin{overpic}[width=1\textwidth]{Figure/blank}
\end{overpic}
\begin{overpic}[width=1\textwidth]{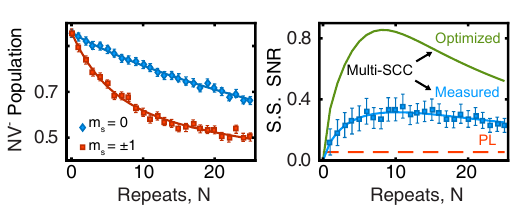}
\end{overpic}
\end{minipage}
\begin{minipage}[h]{0.44\textwidth}
\begin{overpic}[width=1\textwidth]{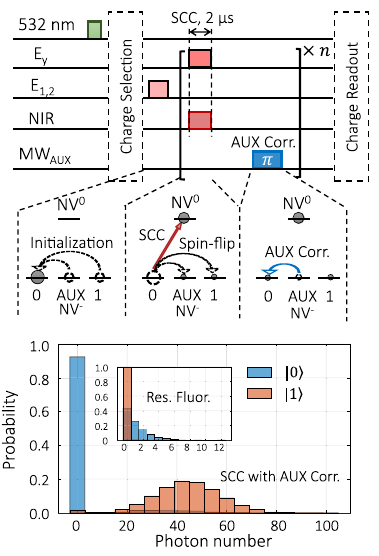}
\put (-72, 97) {(a)}
\put (-72, 49) {(b)}
\put (-72, 30) {(c)}
\put (-2, 97) {(d)}
\put (0, 37) {(e)}
\end{overpic}
\end{minipage}
\caption[NV Center]{\SCC techniques. (a) Schematics illustrating the SCC method at ambient temperature. The left showcases the singlet-SCC, while the right presents the triplet-SCC. Solid lines depict the laser pump, and the dashed lines indicate decay transitions. Adapted from \cite{hopper2018spin}. (b) Control protocol for the SCC. (c) On the left, SCC measurement results are contingent on the initial electron spin state. On the right, a comparison of the signal-to-noise ratio (SNR) between SCC method and conventional fluorescence measurements across repeat times. Figures (b,c) adapted from \cite{Hopper2016}. (d) Pulse sequence and diagram depicting the SCC readout method under low temperatures where resonant transitions are feasible. (e) Distribution of photon numbers in charge readout using the SCC method, acquired from $2\times 10^4$ measurement repetitions with NV spin initially prepared in the $m_s=0$ (blue) and $m_s=1$ (orange) states. Inset: Photon distribution for the resonant fluorescence method under helium temperature.}  \label{fig:SCC}
\end{figure*}

The \scc process stands as another pivotal element in the \scc readout. Three methods have been devised to achieve \scc, each playing a distinct role. The first two methods are executed at ambient temperatures and rely on the spin-dependent shelving process into the singlet metastable state \cite{shields2015Efficient,Hopper2016}. In this process, a pump laser elevates the \NV center from its ground to the excited state. The spin initially in the $m_s=\pm 1$ state shelves into the metastable singlet level through an \isc process, exhibiting a probability approximately an order higher than the $m_s=0$ branch \cite{goldman2015Stateselective}. Subsequently, a high-power laser ionizes either the singlet or triplet populations (Fig. \ref{fig:SCC}(a,b)), employing wavelengths between 900 and 1024 nm for the singlet absorption band or 500 and 637 nm for the triplet absorption band, respectively. The third method operates under low temperatures, as illustrated in Fig. \ref{fig:SCC}(d). Here, a resonant laser pulse selectively excites the $m_s=0$ state, while simultaneously, a high-intensity laser facilitates ionization from the excited state of \NVm to \NVo \cite{irber2021Robust,zhang2021Highfidelity}. 

The conventional fluorescent readout method typically yields a low spin readout fidelity ($F_{\text{read}}$) ranging between 0.03 to 0.05, with a signal accumulation time ($T_{\text{read}}$) of 0.2 to 3 $\mu$s. Ambient temperature SCC methods generally elevate $F_{\text{read}}$ to a range of 0.2 to 0.4, albeit at the expense of longer $T_{\text{read}}$ exceeding 100 $\mu$s \cite{shields2015Efficient,Hopper2016,hopper2018amplified,jaskula2019Improved,hopper2020RealTime}. Consequently, the enhancement of SCC on sensitivity diminishes when the sensing time $T_{\text{accu}}$ is too brief. This trade-off between factors makes SCC more suited for sensing tasks with $T_{\text{accu}}$ greater than 100 $\mu$s, typically resulting in a 2 to 5-fold sensitivity increase (Fig. \ref{fig:SCC}(c)). Notably, SCC exhibits distinct behavior under low-temperature conditions, achieving fidelity $F_{\text{read}}$ surpassing 0.9 with a readout time $T_{\text{read}}$ ranging from 30 to 50 $\mu$s (Fig. \ref{fig:SCC}(e)) \cite{zhang2021Highfidelity}.

A prominent advantage of \scc lies in its lack of specific requirements regarding the magnitude of the magnetic field. In contrast, nuclear-assisted repetitive readout often necessitates operating at magnetic fields exceeding 0.2 T to suppress optical perturbations on the nuclear spin. This magnetic field flexibility renders \scc highly compatible with NV-based low-field quantum sensing techniques like zero-field magnetic resonance (sec. \ref{sec:zerofield}). However, \scc utilizes a relatively high-intensity ionization laser. For instance, \cite{hopper2018amplified} employs 592 nm laser pulses (50 ns) with a peak power of 30 mW. Intense illumination can trigger photochemical reactions on the sample, leading to perturbations or potential photodamage \cite{shi2018SingleDNA}. One potential solution lies in altering the wavelength of the ionization laser. As the excited state of \NVm resides 0.67 eV below the diamond conduction band bottom \cite{aslam2013Photoinduced}, the ionization wavelength can be chosen from a wide range \cite{razinkovas2021Photoionization}. Transitioning to near-infrared ionization lasers still demands high-intensity \cite{Hopper2016,zhang2021Highfidelity}, yet the lower photon energy might mitigate the excitation of sample molecule transitions, warranting further exploration. 

While the efficacy of the \scc technique on NV centers in bulk diamonds and nanoparticles has been well-demonstrated \cite{hopper2018amplified}, there remains a gap in experimental investigations assessing its effectiveness on NV centers situated at depths shallower than 10 nm. Band bending and charge traps near the diamond surface \cite{dhomkar2018Charge,bluvstein2019identifying} could potentially influence the photostability of the NV charge state and the readout fidelity of SCC. Hence, it becomes critical to enhance the charge environment surrounding the NV center through appropriate doping and surface modification (sec. \ref{sec:surface_modification}). Such improvements not only aid in suppressing electric noise but also prolong the NV coherence time.

Beyond the \scc method, the NV spin state can also be read using photoelectric measurement. Direct electrical readout of the charge state is possible. \EDMR of a single electron spin was initially demonstrated on the quantum dot and silicon field-effect transistor \cite{Elzerman2004,Xiao2004}. However, these EDMR experiments demand extremely low temperatures and ultra-high magnetic fields to ensure that the Zeeman splitting exceeds the thermal broadening ($\gamma_e B_0\hbar>k_BT$). Hybrid optical-electrical detection exploits photoionization to relax stringent temperature and magnetic field requirements, proving successful in single quantum sensors \cite{yin2013Optical,Zhang2019}. Enabling photoelectric readout of NV centers has extended this method to liquid nitrogen temperatures \cite{brenneis2015Ultrafast} and even ambient conditions \cite{Bourgeois2015,Gulka2017,Hrubesch2017,siyushev2019photoelectrical,Morishita2020,Gulka2021,zheng2022Electrical}.

Photoionization of the NV center occurs through its photoexcited triplet state, enabling the detection of single NV centers via photocurrent measurement \cite{siyushev2019photoelectrical}. Moreover, spin-selective shelving into the singlet state shields the NV center from photoionization, rendering the photoelectric readout method suitable for discerning its spin states. Recent advancements have even applied this technique to detect single nuclear spins \cite{Gulka2021}, with further comprehensive discussions provided in \cite{bourgeois2020Photoelectric,Bourgeois2021}. This approach presents an enticing opportunity for integrating scalable quantum sensor arrays onto high-throughput test chips. However, the collected photocurrent is ultimately constrained by the carrier recombination lifetime of the diamond material, which is several orders of magnitude shorter than the NV excited state lifetime, thus limiting the photon counting rate. Despite the potential for the number of detected electrons and holes generated by the NV center to exceed the count of detected fluorescence photons, offering the possibility of higher fidelity spin state readouts and improved sensitivity of quantum sensors, significant efforts are essential to suppress background currents from other impurities, enhance carrier collection and detection, and address various technical hurdles. While deploying photoelectric detection in a biocompatible solution environment poses challenges, its utility remains promising for single-molecule structure analysis at cryogenic temperatures.

\subsection{Decoherence mechanism} \label{sec:decoherence}

Bringing the NV center closer to the diamond surface can greatly enhance both the sensitivity of spin detection and spatial resolution. However, shallow NV centers exhibit significantly reduced decoherence times compared to bulk NV centers \cite{myers2014probing,rosskopf2014investigation,Romach2015}, highlighting the pivotal role of the surface in shallow NV center decoherence. The electromagnetic noise emanating from the surface appears to contribute to the decoherence of NV electron spins. As illustrated in \eq \ref{eqn:NV_ham}, the NV center is susceptible to both electric and magnetic fields. This susceptibility can be quantified by a simplified form of the NV ground-state spin Hamiltonian
\begin{align}\label{Ham:decoherence}
H_{\mathrm{NV}}= & \left(\mathcal{D}+d_{\parallel} (E_{z}+\delta_{z})\right) S_{z}^{2} +d_{\perp}(E_y+\delta_y)\qty(S_{x}S_{y}+S_{y}S_{x}) \nonumber \\
 & +d_{\perp}(E_x+\delta_x)\qty(S_{y}^2-S_{x}^2) +\gNV  \vb{B}_0 \cdot \vb{S} . 
\end{align}
Accordingly, three factors significantly affect the coherence time of NV centers: the intrinsic noise of bulk diamond, noise caused by magnetic fields on the surface of diamond, and noise generated by the surface electric field. The spin coherence time of NV center is thus expressed as follows:
\begin{align}
	\frac{1}{T_2^{(*)}}\approx & \frac{1}{T_2^{(*)}\{\text{noise in bulk diamond}\}} +\frac{1}{T_2^{(*)}\{\text{unknown}\}} \nonumber\\
	& +\frac{1}{T_2^{(*)}\{\text{surface magnetic field noise}\}}\nonumber\\
	& +\frac{1}{T_2^{(*)}\{\text{surface electric field noise}\}}.
\end{align}

The noise of the single NV center electronic spin from the bulk diamond mainly consists of electronic spin bath noise, nuclear spin bath noise, spin-lattice relaxation \Tone, and certain other unknown noise \cite{Barry2020Sensitivity}. The electric spin bath consists of paramagnetic defects in the diamond. Moreover, the paramagnetic substitutional nitrogen defects N$_{\text S}^0$ ($S$=1/2), known as P1 centers, may account for the majority of the electronic spin-bath decoherence \cite{bauch2020Decoherence}. The $\mathrm{N}_{\mathrm{S}}^0$ contribution to decoherence obeys $1/T_2^*\left\{\mathrm{N}_{\mathrm{S}}^0\right\}=A_{\mathrm{N}_{\mathrm{S}}^0}\left[\mathrm{N}_{\mathrm{S}}^0\right]$ and $1/T_2\left\{\mathrm{N}_{\mathrm{S}}^0\right\}=B_{\mathrm{N}_{\mathrm{S}}^0}\left[\mathrm{N}_{\mathrm{S}}^0\right]$, where $\mathrm{N}_{\mathrm{S}}^0$ is the concentration of neutral substitutional nitrogen, and $A_{\mathrm{N}_{\mathrm{S}}^0}$ and $B_{\mathrm{N}_{\mathrm{S}}^0}$ characterizes the effective magnetic coupling strength between the NV center and $\mathrm{N}_{\mathrm{S}}^0$ paramagnetic defects. Herein, $A_{\mathrm{N}_{\mathrm{S}}^0}=101 \pm 12 \mathrm{~ms}^{-1} \mathrm{ppm}^{-1}$ and $B_{\mathrm{N}_{\mathrm{S}}^0}=6.25 \pm 0.47 \mathrm{~ms}^{-1} \mathrm{ppm}^{-1}$ lie in the range $\left[\mathrm{N}_{\mathrm{S}}^0\right]=0.75-60 \mathrm{~ppm}$ \cite{bauch2020Decoherence}.

The magnetic noise is conventionally attributed to a monolayer of $S=1/2$ spins on the diamond surface, such as dangling bonds \cite{tisler2009Fluorescence,samsonenko1979Characteristic}, terminating surface atoms \cite{osipov2009Exchange,mcguinness2013Ambient}, or adsorbed molecules like paramagnetic oxygen \cite{bansal1972Kinetics}. This magnetic interference is typically well characterized by a Lorentzian spectral density centered at zero frequency, typically described phenomenologically as \cite{Romach2015}:
\begin{equation}
S(\omega)=\sum_{i} \frac{\gamma_{\text{NV}}^2B_{i}^{2}}{\pi} \cdot \frac{\tau_{c,i}}{1+\left(\omega \tau_{c,i}\right)^{2}},
\end{equation}\label{eqn:noise_PSD}
where $B_i$ is the average magnetic noise strength with different origins, $\tau_{c,i}$ is the corresponding correlation time of the magnetic noise, and the index $i$ indicates a different noise source. For a \td layer of surface with $g$-factor $g=2$ and $S=1/2$ electron spins, by integrating over a surface with uniform spin distribution $\sigma_{\mathrm{surf}}$, the total mean square noise field is calculated as \cite{myers2014probing}
\begin{equation}
B_{\mathrm{rms}}^{2}=\left(\frac{ \mu_{0} \gamma_e\hbar}{4 \pi}\right)^{2} \frac{\pi}{4} \frac{\sigma_{\text {surf }}}{d^{4}},
\end{equation}
where $d$ is the depth of the \NV center. The surface spin density is characterized by different methods, including magnetic noise spectrum estimation \cite{myers2014probing,rosskopf2014investigation,bluvstein2019Extending} and magnetic resonance imaging \cite{luan2015Decoherencea,grinolds2014Subnanometre}. This parameter varies across diamond surfaces subject to different treatments, as presented in Table \ref{tab:surface}. 

The near-surface \NV decoherence is not only dominated by the magnetic noise but also affected by the electric field noise of the surface charge fluctuations. The Hamiltonian Eq. \ref{Ham:decoherence} yields the coupling between the \NV electric dipole $d_{\parallel},d_{\perp}$ and the effective electric field $\boldsymbol{\Pi}=(\mathbf{E}+\boldsymbol{\delta})$, which comprise the electric field $\mathbf{E}$ and strain term $\boldsymbol{\delta}$. When the magnetic field is sufficiently high, satisfying $\gamma_e  B_{0} \gg d_{\perp} \Pi_{\perp} $, where $\Pi_{\perp}$ is the transverse component of $\boldsymbol{\Pi}$, the single and double quantum transition frequencies are given by \cite{bluvstein2019Extending}:
\begin{align}
& f_{0 \rightarrow \pm 1}  \approx D+\frac{d_{\|} \Pi_{\|}}{2\pi} \pm\left(\frac{\gNV}{2 \pi} B_{z}+\frac{1}{4\pi} \frac{\left(d_{\perp} \Pi_{\perp} \right)^{2}}{\gNV  B_{z}}\right)
\\
& f_{-1 \rightarrow+1}  \approx 2\left(\frac{\gNV}{2 \pi} B_{z}+\frac{1}{4\pi} \frac{\left(d_{\perp} \Pi_{\perp} \right)^{2}}{\gNV  B_{z}}\right). \label{eqn:dq_ele}
\end{align}
This effect has been demonstrated through the application of high-dielectric-constant liquids onto the diamond surface \cite{Kim2015}. When contemplating a point charge $q$ positioned on the diamond surface, the electric field experienced at the NV center is influenced by the dielectric constants of both the diamond and the liquid \cite{Kim2015}, which is given by:
\begin{equation}
\mathbf{E}=\frac{1}{4 \pi \varepsilon_{0}} \frac{2}{\kappa_{d}+\kappa_{\mathrm{ext}}} \frac{q}{r^{2}} \hat{r},
\end{equation} 
where $r$ is the distance between the charge and NV, $\hat{r}$ the unit direction from NV to the point charge, $\kappa_{d}$ is the dielectric constant of diamond, $\kappa_{\mathrm{ext}}$ is the dielectric constant of the external medium, and $\varepsilon_{0}$ is the vacuum permittivity. After being covered by a dielectric medium, the electrical field on NV reduces by $\left(\kappa_{d}+1\right) /\left(\kappa_{d}+\kappa_{\mathrm{ext}}\right)$ compared to the air. 
According to \cite{Kim2015}, glycerol (dielectric constant $\kappa_{G}=42$) elongates the $T_2$ time by 4.6 times, and propylene carbonate ($\kappa_{\mathrm{PC}}=64$) elongates it by 2.4 times, thereby greatly enhancing the magnetic field sensitivity (Eq. \ref{equ:mag_sensitivity_fluc}, \ref{equ:mag_sensitivity_pola}). However, this liquid-based enhancement diminishes over several hours due to the formation of a layer with reduced mobility, thereby reducing the effective dielectric constant \cite{xu1998Scanning,capponi2010Structural}.

\begin{table*}
\begin{center}
\setlength{\tabcolsep}{3pt}
    \begin{tabular}[t]{cccp{8cm}}
    \toprule
    \toprule
    Reference  & \parbox{7.5em}{Spin density\\ ($\mu_B$/nm$^2$)} & \parbox{7.5em}{Charge density \\ ($e$/nm$^2$)} & Surface treatment  \\  \midrule
   \cite{myers2014probing}  & 0.04 & - &  PE-CVD grown diamond, delta-doped \Nfif \/ atoms with  $8.6\times 10^{15}$ /cm$^3$, the diamond surface was oxygen terminated in an acid process in 190 $^\circ$C a 1:1:1 mixture of sulfuric, nitric, and perchloric acid for 30 min.  \\ 
    \cite{rosskopf2014investigation}  & 0.1 & - &  MPCVD grown diamond layer on HPHT substrate, delta-doped \Nfif \/ atoms with  $8.6\times 10^{18}$ /cm$^3$, oxygen terminated with three acids.  \\  
                                      & 0.047 & - &  (100)-oriented single crystal, implanted with \Nfif$^{+}$ or \Nfif$^{2+}$ ions at a series of extremely low energies (0.4 – 5 keV) and fluences (10$^{10}$ – 10$^{14}$ /cm$^2$), hydrogen-terminated.   \\ 
                                      & 0.04 & - &  Same as above, except for oxygen terminated with three acids.  \\
                                      & 0.013 & - &  Same as above, except for fluorine-terminated.  \\
    \cite{grinolds2014Subnanometre}  & 0.5 & - & (100)-oriented single crystal, implanted with \Nfor \/ at 6 keV and 3$\times 10^{11}$ /cm$^2$), annealed 2 hours at 800 $^\circ$C in vacuum.  \\ 
    \cite{luan2015Decoherencea}  & 0.28 & - &  Single crystal diamond, implanted with \Nfor \/ at 6 keV and 3$\times 10^{11}$ /cm$^2$), annealed 2 hours at 800 $^\circ$C in vacuum, cleaned in boiling acid mixture of 1:1:1 sulfuric, nitric and perchloric acid and subsequently thoroughly rinsed in distilled water.   \\ 
    \cite{Myers2017}  & - & 1.9 & Single crystal diamond, polished and subsequently etched by ArCl$_2$, implanted with \Nfor \/ at 4 keV and 3$\times 10^{11}$ /cm$^2$) on a CVD grown layer, annealed 2.5 hours at 850 $^\circ$C in vacuum, cleaned in boiling acid mixture of 1:1:1 sulfuric, nitric and perchloric acid and subsequently thoroughly rinsed in distilled water.     \\ 
    \cite{bluvstein2019Extending}  & 0.01-0.01 & - & Single crystal diamond, polished and subsequently etched by ArCl$_2$, implanted with \Nfor \/ at 4 keV and 5.2$\times 10^{10}$ /cm$^2$) on a CVD grown layer, annealed 2.5 hours at 850 $^\circ$C in vacuum, cleaned in 230-240 $^{\circ}$C acid mixture of 1:1:1 sulfuric, nitric and perchloric acid.  \\ 
    \cite{stacey2019Evidence}  & - & 0.001-0.1 & Theory   \\ 
   \bottomrule \bottomrule
   \end{tabular}
\end{center}
\caption{The surface electron spin density of various diamond surfaces subjected to different treatments.. }
\label{tab:surface}
\end{table*}


\subsubsection{Surface treatment}\label{sec:surf_tr}

\begin{figure}[htp]
\begin{overpic}[width=0.83\columnwidth]{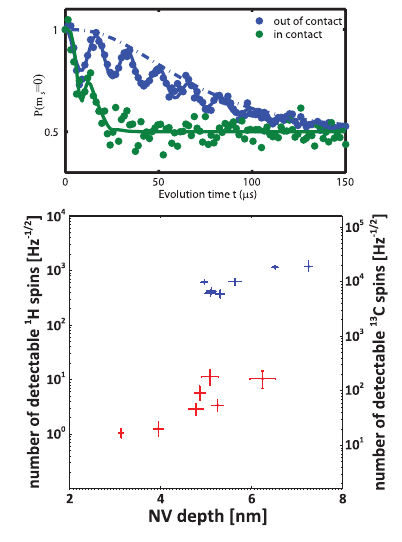}
	\put (-3, 95) {(a)}
	\put (-3, 60) {(b)}
\end{overpic}
\caption[NV Center]{ (a) Surface electron detection utilizing the NV sensor. The green line represents the coherence decay curve observed when the NV sensor is positioned close to another diamond surface, while the blue line shows the decay curve when the NV sensor is positioned farther away from the diamond surface. Oscillations in the curves stem from the \Cthir nuclear spin bath within the diamond. Adapted from \cite{luan2015Decoherencea}. (b) Comparison of nuclear spin sensitivity (\Hone and \Cthir nuclear spins) against NV center depth before (blue) and after (red) surface treatment.  Adapted from \cite{Lovchinsky2016} }  \label{fig:surface_remove_decoherence}
\end{figure}

Several methods are available to extend the coherence time of NV center in \sm experiments. The proximity of the NV sensor to the surface significantly reduces the coherence time compared to when it is far away from the surface (Fig. \ref{fig:surface_remove_decoherence}(a)). The most straightforward way is to remove the possible spins, charges, or strain \cite{chu2014Coherent} of the surface. 
Several common surface treatment methods are presented below:

\begin{enumerate}[label=(\roman*)]
\item \emph{Preparation } The diamond is initially a polished single crystal diamond. Subsequently, the diamond is etched using an Ar-Cl$_2$ plasma, followed by an O$_2$ plasma etch, to clean the diamond surface \cite{sushkov2014Magnetic};
\item \emph{Ion implantation } The diamond is  implanted by nitrogen ions with different dosages and energies. The relationship between  created NV depth with implanted energies is discussed in sec. \ref{sec:NV_creation} in details;
\item \emph{Annealing } The implanted diamond is annealed in vacuum. The temperature change steeply from 400 $^{\circ}$C to 800 $^{\circ}$C or 1200 $^{\circ}$C under high vacuum, $P\lesssim 10^{-6}$ Torr \cite{sushkov2014Magnetic,sushkov2014alloptical,Lovchinsky2016,Lovchinsky2017magnetic};
\item \emph{Acid clean I } A  mixture of three-acid (1:1:1 H$_2$SO$_4$/HNO$_3$/HClO$_4$) treatment on the diamond at 180 $^{\circ}$C to boiling for several hours \cite{muller2014Nuclear,sushkov2014Magnetic,Lovchinsky2017magnetic,schlipf2017molecular,pinto2020Readout};
\item \emph{Oxygen annealing } The diamond is annealed in an oxygen environment for several hours at 465 $^{\circ}$C to make an oxygen-terminated surface \cite{sushkov2014Magnetic,Lovchinsky2016,Lovchinsky2017magnetic};
\item \emph{Acid clean II } Piranha solution (sulfuric acid (98 \%) and hydrogen peroxide (30 \%), volume ratio 3:1) treatment on diamond at 75 $^{\circ}$C for several hours \cite{sushkov2014Magnetic,shi2015Singleprotein,Lovchinsky2016,Lovchinsky2017magnetic}.
\end{enumerate} 
The sensitivity is largely enhanced with procedures (i) -- (vi) compared to procedures (i) -- (iv) as shown in Fig. \ref{fig:surface_remove_decoherence}(b).

In addition to the diamond surface, the layer near the diamond surface holds significant influence over both the stability of charge states and the decoherence time of NV centers. Following growth or ion implantation, this shallow layer often hosts various defects alongside NV centers. For instance, implanted defects, post-annealing, give rise to di-vacancy ($V_2$) complexes or higher-order vacancy chains. Post low-energy nitrogen implantation and thermal annealing, it is estimated that within a nanometer-scale volume, approximately 4 to 6 $V_2$ complexes may surround a single NV center \cite{Oliveira2017}. These complexes demonstrate thermal stability \cite{Yamamoto2013} and exhibit electronic paramagnetism with a spin of $I=1$ \cite{Twitchen1999b}. The presence of these $V_2$ electron spins could notably contribute to decoherence \cite{Hanson2008}. Strategies aimed at curbing the formation of such defects during thermal annealing serve as pivotal steps toward enhancing the quantum properties of NV centers.

During thermal annealing, individually charging vacancies within the defect cluster has been shown to limit vacancy recombination \cite{Oliveira2017}. An alternate method involves nitrogen ion implantation into a space charge layer of free carriers, specifically holes, generated by a thin layer of boron-doped diamond on the diamond surface. This process yields considerable enhancements in various NV center properties: a tenfold increase in \Ttwo time, an extension of \Tone time to over 5 ms, and a twofold improvement in the yield of NV centers situated at depths of 2--8 nm \cite{Oliveira2017}. Additionally, incorporating various donors such as phosphorus, oxygen, and sulfur during the \cvd diamond growth process has been explored \cite{luhmann2019Coulombdriven}. This method utilizes negatively charged vacancies during annealing, while the introduced nitrogen N$^+$ becomes positively charged. This interplay may bolster NV center formation, potentially impeding V--V formation through Coulomb forces. This approach has exhibited a tenfold increase in NV center yield \cite{luhmann2019Coulombdriven}, significantly enhancing charge stability and electron spin coherence time for NV centers located as deep as 15 nm \cite{watanabe2021Shallow}.

\subsubsection{Quantum control for sensor} \label{sec:decoherence:sensor_control}

\begin{figure}[htp]
\begin{overpic}[width=0.9\columnwidth]{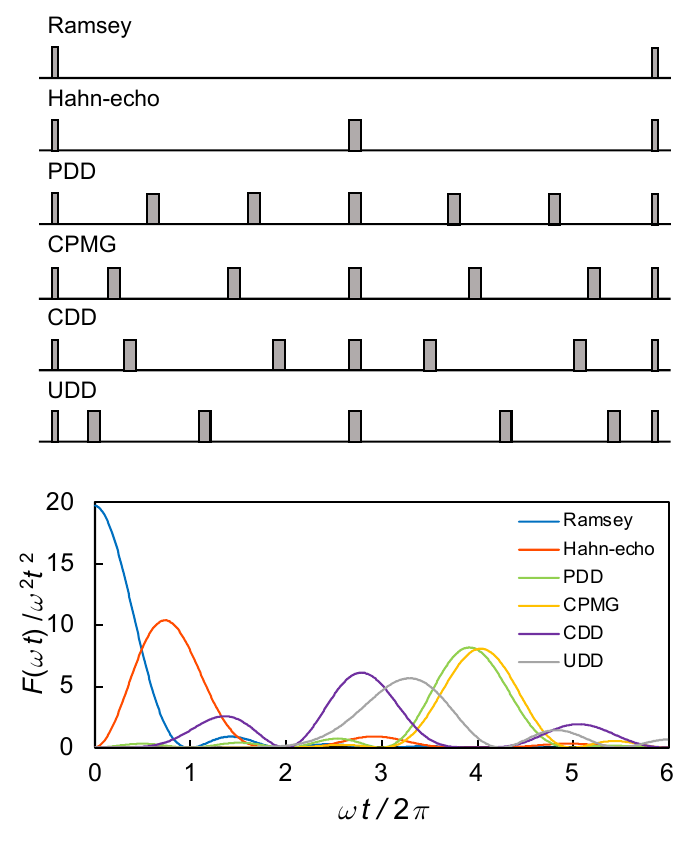}
\put (-2, 97) {(a)}
\put (-2, 42) {(b)}
\end{overpic}
\caption[NV Center]{Dynamical decoupling. (a) The control pulse sequences for different control methods. (b) The filter function for each control method, respectively. The details for the sequence are listed in Tab. \ref{tab:dd_sequence}.}  \label{fig:dd_sequence}
\end{figure}

Passive surface treatment is not the only strategy available to mitigate decoherence noise; active quantum control methods are available, which can further improve coherence time. The effect of fluctuating fields (Eq. \ref{eqn:PSD_define}) can be reduced by dynamical control. 
In particular, the power spectral density $S(\omega)$ can be used to characterize the magnetic signal generated by magnetic fluctuations $\Delta B$, where
\begin{equation}
\left\langle\Delta B(0)\Delta B(t)\right\rangle=\frac{1}{2\pi\gamma_e^2}\int_{-\infty}^{\infty} S(\omega) e^{{i\omega t}}\dd \omega.
\end{equation}
Under the influence of magnetic fluctuation noise, the NV coherence (Eq. \ref{eqn:decoherence_define}) is calculated as:
\begin{equation}
\xi(t)=\exp\left( - \frac{1}{2\pi} \int_{-\infty}^{\infty} \dd \omega S(\omega) \frac{F_{t}(\omega)}{\omega^{2}}    \right), \label{eqn:deocherence}
\end{equation}
where $F_{t}(\omega)$ is the filter function as in Table \ref{tab:dd_sequence} and Fig. \ref{fig:dd_sequence}.

\begin{table*}
\newcommand{\tabincell}[2]{\begin{tabular}{@{}#1@{}}#2\end{tabular}}
\centering
\begin{center}
\setlength{\tabcolsep}{10pt}
    \begin{tabular}{cccl}
    \toprule
    \toprule
    Pulse sequence  & Microwave pulse & Filter function  & Frequency range  \\  \midrule
    $T_1$  & $\displaystyle(\pi)-{t}$ & $2\delta(\omega-\omega_0)^2\omega^2t/\pi$ & $\sim$ GHz \\[0.1in]
    \tabincell{c}{Ramsey } & $\displaystyle\frac{\pi}{2}-{t}-\frac{\pi}{2}$ &  $\text{sin}^2 \frac{\omega t}{2}$ & $\sim$ d.c. \\[0.1in]
    \tabincell{c}{Hahn-echo }  & $\displaystyle\frac{\pi}{2}-\frac{{t}}{2}-\pi-\frac{{t}}{2}-\frac{\pi}{2}$ & $\displaystyle 8 \sin ^{4} \frac{\omega t}{4}$ & $\sim$ MHz \\[0.1in]
    \tabincell{c}{PDD (odd $N$) } & $\displaystyle\frac{\pi}{2}-\frac{{t}}{N}-(\pi-\frac{{t}}{N})^N-\frac{\pi}{2}$ & $\displaystyle 2 \tan ^{2} \frac{\omega t}{2 N+2} \sin ^{2} \frac{\omega t}{2}$ & $\sim$ MHz \\[0.1in]
    \tabincell{c}{CPMG (even $N$) }  & $ \displaystyle\frac{\pi}{2}_x-(\frac{{t}}{2N}-\pi_y-\frac{{t}}{2N})^N-\frac{\pi}{2}_x$ & $ \displaystyle 8 \sin ^{4} \frac{\omega t}{4 N} \sin ^{2} \frac{\omega t}{2} / \cos ^{2} \frac{\omega t}{2 N} $ & $\sim$ MHz \\[0.1in]
    \tabincell{c}{CDD } & $\displaystyle \mathrm{CDD}_{l-1}\left(\frac{{t}}{2}\right)-\pi-\mathrm{CDD}_{l-1}\left(\frac{{t}}{2}\right)-\pi$ & $\displaystyle 2^{2 l+1} \sin ^{2} \frac{\omega t}{2^{l+1}} \prod_{k=1}^{l} \sin ^{2} \frac{\omega t}{2^{k+1}}$ & $\sim$ MHz \\[0.1in]
    \tabincell{c}{UDD }  & $\displaystyle\frac{\pi}{2}-{t}_1-\pi-\ldots-{t}_N-\frac{\pi}{2}$ & $\displaystyle \frac{1}{2}\left|\sum_{k=-N-1}^{N}(-1)^{k} \exp \left[\frac{i \omega t}{2} \cos \frac{\pi k}{N+1}\right]\right|^{2}$ & $\sim$ MHz \\[0.1in]
    Spin-lock  & $\displaystyle\frac{\pi}{2}_x-{\text{MW}}({t})_y-\frac{\pi}{2}_x$ & $2\delta(\omega-\omega_0)^2\omega^2t/\pi$ & $\sim$ MHz \\[0.1in]
   \bottomrule \bottomrule
   \end{tabular}
\end{center}
\caption{The microwave pulse sequence, filter function, and the detection frequency range of various quantum control  sequences \cite{cywinski2008How}. Some of the filter functions are plotted in Fig. \ref{fig:dd_sequence}. }
\label{tab:dd_sequence}
\end{table*}

\begin{figure}[htp]
\begin{overpic}[width=1\columnwidth]{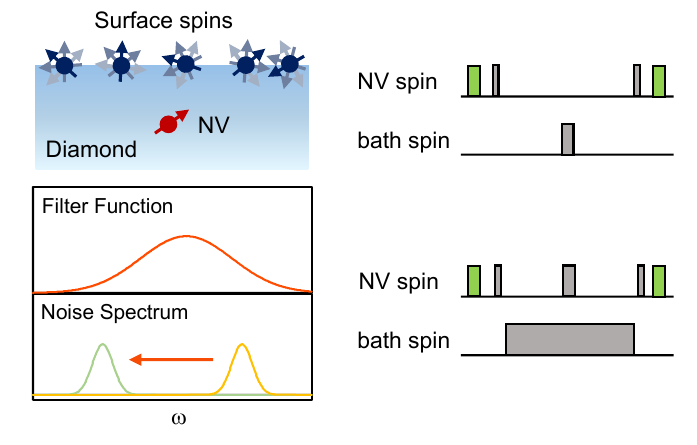}
\put (-2.5, 60) {(a)}
\put (-2.5, 35) {(b)}
\put (46, 60) {(c)}
\put (46, 30) {(d)}
\end{overpic}
\caption[NV Center]{ Coherence elongation by modulating the surface spins. (a) The shallow NV center is affected by the fluctuating field from the surface electronic spin bath, which can cause decoherence and depolarization of the shallow NV center. (b) Modulating the electronic spin bath alters the noise spectrum, creating a mismatch with the filter function, thereby elongating the decoherence time of NV centers. (c) The control sequence of pulsed control of the bath spin. (d) The control sequence of continuous driving of the bath spin.
 }  \label{fig:driving_bath}
\end{figure}
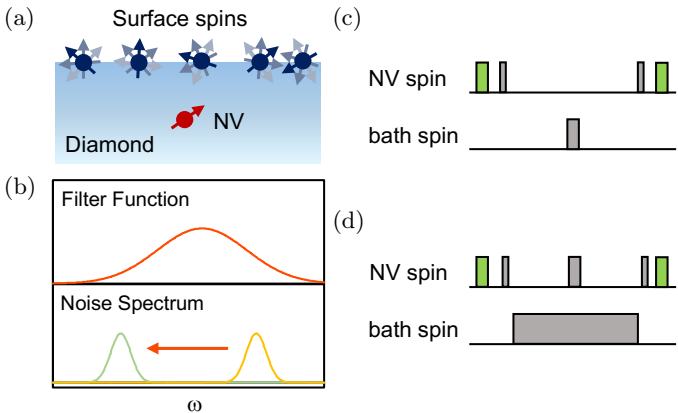

Magnetic noise, stemming from residual paramagnetic impurities (Fig. \ref{fig:driving_bath}(a)) within bulk diamond and on its surface \cite{romach2019measuring,myers2014probing,rosskopf2014investigation,grinolds2014Subnanometre}, significantly contributes to the decoherence of NV centers. Typically employed for d.c. or low-frequency magnetic field measurements, the Ramsey fringe technique exhibits a dephasing time \Ttwostar at the $\sim\mathrm{\mu}$s level. In contrast, the Hahn echo protocol, used to detect a.c. magnetic fields, differs by introducing an additional microwave $\pi$ pulse. This echo sequence remains insensitive to static magnetic fields, as the phase before and after the $\pi$ pulse nullifies each other. The coherence time $T_2$ typically exceeds \Ttwostar by one to two orders of magnitude. 

However, the response of Hahn echo measurements proves sensitive to a.c. magnetic fields, particularly at certain frequencies. The Hahn echo sequence can be expanded into multipulse sequences like periodical dynamical decoupling (PDD), \cpmg, concatenated dynamical decoupling (CDD), and \udd, as detailed in Table \ref{tab:dd_sequence} and depicted in Fig. \ref{fig:dd_sequence}.
By inserting a series of equally spaced $\pi$ pulses into the sequence, the coherence time $T_2$ extends by flipping the NV electron spin, effectively decoupling it from environmental noise. Alternatively, considering an increase in the number of $\pi$ pulses, the filter function (outlined in Table \ref{tab:dd_sequence}) narrows, resulting in lower integrated noise levels (Eq. \ref{eqn:deocherence}). An astute choice of microwave $\pi$ pulse phases further mitigates the impact of non-ideal controls \cite{deLange2010,Wang2012,wang2019Randomizationa}. CPMG, XY8-N, as a phase-modulated pulse sequence, involves eight repeated $\pi$ pulses in the pattern \texttt{XYXYYXYX} (\texttt{X} and \texttt{Y} representing $\pi$ pulses along the $x$ or $y$ axis). This sequence is robust against errors in both pulse duration and phase. In bulk diamond, the longest decoherence time achieved for an NV center using DD stands at 3.3 ms with 512 $\pi$ pulses \cite{herbschleb2019Ultralong}. However, this coherence time remains about half of the $T_1$ time, despite the theoretical limit for coherence time being $2T_1$. This discrepancy might be due to high-frequency noise surpassing the maximum filter frequency constrained by the duration of the $\pi$ pulses \cite{herbschleb2019Ultralong}. Unfortunately, for shallow NV centers, the coherence time is approximately an order of magnitude smaller than the \Tone time \cite{myers2014probing,Romach2015}.

The dynamical decoupling approach aims to actively control NV spins to counter environmental noise. However, its compatibility with sensing operations poses limitations, especially concerning the resonance frequencies of paramagnetic impurities near the surface, which differ from the NV electron spin. Manipulating the noise spectrum itself within the environment (Fig. \ref{fig:driving_bath}(a)) presents an alternative approach to counter NV spin decoherence (Fig. \ref{fig:driving_bath}(b)).

To tackle this, various methods are employed, including pulsed control and continuous driving of the bath spin. The pulsed spin control sequence (Fig. \ref{fig:driving_bath}(c)) involves a $\pi$ pulse on the bath, inserted midway through the NV Ramsey fringe sequence. Analogous to the Hahn-echo sequence on the NV, this $\pi$ pulse flips the electron spins in the bath, neutralizing low-frequency noise.
Continuous spin driving (Fig. \ref{fig:driving_bath}(d)) employs resonant microwave control, continuously driving the bath spin while the dynamical decoupling sequence acts on the NV electron spin. This method extends coherence time by shifting the noise spectrum frequency through bath spin driving. Bath control techniques have improved \Ttwostar and \Ttwo of the NV electron spin by roughly 1.5 times \cite{DeLange2012,bluvstein2019Extending}, though the extent of enhancement varies with NV depth and environmental factors. Stochastic bath driving techniques have shown promise in further extending coherence time \cite{joos2022Protecting}.

The NV center electron spin, constituting a spin-1 system, facilitates the seamless extension of control techniques from the \sq methods ($\Delta m_{s}=1$) to \dq methods ($\Delta m_{s}=2$).
In \dq methods, the impact of temperature fluctuations on noise ($\partial D / \partial T=-71.9$ kHz$\cdot$K$^{-1}$) is nullified, as evidenced in studies \cite{acosta2010temperature,lourette2023Temperature,xu2023Highprecision}. This encompasses the elimination of the axial electric field $E_{\|} $ and axial strain  $\sigma_{\|} $. Simultaneously, the influence of off-axis electric field $E_{\perp}$ and strain $\sigma_{\perp}$ is subdued by the magnetic field $B_{z}$ (Eq. \ref{eqn:dq_ele}).

Moreover, the DQ method substantially augments signal strength for magnetometry, owing to the quantum number difference being twice as large as that in the \sq method, a point validated in a study by Mamin et al. \cite{mamin2014Multipulse}. However, this enhancement comes at a cost: in environments where magnetic noise dominates the spin-bath, the $T_2^*$ and $T_2$ times in the DQ measurement approximately halve compared to the values in the SQ method, as established by Fang et al. \cite{fang2013HighSensitivity} and Mamin et al. \cite{mamin2014Multipulse}. Despite this, accounting for electric field noise and temperature fluctuations, an overall enhancement in magnetometry sensitivity is anticipated.

With the DQ method, the coherence time $T_{2}$ of a single NV center shows a marked increase, rising from $(1.66 \pm 0.16)$ ms in the SQ method to $(2.36 \pm 0.09)$ ms, thereby marking a 2.7-fold improvement \cite{angerer2015Subnanotesla}. For shallow NV centers at 12.8 (3) nm, the sensitivity of DQ exhibits a 1.8-fold increase compared to SQ, which escalates to a 2.5-fold improvement with bath driving \cite{bluvstein2019Extending}. It is imperative to note that the DQ experiment necessitates the simultaneous manipulation of two microwave pulses to drive the two subspaces of the NV ground state, rendering it notably more intricate than the SQ experiment.
Certain considerations must be taken into account during manipulation. Specifically, the separation of the two microwave frequencies is imperative to prevent mutual interference.

\subsection{Signal amplification} \label{sec:NMR:amp}

The magnetic dipolar interaction experiences a substantial decay concerning distance, approximately $\sim r^{-3}$, while the gyromagnetic ratios of nuclear spins are approximately three orders of magnitude smaller than those of electron spins. Consequently, the detection of individual nuclear spins poses a formidable challenge. To establish a coupling between the sensor and the target nuclear spin, intermediary quantum systems prove instrumental in amplifying the nuclear spin signal. These systems encompass various options such as single electron spins \cite{schaffry2011Proposed,bermudez2011ElectronMediated}, an electron spin ensemble \cite{goldstein2011environment}, a spin chain \cite{bose2003Quantum,yao2012Scalable}, and even ferromagnetic particles \cite{trifunovic2015Highefficiency}.

Presently, the sole experimentally realized protocol \cite{schaffry2011Proposed} involves leveraging single dark reporter electron spins positioned on the diamond surface \cite{sushkov2014Magnetic}. The protocol reliant on electron spin ensembles \cite{schaffry2011Proposed} poses a significant challenge due to the difficulty in acquiring multiple highly polarized electron spins that exhibit strong coupling to the NV center.

\begin{figure}[htp]
\begin{overpic}[width=1\columnwidth]{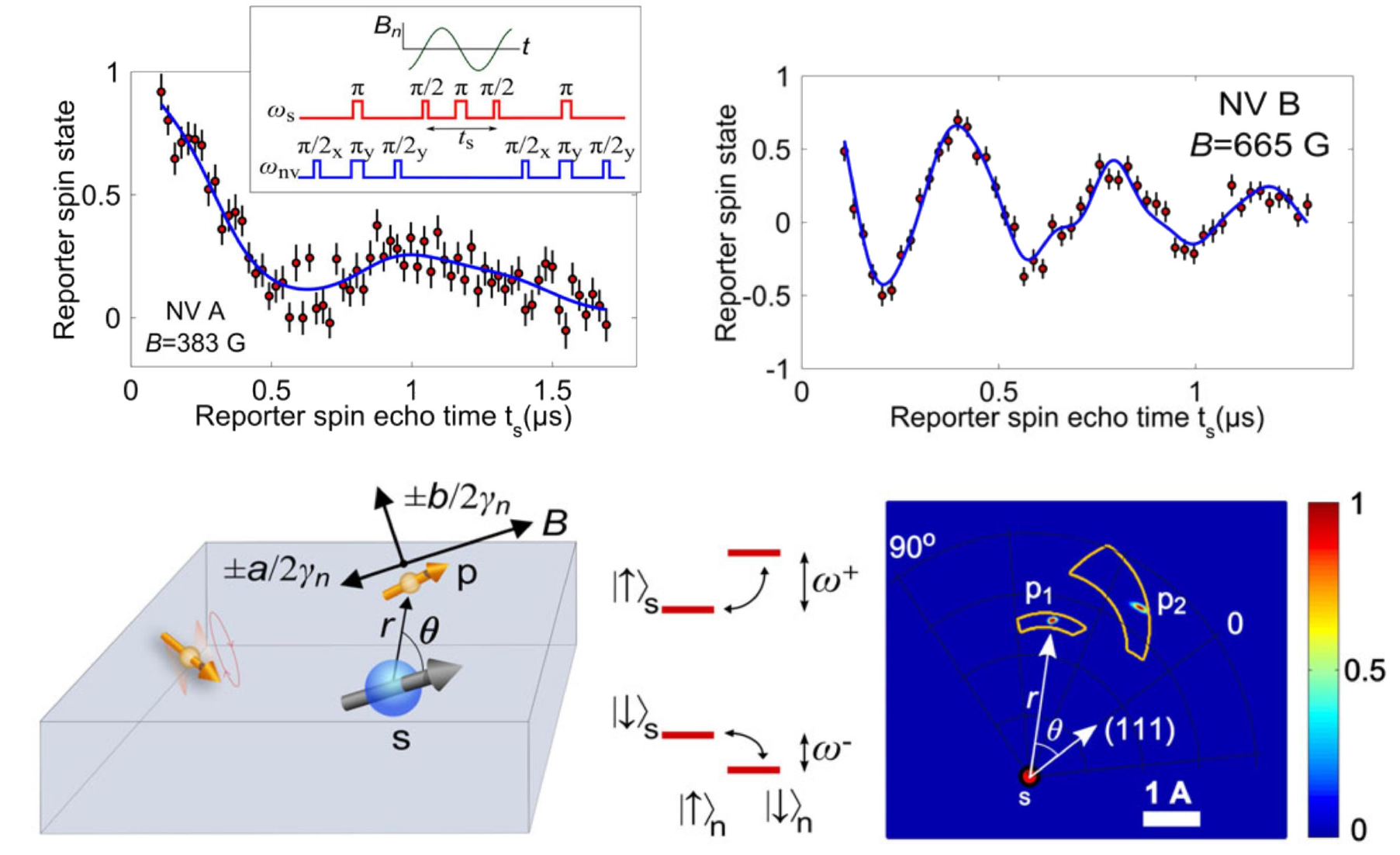}
	\put (0, 60) {(a)}
	\put (50, 60) {(b)}
	\put (0, 26) {(c)}
	\put (42, 26) {(d)}
	\put (58, 26) {(e)}
\end{overpic}
\caption[NV Center]{Single nuclear-spin detection enabled by the reporter electron spins. (a) Measurement of a nuclear spin bath through the reporter spins on the diamond surface. It exhibits the spin-echo modulation induced by a proton spin bath. Inset: the control pulse sequence applied to the reporter and the NV center. (b) The spin-echo modulation of the reporter measured with the NV center. It is estimated by the best fit that the reporter is coupled to two proton spins. (c) Schematic diagram of the hyperfine interaction between the reporter spin and the proton spins (gold arrows). (d) Energy level diagram for the coupled system of the reporter spin and proximal proton spin. (e) Localization of the two nearby proton spins with respect to the reporter spin, which is extracted from a fit to the data shown in (b). The figures are adapted from \cite{sushkov2014Magnetic}. }  \label{fig:reporter}
\end{figure}
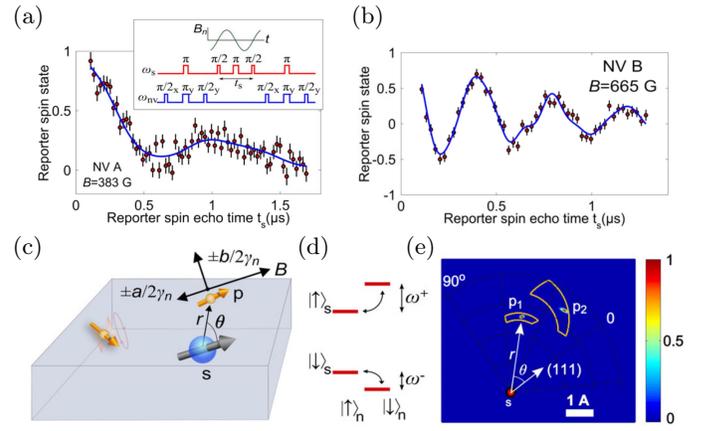

Utilizing dark reporter electron spins positioned on the diamond surface serves as a means to detect nuclear spins \cite{sushkov2014Magnetic}. The term ``reporter electron spin'' pertains to the dangling bond (refer to Table \ref{tab:surface}) on the diamond surface (depicted in Fig. \ref{fig:reporter}(a)), strategically situated in close proximity to both the NV sensor and the target nuclear spins. Employing the DEER sequence (see \se \ref{sec:deer}), the reporter electron spin can be measured and manipulated by $\pi$ pulses, inducing polarization transfer that corresponds to the DEER signal. The NV center exhibits the capability to couple with multiple reporter spins within a range of approximately $\sim$ 10 nm.

Upon subjecting the diamond to a robustly oxidizing reflux mixture of concentrated nitric, sulfuric, and perchloric acids, the DEER signal undergoes clear modification. This suggests that the surface treatment effectively ``resets'' the positions of the surface reporter spins \cite{sushkov2014Magnetic}. The longitudinal relaxation time \Tone of the reporter electron spin is approximately 29 $\mu$s. Through a Hahn-echo measurement of the reporter electron spin, the oscillation induced by a nuclear spin bath is observable in Fig. \ref{fig:reporter}(a). The measured gyromagnetic ratio aligns well with that of the proton nuclear spin, indicating the specific nuclear species. The pronounced modulation in Fig. \ref{fig:reporter}(b) implies the presence of two proton spins in close proximity to the reporter spin. The relative azimuth angles of these two nuclear spins can be further elucidated through detailed calculation, as illustrated in Fig. \ref{fig:reporter}(c, d, e).

\subsection{Conclusion} \label{sec:sensitivity:conclusion}

In conclusion, several methods have been pursued to enhance the sensitivity of the \sqs, \NV center. Notably, improvements have been witnessed in both readout fidelity and coherence time. Achieving a readout fidelity of 97.7\% with helium cryogen \cite{humphreys2018Deterministic} and 86.9\% under ambient temperature \cite{xie2021Beating} through nuclear spin-assisted repetitive readout showcases significant advancements. The latter method, adaptable to ambient conditions, maintains a fidelity of approximately 84\% even for shallow NV centers with depths as minimal as 15 nm \cite{zhao2022Subnanotesla}. However, this fidelity hinges significantly on the depolarization time of the assisted nuclear spin and the stability of the NV charge state \cite{xie2021Beating}.

Efforts involving deterministic charge state control \cite{xie2021Beating} and high magnetic fields \cite{Neumann2010} are anticipated to elevate the readout fidelities of shallow NV centers \cite{Lovchinsky2016,Lovchinsky2017magnetic} from approximately 0.2 to around 0.8 \cite{xie2021Beating}. Nevertheless, this approach faces limitations and may not be universally applicable under conditions characterized by short coherence times. In scenarios wherein $T_{\text{read}}\gg T_{\text{accu}}$, 
the scaling relationship between the spin sensitivity (Eqn. \ref{equ:spin_sensitivity_fluc} and \ref{equ:spin_sensitivity_pola}) and $T_{\text{read}}$ degenerate to 
\begin{equation}
\eta_{\text{fluc,pol}}\propto \sqrt{T_{\text{read}}+\frac{T_{\text{read}}}{n_{\text{avg}}C^2}}
\end{equation}
which will not benefit from repetitive readout. Thus, the pivotal challenge in enhancing sensitivity rests with coherence time. Even with shallow NV centers at depths below 5 nm, the current record for coherence time stands at 100 - 200 $\mu$s \cite{Lovchinsky2016,Lovchinsky2017magnetic,muller2014Nuclear}, still an order of magnitude shorter than $2T_1$ of the NV center. Despite the best $T_2$ time registering at 2.4 ms \cite{herbschleb2019Ultralong} and an optimal fluorescence count rate of approximately $\sim$ 4 Mcounts/s \cite{wang2022Selfaligned}, the achieved proton nuclear spin sensitivity by NV centers under ambient conditions remains at around $\sim$ 0.01 proton\sqrtHz within 5 nm and approximately $\sim$ 1 proton\sqrtHz within 10 nm for both fluctuation and polarization signals. This sensitivity facilitates the identification of a single nuclear spin (defined here as SNR$\geq$3) within 1 ms to 3 s, enabling the real-time observation of a single molecule's dynamics below $\sim$ kHz within a 5 nm proximity. The measurement of electron spins strongly relies on the characteristics of the molecular electron spin. Under ambient conditions, the most optimal depolarization time for the electron spin spans 1 $\mu$s, extending to 10 ms under cryogenic temperatures \cite{bader2016Tuning,graham2017Forging,zadrozny2015Millisecond}. At ambient temperature, the sensitivity of the electron spin stands at $10^{-4}$ \uB\sqrtHz within 5 nm and approximately 0.01 \uB\sqrtHz within 10 nm for fluctuation measurements (as per Eq. \ref{equ:spin_sensitivity_fluc}). This sensitivity allows the identification of a single electron spin (defined here as SNR$\geq$3) within 0.1 ms to 1 ms, facilitating the real-time observation of a single molecule's dynamics and chemical reactions at frequencies below $\sim$ 10 kHz within a 5 nm distance. This stands as a substantial advancement compared to the current sensitivity, which hovers around $\sim$ 10 molecules \cite{barton2020Nanoscale}. Beyond the electron spin's depolarization time, the photo-bleaching effect in optical measurements remains a critical concern. In typical sensing experiments, the laser power density at the optical focal point reaches approximately $\sim$ $10^5$ W$\cdot$cm$^{-2}$, significantly impacting the stability of both electron spins and molecules. The resolution of this issue is pivotal to eliminate laser-induced damage in optical measurements. Addressing this problem could pave the way for highly stable single-molecule EPR studies in aqueous solutions, membranes, and cells, providing crucial insights into molecular structures and local environments. An alternative avenue for enhancement involves parallel high-throughput detection \cite{cai2021Parallel}, employing a large array of NV center probes to conduct single-molecule measurements in parallel. This high-throughput detection methodology enables the practical implementation of single-molecule tracing diagnostics \cite{chen2022Immunomagnetic,millerSpinenhancedNanodiamondBiosensing2020,li2022SARSCoV2}.

At cryogenic temperatures, the typical optimal depolarization time spans 10 ms \cite{bader2016Tuning,graham2017Forging,zadrozny2015Millisecond}, enabling the recognition of a single electron spin within 100 nm. This spatial range permits the creation of adaptable configurations within the electron spin network via lithography \cite{abb2016Twodimensional} or peptide self-assembly \cite{dey2021DNA,gopinath2016Engineering}. Harnessing and manipulating these spins could prove instrumental in crafting a scalable solid-state quantum information processor \cite{yao2012Scalable,schlipf2017molecular} or a large-scale quantum simulator \cite{cai2013Largescale}. Beyond quantum architecture, molecular spintronics stands to benefit significantly from solid-state magnetic resonance detection. NV center-based magnetic resonance detection operates across a broad temperature spectrum, ranging from ambient to cryogenic conditions. This capability extends to the measurement of the state of single-molecule spintronic devices even under elevated cryogenic temperatures \cite{coronado2020Molecular,gaita-arino2019Molecular}.

\section{Spectral resolution}\label{sec:spectral_resolution}

Spectral resolution holds significant importance within both EPR and NMR studies. In EPR spectroscopy, achieving a spectral resolution in the range of MHz or below enables the differentiation of molecules with slight structural variations \cite{martorana2014Probing} or the delineation of local polarity profiles \cite{kurad2003Lipida}. Conversely, in NMR spectroscopy, the precise measurement of distinct chemical environments, such as chemical shift and J-coupling, necessitates attaining high spectral resolution. Consequently, merging single-molecule techniques with high-resolution spectroscopy methods emerges as a pivotal concern. The capacity of spin spectroscopy to discern chemical environments offers unique insights distinct from conventional single-molecule techniques. This ability underscores the significance of enhancing spectral resolution in these methodologies. This chapter aims to introduce various methods geared towards improving spectral resolution. Included among these are the zero-field technique \cite{kong2018Nanoscale,kong2020Kilohertz,li2020Nanoscale}, correlation spectroscopy \cite{aslam2017Nanoscale,laraoui2013Highresolution, kong2015Chemical,staudacher2015Probing}, \qdyne \cite{boss2017Quantum, schmitt2017Submillihertz}, and weak measurement \cite{cujia2019Tracking, pfender2019Highresolution}.

\subsection{Spectral resolution}

The spectral linewidth is limited by different mechanisms for different control sequences. 
\begin{enumerate}[label=(\roman*)]
  \item cw-ODMR: In the cw-ODMR sequence, the microwave frequency undergoes a sweep across the complete \NV resonance range while the 532 nm laser continues its operation concurrently. The resonant frequency, determined by the spectrum's center, necessitates a narrow linewidth for precise frequency measurement. At resonance, the signal achieves maximum contrast as the resonant microwave drive transitions the spin state to $m_s=\pm 1$, consequently reducing the \PhL. The linewidth $\Delta f$ relies on both the microwave control strength and laser power \cite{jensen2013Light}:
  \begin{equation}
    \Delta f=\sqrt{\left(\frac{1}{\pi T_{2}^{\mathrm{eff}}}\right)^{2}+\left(\frac{4 T_{1}^{\mathrm{eff}}}{T_{2}^{\mathrm{eff}}}\right) \cdot \left(\frac{\Omega_{e}}{2\pi}\right)^{2}},
  \end{equation}
  where $\Omega_e$ is the Rabi frequency, $1 / T_{1}^{\text {eff }}=1 / T_{1}+\Gamma_{P}$ and $1 / T_{2}^{\text {eff }}=1 / T^*_{2}+\Gamma_{P}/2$ are  the effective relaxation times, and $\Gamma_{P}$ is the optical pumping rate. At low microwave power, the linewidth $\Delta f \approx 1 /\pi T_{2}^{\mathrm{eff}}$ is limited by \Ttwostar. With high power, the linewidth broadens according to the MW power, $\Delta f \approx  \sqrt{T_{1}^{\mathrm{eff}} / T_{2}^{\mathrm{eff}}} \cdot \Omega_{e}/\pi$.
  \item DEER, Ramsey, Hahn echo, and DD: The control sequences are listed in Table. \ref{tab:dd_sequence} and Fig. \ref{fig:dd_sequence}. The linewidth is limited by both the filter function and decoherence time.
   \item Quantum relaxometry: The cross relaxation approach probes target spins through spectral overlap between them and the NV center. The linewidth is limited by both the NV spin decoherence time and the relaxation time of the target spins.
  \item Spin-lock (Hartmann-Hahn): The spin-lock method employs a resonant microwave to reduce the dephasing of NV electron spin, locking it in a specific direction in the rotating frame. The linewidth is limited by the power instability-induced  linewidth $\delta_{\text{power}}$ and $T_{1 \rho}$, $\Delta f=1/\pi T_{1 \rho}+\delta_{\text{power}}$.
  \item Correlation or Qdyne: The control sequence is shown in Fig. \ref{fig:correlation}. The spectral resolution of native correlation spectroscopy \cite{laraoui2010Magnetometry,laraoui2013Highresolution,kong2015Chemical,staudacher2015Probing} is $\Delta f=1/\pi T_1$, dependent on the NV center electron spin depolarization time \Tone, which is on the scale of $\sim$ ms. However, utilizing the adjacent nitrogen nuclear spin \cite{laraoui2013Highresolution} or classical memory in \Qdyne detection scheme can achieve a spectral resolution of $1 /\pi T_{1}^{\mathrm{memory}}$, depending on the memory type, which can reach submillihertz scale \cite{aslam2017Nanoscale,boss2017Quantum,schmitt2017Submillihertz,Glenn2018}.

\end{enumerate}

\subsection{Methods to enhance the spectral resolution}

\subsubsection{Zero-field electron paramagnetic resonance (ZF-EPR)}\label{sec:zerofield}

Under ambient conditions, ensemble powder EPR spectra or single-molecule EPR spectra experience line broadening due to the magnetic inequivalence among otherwise identical spins or the movement in molecule configuration.  While high-field EPR experiments can partially mitigate this broadening \cite{schweiger2001Principles}, \zfepr spectroscopy presents a method to entirely eliminate line broadening by eradicating the Zeeman term, which depends on the magnetic field orientation \cite{bogle1961Paramagnetic,bramley1983Electron,bramley1986Variablefrequency,silver1963Nuclear}. Despite offering a straightforward route to enhance spectral resolution, \zfepr spectroscopy remains seldom utilized due to significantly lower sensitivity compared to conventional high-field EPR. However, this limitation does not hold true for NV sensor-based EPR, as the signal derives from statistical fluctuations rather than target spin polarization (\se \ref{sec:deer}). 

\begin{figure}[htp]
\begin{overpic}[width=0.85\columnwidth]{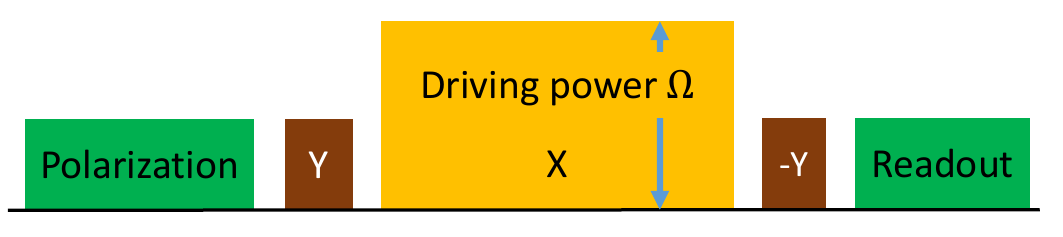}
\end{overpic}
\begin{overpic}[width=0.48\columnwidth]{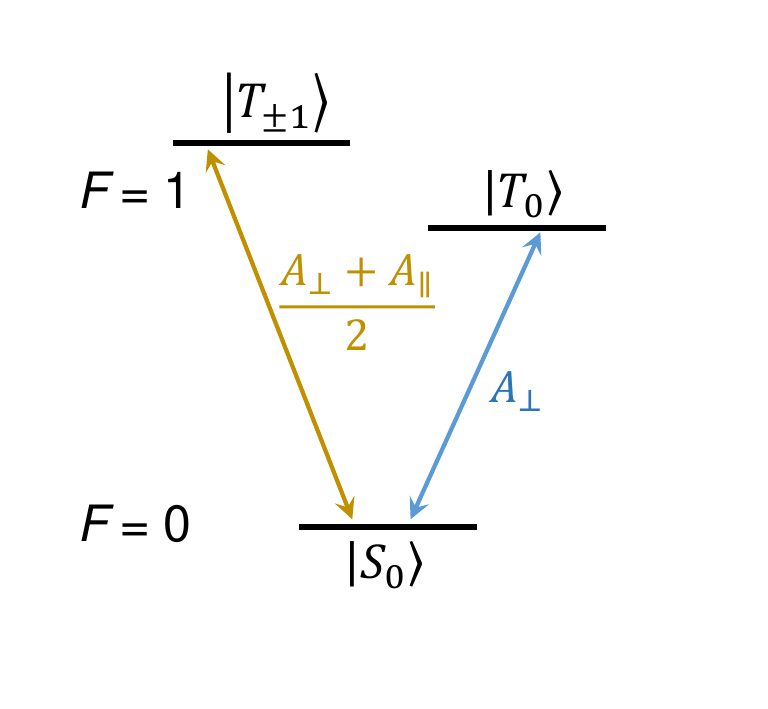}
\end{overpic}
\begin{overpic}[width=0.49\columnwidth]{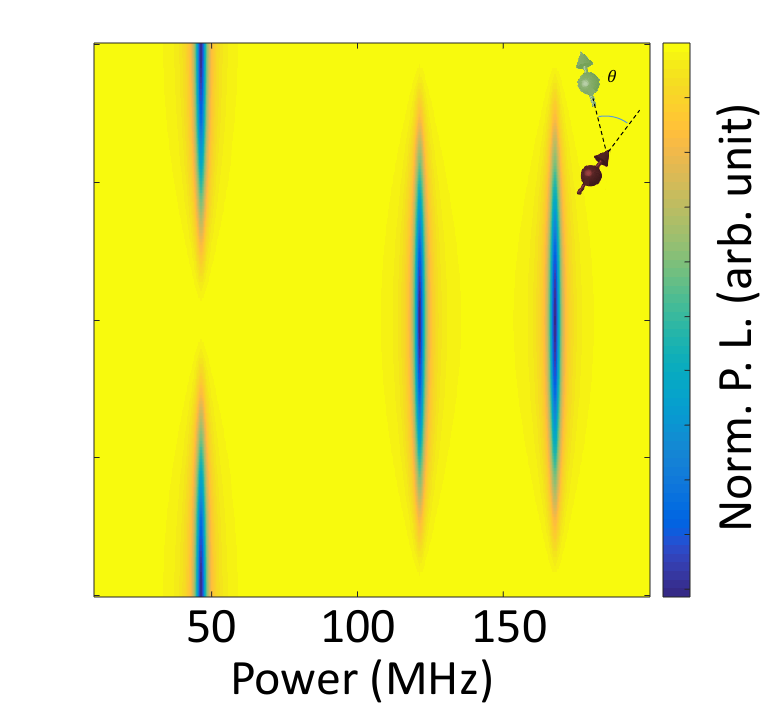}
\end{overpic}
\begin{overpic}[width=0.95\columnwidth]{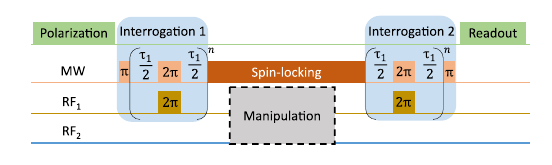}
\end{overpic}
\begin{overpic}[width=0.95\columnwidth]{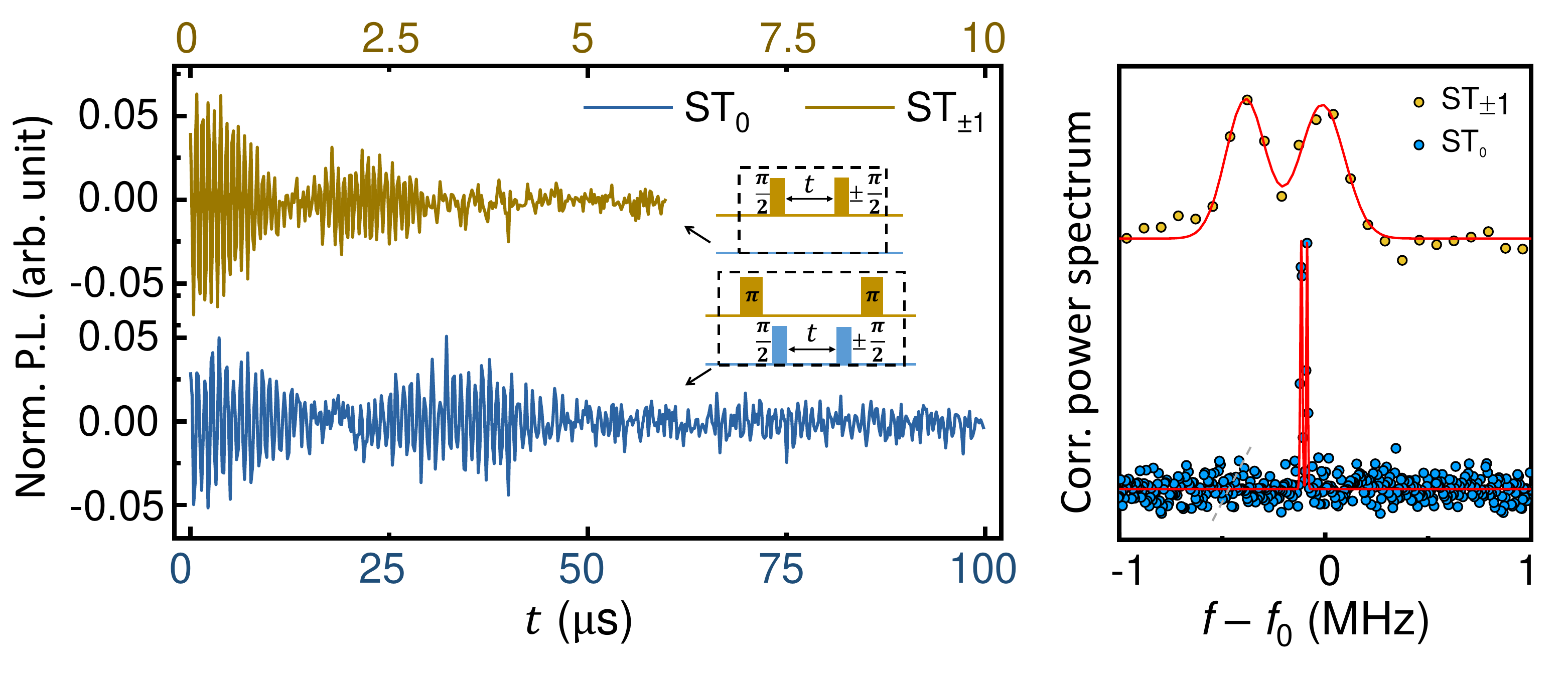}
\put (-3, 135) { (a) }
\put (-3, 112) { (b) }
\put (45, 112) { (c) }
\put (-3, 68) { (d) }
\put (-3, 38) { (e) }
\end{overpic}
\caption[NV Center]{ Zero-field EPR. (a) Power-sweeping sequence for zero-field measurements. (b) Energy levels of the coupling system involving an $S_{e}$=1/2 electron spin and $I$=1/2 nuclear spin at zero field. (c) Dependence of EPR spectra on the orientations of target spins. The spectra are simulations of $^{15}$N Nitroxide spin label. (d) Correlation sequence for zero-field measurements. (e) Comparison between normal (ST$_{\pm1}$) and clock (ST$_0$) transitions. Left and right panels represent the Ramsey measurements and their Fourier transform, respectively. Adapted from \cite{kong2018Nanoscale, kong2020Kilohertz}. }  \label{fig:continuous_driving_spectrum}
\end{figure}

In the context of biological or ambient temperature single-molecule scenarios, controlling the rotational dynamics of the target spin becomes complex. Consequently, the DEER method becomes an unsuitable choice. The resolution of this issue lies in manipulations solely on the NV sensor. The NV sensor undergoes resonant microwave $\vb{B}_1(t)=\vb{B}_1 \cos \mathcal{D} t$ radiation, driving it into dressed states (Fig. \ref{fig:continuous_driving_spectrum}(a)), represented by:
\begin{align}
\begin{aligned}
|-1\rangle_{\mathrm{d}}=\frac{1}{2}|1\rangle-\frac{1}{\sqrt{2}}|0\rangle+\frac{1}{2}|-1\rangle \\
|0\rangle_{\mathrm{d}}=-\frac{1}{\sqrt{2}}|1\rangle+\frac{1}{\sqrt{2}}|-1\rangle \\
|1\rangle_{\mathrm{d}}=\frac{1}{2}|1\rangle+\frac{1}{\sqrt{2}}|0\rangle+\frac{1}{2}|-1\rangle
\end{aligned}\label{eqn:dressed}
\end{align}
in the rotating reference frame with eigenenergies $-\Omega/2$, 0, and $\Omega/2$, respectively, where $\Omega=\gamma_{\text NV}(B_{1,x}^2+B_{1,y}^2)^{1/2}$ is the Rabi frequency and proportional to the perpendicular amplitude of the microwave driving field. The dipole-dipole interaction between the NV center and the target spins induces a flip-flop process at the resonance condition 
\begin{equation}
\Omega=2 \Delta \omega_{i j},
\end{equation}
where $\Delta \omega_{i j}=\omega_{j}-\omega_{i}(i<j)$ is the energy level splitting of the target spins corresponding to an allowed magnetic dipole transition. When the NV center is initially polarized to one of the dressed states given by Eq. \ref{eqn:dressed}, the polarization will be transferred from the NV center to the target spins, and consequently, the populations of dressed states changes. At the zero magnetic field, all the Zeeman terms vanish and the energy level structure of the target spins is determined solely by the intrinsic spin–spin interactions. Thus, for a spin-1/2 electron-nuclear system, the simplified spin Hamiltonian under secular approximation is determined solely by the hyperfine interaction expressed as
\begin{equation}
H_{}=A_{\perp}\left(S_{x,{e}} I_{x}+S_{y,{e}} I_{y}\right)+A_{\|} S_{z,{e}} I_{z}
\end{equation}
where $A_{\perp}$ and $A_{\|}$ are the hyperfine constants. The eigenstates consist of one antisymmetric singlet $\left|\text{S}_{0}\right\rangle$ and three symmetric triplet states $\left|\text{T}_{0}\right\rangle$ and $\left|\text{T}_{\pm 1}\right\rangle$ (Fig. \ref{fig:continuous_driving_spectrum}(b)). The corresponding eigenvalues are
\begin{align}
\begin{aligned}
\omega_{\text{S}_{0}}=-\frac{A_{\perp}}{2}-\frac{A_{\|}}{4},\\
\omega_{\text{T}_{0}}=\frac{A_{\perp}}{2}-\frac{A_{\|}}{4},\\
\omega_{\text{T}_{\pm 1}}=\frac{A_{\|}}{4},
\end{aligned}\label{eqn:dressed_level}
\end{align}
respectively. Precision in tuning the energy levels of NV centers in dressed states to resonate with target spins facilitates polarization transfer and enables measurement of the dark spin by the NV sensor (sec. \ref{sec:relaxometry}). The \zfepr spectrum is attained by varying the driving power. As depicted in Fig. \ref{fig:continuous_driving_spectrum} (c), the peak positions remain independent of target orientations, thereby providing a narrow spectrum.

Furthermore, the application of zero magnetic field EPR effectively mitigates the primary cause of line broadening induced by magnetic noise. The presence of magnetic noise, denoted as $\delta B$, instigates fluctuations in energy levels within the targeted spin, resulting in line broadening. Notably, the $\left|\text{S}_{0}\right\rangle$ and $\left|\text{T}_{0}\right\rangle$ states exhibit no first-order dependence on the magnetic field. Consequently, the frequency fluctuation of the $\left|\text{S}_{0}\right\rangle \rightarrow \left|\text{T}_{0}\right\rangle$ transition (referred to as $\mathrm{ST}_{0}$ hereafter) diminishes to approximately $\sim \gamma_e^2\delta B^{2} / A$. This reduction leads to a noticeable line-narrowing effect. However, observing such a narrowed $\mathrm{ST}_{0}$ spectrum proves challenging using the conventional microwave power-sweeping method, demanding exceptional power stability across the entire microwave circuits. 

Similarly, the widely used DEER method faces limitations due to the interrogation time constrained by the coherence time \Ttwo of the NV center. Drawing inspiration from the correlation spectroscopy of nuclear spins \cite{laraoui2013Highresolution}, a modified correlation detection protocol for \zfepr spectroscopy has been devised (\cite{kong2020Kilohertz}, Fig. \ref{fig:continuous_driving_spectrum} (d)). This modification extends the sensor's lifetime to the spin-locking relaxation time \Tonerou, often significantly longer than \Ttwo for shallow NV centers \cite{kong2018Nanoscale,rosskopf2014investigation}.
Employing Ramsey experiments on the P1 center enables the acquisition of ZF-\epr spectra (Fig. \ref{fig:continuous_driving_spectrum} (d)). The Fourier transformation of these spectra, as shown in Fig. \ref{fig:continuous_driving_spectrum} (e), distinctly illustrates the enhanced spectral resolution. This method underscores the potential to reduce the EPR linewidth of dark electron spins by more than an order, achieving values on the scale of several kilohertz. 

\subsubsection{Correlation and quantum heterodyne (Qdyne) method } \label{sec:correlation} 

Quantum sensing techniques involve subjecting a quantum coherent sensor to a response Hamiltonian $H_{V}(t)$ (Eq. \ref{eqn:sensing_hamiltonian}), influencing the evolution of the probe state within the sensor's coherence time. These protocols (section \ref{sec:quantum_sensing}) enable the extraction of information concerning the Hamiltonian $H_V (t)$. Consequently, the coherence time of the sensor dictates the duration of coherent signal accumulation. The accurate reconstruction of a given Hamiltonian frequency spectrum holds paramount importance for chemical analysis, molecular structure determination, and various other applications. 
The linewidth of this method serves as a pivotal metric for spectral analysis, impacting not only the precision in estimating individual frequency components but also the capability to resolve multiple frequencies simultaneously.

\begin{figure*}[tbhp]
	\begin{overpic}[width=0.98\textwidth]{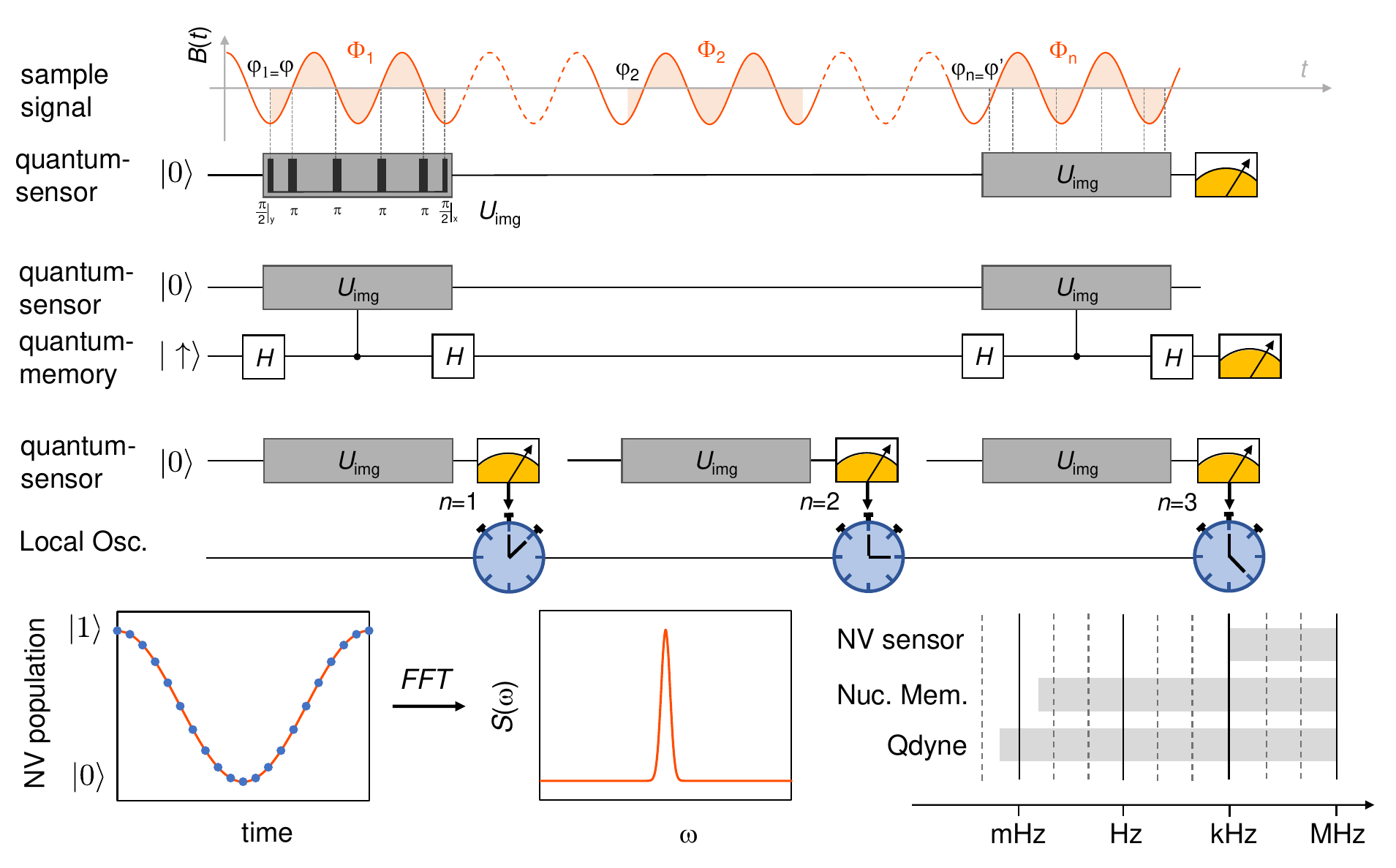}
		\put (-2, 59.5) {(a)}
		\put (-2, 50.5) {(b)}
		\put (-2, 42.5) {(c)}
		\put (-2, 30) {(d)}
		\put (-2, 18) {(e)}
		\put (58, 18) {(f)}
	\end{overpic}
	\caption[NV Center]{ Correlation spectroscopy utilizing the NV sensor, ancilla quantum memory, and Qdyne. (a) An oscillating magnetic field $B(t)=B_{\text{a.c.}}\sin (\omega t+\varphi)$ is probed by a quantum NV sensor, employing a dynamical decoupling sequence $U_{\text{img}}$ aligned with the resonant condition. The intervals between the $\pi$ pulses are $\tau=\pi/\omega$, and the accumulated phase $\Phi_n=2\gamma_eB_{\text{a.c.}}t_s{\cos \varphi_n}/\pi $ relies on the initial oscillating field phase $\varphi_n$ for each sample. (b) The control procedure for correlation spectroscopy involves detecting two phases $\varphi$ and $\varphi'$ using two sensing protocols, culminating in a final measurement recording a correlation signal $\langle \sin\Phi\sin\Phi'\rangle$. (c) Correlation spectroscopy integrates a quantum memory, such as an adjacent nitrogen nuclear spin, and can be combined with the ENDOR sequence. (d) The \Qdyne sequence involves the repetitive accumulation and measurement of phases $\Phi_n$ over the entire measurement time $T_{\text{exp}}$. (e) Heterodyning with an external clock records the NV populations series $\sin\Phi_n$ and consequently the corresponding photon numbers. Through FFT, the time evolution is transformed into the spectrum, enabling the resolution of signal frequency concerning the local oscillator frequency. (f) Spectral resolution comparison among correlation spectroscopy, ancilla quantum memory, and the Qdyne method.  The figures are replotted from \cite{schmitt2017Submillihertz}.
	}  \label{fig:correlation}
\end{figure*}
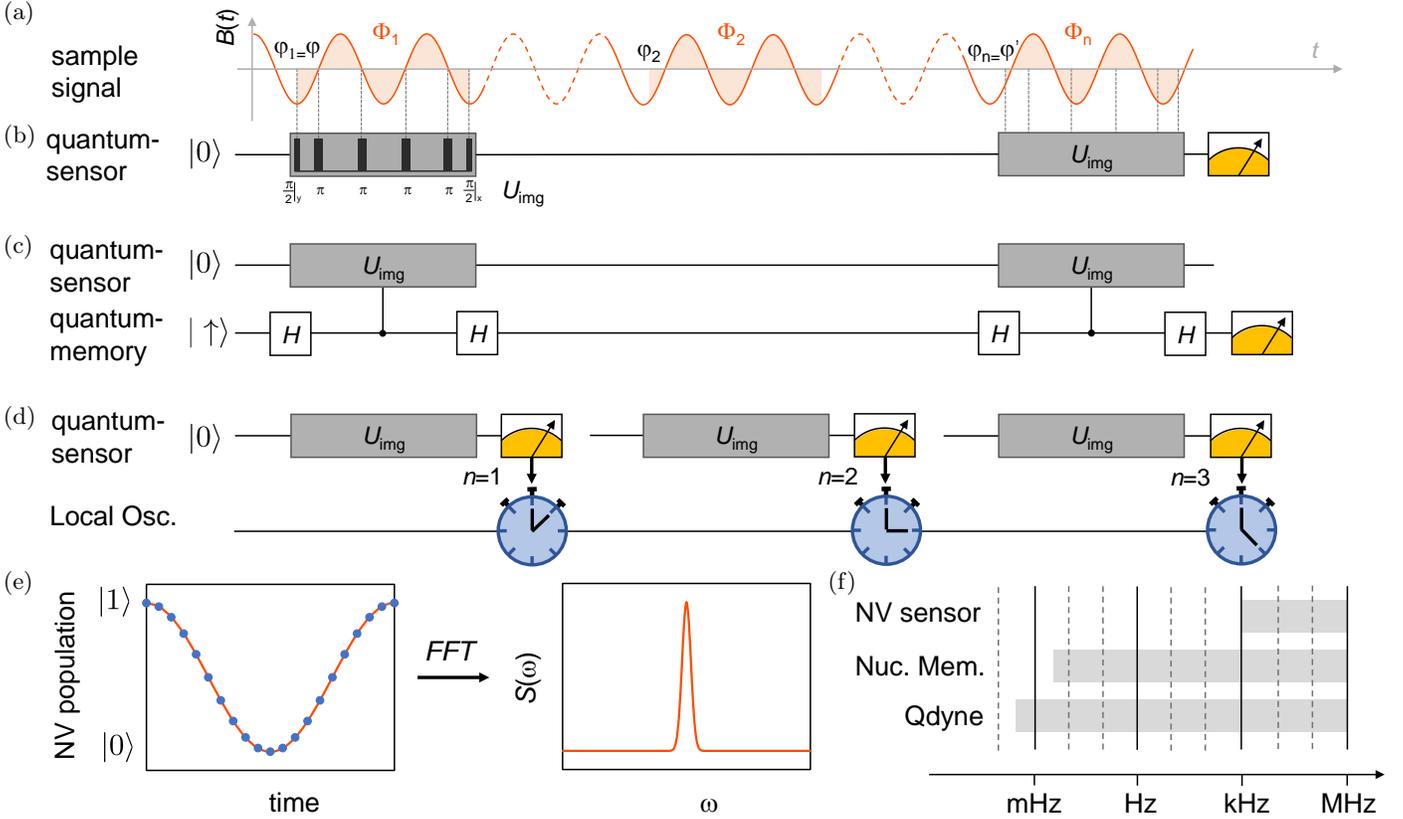

Various methods (refer to \se \ref{sec:decoherence:sensor_control} and \se \ref{sec:surf_tr}) have been employed to extend the coherence time of the NV sensor. However, despite these approaches, including dynamical decoupling and diamond surface fabrication, the decoherence time remains relatively limited. The most prolonged achieved coherence stands at 3.3 ms \cite{herbschleb2019Ultralong} for a single NV center at ambient temperature, approximately half the duration of \Tone.

An alternative path toward achieving high-resolution spectroscopy involves transferring the signal encoded in the phase of the \NV quantum state to electron spin population, nuclear spin, or even classical computer memory. While enhancing the spectral resolution to the memory relaxation time is feasible, it does not augment the sensitivity of the \NV sensor, which relies on coherent phase accumulation. 
An a.c. magnetic field $B(t)=B_{\text{a.c.}}\cos (\omega t+\varphi)$ is assessed via correlation spectroscopy (Fig. \ref{fig:correlation}(a)), where $B_{\text{a.c.}}$ signifies the field strength, $\omega$ denotes the oscillating frequency, and $\varphi$ represents the initial phase of the oscillating magnetic field. The sensing protocol corresponds to the imaginary component measurement (Fig. \ref{fig:sensing_real_img}(d,e)). Under the imaginary component sensing protocol $U_{\text{img}}$, the intervals between control $\pi$ pulses align with the half period of the a.c. field, resulting in an accumulated phase
 $\Phi=(2\gamma_eB_{\text{a.c.}}t_s \cos \varphi)/\pi$, where $t_{\text{s}}$ denotes the interaction time \cite{kotler2011Singleion,maze2008Nanoscale}. This phase $\Phi$ is stored in the NV spin magnetization as $(1-\sin\Phi)/2$ (Eq. \ref{eqn:img_read}). Subsequently, a second sensing protocol is conducted after a waiting time $\Delta t$, accumulating a phase  $\Phi'=(2\gamma_eB_{\text{a.c.}}t_s \cos \varphi')/\pi$.
Employing correlation spectroscopy (Fig. \ref{fig:correlation}(b)), the resulting correlation signal is thus derived \cite{laraoui2013Highresolution} as: 
\begin{equation}
S(t)\propto \langle \sin\varphi \sin\varphi' \rangle\sim \frac{1}{2}\cos \omega\Delta t  ,
\end{equation}
where the bracket stands for time average. Unlike the Hahn-echo or the dynamical decoupling protocols, the spectral resolution is limited only by the NV $T_1$ relaxation time, usually longer than the decoherence time. 

By utilizing the nitrogen nuclear spin as a memory ancilla for storing the quantum state (Fig. \ref{fig:correlation}(c)), the achievable spectral resolution can be improved beyond the NV sensor $T_1$ limit, up to the nuclear longitudinal-relaxation time $T_1^{\text{nuc}}$, which can typically last several minutes \cite{laraoui2013Highresolution, pfender2017Nonvolatile, aslam2017Nanoscale}.
In the encoding and decoding process, the accumulated phase $\Phi$ of the NV sensor is transferred to the superposition state of nuclear spin, which is performed by entangling and disentangling the NV center and nuclear memory with a $\mathrm{C}_{\mathrm{m}} \mathrm{ROT}_{\mathrm{s}}$-gates.
The two Hardmard gates on the nuclear spin  switch between the NV state and nuclear memory information.
After a target spin evolution, the decoding process allows the current target spin state to be correlated with its  initial state stored in the memory nuclear spin.
Finally, the state of nuclear memory is extracted by the single-shot readout method.
The NMR spectrum of a nanoscale liquid, with a chemical resolution of approximately 1 ppm, is demonstrated in \cite{aslam2017Nanoscale}.

Further enhancement in spectral resolution is attainable through narrowband synchronized readout protocols such as the \Qdyne detection scheme \cite{schmitt2017Submillihertz,boss2017Quantum} (Fig. \ref{fig:correlation}(d)) and analogous coherently averaged synchronized readout (CASR) schemes \cite{Glenn2018}. The CASR protocol finds application in micro-sized nuclear spin samples exhibiting sufficiently strong thermal polarization \cite{Glenn2018,munuera-javaloy2022High} or those enhanced via the DNP method \cite{arunkumar2021Micron,bucher2018hyperpolarization}. It involves initiating polarized nuclear spins with an initial $\pi/2$ pulse for coherent signal averaging. Subsequently, the NMR free-induction-decay signal is acquired through a synchronized NV magnetometry pulse sequence interspersed with projective NV spin state readouts.

The \Qdyne technique, applied in nanoscale nuclear spin samples experiencing fluctuations, resembles classical heterodyne detection where an unknown signal is mixed with a local oscillator. However, in the Qdyne protocol (Fig. \ref{fig:correlation}(c)), the nonlinear mixing element is a quantum coherent probe, the NV sensor. Throughout the Qdyne protocol, the NV sensor continuously samples at a rate of $f_{\text{LO}}$, capturing a phase $\Phi_n=(2\gamma_eB_{\text{a.c.}}t_s \cos \varphi_n)/\pi$ with each sample. This phase, measured by the imaginary component for each sample, considering the sample frequency $f_{\text{LO}}$ and sampling time $t_L=1/f_{\text{LO}}$, can be described as:  
\begin{equation}
\Phi_n=\frac{2\gamma_eB_{\text{a.c.}}t_s}{\pi}\cos[2\pi\cdot{n}\,\delta f \, t_L+\varphi]
\end{equation}
where $\delta f=\omega/2\pi-f_{\text{LO}}$ represents the undersampling ``mixed'' beating frequency and $n$ is an integer. By measuring the beating frequency $\delta f$ and the knowledge of $f_{\text{LO}}$, the signal frequency can be determined, although limited by the local oscillator stability. The resolution of the Qdyne method depends on the experiment time $\Texp$, approximating $\sim 1/\pi \Texp$.

\begin{figure}[tbhp]
	\begin{overpic}[width=1\columnwidth]{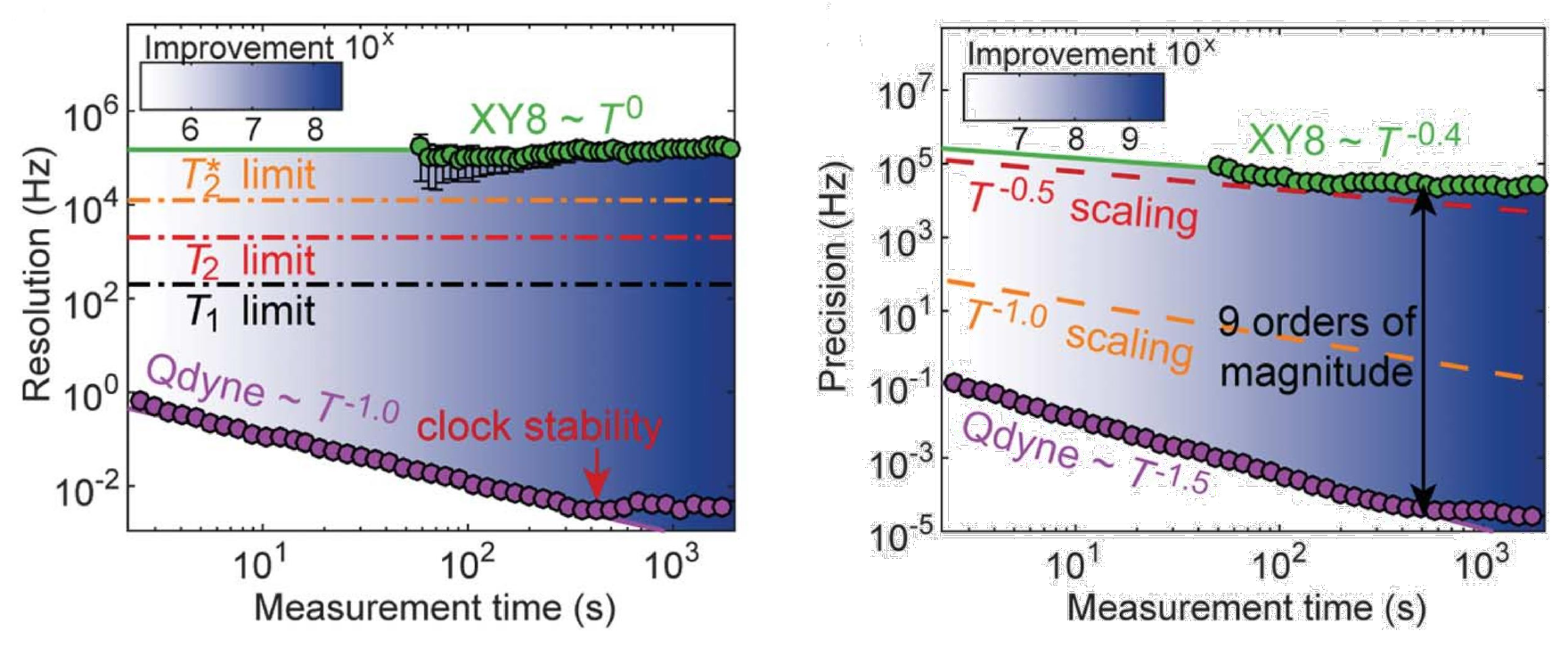}
		\put (0, 38) {(a)}
		\put (52, 38) {(b)}
	\end{overpic}
	\caption[NV Center]{ (a) Comparison of the Qdyne spectral resolution with the XY8 technique. The resolution of the XY8 technique is determined by the interrogation time and remains constant at 300 kHz. The resolution of the Qdyne method improves with the experiment time $T^{-1}$ until the limit of clock stability is reached. (b) Comparison of Qdyne spectral precision and XY8 technique. The frequency precision is fitted by least-squares estimation and thus is always improved compared to the spectral resolution. XY8 technique has a measurement precision approximating the standard quantum limit. In contrast, the spectral accuracy of Qdyne exceeds and scales as $T^{-3/2} $ until reaching the clock stability limit. The figures are adapted from \cite{schmitt2017Submillihertz}.
	}  \label{fig:Qdyne}
\end{figure}

The frequency precision, surpassing the corresponding linewidth, can be obtained by a least-squares fit of the spectra. The precision of estimated frequency of Qdyne scales as \cite{schmitt2017Submillihertz}
\begin{equation}
\Delta f_Q\approx\frac{1}{{g_s}\Texp\sqrt{\Texp T_2}}\propto \Texp^{-3/2}, 
\end{equation}
where $g_s=\gamma_eB_{\text{a.c}}$ is the interaction strength. It should be noted that this precision refers to the sensor and does not take into account the linewidth of the target sample itself. The formula holds until the experiment time reaches the clock stability $T_{\text{LO}}$ and follows the standard quantum limit $\Delta f_{\text{Q}}\sim 1/{g_s}T_{\text{LO}}\sqrt{\Texp T_2}$. Compared to the frequency precision of dynamical decoupling $\Delta f_{\text{DD}}\approx 1/{g_s}T_{\text{2}}\sqrt{\Texp T_2}$ and correlation spectroscopy $\Delta f_{\text{M}}\approx 1/{g_s}T_{\text{2}}\sqrt{\Texp T_{\text{M}}}$, the Qdyne method substantially improves both the frequency precision and spectral resolution by several orders beyond the clock stability $T_{\text{LO}} \gg T_{\text{2}}, T_{\text{M}}$ (Fig. \ref{fig:Qdyne}). The ultimate absolute precisions achieved are 20 $\mu$Hz \cite{schmitt2017Submillihertz} and 70 $\mu$Hz \cite{boss2017Quantum}. The bandwidth of the Qdyne method is only limited by the phase accumulation protocol, dynamical decoupling sequence, peaking at $\sim$ MHz.

\begin{figure}[tbhp]
	\begin{overpic}[width=1\columnwidth]{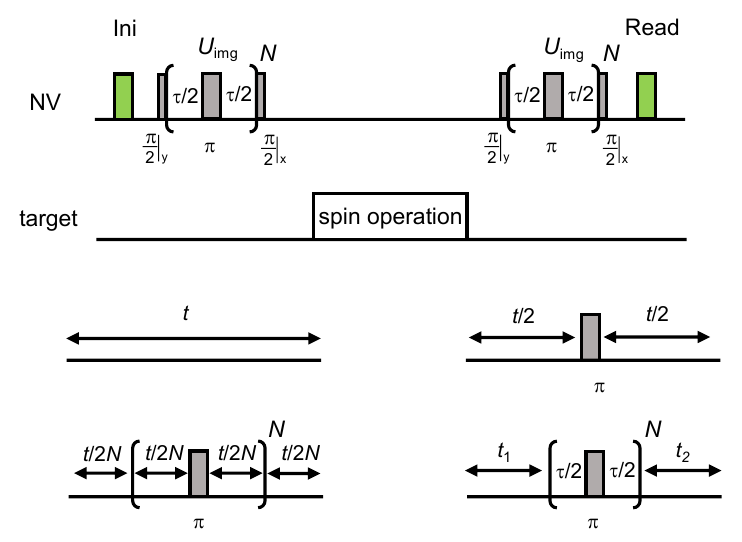}
		\put (0, 68) {(a)}
		\put (0, 31) {(b)}
		\put (55, 31) {(c)}
		\put (0, 14) {(d)}
		\put (55, 14) {(e)}
	\end{overpic}
	\caption[NV Center]{ Combination of the high resolution method with the nuclear spin detection. (a) The overall correlation pulse sequence. The spin operations (b--e) are inserted in the free evolution interval of the correlation sequence. (b) Spin operation of the free effective Hamiltonian $H_0=\omega_0 I_z+a_{\parallel}(S_z+1/2)I_z$. (c) Spin operation of the effective Hamiltonian $H_{\text{eff}}=\omega_0 I_z+a_{\parallel}I_z/2$. (d) Spin operation of the effective Hamiltonian $H_{\text{eff}}={2}a_{\perp}/\pi S_zI_x$. (e) The two-dimensional NMR COSY sequence. The figures (b-c) are adapted from \cite{boss2016One}. }  \label{fig:corr_seq}
\end{figure}
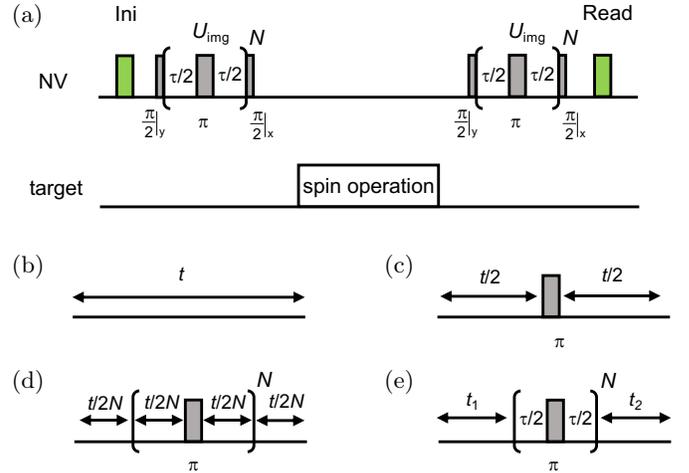

The high resolution method can be readily combined with measurements for different properties of the nuclear spins.  The protocols, which entail a combination of correlation and operation on the target nuclear spins, are illustrated in Fig. \ref{fig:corr_seq}(a). The dynamical decoupling sequence $U_{\text{img}}$ is comprised of a periodical $(\tau/2-\pi-\tau/2)^N$ control sequence with $\pi/2|_y$ and $\pi/2|_x$ before and after respectively. 
The overall effect of the $U_{\text{img}}$ sequence is \cite{boss2016One}
\begin{equation}
\hat{\sigma}_{a}=\cos \left(a_{\perp}N\tau/\pi \right) {S}_{x}+\sin \left(a_{\perp}N\tau/\pi \right) 2 {S}_{z} {I}_{x}.
\end{equation}
Thus the operator $\hat{\sigma}_{a}$ consists of both the electron coherence $S_x$ and nuclear coherence $2S_zI_x$ which is conditional on the electron spin state. By the target nuclear spin operation $U_{\text{op}}$ between two $U_{\text{img}}$ evolution, the final observation is 
\begin{align} 
p(t)=& \cos ^{2}\left(\omega_{1} t\right) \Tr\left({U}_{\text {op }} \hat{S}_{x} {U}_{\text {op }}^{\dagger} \hat{S}_{x}\right) \nonumber \\
 &+\sin ^{2}\left(\omega_{1} t\right) \Tr\left({U}_{\text {op }} \hat{I}_{x} {U}_{\text {op }}^{\dagger} {I}_{x}\right). \label{eqn:nuc_corr}\end{align}
The desired nuclear evolution can be measured by recording the final observation. 
The different sequences in Fig. \ref{fig:corr_seq}(b-e) represent the protocols for free evolution, detection for the parallel component of hyperfine interaction $a_{\parallel}$, the perpendicular component $a_{\perp}$ \cite{boss2016One}, and two-dimensional NMR COSY spectrum \cite{yang2019Structural}.

\subsubsection{Back action effect of measurement} \label{sec:measure_ind_eff}

In traditional NMR or \epr, the influence of measurement on the system is usually negligible due to the weak coupling between the target spin and the detector. However, exceptions arise, such as the damping of magnetic resonance induced by the electric detection circuit in the presence of a nuclear spin ensemble \cite{bloembergen1954Radiation}.
Quantum measurements inherently induce back-action on the measured system, causing the system to collapse into different states determined by the random measurement outcome. In the context of single-molecule techniques, where measurements often involve only a few or even a single spin, overcoming the impact of back action becomes crucial.

The back-action effect is generally negligible for dynamical decoupling and correlation spectroscopy, where the fluctuation signal is measured, making the signal independent of the sample state. Quantum memories enhance spectral resolution at the expense of sensitivity. In contrast, the \Qdyne method achieves exceptional spectral resolution through continuous measurement techniques without compromising sensitivity. 
When monitoring the dynamics of a quantum object by sequential measurements, the back-action induces inevitable disturbance. The strong measurement can even trap the state coherently \cite{pfender2019Highresolution} and thus change the spin oscillation frequency. Weak measurements, as introduced theoretically \cite{aharonov1988How,jordan2005Quantum,jordan2006Qubit,korotkov2001Output,korotkov2001Selective,wiseman2002Weak,2018Mathematical} and demonstrated experimentally with NV centers \cite{blok2014Manipulating,liu2017SingleShot,cujia2019Tracking,pfender2019Highresolution}, superconducting qubits \cite{groen2013PartialMeasurement,murch2013Observing,hatridge2013Quantum} and other systems \cite{piacentini2016Measuring,muhonen2018Coherent}, are a potential solution to approach the limit of negligible disturbance of the system under study. 

\begin{figure}[tbhp]
	\begin{overpic}[width=0.5\columnwidth]{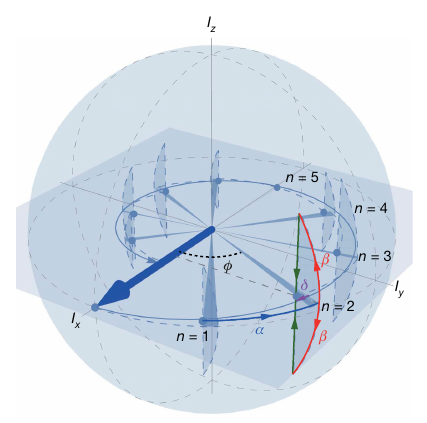}
	\end{overpic}
	\begin{overpic}[width=1\columnwidth]{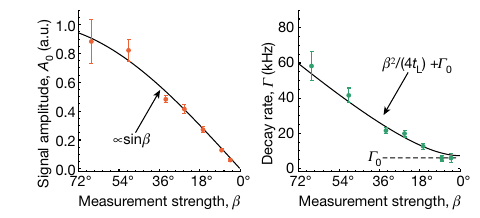}
		\put (2, 90) {(a)}
		\put (2, 40) {(b)}
		\put (49, 40) {(c)}
	\end{overpic}
	\caption[NV Center]{Tracking the precession of single nuclear spins by weak measurements. 
		(a) Bloch sphere representation of the spin evolution of the measured spin.
		(b) The signal amplitude depends  on the measurement strength $\sin\beta$.
		(c) By performing the weak measurements,  the measurement-induced decoherence can be suppressed quadratically with the measurement strength. The figures are adapted from \cite{cujia2019Tracking}. }  \label{fig:corr_nuc}
\end{figure}

The influence of back-action is significant when observing the nuclear spin due to its considerably longer coherence time  (Fig. \ref{fig:corr_nuc}). In observing the nuclear spin, sequential measurements utilizing $U_{\text{img}}$ are interleaved. Initially, the nuclear spin is initialized to the $x-y$ plane \cite{cujia2019Tracking}, or its statistical fluctuations are observed through a measurement-correlation scheme \cite{pfender2019Highresolution}. The rate of dephasing induced by the measurement is determined as:
\begin{equation}
\Gamma_{\beta}=\frac{a_{\perp}^2t_s^2}{{\pi^2} t_L},
\end{equation}\label{eqn:measurement_induced}
where $\tau$ is the interrogation time of $U_{\text{img}}$ and $t_L$ is the sampling time. By choosing the interaction strength $\beta=a_{\perp}t_s/\pi$, the back-action induced dephasing can be eliminated. However, the weak measurement comes at the price of less information on the system. In particular, the trade-off between signal strength and back-action plays an important role in high-resolution sensing. On the condition that interrogation time $t_s$ dominates the measurement time and $t_s\leq \sqrt{{\pi^2}\Gamma_nt_L/a_{\perp}^2}$, where $\Gamma_{n}$ is the intrinsic dephasing rate, the sensitivity is almost not affected by the back-action \cite{cujia2019Tracking}. Another measurement-related dephasing is caused by the residual hyperfine interaction during optical illumination. The hyperfine field on the nuclear spin stochastically changes as the spin state and the charge state of NV center randomly switch during the optical readout process. The decay rate is $\Gamma_{\gamma}\propto a_{\parallel}^2t^2_{\text{optical}}/(2 t_L)$. The best achieved linewidth is 120 Hz \cite{cujia2019Tracking} and 3.8 Hz \cite{pfender2019Highresolution}.

\subsubsection{Confined space } \label{sec:confined_space}

\begin{figure}[tbhp]
  \begin{overpic}[width=\columnwidth]{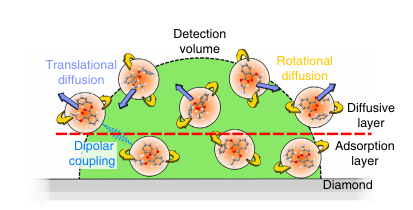}
  \end{overpic}
  \caption[NV Center]{Dynamics of the sample near the diamond surface. A layered model in the detection volume (green semisphere) consisting of a static adsorption film (below red dashed line) where translational diffusion of the molecules (light blue arrows) is restricted. In contrast, the outer section behaves as a bulk liquid, where the molecules self-diffuse. The figure is adapted from \cite{staudacher2015Probing}.   }
  \label{fig:diffusion}
\end{figure}

The precision of correlation spectroscopy and the Qdyne method in resolving spectra relies significantly on sensor properties and the back-action stemming from the measurement protocols. However, the intrinsic relaxation of the target spin remains the dominant factor in determining the linewidth.
In both fluidic systems \cite{staudacher2015Probing} and solid nanoscale systems \cite{shi2015Singleprotein}, molecules experience rapid tumbling and self-diffusion (Fig. \ref{fig:diffusion}). These mechanisms contribute to the averaging of internuclear spin couplings, leading to longer nuclear spin relaxation times.
In nanoscale NMR, molecular diffusion can result in molecules escaping the detection volume, reducing interaction time with the sensor and broadening the linewidth \cite{yang2022phase}. This differs from conventional NMR, where molecular diffusion often leads to narrower linewidths through motional narrowing.
The diffusion induced spectral resolution for NV based nanoscale NMR is
limited by the relaxation time of sensor $T_{\text {sensor }}$, dephasing time of the target sample spins $T_{2,\text{tar}}^{*}$ , and  molecular diffusion induced relaxation \cite{steinert2013Magnetic},
\begin{equation}
  \frac{1}{T_{\text {D }}}=D_{\text {diff }}\left(\frac{3}{4 d}\right)^{2},
\end{equation}
where $d$ is the depth of NV sensor. Previous experiments observing nanoscale spins generally involve immobilized target samples \cite{staudacher2015Probing,yang2022phase,muller2014Nuclear,Staudacher2013nuclear,mamin2013nanoscale} or samples with relatively low diffusion rates \cite{aslam2017Nanoscale,yang2022phase,schmitt2017Submillihertz,staudacher2015Probing,kong2015Chemical,Staudacher2013nuclear,mamin2013nanoscale}. The best spectral resolution achieved, at 200 Hz \cite{aslam2017Nanoscale} with an NV sensor depth of 30--50 nm, corresponds to a chemical shift resolution of 1.3 ppm for \Hone nuclear spins under 3 T. However, due to the greater depth of the NV sensor, the nuclear spin number sensitivity achieved in \cite{aslam2017Nanoscale} is still far from the single-molecule limit.

To mitigate the linewidth broadening challenge, some methods prioritize measuring the polarization signal over the fluctuation signal. However, for nanoscale or single-molecule spin detection (Fig. \ref{fig:nmr_pol_flu}), the polarization might not reach sufficiently high levels. This method is notably advantageous for micrometer-scale samples, such as single cells \cite{xie2018Mesoscopic}.
Another approach involves immobilizing samples on the sensor surface, showing potential in resolving these challenges. Experiments with chemically attached DNA molecules have shown promise in potentially eliminating spin--spin interactions due to molecular liquid motion \cite{shi2018SingleDNA}, although further investigation is necessary.
A third strategy to address diffusion effects is sample confinement within a limited space. Theoretical analyses focused on confined nano-NMR using an NV center \cite{cohen2020confined}. When the experiment time $t$ is less than the diffusion time $T_{\mathrm{D}}$, nuclear spins hardly diffuse out of the detection region, maintaining nearly constant signal intensity. For sensing times $t$ less than the boundary reaching time $t_{V}=V^{2 / 3}/{D}_{\text{diff}}$, the signal decays according to a power law, resulting in sharply peaked spectral lines \cite{staudenmaier2022powerlaw}. Over a prolonged experiment, nuclear spins might reflect back from the boundary and be detected by the NV sensor, eventually causing the signal to degrade into a constant.
Confining the sample significantly reduces diffusion effects, thereby enhancing spectral resolution. Experiments employing metal-organic frameworks (MOFs) with angstrom-sized pores demonstrated remarkable outcomes. In \Pone NMR, these experiments achieved a minimum linewidth of 3 kHz, whereas the spectrum was undetectable without MOF confinement \cite{liu2022Using}.

\subsection{Conclusion}\label{sec:spectral_resolution:conclusion}

The NV sensing protocol has successfully achieved a spectral resolution on the order of mHz \cite{schmitt2017Submillihertz, boss2017Quantum}, attaining a best-reported linewidth of approximately Hz for NMR \cite{cujia2019Tracking, pfender2019Highresolution} and kHz for EPR \cite{kong2018Nanoscale}. This level of precision proves adequate for distinguishing \Hone and \Cthir chemical shifts at external magnetic fields below 1 T \cite{kong2015Chemical, Smits2019two, Glenn2018}, as well as resolving chemical species under magnetic fields of around 20 mT \cite{haberle2015Nanoscale, Devience2015}.

The primary challenge in achieving chemical shift resolution at the single-molecule level arises from diffusion-induced nuclear spin relaxation. Utilizing a confined structure to minimize surface diffusion of liquid on diamond \cite{staudenmaier2022powerlaw, liu2022Using} presents a viable approach to reducing linewidths for the realization of single-molecule chemical resolved NMR. When combined with two-dimensional NMR  protocols, this approach enables the resolution of single-molecule structures by leveraging the nuclear Overhauser effect, residual dipolar coupling, or paramagnetic relaxation enhancement. Furthermore, the enhancement in spectral resolution is advantageous for single-molecule EPR, allowing for more precise measurements of bond distances and chemical environments, along with accurate determination of the g factor.

\section{Spatial resolution}\label{sec:spatial_resolution}

 \mri typically offers sub-millimeter spatial resolution, rendering it inadequate for imaging subcellular biological structures. While \TEM, \MFM, and nano-SQUID can achieve nanometer-scale resolution, they are not suitable for imaging biological structures in living samples \cite{degen2009Nanoscale, finkler2010SelfAligned, rugar2004Single}. Although super-resolution optical microscopy, \stm, and \afm can achieve high-resolution imaging in biological environments, simultaneous integration of chemical recognition capabilities poses significant challenges.

The NV center presents a possibility to achieve nano and sub-nano level spatial resolution alongside chemical recognition under in vivo biological conditions. The chemical recognition capabilities and spectroscopic methods inherent in NV-based imaging techniques will play a pivotal role in physics, biology, and chemistry. This chapter introduces two methods: scanning probe magnetic imaging \cite{haberle2015Nanoscale,wang2019Nanoscale, maletinsky2012robust,grinolds2013Nanoscale,rugar2015Proton}, and gradient field-based magnetic imaging \cite{grinolds2014Subnanometre,Arai2015,muller2014Nuclear,abobeih2019Atomicscale}.

\subsection{Scanning probe magnetic imaging}

\begin{figure}[htp]
\begin{overpic}[width=1\columnwidth]{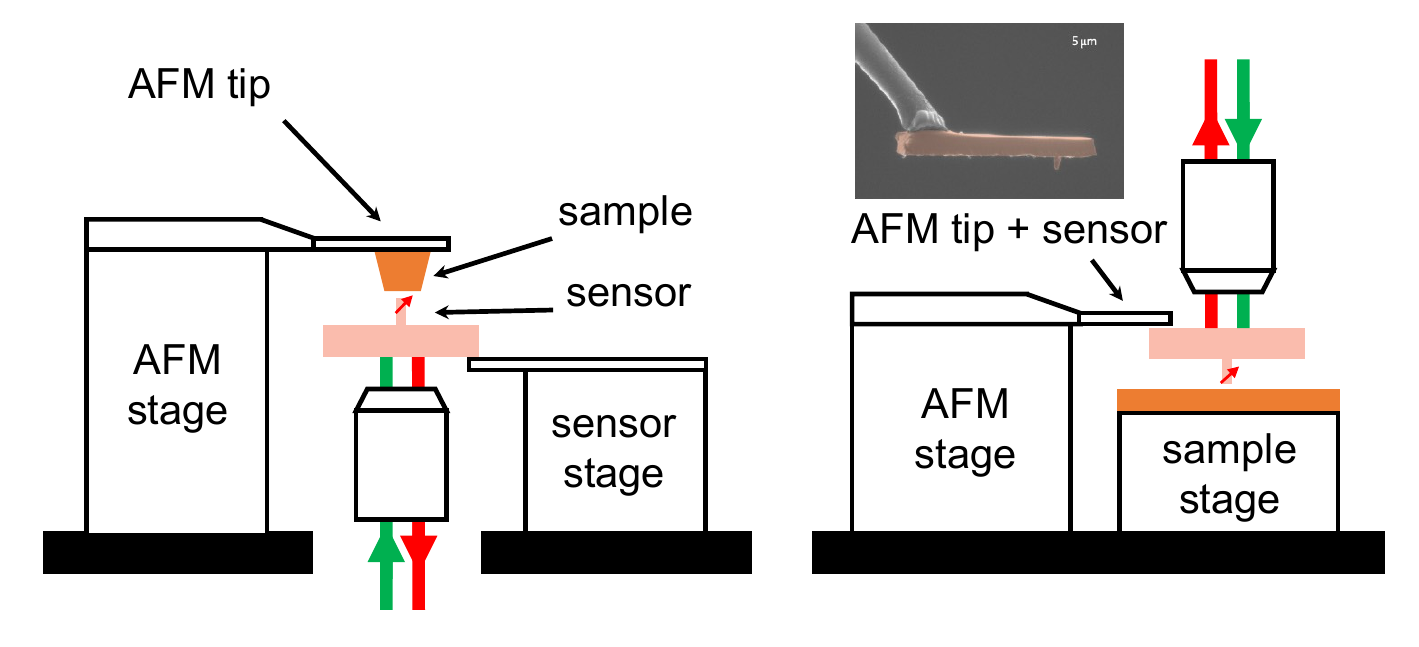}
\put (0, 43) {(a)}
\put (53, 43) {(b)}
\end{overpic}
\caption[NV Center]{ Schematics of NV scanning probe microscopy. (a) The target sample is attached to the cantilever of the conventional AFM probe. The sample scans above the diamond which contains a shallow NV center. The NV sensor is read out by the confocal microscope. (b) The NV sensor is fabricated at the end of a diamond nanopillar (as shown in the inset figure). The diamond nanopillar is utilized as a magnetic probe to scan over the target sample.
 The inset is adapted from \cite{maletinsky2012robust}. }  \label{fig:scanning}
\end{figure}

Scanning microscopy stands as one of the most potent tools for exploring the nanoscale world, encompassing various techniques such as optical microscopy \cite{betzig1993Single}, \sem \cite{glaeser1985Electron}, \stm \cite{binnig1982Tunneling}, spin-polarized \stm \cite{wiesendanger2009Spina}, \afm \cite{giessibl2003Advances,gross2009Chemical} and \mrfm \cite{degen2009Nanoscale,rugar2004Single}. Integrating magnetic resonance measurements with \AFM via the fabrication of individual nanoscale diamond probes allows high spatial resolution magnetometry under less stringent conditions.
Combining the \NV sensor with \AFM necessitates a confocal optical microscope system (Fig. \ref{fig:scanning}). A green laser from this system is utilized for addressing, initializing, and reading out the NV center. The spin state-dependent fluorescence is collected by an objective, enabling the acquisition of the sample's magnetic image  by scanning it across the NV sensor while simultaneously measuring the desired signal.
The leapfrog \AFM scanning mode, chosen due to the diamond's hardness minimizing sample abrasion, ensures result reproducibility \cite{rugar2015Proton, wang2019Nanoscale}. The fluctuating magnetic field follows an inverse sixth power law ($\sim 1/r^6$, \eq \ref{eqn:ion_field}), limiting achievable resolution to the depth $d$. Achieved spatial resolution ranges from 8.3 nm to 20 nm, contingent upon the distance between the NV center and the sample \cite{haberle2015Nanoscale, rugar2015Proton, wang2019Nanoscale, maletinsky2012robust}.
Two forms of NV center-based scanning microscopy exist: one involves attaching the target sample to the scanning AFM tip over the NV sensor (Fig. \ref{fig:scanning}(a)), while the other employs high-purity diamond nanopillars containing NV centers as the AFM tip \cite{maletinsky2012robust} (Fig. \ref{fig:scanning}(b)).

\begin{figure}[htp]
	\begin{overpic}[width=1\columnwidth]{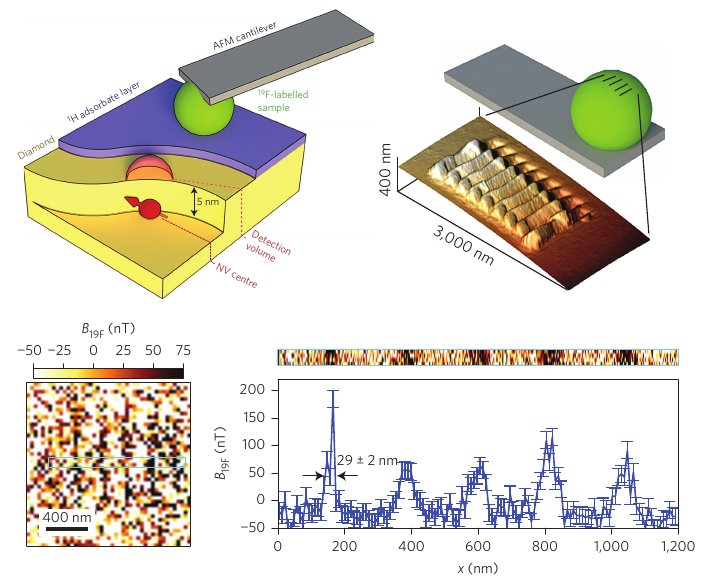}
		\put (0, 76) {(a)}
		\put (0, 36) {(c)}
		\put (30, 36) {(d)}
		\put (54, 76) {(b)}
	\end{overpic}
\caption[NV Center]{  Nanoscale MRI of fluorine (\Fnine) and protons (\Hone). (a) Utilizing a single shallow NV center (red spin), nuclear spins are detected within a nanoscale sensing volume above the diamond surface (red hemisphere). Imaging involves scanning the target sample through this sensing volume. (b) A calibration grating is etched into the sample (green sphere representing \Fnine-rich Teflon) using the AFM tip. (c) The \Fnine NMR imaging of the sample (calibration grating). (d) The accumulated signals obtained through 4 $\times$ 3 binning of the linescan in (c). The figures are adapted from \cite{haberle2015Nanoscale}.  }
  \label{fig:scanning_NMR}
\end{figure}
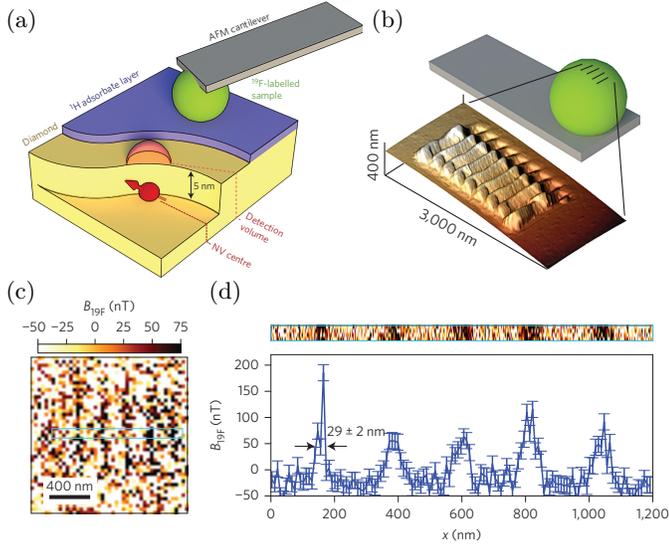

The first scheme, compatible with the typical AFM setup, enhances spatial resolution beyond optical diffraction limits. Here, the microscale sample attaches to the cantilever of an \AFM tip and scans above the diamond containing a shallow NV center (Fig. \ref{fig:scanning_NMR}(a)). This configuration enables the implementation of scanning magnetic resonance imaging. For instance, in one setup, a nanoscale fluorine sphere scans over an \NV sensor while the sensor conducts NMR DD measurements (\se \ref{sec:NMR:DD}) simultaneously to measure the NMR spectra \cite{haberle2015Nanoscale}. By analyzing the NMR spectra of \Fnine nuclear spins, the power spectral density of the nuclear magnetic signal reconstructs from the fluorescence of the NV sensor, facilitating \Fnine density measurement \cite{haberle2015Nanoscale} as illustrated in Fig. \ref{fig:scanning_NMR}(b). In another experiment, scanning \Hone NMR observes the \Hone nuclear spin sample, represented by a sharp polymer tip drawn from glass (Fig. \ref{fig:scanning_NMR}(c,d)) \cite{rugar2015Proton}. The method achieves a spatial resolution of $\sim$ 10 nm and a sensitivity of $\sim$ 200 nT$\cdot$Hz$^{-1/2}$. A pertinent application involves measuring ferritins in liver cancer cells. Here, the cell, fixed in a solid state, is subsequently sectioned into a cube and placed on a tuning fork scanning probe attached to an \AFM, exposing the cell's flat cross-section to air. To minimize cell  sample abrasion and ensure result reproducibility, scanning occurs in a slow ``leapfrog'' mode, albeit limiting sensitivity. This approach achieves intracellular ferritin imaging with a spatial resolution of about 10 nm.

The second approach situates a single NV center at the termination of a high-purity diamond nanopillar (Fig. \ref{fig:scanning}(b)), amalgamating extended spin coherence time, augmented photon collection efficiency due to optical waveguiding \cite{babinec2010diamond}, and superior spatial resolution.
Beyond the pointed diamond probes, flat probes find application in scanning probe microscopy through far-field optical leveling techniques \cite{ernst2019Planara}. These flat probes streamline the diamond engineering process and mitigate potential engineering damage.

Imaging via short-distance scanning probes achieves nanoscale, and sometimes sub-nanoscale, resolution. The inception of nanoscale NV magnetic imaging occurred through the NV sensor integration into a diamond nanocrystal affixed to the \AFM tip \cite{Balasubramanian2008}. The 40 nm diameter of the nanocrystal ensures close proximity between the NV center and the sample, yielding a spatial resolution in the range of several tens of nanometers.

While these methods achieve impressive spatial resolution, their detection sensitivity remains considerably limited, falling short of single-molecule applications. An alternative approach involves employing nano-fabricated diamond optical waveguides to enhance the NV sensor's optical collection efficiency. This optical waveguide constitutes a nano-pillar formed on a thin film of high-purity diamond single crystals using nano-fabrication techniques (Fig. \ref{fig:scanning}(b) inset). NV sensors crafted within these nano-pillars possess minute end surfaces (ranging between $\sim$ 100 nm and $\sim$ 1 $\mu$m), mitigating the typically short coherence times observed in NV centers within nanodiamonds. The design of optical waveguides in the nano-pillars facilitates stronger coupling between the excitation light and the embedded NV center, thereby amplifying optical excitation efficiency.  Moreover, this design alters the fluorescence emission light field of the NV center, resulting in a substantial increase of about 5 to 10 times in the collected far-field photon numbers by the objective (\se \ref{sec:collection_efficiency}). This heightened signal collection efficiency and enhanced excitation efficiency hold promise for improving measurement sensitivity while concurrently reducing the intensity of the excitation light and its impact on the sample.

\begin{figure}[htp]
	\begin{overpic}[width=1\columnwidth]{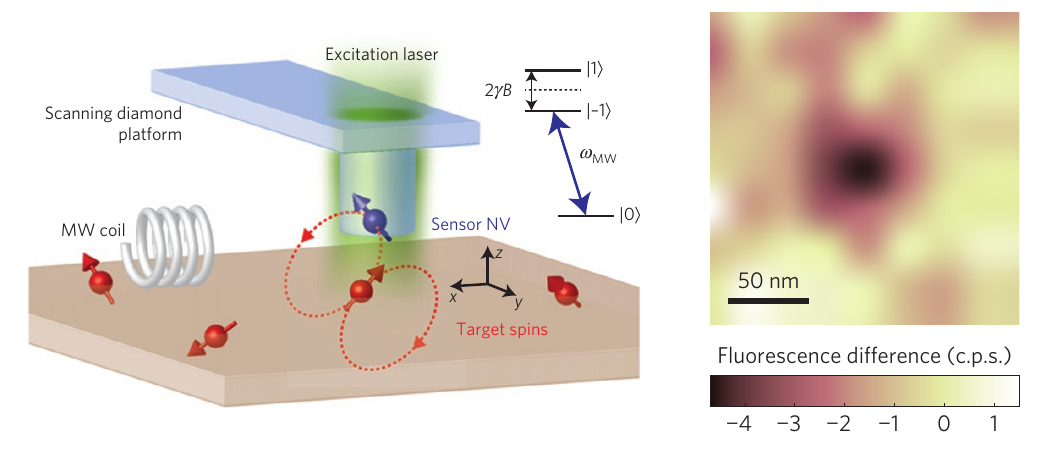}
	\put (0,39) {(a)}
	\put (62,39) {(b)}
\end{overpic}
\caption[NV Center]{
(a) A nanopillar embedding an NV center is employed to scan the target sample. (b) The NV sensor observes the magnetic field imaging of a single electron spin. A substantial decrease in fluorescence near the center signifies the detection of a single electron spin. These figures are adapted from \cite{grinolds2013Nanoscale}.  }  \label{fig:scanning_AFM}
\end{figure}

Through highly efficient optical waveguiding (Fig. \ref{fig:scanning_AFM}(a)), the sensitivity of the scanning NV sensor reaches 96 \nTHz, primarily constrained by surface noise and background fluorescence \cite{grinolds2013Nanoscale}. Given that the magnetic field of a single electron spin measures approximately 9 nT (1 $\mu$T) when the distance between the sensor and the target spin spans about 50 nm (10 nm), detecting a single electron spin becomes feasible (Fig. \ref{fig:scanning_AFM}). To map the magnetic field intensity of the target single electron spin, the NV magnetic probe averages multiple scans across an area of approximately $\sim$ 200 $\times$ 200 nm$^2$. In these scanning experiments, the spin sensitivity can attain 11 $\mu_B\cdot$Hz$^{-1/2}$ at ambient temperature \cite{grinolds2013Nanoscale}. While this sensitivity stands 15 times lower compared to ultra-low temperature \MRFM experiments \cite{rugar2004Single}, future advancements aim to surpass or meet this level by reducing the sensor-to-target distance.

\subsection{Gradient field based magnetic imaging} \label{sec:gradient}

Improved spatial resolution becomes attainable through the exploitation of magnetic gradients, altering the resonant frequencies of spins across various spatial locations. Magnetic gradient techniques enable precise addressing of nuclear spins in conventional NMR at resolutions down to $\sim\mu$m. This technological advancement has propelled the use of MRI in medical and biological  sciences, facilitating microscopic insights into organisms \cite{lee2001One,mansfield2004Snapshot,lauterbur1973Image,plewes2012Physics}. Nevertheless, conventional NMR's spatial resolution struggles to surpass $\sim\mu$m due to limited sensitivity and gradients \cite{glover2002Limits}. Leveraging the NV center's high sensitivity and atomic-scale positioning presents a distinctive opportunity for magnetic imaging at nanoscale or even single-molecule levels.

\subsubsection{External applied gradient magnetic fields}

\begin{figure}[htp]
\begin{overpic}[width=1\columnwidth]{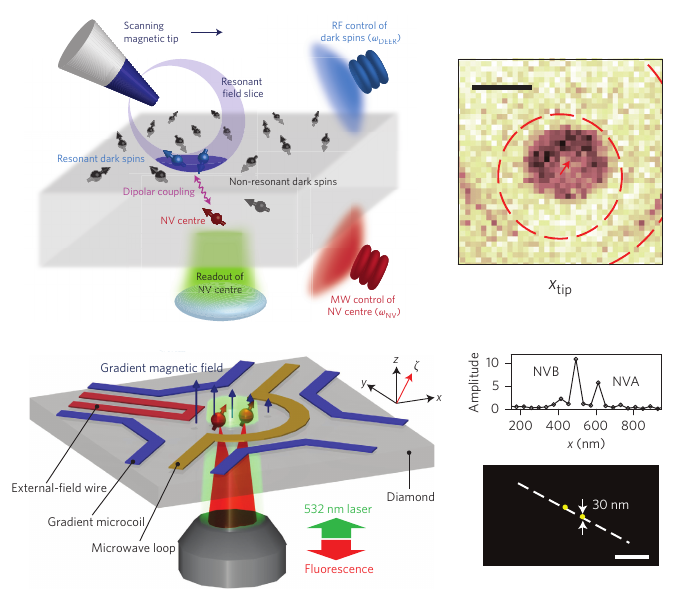}
\put (2, 82) {(a)}
\put (62, 82) {(b)}
\put (69, 76) {25 nm}
\put (2, 35) {(c)}
\put (64, 35) {(d)}
\end{overpic}
\caption[NV Center]{ Gradient-based MRI with NV sensor. 
(a) MRI of dark spins in a diamond utilizing a scanning gradient and a single NV sensor. A scanning magnetic tip is positioned 100 nm above the diamond surface, enabling the measurement of the spatial distribution of dark spins density by the NV sensor.
(b) 0.8 nm resolution NV-MRI of a single dark spin. Figures (a,b) adapted from \cite{grinolds2014Subnanometre}.
(c) Schematic of Fourier magnetic imaging. An adjustable magnetic field gradient for phase encoding is established by passing an electric current through a pair of gradient microcoils.  Currents of 1 A through two gradient microcoil pairs generate magnetic field gradients up to 0.2 $\mu$T$\cdot$nm$^{-1}$.
(d) Nanoscale resolution MRI of the NV center.
Top: cross-section depicting two distinct peaks along the direction indicated by the white dashed line in the \td image. These peaks correspond to two separate NV centers spaced 121(9) nm apart.Bottom: two-dimensional real-space images of the Fourier transform of these two NV centers. Figures (c,d) adapted from \cite{Arai2015}.
 }  \label{fig:gradient}
\end{figure}
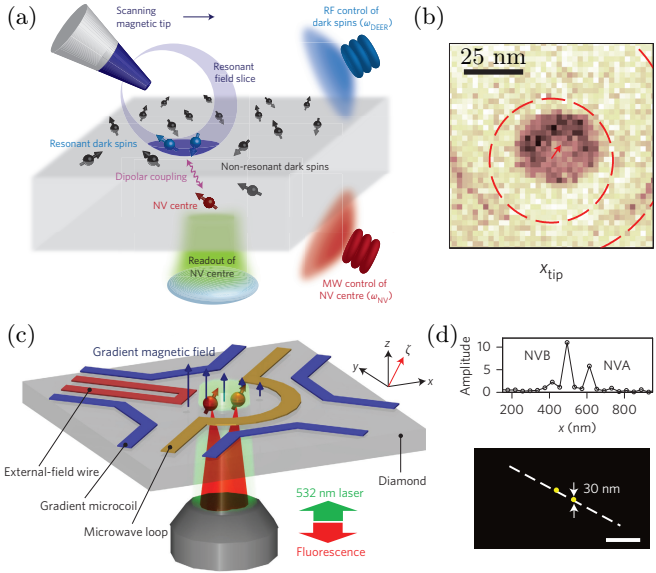

By utilizing an external magnetic field gradient, higher spatial resolution is possible, which is independent of sensor-sample distance. The magnetized AFM tip can be positioned infinitesimally close to the sample, thereby generating a strong magnetic field gradient. The magnetic tip can be produced by depositing a layer of magnetic material on the \afm tip \cite{grinolds2011Quantum}. A typical magnetic probe can produce a magnetic field gradient of approximately 1 mT$\cdot$nm$^{-1}$ at a distance of approximately 20 nm \cite{grinolds2014Subnanometre}. By placing the magnetic tip above an NV center, as shown in Fig. \ref{fig:gradient}(a), the magnetic field gradient is combined with the NV-based magnetic resonance microscopy, thereby enabling three-dimensional imaging of the target spins. The driving microwave frequency and the inhomogeneous magnetic field determine a narrow spatial volume, a ``resonant slice'' \cite{rugar2004Single}, where the spins resonate with the driving microwave frequency. By scanning the magnetic tip precisely, the high-resolution three-dimensional map is obtained. Specifically, the \psf of the measured spins can be directly determined experimentally using the spatial map of the resonant \nv, which is independent of the magnetic field models or iterative deconvolution schemes. The density distribution of the spins to be measured can be reconstructed by convolving the measurement results by the \psf. Thus, using the NV sensor , \mri measurement for the target spins can be accomplished with the \epr or \nmr methods.

The spatial resolution of the gradient-based \mri is given by \cite{grinolds2014Subnanometre},
\begin{equation}
\delta r=\frac{2\pi}{t_s\gamma_e|\vb{n}\cdot\nabla B_{\text{tip}}|},
\end{equation}\label{eqn:gradient_resolution}
where $\gamma_e$ denotes the gyromagnetic ratio of the target spin, $t_s$ represents the interrogation time, $\vb{n}$ depicts the unit vector of the target spin's quantization axis, and $\nabla B_{\text{tip}}$ indicates the magnetic field gradient at the target spin position. For electron spin measurements, a spatial resolution of 2.5 {\AA} is achievable when the magnetic gradient is 1.2 mT $\cdot$ nm$^{-1}$ and the interrogation time is 120 ns \cite{grinolds2014Subnanometre}. This spatial resolution can be further enhanced by commercial magnetic recording heads, which can generate up to $\sim$ 10 mT$\cdot$nm$^{-1}$ magnetic field gradient \cite{bodenstedt2018Nanoscale,jakobi2017Measuring}. The amalgamation of scanning gradient fields with NV magnetic resonance microscopy holds promise for achieving sub-nanoscale resolution in three-dimensional NV-based MRI. However, the current spatial resolution of NV-based MRI  confronts limitations on two fronts. Firstly, maintaining stability in the relative position between the magnetic tip and the NV center is critical, yet even well-designed setups experience position drift of several nanometers due to 1 K temperature vibrations \cite{grinolds2014Subnanometre, wang2019Nanoscale}. Secondly, the interrogation time is restricted by \Ttwostar, which is consistently much shorter than \Ttwo for shallow NV centers. However, the intricate \dyde control protocols are incompatible with static magnetic gradients.

The adjustable magnetic field gradient can be integrated with the dynamic control protocol of the NV center, utilizing microcoils that adjust currents to generate these gradients. Enhancements in spatial resolution and measurement efficiency can be achieved by extending the interrogation time limit from \Ttwostar to \Ttwo. Encoding both the magnetic field gradient and the spatial information of spins in phase , the real-space image of the spin distribution can be reconstructed via Fourier transformation. In experimental setups (Fig. \ref{fig:gradient}(c)), the NV magnetic resonance microscope is positioned below a pair of gradient microcoils—metallic wires capable of producing adjustable magnetic field gradients through delivered currents. However, limitations arise due to electric resistance heating, restricting the achieved magnetic field strength to 0.07 \mTnm at an interrogation time $\tau=104$ $\mu$s. This limitation results in a spatial resolution of < 5 nm and a dynamic range of approximately 500 \cite{Arai2015}. Utilizing Fourier magnetic imaging, the magnetic resonance frequency encoding selectively addresses electron spins \cite{zhang2017Selective}.

\subsubsection{Spin based gradient magnetic fields}

\begin{figure}[htp]
\begin{overpic}[width=1\columnwidth]{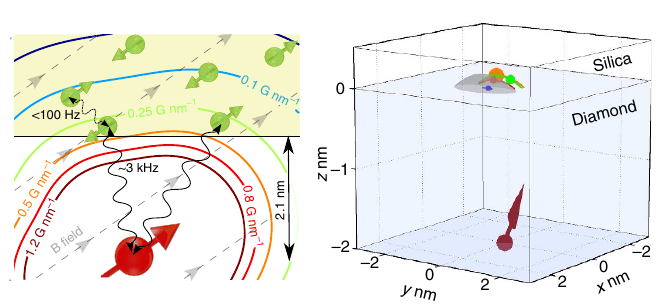}
\put (0, 42) {(a)}
\put (50, 42) {(b)}
\end{overpic}
\caption[NV Center]{  Nanoscale resolution MRI of single nuclear spins using the gradient magnetic field of the NV center.
(a) A shallow NV center in diamond (2 nm from the surface) couples with nearby nuclear spins due to hyperfine interaction. The contour lines depict the strength of the effective magnetic gradient experienced by the nuclear spins.
(b) The optimal locations of the four most significant contributing nuclear spins, determined through basis pursuit, are illustrated as colored arcs. Each spin's representative position is marked with a filled ball, and the positional uncertainty in the x and z directions is indicated by the ball's size. The y-axis resolution ranges from 0.1 nm for the nearest spin to 0.5 nm for the furthest spin. Figures (a,b) are adapted from \cite{muller2014Nuclear}.
 }  \label{fig:MRI}
\end{figure}

The NV center itself possesses a magnetic electron spin, generating a robust magnetic field gradient when in close proximity to a sample. This gradient, produced by the NV center, serves as a field gradient in \mri  applications. For an NV center positioned at a depth $d$, the effective magnetic field gradient (Fig. \ref{fig:MRI}(a)) is described as follows:
\begin{equation}
\abs{\pdv{(H_{\text{dip}}/\gamma_{\text{tar}})}{\vb{r}}}=\frac{3\mu_0\gamma_e\hbar}{4\pi r^4}\sqrt{\cos^2\theta(1+5\cos 2\theta)}
\end{equation}
which is $\sim$ 0.1 mT$\cdot \text{nm}^{-1}$ for 2 nm depth \NV center.

The NV spin gradient proves useful in detecting nuclear spins, where the target sample is deposited onto the diamond's surface (Fig. \ref{fig:MRI} (a)). Diverse NV-based NMR techniques, such as DD, ENDOR, and HHDR, can be applied. In the presence of high-order DD, the NMR signal exhibits apparent inhomogeneous broadening (Fig. \ref{fig:MRI} (a)). The spatial information of nuclear spins becomes encoded in the frequency shift induced by the NV hyperfine interaction. Although the frequencies of individual nuclei cannot be resolved in Ref. \cite{muller2014Nuclear}, extracting them is feasible using the super-resolution capabilities of basis pursuit de-noising \cite{chen1998Atomic}. The experimental results display the spectral decomposition of contributing nuclear spins and their hyperfine coupling parameters. It's indicated that four nuclei constitute over 50\% of the signal, a size comparable to many chemical functional groups and smaller than a small amino acid. Despite the added complexity, estimating the location of two nuclei is achievable with an uncertainty below 0.2 nm (Fig. \ref{fig:MRI}(b)) \cite{muller2014Nuclear}.

Imaging nuclear spins in molecules rather than the sensor background poses an immense challenge due to limitations imposed by diamond and shallow NV centers. Although demonstrated solely in amorphous silica \cite{muller2014Nuclear}, the  potential applications of NV-based \mri have enabled the analysis of multiple theoretical schemes and demonstrations of the sensor background signal. Several sensing protocols have been proposed to resolve the spatial positions of nuclear spins using one-dimensional or multi-dimensional \nmr methods \cite{ajoy2015AtomicScale,kost2015Resolvinga,perunicic2016Quantuma,wang2016Positioning}. Beyond the NV gradient, resolving the interaction among nuclear spins allows determination of the distance and direction between nuclear spin pairs \cite{shi2014Sensing,yang2019Structural,kong2020Artificial,zhao2011Atomicscalea}. Proof-of-concept experiments have been conducted to demonstrate \mri of the background \Cthir nuclear spins in a diamond sensor. Absolute three-dimensional positions of nuclear spins can be determined by manipulating the direction of the external magnetic or spin control field \cite{zopes2018ThreeDimensional,zopes2018Threedimensionalb}. However, the coherence time limitation restricts the resolution of these methods, making them suitable only for locating strongly coupled nuclear spins numbering less than ten. Cryogenic temperature multidimensional \nmr enables \mri of 27 nuclear spins with sub-angstrom resolution \cite{abobeih2019Atomicscale}. Recently, a new parallel detection technique, combining weak quantum measurements, phase coding, and simulated annealing, has enabled imaging of over 20 \Cthir nuclear spins within 2.4 nm (equivalent to 5--6 nm for \Hone \nmr) at ambient temperature, achieving sub-angstrom resolution. These proof-of-concept imaging techniques using one-dimensional or multidimensional NMR spectroscopy mark a significant breakthrough toward future

The gradient of the NV or electron spin holds promise for structural analysis through EPR. In traditional EPR, DEER measurements analyze distances between two electron spins, such as in proteins with dual spin labels. Here, coupling depends on both the distance and direction between the labels. Ensemble averaging over all possible directions allows the extraction of distance, a powerful tool for biomolecular structure analysis, especially for biomacromolecules. The NV sensor offers the potential to measure this spin-spin coupling at a single-molecule level. Extracting distance involves repeating measurements with varying directions of the external magnetic field. In traditional EPR, measured distances often have significant variance due to spin label decoherence. Additionally, molecular heterogeneity contributes to variance, indistinguishable from decoherence. NV-EPR might offer more precise distance information due to the unique capabilities of NV-based measurements. This heightened precision could enable the resolution of single-molecule conformational transitions  \cite{munuera-javaloy2022Detectiona}. 

\subsection{Conclusion}

In the pursuit of single-molecule MRI using NV centers, the aim is to attain structural resolution for individual molecules. Unlike established techniques like X-ray crystallography or cryo-electron tomography, NV-based MRI only needs a single copy of a molecule. This allows direct observation of conformational variations among individual molecules, potentially unveiling new insights into their structure and functions. Both liquid and solid-state NMR measurements are feasible with NV centers for structural analysis. Liquid-phase single-molecule NMR  (discussed in sec. \ref{sec:spectral_resolution:conclusion}) relies on effects like the nuclear Overhauser effect, residual dipolar coupling, or paramagnetic relaxation enhancement. Conversely, solid-state single-molecule NMR relies on dipolar interactions between NV centers and nuclear spins or among nuclear spins themselves. The proof-of-concept methods outlined in this chapter, employing external gradients or NV-coupled techniques, have achieved resolutions within the angstrom range, meeting the structural resolution requirements. Through cryogenic temperatures or parallel detection methods, proof-of-concept experiments have shown fixed relative positions between NV centers and molecules with sub-angstrom resolution \cite{abobeih2019Atomicscale,cujia2022Parallel}.

Despite some proof-of-concept demonstrations in structural analysis, practical single-molecule measurements under biological conditions  pose challenges. The primary hurdles involve sensitivity limitations and stability (\se \ref{sec:sensitivity:conclusion}). In real biological settings, maintaining a stable relative position between sample molecules and NV sensors is crucial for achieving high-resolution single-molecule structural analysis.

Once the sensitivity challenge is addressed, single-molecule imaging with less stringent resolution can also be accomplished, complementing single-molecule structural analysis. In comparison to traditional optical super-resolution microscopy, NV-based single-molecule imaging offers simultaneous analysis of individual molecules within in vivo biological environments. This includes chemical recognition, single-molecule dynamics, chemical environment, and chemical reactions. While achieving nuclear spin chemical resolution requires specific conditions, recognizing chemical species within magnetic fields of around $\sim$ 10 mT is readily feasible.

\section{Sensor fabrication}\label{fabrication}

This chapter provides an extensive overview of processing techniques pertinent to NV sensors, encompassing the establishment of NV centers (\se \ref{sec:NV_creation}), surface modification (\se \ref{sec:surface_modification}), and diamond structure engineering (\se \ref{sec:diam_engineer}). Given the pivotal role of the NV sensor in NV-based quantum sensing, these techniques are anticipated to address previously highlighted challenges. Initially, NV centers are formed in diamond through ion implantation or delta doping, followed by subsequent annealing. Furthermore, the advantage of NV lies in its ability to sustain high sensing sensitivity in close proximity to the surface. Thus, surface treatment methods aid in enhancing coherence time, stabilizing charge states, and facilitating sample assembly. Lastly, diamond engineering holds significance. On one hand, the optical structure enhances optical collection efficiency, thereby elevating readout sensitivity. On the other hand, diamond can be shaped into a probe suitable for relevant sensing tasks.

\subsection{NV creation} \label{sec:NV_creation}

The realization of \ssmr relies on NV centers situated several to tens of nanometers below the diamond surface. An NV center comprises a nitrogen atom substituting for a carbon atom alongside a vacancy. Introduction of the nitrogen atom commonly occurs through nitrogen ion implantation or delta doping. The vacancy, naturally introduced through ion implantation, can also be incorporated via the delta doping method using ion implantation, laser writing, or electron irradiation. Subsequent vacuum annealing becomes imperative to amalgamate the vacancy and nitrogen, culminating in the formation of an NV center and bolstering its properties.

\begin{figure}[htp]
\begin{overpic}[width=0.95\columnwidth]{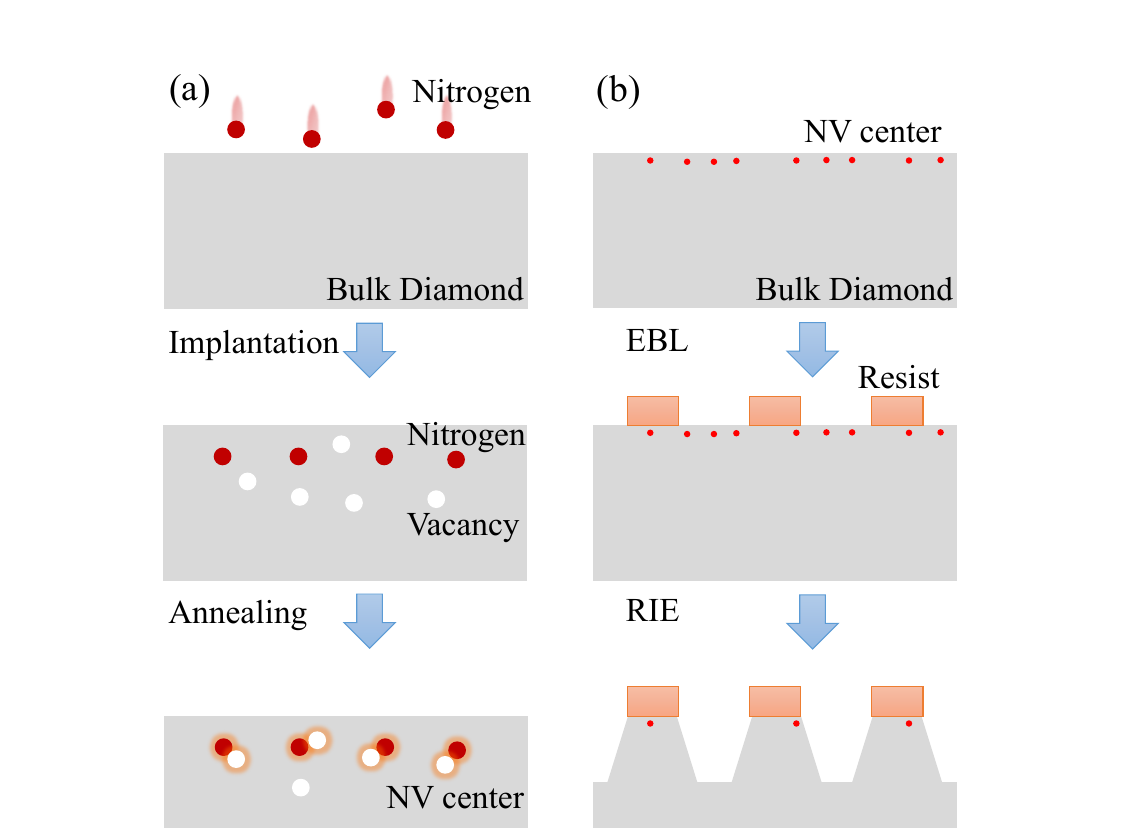}
	\put (-4, 98) {(a)}
	\put (46, 98) {(b)}
\end{overpic}
\caption[NV Center]{ (a) Process for creating NV center via ion implantation. NV centers are formed through nitrogen ion implantation (with energies ranging from 2 keV to 50 keV and a density of $10^8$ to $10^{11}$ cm$^{-2}$), followed by high-temperature annealing (ranging from 600 \celsius to 1200 \celsius). (b) Delta-doping procedure: A thin nitrogen-doped layer (light red) is produced by introducing N$_2$ gas during the diamond growth process. Subsequently, irradiation via laser, ions, or electrons induces vacancies. Annealing is then performed to generate NV centers.    }  \label{fig:ion_implantation}
\end{figure}

Ion implantation is a well-established and pivotal technique for altering the physical or chemical properties of a target sample by accelerating ions of an element into it. Widely employed in semiconductor device fabrication, this method plays a crucial role in quantum device creation to generate shallow NV centers \cite{pezzagna2010Creation,pezzagna2011Creation,spinicelli2011Engineered}.

\begin{figure}[thp]
\begin{overpic}[width=0.75\columnwidth]{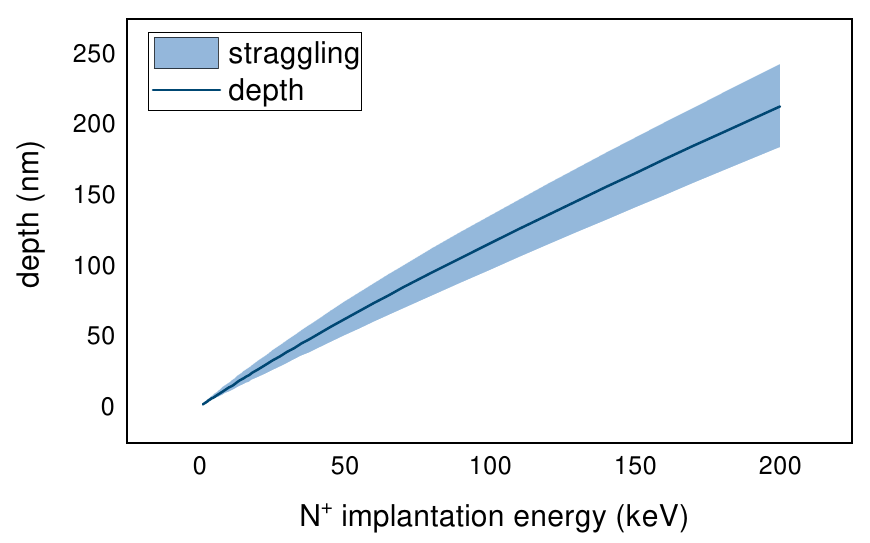}
\put (67.5, 11.5) {\includegraphics[width=0.11\textwidth]{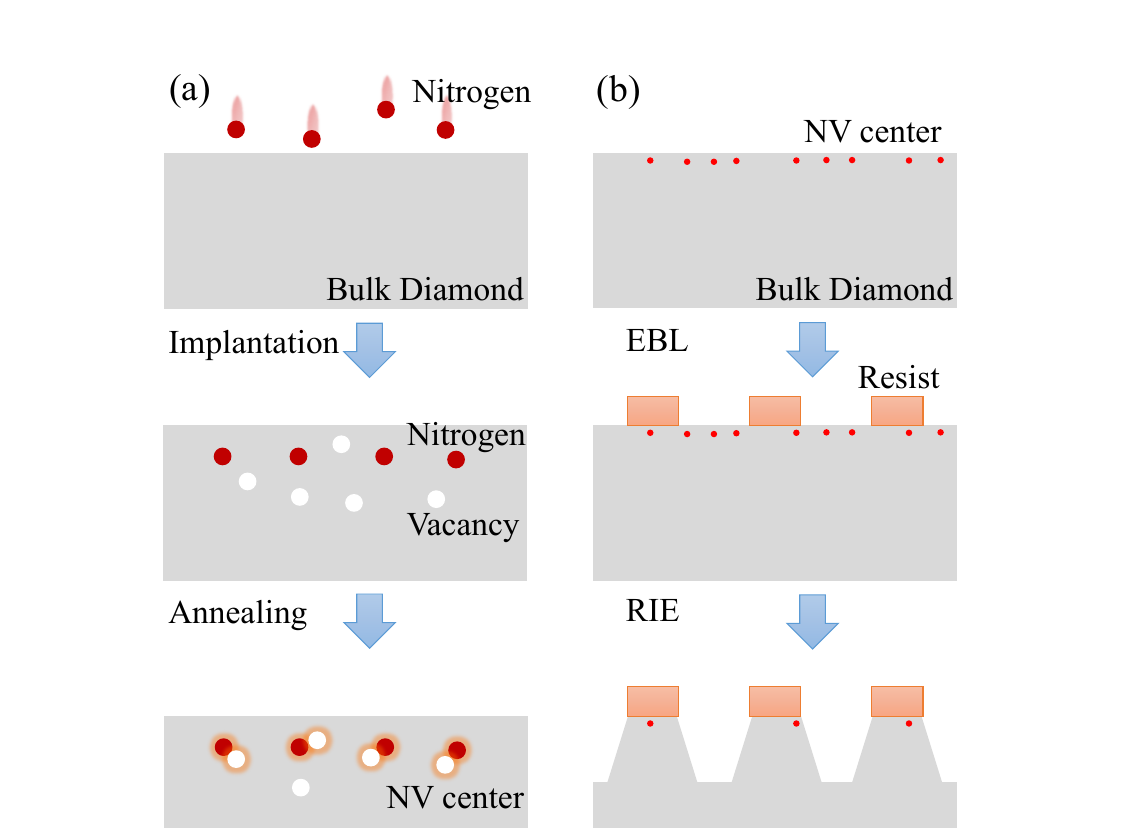}}
\end{overpic}
\begin{overpic}[width=0.8\columnwidth]{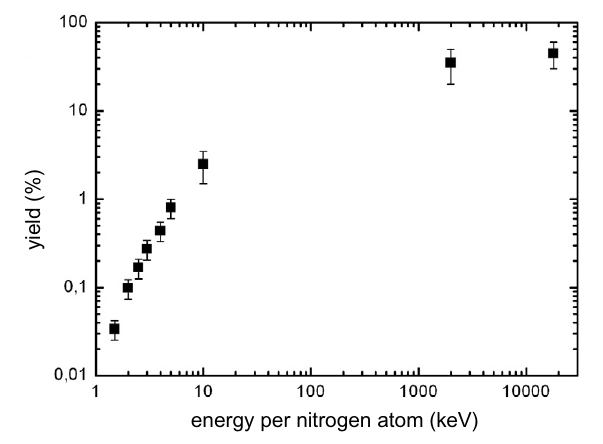}
	\put (-2, 128) {(a)}
	\put (-2, 70) {(b)}
\end{overpic}
\caption[NV Center]{ (a) The depth of the implanted ions versus the implantation energy. The blue shadow indicates straggling of ions. (b) The yield of the NV centers versus the implantation energy. Figure (b) is adapted from \cite{pezzagna2010Creation}. }\label{fig:srim}
\end{figure}

In the process, nitrogen ions are introduced into the diamond via low-energy ion bombardment, concurrently inducing vacancy creation (Fig. \ref{fig:ion_implantation}(a)). To create the shallow NV center essential for single-molecule spectroscopy, N$^+$ ions typically bombard the diamond with energies ranging between 2 and 8 keV, resulting in mean implantation depths of approximately 2 nm to 12 nm. These depths are calculated through \srim simulations (Fig. \ref{fig:srim}(a)). As the implanted ions collide with the diamond, their energy dissipates,  leading to positional deviations in both lateral and longitudinal directions, a phenomenon referred to as ``straggling''. This straggling effect imposes limitations on the spatial resolution achievable in NV sensor fabrication. Nevertheless, various strategies exist to regulate this effect.

The focused ion beam method \cite{meijer2005Generation,lesik2013Maskless,schroder2017Scalable} offers ion implantation with spatial precision smaller than 50 nm. Enhanced control over implantation location is achievable through the utilization of a nanoporous mask deposited on the diamond during ion implantation. Nanochannels in mica \cite{Neumann2010b,pezzagna2011Creation}, SiO$_2$ masks \cite{toyli2010ChipScale}, and polymethyl methacrylate (PMMA) masks \cite{spinicelli2011Engineered} are employed to create NV centers with spatial precision on the order of ten nanometers. Combining ion implantation with \afm facilitates the production of NV centers at specific positions with high spatial resolution or the preparation of NV center arrays with distinct patterns. This is achieved by fabricating a nanopore on a piezo-sensitive AFM tip made of silicon nitride (Si$_3$N$_4$). The tip maneuvers to a specific position, and the nanopore precisely focuses the ion beam to a location with a lateral spatial resolution of a few nanometers \cite{pezzagna2010Nanoscale,meijer2008implanting,riedrich-moller2015Nanoimplantation}.
Considering the longitudinal direction, the ion channeling effect becomes a crucial consideration. Optimal conditions for a [100] diamond surface necessitate a 12$^{\circ}$-inclined beam \cite{raatz2019Investigationa}.

However, the aforementioned implantation methods rely on a random ion source, posing challenges in achieving deterministic single ion implantation \cite{meijer2006Concepta}. This limitation, in turn, restricts the accuracy achievable in quantum sensor fabrication. Fortunately, the advent of a deterministic ion source has been realized by loading a single laser-cooled ion into a linear Paul trap \cite{alves2013Controlleda,jacob2016Transmissionb,zhou2013Transmissionb}. Through laser cooling, a single ion is precisely confined to a specific location.
By attaining deterministic ion implantation at precise locations, this method ensures minimal charging and irradiation, consequently minimizing damage to the diamond substrate \cite{groot-berning2021Fabricationa}. Presently, the production of individual NV centers deterministically via this method faces limitations due to the creation yield dependency on the need for NV to be bound to a defect alongside nitrogen. Despite this constraint, this approach signifies a crucial stride towards realizing a truly deterministic single NV sensor creation process in the future \cite{groot-berning2021Fabricationa}.

Another bottom-up approach involves delta-doping (Fig. \ref{fig:ion_implantation}(b)), wherein a minimal quantity of nitrogen is doped during the diamond growth using plasma-enhanced chemical vapor deposition (PECVD). In this process, a mixture of CH$_4$ and H$_2$ undergoes ionization into plasma by microwaves within the growth chamber. Nitrogen injection for 10 minutes while the [100]-surface diamond slowly grows at a rate of 8 nm/hour results in the growth of a nitrogen layer only a few nanometers thick \cite{Ohno2012}. Research efforts \cite{fukui2014Perfecta,lesik2015Preferential,ozawa2017Formation} have showcased that the diamond growth process can be conducted on surfaces of orientations other than the [100] surface. Additional defects are introduced through electron irradiation \cite{McLellan2016,Ohno2012}, ion implantation \cite{huang2013Diamond,ohno2014Threedimensional}, or laser writing \cite{chen2017Laser}. Growth of thin, uniform epitaxial diamond layers facilitates the formation of high-quality, shallow NV centers. Controlling the thickness and density of the nitrogen layer is contingent on the duration of nitrogen injection and the gas flow rate. Maintaining a slow growth rate minimizes defects and strain, ensuring extended coherence times for NV centers. Notably, NV centers produced via delta doping consistently exhibit long \Ttwo coherence times \cite{Ohno2012,ohashi2013Negatively,chandran2016Fabrication}. Moreover, preferential orientation of NV centers can be achieved by carefully selecting growth surfaces \cite{Lesik2014,Fukui2014,Michl2014,Miyazaki2014,Tahara2015,ozawa2017Formation}. However, the delta doping method, lacking vacancies, relies on the generation of vacancies through ion or electron irradiation to enhance NV center creation yield, potentially reintroducing other defects \cite{Oliveira2016,McLellan2016}.

Following ion implantation, the diamond undergoes annealing in a vacuum or forming gas. During this procedure, crystal damage is repaired, and vacancies move randomly until captured by nitrogen defects, culminating in the formation of NV centers. To prevent etching of the diamond surface, the vacuum maintains a pressure consistently below $5 \times 10^{-7}$ mbar \cite{ofori-okai2012Spin,wang2016Coherence,appel2016Fabrication}. Annealing temperatures typically range from 800 \celsius to 1200 \celsius, although studies have shown insignificant differences within this range \cite{appel2016Fabrication}. Alternatively, employing a forming gas—composed of 4\% H$_2$ in Ar at 8 GPa—serves as a protective measure for the diamond surface during annealing \cite{orwa2011Engineering,santori2009Verticala}. For optimal annealing under high pressure, temperatures between 900 \celsius and 1000 \celsius prove effective, whereas temperatures exceeding 2000 \celsius diminish the NV yield \cite{orwa2011Engineering}. The conversion efficiency from implanted nitrogen to NV centers typically remains below 1\% for implantation energies below 5 keV, resulting in depths around 8 nm (Fig. \ref{fig:srim}(b)) \cite{ofori-okai2012Spin,pezzagna2010Creation}. Yield improvements of twofold have been achieved using boron-doped diamond structures \cite{Oliveira2017}, while dopants like phosphorus, oxygen, or sulfur have enhanced yields by tenfold, correlating with improved coherence times \cite{luhmann2019Coulombdriven}.

\subsection{Surface modification} \label{sec:surface_modification}

Surface treatment holds paramount importance in \sm applications. Diamond is renowned for its chemical inertness, and its surface can be functionalized with diverse groups and attached to various molecules. Numerous coating methods based on surface chemistry have emerged in the realms of physics and biology for both diamonds and nanodiamonds \cite{krueger2008New,krueger2012Functionality,reina2019Chemical,jung2021Surface,james2021review,crawford2021Surface}. In nanoscale sensing applications, the performance of shallow NV centers is notably influenced by the diamond surface. Consequently, surface treatments for diamonds prove beneficial in sample assembly and enhancement of NV sensor properties.

\subsubsection{Chemical treatment}\label{sec:chem_treat}

Apart from dissolving and depositing samples on the diamond surface \cite{muller2014Nuclear,shi2015Singleprotein,schlipf2017molecular}, alternative approaches involve covalent methods (such as carbodiimide crosslinker chemistry and click chemistry) and noncovalent methods \cite{shi2018SingleDNA} for immobilizing samples on the diamond surface. The initial step involves surface purification, often achieved through wet chemistry methods employing harsh oxidation for surface cleaning. A common procedure involves treating the surface with a boiling mixture of nitric, perchloric, and sulfuric acids in a 1:1:1 volume ratio to restore a pristine diamond surface \cite{brown2019Cleaning}. This process effectively removes unwanted graphitic and pyrolytic carbon domains, substantially enhancing coherence times by over an order of magnitude \cite{Lovchinsky2016}. Additionally, strongly oxidizing mineral acids, such as the ``piranha solution'' (a mixture of sulfuric acid and hydrogen peroxide in a 3:1 volume ratio) \cite{Rondin2010}, or a mixture of sulfuric and nitric acids (3:1 volume ratio) \cite{tu2006Sizedependent}, have found wide usage. These wet chemical treatments introduce diverse oxygen-containing groups, including carboxyl groups, to the surfaces \cite{schlipf2017molecular,Lovchinsky2016}. Carboxylated diamond surfaces serve as versatile starting points for subsequent modifications required in biological and chemical applications. Silylation \cite{grotz2011Sensing} or acylation \cite{terada2018OnePot} of the diamond surface becomes feasible through the carboxyl surface. Furthermore, activation of carboxyl groups using EDC and NHS allows for the preparation of an aminated diamond surface \cite{sushkov2014alloptical,Lovchinsky2016}.

\begin{figure}[htp]
\begin{overpic}[width=1\columnwidth]{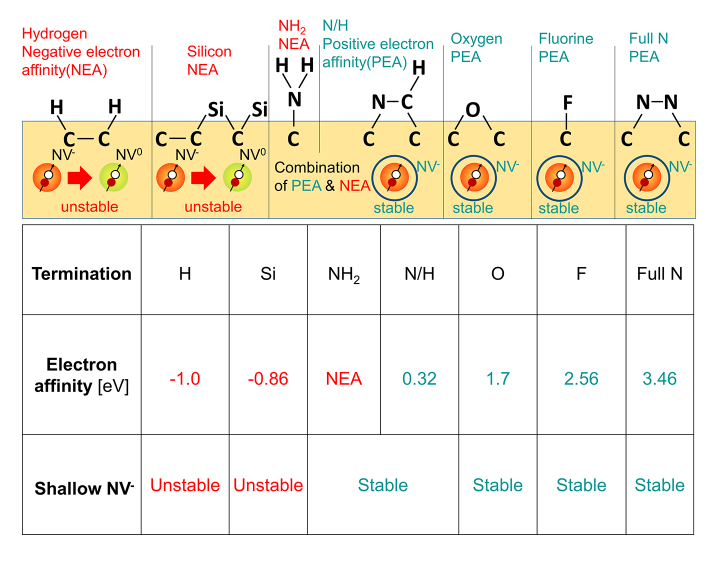}
\end{overpic}
\caption[NV Center]{ The electron affinity (eV) and the stability of NV charge state depend on the diamond surface termination for shallow NV center. The figure is adapted from \cite{kawai2019NitrogenTerminated}.
 }  \label{fig:affinity}
\end{figure}

The density of carboxyl groups on a carboxylated diamond surface typically remains below a few percentage points \cite{wolcott2014Surface}. Utilizing plasma treatment to achieve hydrogen termination on diamond surfaces presents a more effective chemical bonding method. Through hydrogen termination, functional groups including carboxylic acid (COOH$^{-}$), hydroxyl (OH$^{-}$) \cite{navas2018Oxygen}, hydrogen (H$^{+}$) \cite{nichols2005Photochemical}, amino (NH$_{2}^{-}$) \cite{zhu2016Aminoterminated}, and halide (F$^-$) \cite{rietwyk2013Work} can be affixed to the diamond surface. In general, hydrogen-terminated diamonds \cite{shpilman2008Hydrogen,cui2015Reduced} exhibit a more uniform surface compared to oxygen-terminated diamonds, making them preferable for subsequent chemical modifications.
Electrochemical methods \cite{krysova2016Efficiency,pinson2005Attachment} or photochemical reactions \cite{miller1996Photochemical,strother2002Photochemical,yang2002DNAmodified} offer pathways to synthesize a monolayer comprising 2$\times$10$^{14}$ cm$^{-2}$ amino groups or a monolayer ranging between 10$^{12}$ and 10$^{13}$ cm$^{-2}$ DNA from an H-terminated diamond surface \cite{yang2007Photochemical}.

\begin{figure}[htp]
\begin{overpic}[width=1\columnwidth]{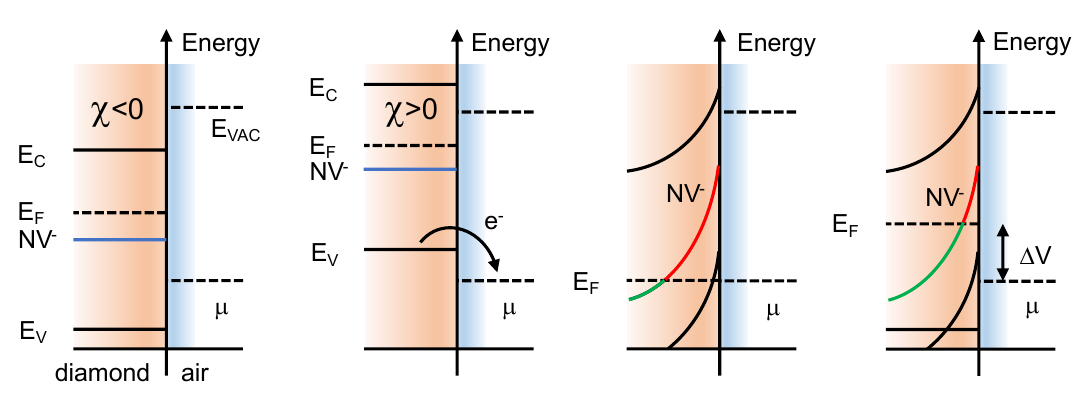}
	\put (0, 34) {(a)}
	\put (30, 34) {(b)}
	\put (55, 34) {(c)}
	\put (78, 34) {(d)}
\end{overpic}
\caption[NV Center]{ Energy band schematic of diamond. (a) Diamond surface with positive electron affinity. The conduction band $E_{\text{C}}$ is below $E_{\text{vac}}$ (O-termination for example). The negatively charged \NVm lies beneath the Fermi level $E_{\text{F}}$. (b) Diamond surface with negative electron affinity: the band levels are shifted up, and the electron in diamond can transfer into the acceptor states on the surface.  (c) The band is bending due to the effect of electron transfer. The \NV center prefers \NVo charge state near the surface. (d) A voltage is applied to lift the Fermi energy to control the \NV charge state.
 }  \label{fig:energy_band}
\end{figure}
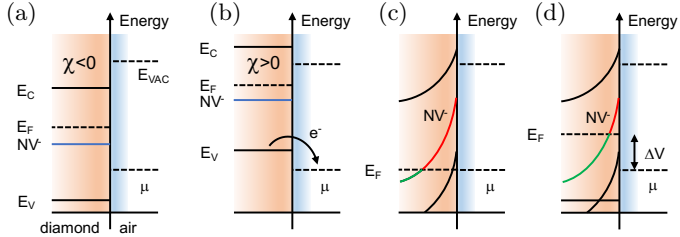

The stability of shallow NV centers critically hinges on the electron affinity polarity of the diamond surface \cite{kawai2019NitrogenTerminated}. Hydrogen termination, possessing negative electron affinity (Fig. \ref{fig:affinity}), induces surface charge transfer, rendering shallow NV centers unstable (Fig. \ref{fig:energy_band}(b)). Conversely, oxygen or fluorine termination, with positive electron affinity, stabilizes shallow NV centers \cite{rietwyk2013Work,Osterkamp2015}. Despite hydrogen-terminated surfaces exhibiting higher chemical modification efficiency, their unsuitability for NV-based quantum sensing due to instability prompts the exploration of novel modification methods or electrical control (\se \ref{sec:ele_control}). Recent advancements involve chemically modifying hybrid diamond surfaces through atomic layer deposition of aluminum oxide \cite{liu2022Surface,xie2022Biocompatible} and integration of graphene \cite{hao2023Sensing}. The aluminum oxide monolayer on the diamond surface not only simplifies technical complexities in chemical and optoelectronic applications but also serves as structural support in various catalytic processes \cite{liu2022Surface,xie2022Biocompatible}.

\subsubsection{ Electric Control } \label{sec:ele_control}

The NV center, comprised of a substituting  nitrogen atom and a vacancy, exhibits remarkable quantum properties when capturing an electron, rendering a stable charge state crucial for NV-based quantum sensing. Surface electric conditions wield a significant influence on charge state stability, particularly affecting shallow NV centers, which frequently encounter charge instability \cite{dhomkar2018Charge,bluvstein2019identifying}. Within a few tens of microseconds post-laser shutdown, the \NVo population rapidly escalates, causing a baseline drop that interferes with \Tone and \Ttwo measurements, thus compromising sensitivity. Mitigating this instability involves applying a negative voltage to the diamond surface (Fig. \ref{fig:energy_band}(d)), thereby stabilizing the NV center's charge state and its ambient environment \cite{Hauf2014}. A robust built-in electric field, formed within this stable charge environment, sustains NV charge stability even during laser illumination \cite{bian2021Nanoscale}.

Control over electron affinity comes via the chemical treatment of the diamond surface, while manipulation of negative voltage involves electrodes fabricated on the diamond surface. Electrically tuning the Fermi level facilitates rapid and stable manipulation of the NV charge state. Electric fields can be induced by various means, including electrolyte electrodes \cite{Grotz2012}, p-i-n diode structures \cite{Doi2014}, and in-plane configurations \cite{Hauf2014,Schreyvogel2015,pfender2017Protecting}. Alternatively, all-diamond in-plane gates can laterally deplete the conductive channel on a hydrogen-terminated diamond surface instead of using an electrolytic top gate \cite{Hauf2014,pfender2017Protecting}. While the addition of electrodes augments operational complexity in diamond sensors, it substantially improves quantum sensing by enhancing charge stability in the NV center and its surroundings. Instead of electrodes directly fabricated on the diamond, a conductive scanning tip provides nanoscale spatial precision for charge manipulation \cite{bian2021Nanoscale}. A stabilized charge environment not only secures the charge state but also extends coherence time by minimizing surface electric field noise effects on shallow NV centers \cite{zheng2022Coherence}.

Another advantage of charge stabilization lies in reducing sample line broadening under examination. Non-resonant optical readouts of NV centers induce stochastic electron spin and charge state switching, along with photoionization of surface defects, introducing random phase shifts to nearby nuclear spin precessions. While these disturbances minimally impact spectral line broadening at a micrometer scale sensing volume \cite{Glenn2018}, the scenario differs for detecting single or a few nuclear spins, where the coupling between nuclear spins and NV centers surpasses their intrinsic dephasing rate 1/\Ttwostar \cite{cujia2019Tracking}. Fluctuations in NV center charge state and diamond surface defects can impede chemical shift measurement resolution. Electrically stabilizing the NV center's charge state and environmental charges emerges as a promising remedy for this issue. 

\subsection{Diamond  engineering}\label{sec:diam_engineer}

\begin{figure}[htp]
\begin{overpic}[width=0.95\columnwidth]{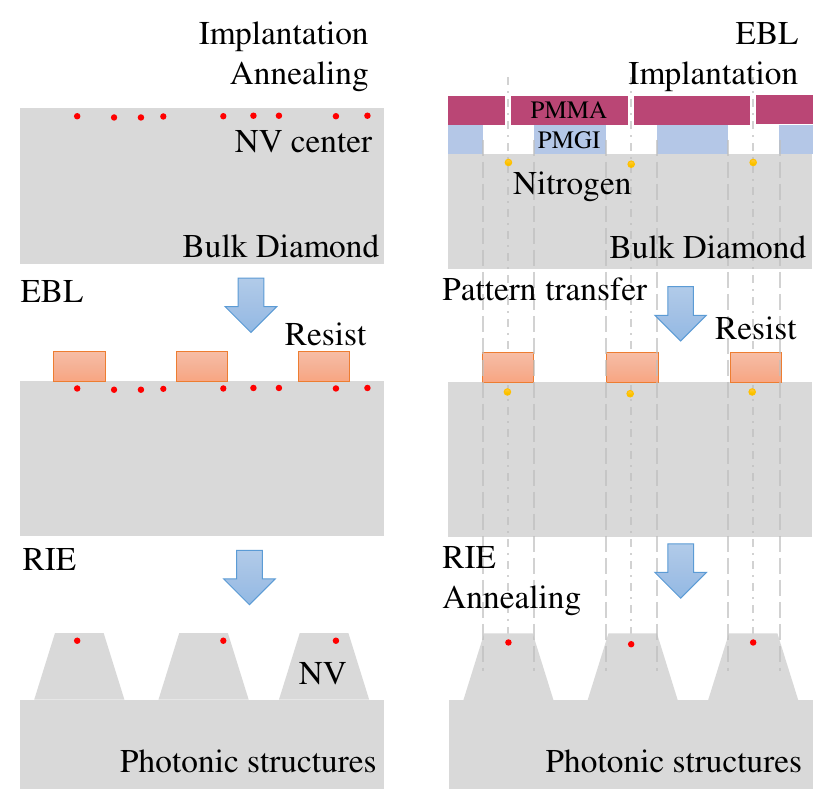}
	\put (2, 88) {(a)}
   	\put (52, 88) {(b)}
\end{overpic}
\caption[NV Center]{ (a) The diamond fabrication process without alignment. NV centers are created through maskless ion implantation followed by annealing. Subsequently, nanopillar photonic structures are fashioned using \ebl and \rie techniques. 
(b) Illustration of the self-alignment process employed for fabricating photonic structures. A double-layer mask comprising PMGI and PMMA, produced via a single-step \ebl process, confines the positions of both the etching resist and ion implantation region. Nitrogen ions are implanted into the diamond through the PMMA layer mask. Subsequently, the PMGI layer masks are transferred to the etching resist through a liftoff process. This is followed by using RIE techniques to produce the photonic structures, which are then annealed to convert the implanted nitrogen ions into NV centers. Adapted figures from \cite{wang2022Selfaligned}.
}\label{fig:pillar}
\end{figure}

Solid-state quantum sensors necessitate nano-fabrication techniques to address requirements for both biological \cite{shi2018SingleDNA,wang2019Nanoscale} and physical \cite{casola2018Probing} sensing tasks (Fig. \ref{fig:pillar}(a)). The NV sensor's functionality involves converting external magnetic fields into fluorescence signals. However, owing to diamond's high refractive index ($n\approx$ 2.4), NV center fluorescence faces hindrance from total internal reflection at the interface, a significant limiting factor for magnetic detection sensitivity. Simulations indicate that less than 5\% of the fluorescence from diamonds with [100] crystal orientation can escape the diamond-air interface \cite{siyushev2010Monolithic}.

Standard ``top-down'' fabrication methods have been employed to enhance fluorescence collection efficiency from NV centers. These methods include nanopillars (1 - 5 Mcps) \cite{babinec2010diamond,hausmann2010Fabrication,Choy2011,neu2014Photonica,Momenzadeh2015}, bullseye grating ($\sim$ 2.7 Mcps) \cite{Li2015}, and inverted nanocones ($\sim$ 2.7 Mcps) fashioned with conical Faraday cages for etching direction guidance \cite{jeon2020Bright}. \FIB milling has been utilized for structuring with sizes $\gtrsim\mu$m, like \sil ($\sim$ 1 Mcps) \cite{hadden2010Strongly,jamali2014Microscopic}. Additionally, Al$_2$O$_3$ anti-reflective coating applications have increased fluorescent counts to 1.2 Mcps \cite{robledo2011Highfidelity}. Larger structures, e.g., $\sim$ 1 mm \sil, are created through laser and mechanical processing stages \cite{siyushev2010Monolithic}. Commonly used fabrication methods include:
\begin{enumerate}[label=(\roman*)]
  \item  Standard ``Top-Down'' Method: This method involves fabricating etching resists masks (such as PMMA or silicon) on diamond surfaces through photolithography or \ebl. Subsequently, reactive plasma technology etches the diamond surface to create desired structures \cite{babinec2010diamond,hausmann2010Fabrication,Choy2011,neu2014Photonica,Momenzadeh2015}.
  \item \FIB Milling: Employing gallium or oxygen ions, this technique directly cuts diamond into desired structures. However, the use of ions in this process can lead to sensor contamination. It finds application in instances such as cutting the diamond holder \cite{appel2016Fabrication} or milling SIL where the ions are distant from the NV center during fabrication.
  \item Laser and Mechanical Processing Stages: This technique mills larger diamond structures, such as $\sim$ 1 mm diameter \SIL, achieving flatness better than 10 nm (rms) \cite{siyushev2010Monolithic}.
  \item ``Lift-Off'' Technique: High-energy ion implantation creates a buried damage layer within the diamond. After annealing, this layer transforms into graphitic material, easily removable by chemical etching. This method is suitable for producing single-crystal thin diamond membranes \cite{parikh1992Singlea,olivero2005IonBeamAssisted}. 
 \item \ICP-\RIE: Utilizing ArCl$_2$ plasma, this technique etches diamond into thin membranes without roughening the surface \cite{friel2009Control}.
\end{enumerate}

Nanopillars easily integrate with \afm, enhancing scanning magnetic resonance technologies for imaging at nanoscale resolutions \cite{appel2016Fabrication,grinolds2014Subnanometre,wang2019Nanoscale,maletinsky2012robust,thiel2019Probing}. The ICP-RIE technique is employed to etch away 3--4 $\mu$m from the diamond surface, minimizing mechanical polishing-induced surface damage. Subsequently, NV centers are created via ion implantation and annealing processes (\se \ref{sec:NV_creation}). The diamond is further etched to a few micrometers in thickness from the original 50 $\mu$m, employing cycling ArCl$_2$ and O$_2$ in the ICP-RIE process. This progression culminates in the creation of diamond sensors featuring a 200 nm diameter nanopillar on a 500 nm cantilever, accomplished through mutually aligned \ebl and ICP-RIE processes \cite{wang2022Selfaligned}. Finally, micromanipulators are used to pick up the diamond sensors and connect them to AFM tips.

The engineering of large-scale photonic devices for multiple quantum sensors relies heavily on accurately aligning NV centers with the optical structures. This alignment significantly impacts the optical photon collection efficiency. One approach involves pre-determining the position of the photonic structure through confocal imaging, but this method tends to be intricate and inefficient. Another strategy entails precise ion implantation onto the photonic structure, necessitating a specially designed implantation system \cite{meijer2008implanting,raatz2019Investigationa}.
A recent technique based on a self-aligning strategy (Fig. \ref{fig:pillar}(b)) has demonstrated significant improvement in device performance \cite{wang2022Selfaligned}. This method streamlines the process by integrating independent steps—ion implantation and photonic structure lithography—into a single pattern. By doing so, it eliminates alignment inaccuracies and greatly simplifies the design and fabrication processes.
This approach, which achieves NV center and photonic structure alignment, has resulted in a saturated photon count rate of 4.7 Mcps \cite{wang2022Selfaligned}. 
 

\section{Conclusion and outlook}\label{conclusion}

This chapter draws conclusions (Fig. \ref{fig:compatibility}) and presents a blueprint (Fig. \ref{fig:all_application}) for single-molecule magnetic resonance spectroscopy. It considers various perspectives, encompassing current achievements, technical gaps, and potential applications (Table \ref{tab:all_summary}). Over recent decades, the \NV center in diamond has emerged as a pivotal physical system in quantum sensing, particularly due to its distinct advantages in magnetometry. Achieving a magnetic field sensitivity of approximately 1 \nTHz, along with a spatial resolution of about 10 nm, showcases its remarkable capabilities \cite{zhao2022Subnanotesla}. Among magnetometry sensors, \sqs exhibits superior performance in sensitivity and spatial resolution, enabling the detection of individual spins and molecules \cite{mitchell2020Colloquium}. Notably, the best sensitivity achieved for \ssmr stands at 0.3 $\mu_B$\sqrtHz \cite{shi2018SingleDNA} for NV center-based \EPR and 0.06 proton\sqrtHz \cite{muller2014Nuclear} for NV-based \NMR under ambient conditions. Moreover, \ssmr's spectral resolution reaches approximately mHz \cite{schmitt2017Submillihertz,boss2017Quantum}, with the best linewidth around Hz for NMR \cite{cujia2019Tracking,pfender2019Highresolution} and kHz for EPR \cite{kong2018Nanoscale}. This technique also achieves sub-nm resolution for spatial measurements, capable of both electron spin \cite{grinolds2014Subnanometre} and nuclear spin \cite{muller2014Nuclear}, and possibly even sub-\AA resolution in future studies \cite{abobeih2019Atomicscale}.   \\[0.8pt]

\subsection{Achievements: quantum sensing outside the diamond}

Significant strides have been made in microscale and nanoscale magnetic resonance spectroscopy using \sqs over the last decade. Notably, advancements in single-molecule EPR and NMR spectroscopy are reviewed. Five attempts utilizing \sqs aimed at achieving magnetic spectroscopy at the single-molecule level. At room temperature, successful \sm solid-state \nmr spectroscopy has been achieved \cite{Lovchinsky2016}, enabling the recognition and sensing of chemical environments. Both solid-state \cite{shi2015Singleprotein} and liquid-state \cite{shi2018SingleDNA} \sm \epr spectroscopy have been achieved, offering insights into g-factors and molecular motion. At cryogenic temperatures, measurements and controls of \sm \epr in spin-labeled peptides network \cite{schlipf2017molecular} and endofullerene N@C$_{60}$ \cite{pinto2020Readout} not only demonstrate single-molecule sensing but also hint at controllable quantum devices. Additional nanoscale sensing achievements include NMR spectroscopy of $\sim$ 10 nm scale organic molecules \cite{Staudacher2013nuclear, mamin2013nanoscale}, nanoscale chemical-shift resolving \cite{aslam2017Nanoscale}, \nqr on single-layer h-BN \cite{Lovchinsky2017magnetic}, and nuclear spin position reconstruction \cite{yang2018Detection,muller2014Nuclear}. These milestones pave the way for advancements in \ssmr.
 
A range of magnetic resonance spectroscopy techniques suitable for \sm sensing has been proposed or experimentally confirmed, including DEER (sec. \ref{sec:deer}), quantum relaxometry (sec. \ref{sec:relaxometry}), DD (sec. \ref{sec:NMR:DD}), ENDOR (sec. \ref{sec:NMR:ENDOR}), HHDR (sec. \ref{sec:NMR:HH}), and two-dimensional NMR (sec. \ref{sec:NMR:twod}). Notable research strides have progressively enhanced the sensitivity (sec. \ref{sensitivity}), spectral resolution (sec. \ref{sec:spectral_resolution}), and spatial resolution (sec. \ref{sec:spatial_resolution}) of \sqs. Additionally, chemical modification methods have been employed to optimize sensor performance and molecule stability (sec. \ref{sec:surface_modification}). Despite these advancements, technical gaps remain, hindering the widespread adoption of \ssmr across disciplines. 

\begin{figure*}[bhtp]
\begin{overpic}[width=0.85\textwidth]{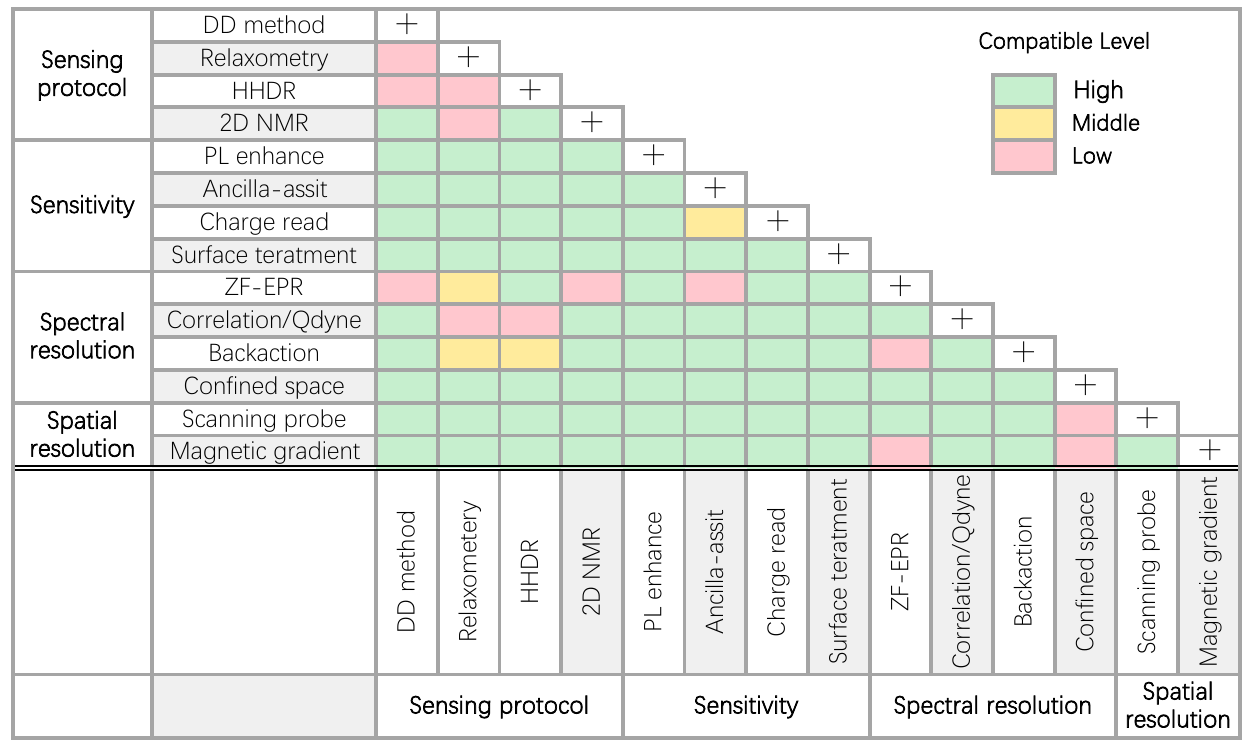} 
\end{overpic}
\caption[NV Center]{ Various sensing protocols exhibit differing compatibility levels. This evaluation categorizes the compatibility between diverse technologies into three tiers: high, medium, and low. High compatibility signifies technologies that can enhance each other's performance, while medium compatibility indicates the absence of conflicts. Low compatibility suggests either an inability to coexist or a severe compromise in performance when combined. Comprehensive assessment of the compatibility of different
technologies is available in Table \ref{tab:compat}.
 }  \label{fig:compatibility}
\end{figure*}

\begin{figure*}[bhtp]
\begin{overpic}[width=1\textwidth]{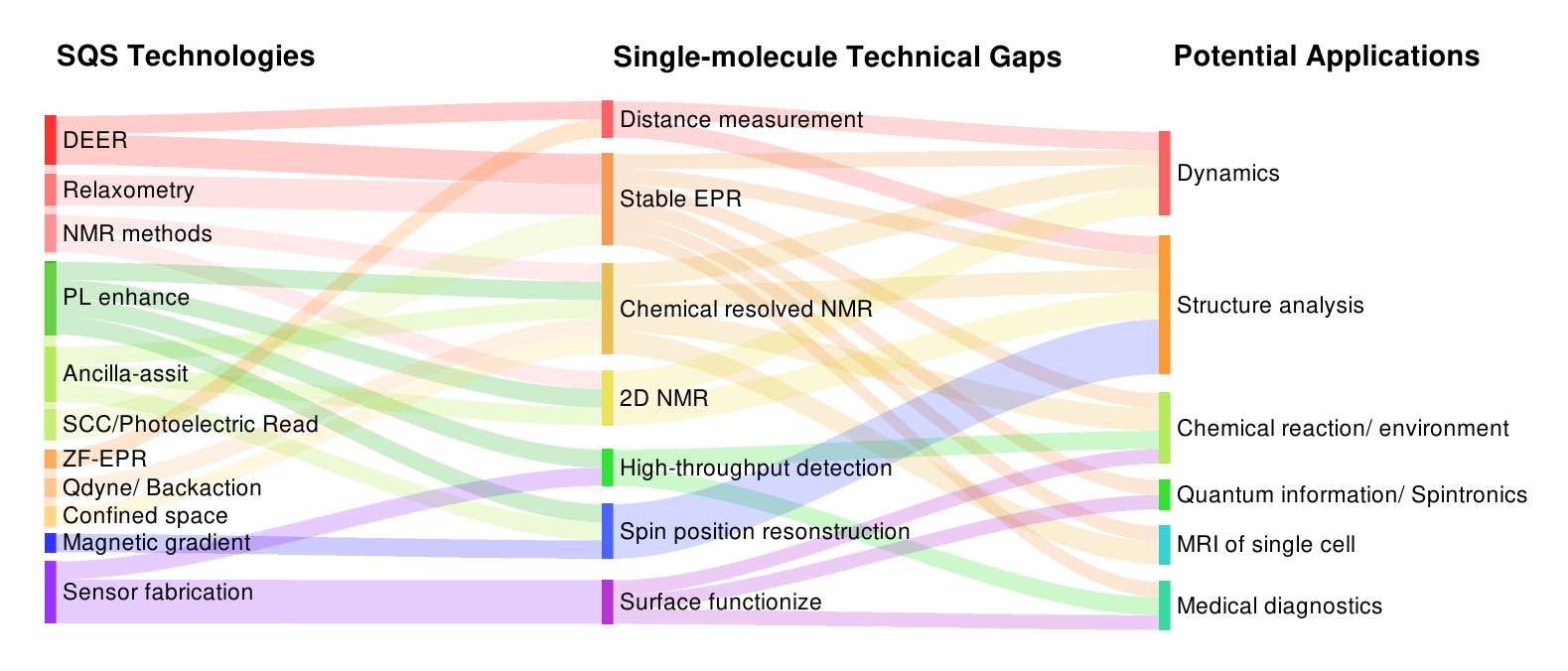} 
\end{overpic}
\caption[NV Center]{There has been significant advancement in microscale and nanoscale magnetic resonance spectroscopy through the utilization of \sqs. Despite these strides, the single-molecule field's progress remains constrained, necessitating further development. The primary hindrances impeding advancement revolve around technical gaps that require resolution to enable substantial applications. By amalgamating various \sqs technologies, potential solutions arise for addressing several of these technical gaps. The figure delineates the current \sqs technologies in the first column, organized into distinct colored groups representing different technology categories. Each technology within these categories is discernible by a unique shade corresponding to its color group. Specifically, the red color group encompasses quantum sensing technologies such as DEER (sec. \ref{sec:deer}), quantum relaxometry (sec. \ref{sec:relaxometry}), and NMR methods (sec. \ref{sec:NMR}). Green indicates sensitivity enhancement technologies like PL enhancement (sec. \ref{sec:collection_efficiency}), ancilla-assisted readout (sec. \ref{sec:anc_ass_read}), and SCC or photoelectric readout (sec. \ref{sec:SCC}). The orange group signifies spectral resolution enhancement technologies encompassing ZF-EPR (sec. \ref{sec:zerofield}), Qdyne, or weak measurement (sec. \ref{sec:correlation} and \ref{sec:measure_ind_eff}), along with confined space techniques (sec. \ref{sec:confined_space}). Additionally, blue denotes spatial resolution enhancement technology (sec. \ref{sec:gradient}), while purple signifies probe fabrication technology (sec. \ref{fabrication}). The second column illustrates the technical gaps within single-molecule technology that can potentially be addressed by combining various \sqs techniques. The third column outlines the potential applications achievable by overcoming these technical gaps.
}  \label{fig:all_application}
\end{figure*}

The amalgamation of existing technologies presents several solutions for bridging technical gaps. However, these technologies aren't entirely compatible with one another (Fig. \ref{fig:compatibility}). Incompatibilities stem from different underlying mechanisms in each technique. Techniques boasting both single-molecule sensitivity and high spectral resolution are pivotal for structural analysis, molecular dynamics, and biological processes. However, striving for high-resolution spectra often compromises sensitivity. For instance, the ZF-EPR method significantly enhances spectral resolution but poses challenges when combined with nuclear-assisted techniques aimed at increasing sensitivity and reliant on a magnetic field. Nevertheless, it can still be amalgamated with PL enhancement and SCC techniques. While the Qdyne technique does not easily mesh with the \NV charge state readout method, it pairs effectively with the ancilla-assisted method, enhancing the NV sensor's spectral resolution without sacrificing sensitivity. Notably, linewidth broadening of the target sample plays a crucial role in determining spectral resolution. Further narrowing of the linewidth can be achieved by combining \ssmr with homonuclear decoupling techniques \cite{aslam2017Nanoscale,Maurer2012} or nanoscale MAS techniques \cite{Wood2018} for solid-state sensing, as well as confined space methods for liquid-state sensing. Innovative spectral narrowing techniques are imperative for high spectral resolution in single-cell MRI. Subsequently, we shall primarily address technology gaps resolvable by compatible techniques, along with associated scientific goals and future applications (Fig. \ref{fig:all_application}).
 \\[0.8pt]

\subsection{Technical gaps on sensors}

The quantum sensor acts as the core of quantum sensing, directly impacting critical parameters like sensitivity and resolution (Eq. \ref{equ:spin_sensitivity_pola} and Eq. \ref{equ:spin_sensitivity_fluc}). The sensitivity of NV sensors relies on their distance from the surface, coherence time, and charge state stability. Enhancing the production yield for shallow NV centers at fixed points is pivotal for facilitating effective preparation of \sqs, given the current low yield. Furthermore, improving the coherence time of shallow NV centers, often inferior to those deeper than 30 nm \cite{zhao2022Subnanotesla}, remains imperative. Surface electric and magnetic noise majorly impact the coherence and charge stability of shallow NV centers. Developing methods that mitigate surface noise, such as surface passivation techniques (\se \ref{sec:chem_treat}),  would augment NV sensor performance in single-cell MRI and high-resolution solid-state sensing. Additionally, enhancing the stability of NV charge states under ambient conditions is crucial, especially for liquid-state sensing. Techniques combining ancilla-assisted readout \cite{shi2018SingleDNA} with SCC \cite{boss2017Quantum} show promise in addressing this challenge, maintaining high sensitivity and resolution while stabilizing NV charge states.

\subsection{Specific technical gaps on single-molecule EPR or NMR}
 There are several specific technical gaps for single-molecule \epr spectroscopy, including detection with low laser power, optical stability of spin labels and distance measurements. \\[0.8pt]

\noindent \emph{Detection with low laser power} Within the optical confocal system, the power density of the laser on the focal point can reach approximately $10^5$ W$\cdot$cm$^{-2}$, significantly impacting the stability of molecules and spin labels at room temperature. Hence, mitigating laser-induced damage stands as a pivotal concern for the effective application of \sm \epr methods. Addressing this necessitates the development of new detection techniques, with one potential solution being the utilization of SCC or photoelectric detection methods (sec. \ref{sec:SCC}) to curtail the average laser power. \\[0.8pt]

\noindent \emph{Optical stability of spin labels} An alternative approach to enhance laser stability involves developing radicals that exhibit stability intracellularly under laser irradiation and allow flexible labeling at specific sites. The tetrathiatriarylmethyl (TAM) radical \cite{Matsumoto2004Phar} and transition metallic ions \cite{zadrozny2015Millisecond} emerge as promising candidates.    \\[0.8pt]

\noindent \emph{Distance measurements} Employing site-specific spin labeling serves as a conventional \epr technique. Labeling two specific sites of a molecule and discerning the distance between the labeled spins facilitates direct access to structural information critical for molecular analysis, studying dynamics, and understanding conformational changes in single molecules. Enhancing spatial resolution may be achievable through methodologies like ZF-EPR (sec. \ref{sec:zerofield}) under single-molecule conditions. \\[0.8pt]

There are also noteworthy technical gaps in single-molecule or nanoscale \nmr spectroscopy, encompassing aspects such as sensing and resolving a single nuclear spin, achieving single-molecule NMR with chemical resolution, and executing single-molecule two-dimensional NMR. \\[0.8pt]

\noindent \emph{Sensing and resolving a single nuclear spin} To scrutinize molecular structure and dynamics at the single-molecule level, the direct detection and resolution of individual nuclear spins via \sm \nmr techniques stand as imperative requisites. While achieving single nuclear spin sensitivity has been demonstrated \cite{sushkov2014Magnetic,muller2014Nuclear,Lovchinsky2017magnetic}, detecting a single nuclear spin within an external molecule remains challenging, necessitating more universal protocols and the ability to resolve individual nuclear spins. \\[0.8pt]

\noindent \emph{Single-molecule NMR with chemical resolution} A primary limitation of current single-molecule NMR is its limited resolution, constraining the obtainable information on chemical structure and nuclear spin relaxation rates. The resolution of \ssmr itself suffices for discerning chemical shifts (sec. \ref{sec:correlation}), yet challenges persist in suppressing molecular diffusion and mitigating the effects of measurement-induced back-action. Solutions to these issues might involve restricting sensing space (sec. \ref{sec:confined_space}) and implementing weak measurement methods (sec. \ref{sec:measure_ind_eff}). Improving sensitivity is also pivotal for analyzing complex spectral structures. \\[0.8pt]

\noindent \emph{Single-molecule two-dimensional NMR} Upon addressing the aforementioned challenges, employing two-dimensional NMR spectroscopy becomes feasible for analyzing molecular structures and dynamics \cite{smits2019Twodimensional,yang2019Structural,abobeih2019Atomicscale}. Multi-dimensional \nmr spectroscopy stands as a robust method for determining molecular and protein structures. Previous demonstrations \cite{muller2014Nuclear,abobeih2019Atomicscale,yang2018Detection} have showcased the efficiency of \sqs in spatial reconstruction, potentially enabling three-dimensional \sm imaging of large nuclear spin structures with atomic resolution in the future.

\subsection{Overall technical gaps}

\noindent \emph{\SM magnetic resonance spectroscopy on in-situ living conditions}  Magnetic resonance spectroscopy stands among the structural biology techniques, akin to cryo-electron microscopy and X-ray diffraction, enabling molecule detection under loose conditions like in vivo and in situ. This capability extends to \ssmr, which inherits these features. While single-molecule solid-state quantum sensing typically employs shallow NV centers in bulk diamond, nanodiamonds exhibit no inherent limitations for magnetic resonance techniques at the nano and single-molecule scale. Notably, nanoscale NMR has been conducted using nanodiamonds of approximately 30 nm \cite{holzgrafe2020Nanoscale}. Given their compatibility with in vivo conditions \cite{mcguinness2011Quantuma}, implementing \ssmr in aqueous solutions, membranes, and cells represents a significant milestone. \\[0.8pt]

\noindent \emph{New methods for achieving high spectral resolution}  Extracting molecular structure and local environment information directly relies on magnetic resonance spectra's structures and line shapes. Therefore, designing new methods and pulsed sequences becomes pivotal to narrow spectral widths and augment the quantum spectrometer's dynamic range. Unlike merely observing single-molecule signals, achieving greater spectral resolution poses a heightened challenge necessitating enhanced sensitivity. \\[0.8pt]

\noindent \emph{Parallel high-throughput detection methods} Expediting sensing procedures entails simultaneous detection of multiple NV centers through high-throughput detection protocols. Current methods enable parallel measurement of NV centers \cite{cai2021Parallel}; however, improvements in NV center numbers and measurement efficiency remain imperative. Achieving parallel detection of over one hundred NV centers marks a significant milestone, requiring efficient and precise creation of two-dimensional arrays of single NV centers with high yield \cite{shi2018SingleDNA,cai2021Parallel}. \\[0.8pt]

\subsection{Scientific objectives and potential applications}

Over the past decades, advancements in single-molecule technologies spanning optical, electric, magnetic, and force-based methods have yielded significant multidisciplinary applications. However, these technologies primarily visualize single molecules under ambient temperatures, particularly in vivo conditions, restricting the direct implementation of chemical recognition or spectroscopic research methods to vacuum or low-temperature settings. Enter \sqs, specifically the NV center in diamond, offering the potential and initial strides toward implementing chemical recognition or spectroscopic analysis at the single-molecule level across diverse conditions. This avenue opens the opportunity for directly observing each molecule's distinct characteristics and behaviors. \\[0.8pt]

\noindent \emph{Dynamics and structure of single molecule} Understanding the dynamics and structure of molecules at the single-molecule level within physiological conditions like living cells and cellular membranes stands as a crucial pursuit. Fundamental biological mechanisms such as cellular signaling, metabolism, and immune responses rely on detecting physiological species like free radicals and ions, demanding sensitive and biocompatible sensing methodologies. Establishing a biologically specific and quantitative platform capable of precise sensing and imaging in the local physiological environment is paramount.

Investigations into molecular dynamics benefit significantly from implementing distance measurement and stable EPR techniques. Leveraging distance measurement, a traditional ensemble EPR technology, at the single-molecule level alongside stable \epr technology facilitates probing the dynamics of single molecules, including molecular conformational changes and tumbling. Enabling \sm chemical shift-resolved NMR spectroscopy, coupled with high-resolution NMR, offers insights into nuclear spin relaxation times for tracking dynamics within well-defined functional groups. Additionally, direct measurement of chemical bond reorientation becomes feasible through two-dimensional NMR exchange experiments.

The high sensitivity and resolution offered by \ssmr across diverse environments present opportunities for structural analysis in solid, liquid, or in-situ conditions, including living systems. Nanotechnology confinement or chemical modification allows assembling molecules with SQS probes. Stable EPR and distance measurement methods aid in gauging distances between molecular sites bearing free-radical electron spin labels in both solid and liquid states. Augmented by chemical-resolved two-dimensional NMR and spin position reconstruction, \sqs enables the analysis of \sm molecular structures, as evidenced by demonstrative experiments \cite{muller2014Nuclear,abobeih2019Atomicscale,yang2018Detection}. This potential for analysis signifies a promising direction for future applications.  \\[0.8pt]

\noindent \emph{Chemical environment and chemical reaction } The NV center offers another advantage: the capability to conduct quantitative measurements across various physical quantities, encompassing magnetic and electric fields, as well as temperature. These measurements enable simultaneous assessment of a diverse range of physiological variables, aligning with the analysis of single-molecule dynamics and structure. Parameters such as ion concentration, free radicals, biomolecule concentration, pH, and temperature can be concurrently evaluated.

Augmenting \sm EPR, \sm NMR, and other foundational measurement techniques, the functionalization of diamond surfaces and the implementation of high-throughput detection methods contribute significantly by enhancing measurement efficiency, thus rendering these technologies notably practical. Such physiological measurements extend the capacity for quantitative and single-molecule detection within biological and biochemical processes. This advancement holds promise in addressing critical questions in the fields of biology and medicine, specifically pertaining to cell signaling, development, and differentiation. \\[0.8pt]

\begin{table*}[hbtp]
\begin{tabular}[t]{@{}lll@{}}
\toprule
 \toprule
   & \parbox{18em}{Single-molecule \epr Spectroscopy}         & \parbox{27em}{Single-molecule/Nanoscale NMR Spectroscopy}       \\ \midrule
\parbox[t]{10em}{\raggedright Achievements: quantum sensing outside diamond} & \parbox[t]{18em}{\raggedright 1. Single protein EPR  \cite{shi2015Singleprotein}\\ 2.  Single DNA EPR \cite{shi2018SingleDNA} \\ 3. Single-molecule EPR on assemblable molecules \cite{schlipf2017molecular,pinto2020Readout}\\ 4. All-optical quantum relaxometry \cite{sushkov2014alloptical} }      & \parbox[t]{27em}{\raggedright 1. Single protein NMR \cite{Lovchinsky2016}\\ 2. Nanoscale NMR  \cite{Staudacher2013nuclear,staudacher2015Probing,mamin2013nanoscale} \\ 3. Nanoscale NMR with chemical resolution \cite{aslam2017Nanoscale}\\ 4. \NQR on atomically thin material \cite{Lovchinsky2017magnetic}\\ 5. Nanoscale \mri \cite{yang2018Detection,muller2014Nuclear} }         \\[1.0in]
\midrule
\parbox[t]{10em}{\raggedright Technical gaps on sensors  } & \multicolumn{2}{l}{\parbox[t]{48em}{\raggedright 1. Fabrication of shallow NV centers with high yield at fixed locations \\ 2. Quantum control, chemical modification, and electronic control methods to improve the coherence time
\\ 3. Chemical modification and electric control methods to enhance the charge state stability of \NV \\ 4. Diamond surface functionalization \\ 5.  \sqs with good quality fabricated in nanodiamond }} \\[0.6in]
\midrule
\parbox[t]{10em}{\raggedright Specific technical gaps} & \parbox[t]{18em}{\raggedright   1. Detection  with low laser power \\ 2. Optical stability of spin labels \\3. Distance measurements }         & \parbox[t]{27em}{\raggedright 1. Sensing and resolving a single nuclear spin\\ 2. Single-molecule NMR with chemical resolution\\ 3. Single-molecule two-dimensional NMR}        \\[0.29in]
\midrule
\parbox[t]{10em}{\raggedright Overall technical gaps} & \multicolumn{2}{l}{\parbox[t]{40em}{\raggedright 1. \SSMR on in-situ living conditions \\ 2. New methods for high spectral resolution\\ 3. Parallel high-throughput detection methods}} \\[0.30in]
\midrule
\parbox[t]{10em}{\raggedright Potential Applications } & \multicolumn{2}{l}{\parbox[t]{40em}{\raggedright 1. Dynamics and structure of single molecule\\ 2. Chemical environment and chemical reaction. \\ 3. Quantum information and spintronics\\ 4. MRI of single cell \\ 5. Medical diagnostics }} \\[0.43in]
 \bottomrule
\end{tabular}
\caption{Summary for the achievements, technical gaps and potential applications 
 for \ssmr.}\label{tab:all_summary}
\end{table*}

\noindent \emph{Quantum information and spintronics }
Besides biochemical measurements, the hybrid system of biological molecules and quantum sensors offers an avenue for quantum information applications. The proficiency of biology and chemistry in designing and synthesizing new molecules, from small entities to supramolecular assemblies, enables the construction of stable, controllable quantum systems on the nanoscale. Scalable quantum technologies demands an unprecedented blend of precision and complexity. The search for an elementary building block conducive to a scalable quantum network remains challenging. Notably, electron and nuclear spins in solids present remarkable coherence times of up to 6 hours \cite{zhong2015Optically}, with coherent control at gigahertz rates \cite{fuchs2009Gigahertz}, and both optical \cite{gruber1997Scanning} and electronic \cite{Bourgeois2015} readouts. However, scaling spin systems to larger arrays is a formidable technical hurdle as the distances for coupling electron spins via magnetic dipole interaction (below 30 nm \cite{Dolde2013}) currently surpass reliable top-down nanotechnology. A programmable molecular structure, such as sequence-controlled self-assembly of peptides on \sqs surfaces \cite{abb2016Twodimensional}, can effectively span these length scales.    \\[0.8pt]

\noindent \emph{MRI of single cell }
Intermolecular interactions measurable through \epr and \nmr based on \sqs hold significance in addressing biological concerns like phase transitions and intracellular organic phase separation. Moreover, enhanced sensitivity of NV centers broadens a single NV center's detection range from tens of nanometers to beyond a micron, promising single-cell MRI with a spatial resolution of $\sim$100 nm \cite{xie2018Mesoscopic}. \\[0.8pt]

\noindent \emph{Medical diagnostics }
The \sqs-based EPR technique, especially quantum relaxometry, has demonstrated partial success in tumor issue sensing \cite{chen2022Immunomagnetic,wang2019Nanoscale} and achieving ultrasensitive quantum diagnostics \cite{millerSpinenhancedNanodiamondBiosensing2020,li2022SARSCoV2}. Current diagnostic methods primarily concentrate on \sqs in nanoparticles. Future advancements, coupled with high-quality nanofabricated sensors and efficient diamond surface functionalization, could lead to single-molecule diagnostics. This could evolve into a practical technology by integrating high-throughput detection methods.  \\[0.8pt]

\noindent \emph{Other applications }
Beyond \ssmr in biology, quantum sensing based on single NV centers has found substantial application in condensed matter physics \cite{casola2018Probing}. Condensed matter physics hosts numerous magnetic or electrically charged quasiparticles \cite{venema2016quasiparticle}. Nanoscale quantum sensing techniques hold promise in realizing a real-space detection platform \cite{gross2017Realspace} for spatial distribution, spectral properties, and real-space correlation functions of these quasi-particles \cite{stano2013Local}. Additionally, albeit compromising spatial resolution, micron to millimeter-sized ensemble NV centers can serve as quantum sensors, delivering high sensitivity across a broad range of applications from life sciences to industry \cite{Barry2020Sensitivity}.

\begin{acknowledgements}
The authors thank Tianyu Xie, Zhiping Yang, Qi Zhang, Fei Kong, Jia Su, and Zhiyuan Zhao for their helpful advice and for revising the manuscript. The authors thank Friedemann Reinhard and Mengqi Wang for feedback and suggestions on the manuscript. 
This work was supported by the National Natural Science Foundation of China (grant no. T2125011), the CAS (grant nos. YSBR-068, GJJSTD20200001), Innovation Program for Quantum Science and Technology (grant No. 2021ZD0302200, 2021ZD0303204), the Anhui Initiative in Quantum Information Technologies (grant no. AHY050000), New Cornerstone Science Foundation through the XPLORER PRIZE, and the Fundamental Research Funds for the Central Universities, the DFG (GRK 2642, FOR 2724), the BW Foundation via project SPOC, the BMBF via the project Spinning and QSens as well as the EU via project Amadeus.
\end{acknowledgements}

\appendix
\clearpage\newpage


\section{List of symbols and abbreviations} \label{sec:symbols}

\begin{longtable}{p{0.17\columnwidth}p{0.77\columnwidth}}
AFM & atomic force microscopy \\
{cw} & {continuous-wave}  \\
{cw-ODMR} & {continuous-wave optically detected magnetic resonance}  \\
{cps} & {counts per second}  \\
{CVD} & {chemical vapor deposition}  \\
{DD} & {dynamical decoupling}  \\
{DEER} & {double electron-electron resonance}  \\
{DNP} & {dynamical nuclear polarization}  \\
{DQ} & {double-quantum}  \\
{EBL} & {electron beam lithography}  \\
{EDMR} & {electrically detected magnetic resonance}  \\
{ENDOR} & {electron nuclear double resonance}  \\
{EPR} & {electron paramagnetic resonance}  \\
FIB   &  focused ion beam  \\
 ERL & energy resolution limit   \\
 FRET   & fluorescence resonance energy transfer \\
{HH} & {Hartmann--Hahn double resonance}  \\
{ICP} & {inductively coupled plasma}  \\
{ISC} & {inter-system crossing}  \\
{MFM} & {magnetic force microscopy}  \\
{NMR} & {nuclear magnetic resonance}  \\
{NV} & {nitrogen-vacancy}  \\
{ODMR} & {optically detected magnetic resonance}  \\
{ONP} & {optically nuclear polarization}  \\
{PL} & {photoluminescence}  \\
{QND} & {quantum non-demolition}  \\
{Qdyne} & {quantum heterodyne}  \\
{RIE} & {reactive ion etching}  \\
{RF} & {radio frequency}  \\
{SCC} & {spin-to-charge}  \\
{SNR} & {signal-to-noise ratio}  \\
{SIL} & {solid immersion lens}  \\
{SQS} & {solid-state quantum sensor}  \\
{SQ} & {single-quantum}  \\
{SSMR} & {spin-based single-molecule magnetic resonance technologies}  \\
{SRIM}  & stopping and range of ions in matter \\
{UHV} & {ultra-high vacuum}  \\
{ZF-EPR} & {zero-field electron paramagnetic resonance}  \\
{ZPL} & {zero phonon line}  \\
 \end{longtable}

\begin{longtable*}{p{0.12\textwidth}p{0.23\textwidth}p{0.56\textwidth}}
\toprule
 \toprule
 Symbol   &   Values              & Quantity \\
 \midrule
 $\gamma_e$ &  $-2\pi$ 28.03 GHz$\cdot$T$^{-1}$      &  gyromagnetic ratio of the electron spin        \\ 
$\gamma_{\text{NV}}$ & $-2\pi$  28.04 GHz$\cdot$T$^{-1}$    & gyromagnetic ratio of the NV center spin \cite{Loubser1978}                        \\ 
$\gamma_{\text{N}}$ & \phantom{$-$}$2\pi$  3.077 MHz$\cdot$T$^{-1}$    & gyromagnetic ratio of  \Nfor          \\ 
    $N_A$         &   \phantom{$-$}6.02214076$\times$10$^{23}$      mol$^{-1} $   &  Avogadro constant  \\
$\mathcal{D}$    &  \phantom{$-$}$2\pi D$  & zero field splitting in the ground state of the NV center     \\
\,\, $D$   &  \phantom{$-$}2.87 GHz + $ C_{\text{T}}\Delta T$  & zero field splitting  depending on the temperature $\Delta T$ (with respect to 298 K) \\
\,\, $C_{\text{T}}$ &    $-71.9(0.3)$  kHz$\cdot$K$^{-1}$ & the temperature coefficient   \cite{acosta2010temperature,lourette2023Temperature,xu2023Highprecision}    \\
$d_{\parallel}$   & \phantom{$-$}$2\pi$ $( 0.35\pm 0.02)$  Hz{$\cdot$}cm{$\cdot$}V$^{-1}$ & parallel electric dipole moment  of the NV center  \\
$d_{\perp}$   &   \phantom{$-$}$2\pi$ $(17\pm 3)$  Hz{$\cdot$}cm{$\cdot$}V$^{-1}$ & perpendicular electric dipole moment of the NV center        \\
   $\mathcal{P}$      &   $-2\pi$ 4.946  MHz & the \Nfor quadrupole coupling \cite{xie2021Identity}      \\
    $A_{\parallel}$     &   $-2\pi$ 2.165  MHz &  the parallel hyperfine coupling between the NV spin and adjacent \Nfor \cite{xie2021Identity}   \\
    $A_{\perp}$     &   $-2\pi$ 2.633   MHz  & the perpendicular hyperfine coupling between the NV spin and adjacent \Nfor  \cite{xie2021Identity}   \\
    $a_{\parallel}$     &    & the parallel hyperfine coupling between the NV spin and nuclear spin    \\
    $a_{\perp}$     &      &  the perpendicular hyperfine coupling between the NV spin and nuclear spin   \\
$\vartheta$ &   & The phase of the control microwave    \\
    \,\, $|_x$ & \phantom{$-$}0\degree   &  \\
     \,\, $|_y$ & \phantom{$-$}90\degree   & \\
{$g_s$}  &   & {the response coefficient of the physical quantity $V (t)$, unit in rad$\cdot$Hz} \\
$H$ &  &  Hamiltonian, unit in rad$\cdot${Hz}   \\
$C$  &   & PL contrast \\
{$T$}   &   & {temperature, unit in Kelvin}\\
{$t$}   &   & time, unit in s\\
\,\, \Tonerou  &   & decay time of the spin-lock sequence\\
\,\, \Tone  &   & depolarization time\\
\,\, \Ttwostar  &   & dephasing time of the Ramsey fringe\\
\,\, \Ttwo  &   & decoherence time under the Hahn-echo sequence\\
\,\, {$t_s$}  &   & {interaction time of sensor and the target}\\
\,\, {$t_L$}  &   & {sampling time $1/f_{\rm{LO}}$}\\
$\Gamma$   &   & relaxation rate, unit in Hz\\
\,\, $\Gamma_{{1}}$ &   &  {depolarization rate $1/T_1$} \\
\,\, {$\Gamma_{2}$} &   &  {dephasing rate $1/T_2^{\ast}$} \\
\,\, $\Gamma_n$    &    & intrinsic dephasing rate of nuclear spin\\
\,\, $\Gamma_{{P}}$ &      & optical {pumping rate}       \\
     $P$ &   & the polarization of the spins \\
{$f$}   &    &   {frequency, unit in Hz}\\
\,\,  $f_{\text{LO}}$   &      &  sampling frequency \\
\,\,  $\Delta f$        &   &  spectral resolution \\
    $B_0$    &               & external magnetic field, unit in Tesla         \\
    $\boldsymbol{\Pi}=\vb{E}+\bm{\delta}$ &  \phantom{$-$}{$(\Pi_x,\Pi_y,\Pi_z)$}    &  {effective electric field}  \\
\,\, {$\vb{E}$} &  \phantom{$-$}{$(E_x,E_y,E_z)$}    &  { electric field}  \\
\,\, {$\bm{\delta}$} &  \phantom{$-$}{$(\delta_x,\delta_y,\delta_z)$}    &  {strain field}  \\
    $D_{\text{diff}}$ &     & difffusion coefficient, unit in m$^2\cdot$Hz \\
    $S(\omega)$ &           & noise spectral density, unit in rad$\cdot$Hz \\
    $F_t(\omega)$ &      &filter function          \\
$\mathscr{F}_{\omega}(f(t))$  &  \phantom{$-$}$\int_{-\infty}^{+\infty} f(t) e^{-i \omega t} \dd t$   &  Fourier transformation    \\
$\mathscr{F}^{-1}_{t}(f(\omega))$  &  \phantom{$-$}$\frac{1}{2 \pi} \int_{-\infty}^{+\infty} f(\omega) e^{i \omega t} d \omega$   &  inverse Fourier transformation    \\
$U_{\text{img}}$ &  \phantom{$-$}$\pi_y-\mathrm{DD}-\pi_x$ &  the evolution of the imaginary component readout      \\
  $\vb{S}_{\rm NV}$   &  \phantom{$-$}$(S_{x,\rm{NV}},S_{y,\rm{NV}},S_{z,\rm{NV}})$ &     the NV$^-$ electronic spin operator  \\
  \,\, $S_{x,y,z,\rm{NV}}$   &  \phantom{$-$}$\frac{1}{\sqrt{2}}\left(\begin{array}{lll}0 & 1 & 0 \\ 1 & 0 & 1 \\ 0 & 1 & 0\end{array}\right)$,$\frac{1}{\sqrt{2}}\left(\begin{array}{ccc}0 & -i & 0 \\ i & 0 & -i \\ 0 & i & 0\end{array}\right)$,$\left(\begin{array}{ccc}1 & 0 & 0 \\ 0 & 0 & 0 \\ 0 & 0 & -1\end{array}\right)$  &    \\
  $\vb{I}$, $\vb{S_e}$ &  \phantom{$-$}$(I_x,I_y,I_z)$, $(S_{e,x},S_{e,y},S_{e,z})$ &    the spin-1/2 spin operator (nuclear spin or electron spin)  \\
  \,\, $I_{x,y,z}$, $S_{e,x,y,z}$ &  \phantom{$-$}$\frac{1}{2}\left(\begin{array}{ll}0 & 1 \\ 1 & 0\end{array}\right)$,$\frac{1}{2}\left(\begin{array}{cc}0 & -i \\ i & 0\end{array}\right)$,$\frac{1}{2}\left(\begin{array}{cc}1 & 0 \\ 0 & -1\end{array}\right)$  &      \\
      \bottomrule
\end{longtable*}

\text{\/}

\newpage
\clearpage


\section{Derivations}

\subsection{Detection of nuclear spins via dynamical decoupling control }\label{sec:sta_fluc_signal}

In a spin bath system comprising an \NVm spin and \Cthir nuclear spins, using the secular approximation, we can express the Hamiltonian as
\begin{align}\label{eqn:nmr_ham_equ}
H=& -\gamma_{\mathrm{n}} \mathbf{B} \cdot \sum_{i=1}^N \mathbf{I}_{i}+\mathbf{S} \cdot \sum_{i=1}^N \mathbb{A}_{i} \cdot \mathbf{I}_{i}
\\\nonumber
 & + \sum_{i<j} \frac{\mu_0}{4\pi}\frac{\gamma_i\gamma_j\hbar}{r_{ij}^3}\left[\mathbf{I}_{i} \cdot \mathbf{I}_{j}-\frac{3\left(\mathbf{I}_{i} \cdot \mathbf{r}_{i j}\right)\left(\mathbf{r}_{i j} \cdot \mathbf{I}_{j}\right)}{r_{i j}^{2}}\right]\\\nonumber
 \approx & \omega_L \sum_{i=1}^N I^z_{i}+S_z \cdot \sum_{i=1}^N \left(a^{\|}_{ j} I_{j}^{z}+a^{\perp}_{ j} I_{j}^{\perp}\right).
\end{align}
The interactions between the nuclear spins, $H_{\text{dip}}^{ij}=\mu_0\gamma_i\gamma_j\hbar/4\pi r^3_{ij}$, can be disregarded when the experimental timescale is smaller than $1/H_{\text{dip}}^{ij}$, which generally holds for shallowed NV centers.

\begin{figure}[htp]
\begin{overpic}[width=1\columnwidth]{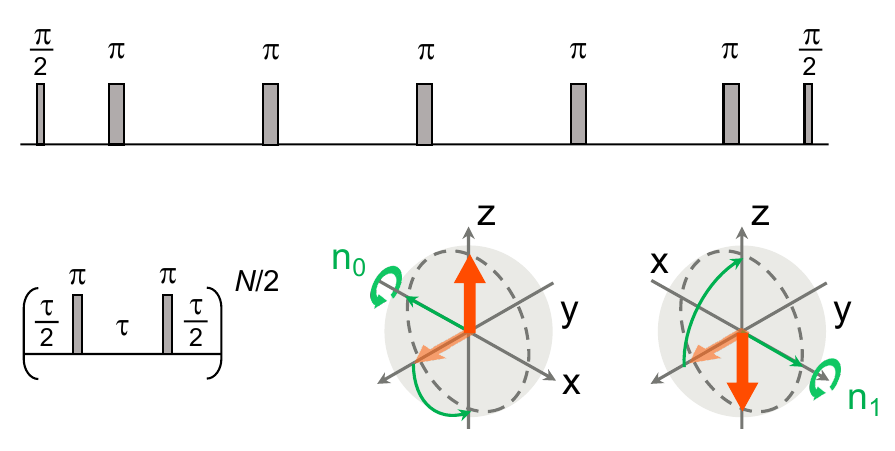}
\put (-2, 50) { (a) }
\put (-2, 26) { (b) }
\put (32, 26) { (c) }
\end{overpic}
\caption[NV Center]{ The effective evolution for nuclear spins under dynamical decoupling controlling. (a) Pulse sequence for a typical dynamical decoupling sequence. The basic unit $\tau/2-\pi-\tau-\pi-\tau/2$ will repeat for a longer sequence. Complicated sequences are reported in Fig. \ref{fig:dd_sequence}. (b) Pulse sequence for manipulating nuclear spins. (c) The nuclear spin evolution for DD under resonant condition $\tau=\tau_0$ shown in the Bloch sphere. Consequently, the nuclear spin will rotate around antiparallel axes $\hat{\mathbf{n}}_{0}$ and $\hat{\mathbf{n}}_{1}$ for different initial \NVm electron spin states $m_s=0$ and $m_s=1$. }  \label{fig:app_nmr_DD}
\end{figure}

Firstly, the \NV sensor is initialized into the $\ket{m_s=0}$ state, followed by a $\pi/2$ pulse transforming it into the superposition state $\ket{\psi_0}=\ket{+}\equiv(\ket{0}+\ket{1})/\sqrt{2}$. Subsequently, a series of resonant $\pi$ microwave pulses (Fig. \ref{fig:app_nmr_DD}) are applied to the \NVm sensor with a basic unit as $\tau/2-\pi-\tau-\pi-\tau/2$, where $\tau$ represents the free evolution time. The overall evolution of the NV-nuclear spin system,   
\begin{align}
U_{\text{DD}}=& \dyad{0}{0}\prod_{i=1}^N \exp \left[-i \phi\left(\hat{\mathbf{I}} \cdot \hat{\mathbf{h}}^{0}_j\right)\right] 
\\\nonumber & +  \dyad{1}{1} \prod_{i=1}^N \exp \left[-i \phi\left(\hat{\mathbf{I}} \cdot \hat{\mathbf{h}}^{1}_j\right)\right],
\end{align}
denotes a rotation of the nuclear spin by an angle $\phi_i$ around an axis $\hat{\mathbf{h}}^{m_s}_i$, depending on the initial \NV spin state $m_s$ \cite{taminiauDetectionControlIndividual2012}. The two axes, $\hat{\mathbf{h}}^{0}_i$ and $\hat{\mathbf{h}}^{1}_i$, lie in the antiparallel direction shown in Fig. \ref{fig:app_nmr_DD}(b) when the free evolution time meets the resonant condition:
\begin{equation}
\tau=\tau_0=\frac{2\pi}{2(\omega_L+a^{\|}_i/2)},
\end{equation}
and the rotation angle,
\begin{equation}
\phi\approx\frac{a^{\perp}_j}{\pi}N\tau,
\end{equation}
with $\omega_L\gg a^{\|}_j,a^{\perp}_j$.

The statistical fluctuation signal is independent of the nuclear spin state. For any nuclear spin ensemble state,
\begin{equation}
\ket{J}=\alpha_j\ket{h_j^+}+\beta_j\ket{h_j^-},
\end{equation}
the \NVm spin and nuclear spin system will evolve as,
\begin{align}
\nonumber
\ket{\Psi}  = & \prod_{j=1}^{N} U_{\text{DD,j}}^K\cdot \ket{+}_e\bigotimes_J\ket{J} \\ \nonumber
    =  & \ket{0}_e \bigotimes_J (\alpha e^{-i\phi_j}\ket{+} + \beta e^{i\phi_j}\ket{-} ) \nonumber \\ \nonumber
    & + \ket{1}_e \bigotimes_J (\alpha e^{i\phi_j}\ket{+} + \beta e^{-i\phi_j}\ket{-})\\
    \equiv & \ket{0}_e\ket{J_0}+\ket{1}_e\ket{J_1}.
\end{align} 
Thus, the signal is the projection on $\ket{+}_e\bra{-}_e$,
\begin{align}
\nonumber
& \text{Tr}(\ket{+}_e\bra{+}_e\ket{\Psi}\bra{\Psi})\\\nonumber
=&\frac{1}{2}+\frac{1}{2}|\bra{J_0}\ket{J_1}|^2\\\nonumber
=& \frac{1}{2}+\frac{1}{2}\prod_{j=1}^N\cos 2\phi_j \\\nonumber
      \approx & 1-\sum_{j=1}^N\frac{\phi^2_j}{4}\\
       = & 1-\sum_{j=1}^N \left(\frac{a_{\|}N\tau}{\pi}\right)^2 \label{eqn:pol_signal}
\end{align}

\subsection{Noise calculation}\label{sec:noise_cal}

Assuming the fluorescence photon numbers of $m_s=0$ and $m_s=\pm 1$ state are $n_0$ and $n_1$, and $p=|\braket{\Psi_f}{0}|^2=\cos^2 \phi/2$ is the probability in the $m_s=0$ state. According to Poisson distribution, the probability to find a photon number is given by
\begin{equation}
P(n)=p\frac{e^{-n_0}n_0^n}{n!}+(1-p)\frac{e^{-n_1}n_1^n}{n!}
\end{equation}
Accordingly, the variance of $n$ is expressed as
\begin{align}
\langle n \rangle & =\sum_{n=0}^{\infty}n P(n) \\ \nonumber
& = pn_0+(1-p)n_1
\end{align}

\begin{align}
\langle \Delta n^2 \rangle & =\sum_{n=0}^{\infty}n^2 P(n)-\langle n\rangle^2 \\ \nonumber
& = p(1-p)(n_0-n_1)^2+pn_0+(1-p)n_1
\end{align}
Considering the real and image magnetometry for $\phi\sim 0$ and $\phi\sim \pi/2$ for the real-component readout, $\phi\sim 0$, thus:
\begin{align}
\Delta\phi^2 &=\frac{\sqrt{\langle \Delta n^2 \rangle}}{|d\langle \hat{n}\rangle /d\phi^2|}\\\nonumber 
&= \frac{4\sqrt{n_0}}{n_0-n_1} \\\nonumber
\end{align}

For $\phi\sim \pi/2$, the normalized noise is
\begin{align}
\Delta\phi &=\frac{\sqrt{\langle \Delta n^2 \rangle}}{|d\langle \hat{n}\rangle /d\phi|}\\\nonumber &= \sqrt{1+\frac{2(n_0+n_1)}{(n_0-n_1)^2}} \\\nonumber
\end{align}


\section{Technology compatibility}

Here we primarily evaluate the compatibility of different
technologies, supporting Fig. \ref{fig:compatibility} in the main text. The focus is placed on middle and
low compatibility scenarios depicted in Fig. \ref{fig:compatibility}. Compatible and previously achieved
cases are not discussed here. For simplicity, the compatibility of PL enhancement and
surface treatment with other methods are both omitted. In addition, certain judgments
regarding compatibility are experientially based on the authors’ decisions.

\begin{longtable*}{p{0.13\textwidth}|p{0.16\textwidth}|p{0.5\textwidth}}
\toprule
 \hline\hline
 \multicolumn{1}{c|}{Methods}  &   \multicolumn{1}{c|}{Methods}    &   \multicolumn{1}{c}{Compatible Level}     \\
       \hline
\multicolumn{1}{c|}{\multirow{13}{*}{DD method}} & Relaxometry  & Low (The relaxometry method relies on the polarization of target spins, whereas the DD method utilizes coherence. Furthermore, the experimental sequences employed for both methods differ significantly.) \\
\multicolumn{1}{c|}{}               & HHDR       & Low (HHDR necessitates continuous driving of the target spins, whereas the DD method operates on a pulsed sequence.) \\
\multicolumn{1}{c|}{}               & 2D NMR      & \cite{abobeih2019Atomicscale,yang2019Structural} \\
\multicolumn{1}{c|}{}               & Ancilla-assit   & \cite{Lovchinsky2016}\\
\multicolumn{1}{c|}{}               & Charge read    & \cite{jaskula2019Improved}\\
\multicolumn{1}{c|}{}               & ZF-EPR      & Low (The DD method necessitates a magnetic field to eliminate the degeneracy of the electron spin state, a requirement in conflict with the zero-field prerequisite of ZF-EPR.) \\
\multicolumn{1}{c|}{}               & Correlation/Qdyne & \cite{laraoui2013Highresolution,schmitt2017Submillihertz,boss2017Quantum} \\
\multicolumn{1}{c|}{}               & Backaction    & \cite{cujia2019Tracking} \\
\multicolumn{1}{c|}{}               & Confined space  & \cite{liu2022Using} \\
\multicolumn{1}{c|}{}               & Scanning probe  & \cite{luan2015Decoherencea}\\
\multicolumn{1}{c|}{}               & Magnetic gradient & \cite{Arai2015,muller2014Nuclear} \\ 
\cline{1-3} 
\multicolumn{1}{c|}{\multirow{15}{*}{Relaxometry}} & HHDR       & Low (HHDR and relaxometry methods utilize different control methods.) \\
\multicolumn{1}{c|}{}               & 2D NMR      & Low (Relaxometry lacks the capacity to disclose information such as the chemical shifts and J-coupling of target spins, which are of concern in 2D NMR, due to the high spectral resolution required.)\\
\multicolumn{1}{c|}{}               & Charge read    & \cite{andersenElectronphononInstabilityGraphene2019}\\
\multicolumn{1}{c|}{}               & ZF-EPR      & Middle(Although the combination of these two methods is not challenging in principle, it does not offer any particular advantages.) \\
\multicolumn{1}{c|}{}               & Correlation/Qdyne & Low (Correlation/Qdyne and relaxometry methods utilize different control methods.) \\
\multicolumn{1}{c|}{}               & Backaction    & Middle (Backaction method is not helpful in relaxometry measurements.) \\
\multicolumn{1}{c|}{}               & Scanning probe  & \cite{wang2019Nanoscale} \\
\multicolumn{1}{c|}{}               & Magnetic gradient & \cite{bodenstedt2018Nanoscale} \\ 
\cline{1-3} 
\multicolumn{1}{c|}{\multirow{5}{*}{HHDR}}               & ZF-EPR      & \cite{kong2018Nanoscale} \\
\multicolumn{1}{c|}{}               & Correlation/Qdyne & Low (Correlation/Qdyne rely on the transverse coupling between the target spins and sensor, $S_zI_x$, whereas HHDR utilizes longitudinal coupling, $S_zI_z$.) \\
\multicolumn{1}{c|}{}               & Backaction    & Middle (Backaction does not contribute significantly to HHDR measurements.) \\
\cline{1-3} 
\multicolumn{1}{c|}{\multirow{2}{*}{2D NMR    }} & ZF-EPR      & Low (The detection targets are different.) \\
\multicolumn{1}{c|}{}               & Correlation/Qdyne & \cite{abobeih2019Atomicscale,yang2019Structural} \\
\cline{1-3} 
\multicolumn{1}{c|}{\multirow{4}{*}{ Ancilla-assit   }} & Charge read & Middle (currently, sensitivity cannot be improved by combining these two methods.)\\
\multicolumn{1}{c|}{}               & ZF-EPR      & Low (ZF-EPR requires zero magnetic field, while ancilla-assited readout requires a certain magnetic field for proper readout fidelity.) \\
\multicolumn{1}{c|}{}               & Correlation/Qdyne       & \cite{aslam2017Nanoscale} \\
\cline{1-3} 
\multicolumn{1}{c|}{\multirow{3}{*}{ZF-EPR}} & Correlation/Qdyne   & \cite{kong2020Kilohertz}  \\
\multicolumn{1}{c|}{}               & Backaction    & Low (Performing weak measurements that prevent back-action of nuclear spins at zero magnetic field is challenging.)\\
\multicolumn{1}{c|}{}               & Magnetic gradient & Low (The magnetic field requirements are in conflict with the ZF-EPR method.) \\ 
\cline{1-3} 
\multicolumn{1}{c|}{\multirow{2}{*}{ Correlation/Qdyne    }} & Backaction & \cite{cujia2019Tracking} \\
\multicolumn{1}{c|}{}               & Confined space  & \cite{liu2022Using}\\
\cline{1-3} 
\multicolumn{1}{c|}{\multirow{2}{*}{ Confined space    }} & Scanning probe & Low (The confined-space method confines the target sample to a nanoscale region, thereby obviating the need for scanning.) \\
\multicolumn{1}{c|}{}               & Magnetic gradient & Low (This confined space method already limits the target sample to a nanoscale region, eliminating the need for a magnetic gradient to achieve high spatial resolution.) \\ 
       \hline
 \caption{Evaluation of the compatibility
of different technologies depicted in Fig. \ref{fig:compatibility}.\label{tab:compat}}
\end{longtable*}

\clearpage\newpage

\bibliography{thebib.bib}

\end{document}